\documentclass{aastex6}

\usepackage{natbib}
\usepackage{graphicx}

\usepackage{graphics}
\usepackage{amssymb}
\usepackage{amsbsy}
\usepackage{amsfonts}
\usepackage{bm}
\usepackage{mathtools}
\usepackage{empheq}

\usepackage{amsmath}

\usepackage[makeroom]{cancel}
\usepackage{tikz}

\newcommand{\bhat}{\boldsymbol{\hat{b}}}
\newcommand{\bu}{\boldsymbol{u}}
\newcommand{\bV}{\boldsymbol{v}}

\newcommand{\bE}{\boldsymbol{E}}
\newcommand{\bb}{\boldsymbol{B}}
\newcommand{\bj}{\boldsymbol{j}}

\newcommand{\bk}{\boldsymbol{k}}

\newcommand{\pr}{\partial}
\newcommand{\nn}{\nonumber}

\newcommand{\bx}{\boldsymbol{x}}

\newcommand{\kpar}{k_\parallel}

\newcommand{\vpar}{v_\parallel}
\newcommand{\vperp}{v_\perp}

\newcommand{\alphapa}{\alpha_\parallel}
\newcommand{\alphape}{\alpha_\perp}

\newcommand{\bee}{\begin{eqnarray}}
\newcommand{\eee}{\end{eqnarray}}

\newcommand{\sign}{\textrm{sign}}

\begin{document}

\title{An introductory guide to fluid models with anisotropic temperatures\\
  Part 2 - Kinetic theory, Pad\'e approximants and Landau fluid closures} 

\author{P. Hunana\altaffilmark{1,2}, A. Tenerani\altaffilmark{3}, G. P. Zank\altaffilmark{4,5}, M. L. Goldstein\altaffilmark{6}, G. M. Webb\altaffilmark{4}, \\
      E. Khomenko\altaffilmark{1,2},  M. Collados\altaffilmark{1,2}, P. S. Cally\altaffilmark{7}, L. Adhikari\altaffilmark{4}, M. Velli\altaffilmark{8} }


\altaffiltext{1}{Instituto de Astrof\'isica de Canarias (IAC), La Laguna, Tenerife, 38205, Spain; peter.hunana@gmail.com}
\altaffiltext{2}{Universidad de La Laguna, La Laguna, Tenerife, 38206, Spain}
\altaffiltext{3}{Department of Physics, The University of Texas at Austin, TX 78712, USA}
\altaffiltext{4}{Center for Space Plasma and Aeronomic Research (CSPAR),
  University of Alabama, Huntsville, AL 35805, USA}
\altaffiltext{5}{Department of Space Science, University of Alabama, Huntsville, AL 35899, USA}
\altaffiltext{6}{Space Science Institute, Boulder, CO 80301, USA}
\altaffiltext{7}{School of Mathematics, Monash University, Clayton, Victoria 3800, Australia}
\altaffiltext{8}{Department of Earth, Planetary, and Space Sciences, University of California, Los Angeles, CA 90095, USA}



\begin{abstract}
  In Part 2 of our guide to collisionless fluid models, we concentrate on Landau fluid closures. These closures were pioneered by
  Hammett and Perkins and allow for the rigorous incorporation of collisionless Landau damping into a fluid framework. It is Landau damping
  that sharply separates traditional fluid models and collisionless kinetic theory, and is the main reason why the usual fluid models do not converge to
  the kinetic description, even in the long-wavelength low-frequency limit. We start with a brief introduction to kinetic theory, where we discuss in detail
  the plasma dispersion function $Z(\zeta)$, and the associated plasma response function $R(\zeta)=1+\zeta Z(\zeta)=-Z'(\zeta)/2$.  We then consider a 1D
  (electrostatic) geometry and make a significant effort to map all possible Landau fluid closures that can be constructed at the 4th-order moment level.
  These closures for parallel moments have general validity from the largest astrophysical scales down to the Debye length, and we verify their validity by
  considering examples of the (proton and electron) Landau damping of the ion-acoustic mode, and the electron Landau damping of the Langmuir mode.
   We proceed by considering 1D closures at higher-order moments than the 4th-order, and as was concluded in Part 1, this is not possible without Landau fluid
  closures. We show that it is possible to reproduce linear Landau damping in the fluid framework to any desired precision, thus showing the convergence of
  the fluid and collisionless kinetic descriptions.  We then consider a 3D (electromagnetic) geometry in the gyrotropic (long-wavelength low-frequency) limit
   and map all closures that are available at the 4th-order moment level. In the Appendix \ref{section:Pade}, we provide  comprehensive tables with Pad\'e
  approximants of $R(\zeta)$ up to the 8th-pole order, with many given in an analytic form.
\end{abstract}  

\maketitle
\newpage
\tableofcontents

\clearpage
\section{Introduction}
The incorporation of kinetic effects, such as Landau damping, into a fluid description naturally requires some knowledge of kinetic theory.
There are many excellent plasma physics books available, for example \cite{Akhiezer}, \cite{Stix1992}, \cite{Swanson}, \cite{Gary},
\cite{GurnettBhattacharjee2005}, \cite{Fitzpatrick} and many others. These books cover numerous topics
in kinetic theory that need to be addressed, if a plasma physics book wants to be considered complete.
However, the topics that are required for the construction of advanced fluid models
are often covered only briefly, or not covered at all. For example, the Pad\'e approximation of the Maxwellian plasma dispersion function $Z(\zeta)$
or the plasma response function $R(\zeta)=1+\zeta Z(\zeta)$, which is a crucial technique for the construction of collisionless fluid closures valid for all $\zeta$,
is not addressed by any of the cited plasma books. 

A researcher interested in collisionless fluid models that incorporate kinetic effects has to follow for example
\cite{HammettPerkins1990,Hammett1992,Snyder1997,PassotSulem2003b,Goswami2005,PassotSulem2006,PassotSulem2007,PSH2012,SulemPassot2015} and references therein. 
The first three cited references are written in the guiding-center reference frame (gyrofluid), which is a very powerful approach
that enables the derivation of many results in an elegant way. However, the calculations in
guiding-center coordinates can be very difficult to follow. The other cited references are written in the usual laboratory reference frame (Landau fluid),
but, the kinetic effects considered are of an even higher-degree of complexity and the papers can be very difficult to follow as well.
There are other subtle differences between gyrofluids and Landau fluids and the vocabulary is not strictly enforced.

Additionally, the cited papers assume that the reader is already fully familiar with the nuances of the kinetic description,
such as the definition of the plasma dispersion function $Z(\zeta)$ and the very confusing sign of the parallel wavenumber  $\sign(\kpar)$,
that almost every plasma book appears to treat slightly differently.  This guide, which is a companion paper to
`` An Introductory Guide to Fluid Models with Anisotropic Temperatures. Part 1: CGL Description and Collisionless Fluid Hierarchy'', 
attempts to be a simple introductory paper to the collisionless fluid models, and we focus on the Landau fluid approach.
The text is designed to be read as ``lecture notes'', and may be regarded as a detailed exposition of \cite{HunanaPRL2018}. 
 We focus on collisionless closures and use a technique pioneered by \cite{HammettPerkins1990}. Alternative approaches, including incorporation
  of collisional effects were presented for example by \cite{JosephDimits2016,JiJoseph2018,Jorge2019,Chen2019,Wang2019} and references therein.

\begin{figure*}[!htpb] 
  \includegraphics[width=0.42\linewidth]{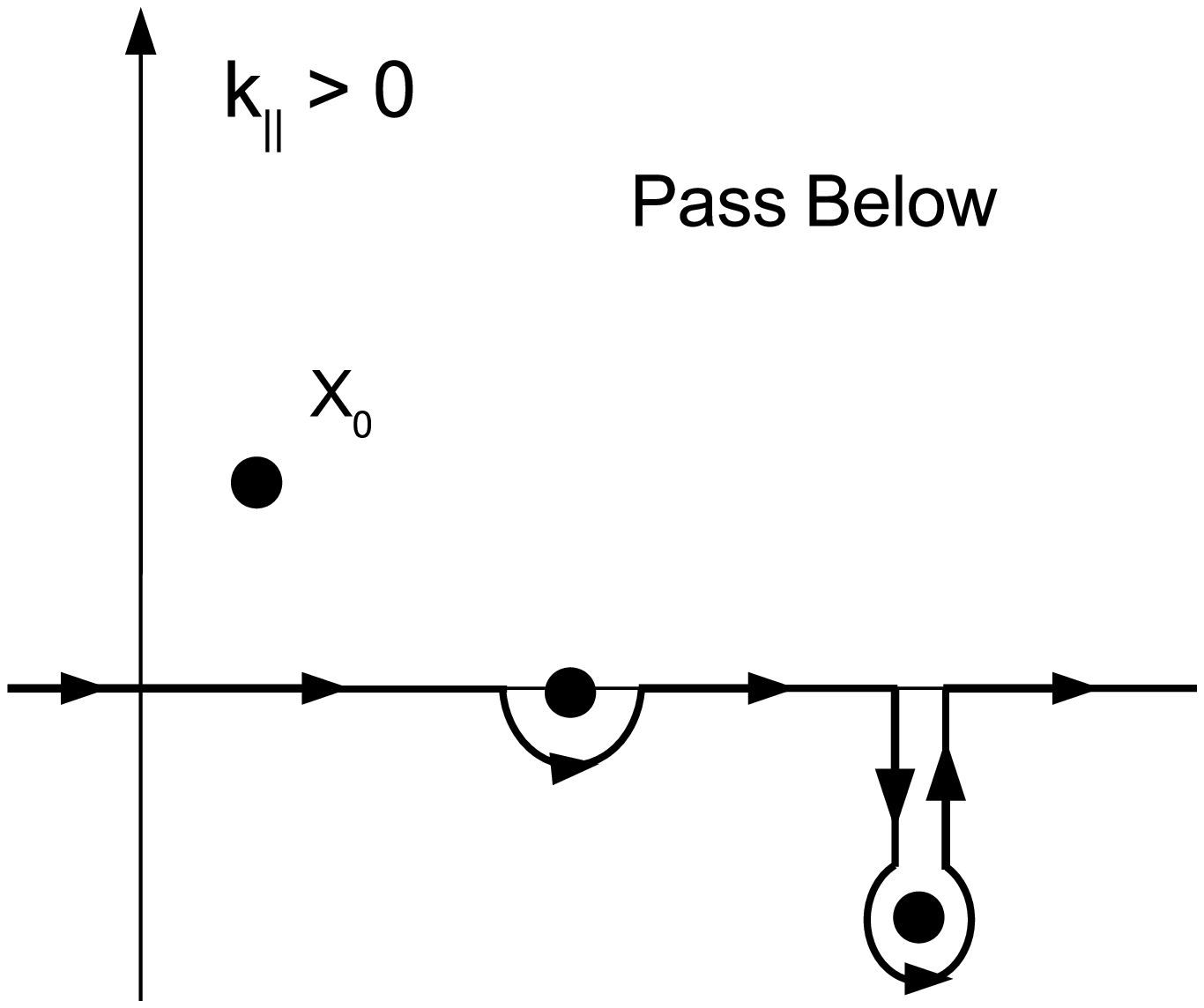}\hspace{0.1\textwidth}\includegraphics[width=0.42\linewidth]{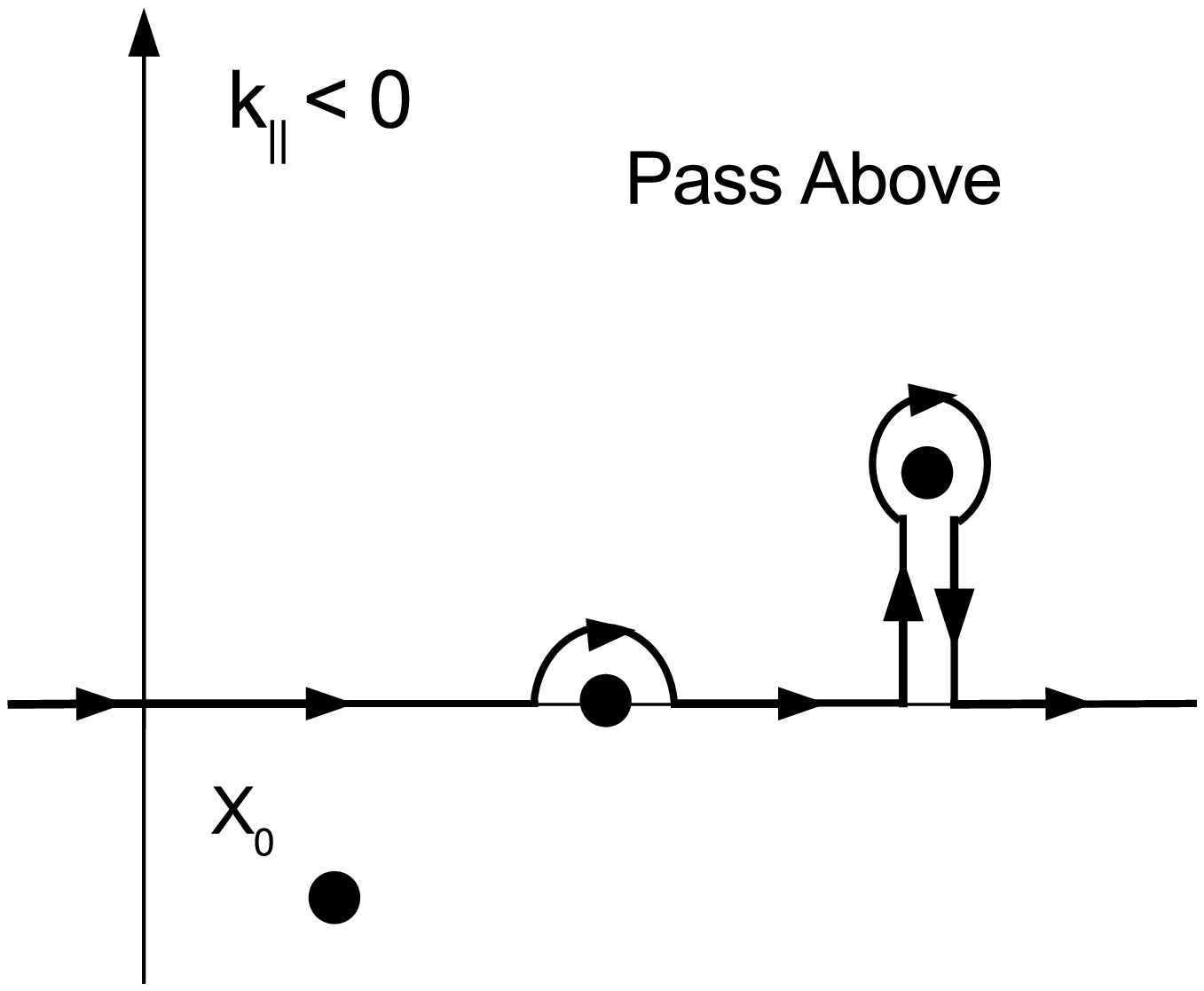}   
  \caption{Left panel: Landau contours for $k_\parallel>0$. Right panel: Landau contours for $k_\parallel<0$.} \label{fig:Landau}
\end{figure*}

In Section \ref{section:KT}, we introduce kinetic theory briefly, and we consider only aspects that are necessary
for the construction of advanced fluid models that contain Landau damping. We focus on the integral $\int \frac{e^{-x^2}}{x-x_0}dx$ that we call the Landau integral,
 see Figure \ref{fig:Landau}. 
We discuss how this integral is expressed through
the plasma dispersion function $Z(\zeta)$ and we discuss in detail the perhaps only technical (but very important)
difference between defining $\zeta=\omega/(k_\parallel v_{\textrm{th}})$ and $\zeta=\omega/(|k_\parallel| v_{\textrm{th}})$. Only the latter choice allows
one to use the original plasma dispersion function of \cite{FriedConte}, and the former choice requires that the $Z(\zeta)$ is redefined. 

In Section \ref{section:1D}, we consider a 1D electrostatic geometry. We discuss the concept of
the Pad\'e approximation to the plasma dispersion function $Z(\zeta)$ and the plasma response function $R(\zeta)$. We introduce a new
classification scheme for approximants $R_{n,n'}(\zeta)$ that we believe is slightly more natural than the classification scheme introduced
by \cite{Martin1980} or the scheme of \cite{HedrickLeboeuf1992}. Nevertheless, we provide conversion relations that allow to convert one notation into the other. 
We verify the numerical values in Table 1 of \cite{HedrickLeboeuf1992} analytically and find a typo in one
coefficient of the quite important $Z_{3,1}(\zeta)$ approximant previously used to construct closures.
 In Figures \ref{fig:6}, \ref{fig:7} we compare precision of various approximants $R_{n,n'}(\zeta)$ with the exact $R(\zeta)$.
We proceed by mapping all plausible Landau fluid closures that can be constructed at the level of 4th-order moment.
For a brief summary of possible closures, see (\ref{eq:Table1S})-(\ref{eq:Table1T}).
For the sake of clarity, all closures are provided in Fourier space as well as in real space. Writing the closures in real space emphasizes the non-locality
of collisionless closures, since all closures contain the Hilbert transform, which in real space should be calculated correctly by integration along the magnetic field lines.  As discussed in detail by \cite{Passot2014},
  neglecting the distortion of magnetic field lines and calculating the Hilbert transform with respect to mean magnetic field $B_0$
can lead to spurious instabilities.
We compare the precision of the obtained closures by calculating the 
dispersion relation of the ion-acoustic mode at wavelengths that are much longer than the Debye length. For some closures, an interesting property is observed in that
the resulting fluid dispersion relation is analytically equivalent to the kinetic dispersion relation, once $R(\zeta)$ is replaced by the $R_{n,n'}(\zeta)$    
approximant, and such closures are viewed as ``reliable'', or physically-meaningful. Subsequently, all unreliable closures were eliminated; see the discussion
below (\ref{eq:Table1T}). The closure with the highest power series precision is the $R_{5,3}(\zeta)$ closure.

We note that electron Landau damping of the ion-acoustic
mode can be correctly captured, even if the electron inertia in the electron momentum equation is neglected (the ratio $m_e/m_p$ still enters the electron heat flux and
the 4th-order moment $\widetilde{r}$). The dispersion relation of such a fluid model is of course not analytically equivalent to the kinetic dispersion relation
(after $R(\zeta)$ is replaced by the $R_{n,n'}(\zeta)$), however, such a fluid model provides great benefit for direct numerical simulations,
since the electron motion does not have to be resolved. 
In Figure \ref{fig:IAdamping1} we plot solutions for selected fluid models without the electron inertia.
In Figure \ref{fig:IAdamping2}, the electron inertia is retained, and we replot the fluid model with the $R_{5,3}(\zeta)$ closure to show that the differences are
negligible. We also plot additional closures and discuss a regime where the electron temperature is much larger than the proton temperature, and where
closures with higher asymptotic precision yield better accuracy.
We then investigate the precision of the obtained closures by using the example of the Langmuir mode, see Figures \ref{fig:Langmuir1} and \ref{fig:Langmuir2}.
These calculations were only noted but not presented in \cite{HunanaPRL2018}.

The case of 1D geometry is then pursued further, and
selected closures with 5th-order and 6th-order moments are constructed. For an impatient reader, the entire text can be perhaps summarized with Figure \ref{fig:IAmode},
where the 
Landau damping of the ion-acoustic mode is plotted for dynamic closures with the highest power-series precision that can be constructed at a given fluid moment level.
For the 3rd-order moment (the heat flux) it is $R_{4,2}(\zeta)$, for the 4th-order moment it is $R_{5,3}(\zeta)$, for the 5th-order moment it is $R_{6,4}(\zeta)$, 
and for the 6th-order moment it is $R_{7,5}(\zeta)$ (we also briefly checked that for the 7th-order moment it will be $R_{8,6}(\zeta)$). 
In Figure \ref{fig:Lmode}, we also plot solutions for the Langmuir mode with the $R_{7,5}(\zeta)$ closure.
Additionally, it was verified that all these closures are ``reliable''.

The remarkable result that the reliable 1D closures reproduce the exact kinetic dispersion relation once $R(\zeta)$ is replaced by $R_{n,n'}(\zeta)$ leads 
us to the conjecture that there exist reliable fluid closures that can be constructed for even higher-order moments,
i.e. satisfying the kinetic dispersion relation exactly, once $R(\zeta)$ is
replaced by the $R_{n,n'}(\zeta)$ approximant. Furthermore, for a given n-th order fluid moment, the reliable closure with the highest power-series precision is the
dynamic closure constructed with $R_{n+1,n-1}(\zeta)$. Indeed, for higher order fluid moments one should be able to construct closures with higher order $R_{n+1,n-1}(\zeta)$
approximants that will converge to $R(\zeta)$ with increasing precision. Thus, one can reproduce the linear Landau damping in the fluid framework to any desired
precision, which establishes the convergence of fluid and kinetic descriptions.

In Section \ref{section:3D}, we consider a 3D electromagnetic geometry in the gyrotropic limit, and 
map all plausible Landau fluid closures at the 4th-order moment level.  
In a 3D electromagnetic geometry, the most difficult part of the calculations actually consists in obtaining
the perturbed distribution function $f^{(1)}$, since in the laboratory reference frame that we use here, one needs to first calculate the fully kinetic
integration around the unperturbed orbit. Only then, the correct gyrotropic limit (where the gyroradius and the frequency $\omega$ are small) can be obtained.
The integration around the unperturbed orbit can be found in many plasma books, and can be found  in the Appendix \ref{sec:Orbit}.
An alternative and very illuminating derivation of $f^{(1)}$ is by using the guiding-center reference frame. By writing the collisionless Vlasov
equation in the guiding-center limit and by prescribing from the beginning that the magnetic moment has to be conserved
at the leading order, the same $f^{(1)}$ is obtained in a perhaps more intuitive way. The various terms in $f^{(1)}$ can be identified with the
conservation of the magnetic moment, the electrostatic Coulomb force (which yields Landau damping) and the magnetic mirror force (which yields
transit-time damping). Usually Landau damping and its magnetic analogue, transit time damping, are summarily described as Landau damping,
and we note that 3D Landau fluid models contain both of these collisionless damping mechanisms.

We show that the closures for the $q_\parallel$ and $\widetilde{r}_{\parallel\parallel}$ moments are the same
as for the $q$ and $\widetilde{r}$ moments in 1D geometry.
The closure for $\widetilde{r}_{\perp\perp}$ in the gyrotropic limit is simply $\widetilde{r}_{\perp\perp}=0$.
One therefore needs to consider only closures for the $q_\perp$ and $\widetilde{r}_{\parallel\perp}$ moments.
For a summary of the $q_\perp$ and $\widetilde{r}_{\parallel\perp}$ closures, 
see (\ref{eq:RparPerpF1})-(\ref{eq:RparPerpF2}). We did not compare the dispersion relation of the resulting fluid models with the fully kinetic
dispersion relation in the gyrotropic limit and therefore we can not conclude which closures are ``reliable''. Nevertheless, by briefly considering
parallel propagation along $B_0$, one closure was eliminated since it produced a growing higher-order mode.   
There is only one static closure available for the perpendicular heat flux $q_\perp$, which is constructed with the $R_1(\zeta)$ approximant.
As discussed later in the Appendix, the simple $R_1(\zeta)=1/(1-i\sqrt{\pi}\zeta)$ is a quite imprecise approximant of $R(\zeta)$. This has the important
implication that 3D Landau fluid simulations should not be performed with static heat fluxes, and time-dependent heat flux equations have to be considered. 
The closure with the highest power-series precision for $\widetilde{r}_{\parallel\perp}$ in the gyrotropic limit is constructed with $R_{3,0}(\zeta)$.
 In the Appendix \ref{section:Pade}, we provide tables of Pad\'e approximants
of $R(\zeta)$ up to the 8-pole approximation, and many solutions are provided in an analytic form. 

\clearpage
\section{A brief introduction to kinetic theory} \label{section:KT}
In this section we introduce some building blocks of kinetic theory starting from the simple case of wave propagation along a mean magnetic field $B_0$
in a homogeneous plasma. Such an approach allows us to introduce the plasma dispersion function and the hierarchy of linearized kinetic moments,
preparing the ground for the next section where various hierarchy closures will be described in detail.
The collisionless Vlasov equation in CGS units reads
\begin{equation} \label{eq:Vlasov}
\frac{\pr f_r}{\pr t}+ \bV\cdot\nabla f_r +\frac{q_r}{m_r}(\bE+\frac{1}{c}\bV\times\bb)\cdot\nabla_v f_r =0.
\end{equation}
It is often illuminating to work in the \emph{cylindrical} coordinate system,
where the particle velocity $\bV=(v_x,v_y,v_z)$ is expressed as
\begin{eqnarray}
\bV =  \left( \begin{array}{c}
    v_\perp \cos \phi \\
    v_\perp \sin \phi \\
    v_\parallel
  \end{array} \right), \label{eq:KinVel}
\end{eqnarray}  
and the gyrating (azimuthal) angle $\phi=\arctan (v_y/v_x)$. The reason is, that it very nicely clarifies the meaning of \emph{gyrotropy}, where the distribution function
and the expressions that follow, are independent of the angle $\phi$. The velocity gradient in the cylindrical coordinate system reads
\begin{equation}
\nabla_v = \hat{\bV}_\perp \frac{\pr}{\pr v_\perp} +\hat{\boldsymbol{\phi}} \frac{1}{v_\perp}\frac{\pr}{\pr \phi} +\hat{\bV}_\parallel \frac{\pr}{\pr v_\parallel},
\end{equation}  
where the unit vectors
\begin{eqnarray}
\hat{\bV}_\perp =  \left( \begin{array}{c}
    \cos \phi \\
    \sin \phi \\
    0
\end{array} \right);\qquad
\hat{\boldsymbol{\phi}} = \left( \begin{array}{c}
    -\sin \phi \\
    \cos \phi \\
    0
\end{array} \right);\qquad
\hat{\bV}_\parallel =  \left( \begin{array}{c}
    0 \\
    0 \\
    1
\end{array} \right),
\end{eqnarray}
so the velocity gradient is
\begin{eqnarray}
\nabla_v =  \left( \begin{array}{c}
    \cos \phi \frac{\pr}{\pr v_\perp} - \frac{\sin\phi}{v_\perp} \frac{\pr}{\pr\phi} \\
    \sin \phi \frac{\pr}{\pr v_\perp} + \frac{\cos\phi}{v_\perp} \frac{\pr}{\pr\phi}\\
    \frac{\pr}{\pr v_\parallel}
  \end{array} \right).
\end{eqnarray}
A straightforward calculation with $\bb_0 = (0,0,B_0)$ yields
\begin{eqnarray}
\bV \times \bb_0 =  B_0 \left( \begin{array}{c}
    v_\perp \sin \phi \\
    - v_\perp \cos \phi \\
    0
  \end{array} \right),
\end{eqnarray}
which further implies
\begin{eqnarray}
  (\bV\times\bb_0 )\cdot \nabla_v &=& v_\perp \sin\phi B_0 \Big( \cos \phi \frac{\pr}{\pr v_\perp} - \frac{\sin\phi}{v_\perp} \frac{\pr}{\pr\phi} \Big)
  - v_\perp \cos\phi B_0 \Big(\sin \phi \frac{\pr}{\pr v_\perp} + \frac{\cos\phi}{v_\perp} \frac{\pr}{\pr\phi} \Big) \nn\\
  &=& -B_0 \sin^2\phi \frac{\pr}{\pr\phi} -B_0 \cos^2\phi \frac{\pr}{\pr\phi} = -B_0 \frac{\pr}{\pr\phi}. \label{eq:KinGyro}
\end{eqnarray}  
Now we need to expand the Vlasov equation (\ref{eq:Vlasov}) around some equilibrium distribution function $f_0$, i.e., the entire distribution function is
separated to two parts as $f=f_0+f^{(1)}$. For the distribution function, we drop the species index $r$.
The magnetic field is separated as $\bb=\bb_0+\bb^{(1)}$, where $\bb_0=B_0\boldsymbol{\hat{z}}$, and the electric field as $\bE=\bE_0+\bE^{(1)}$, but since
there is no large-scale electric field in your system, the $\bE_0=0$.

The most important principle that is usually not emphasized enough, is that the kinetic velocity $v$ is an independent quantity, and is not linearized. 
The entire Vlasov equation reads
\begin{equation} \label{eq:VlasovExp}
\frac{\pr (f_0+f^{(1)})}{\pr t}+ \bV\cdot\nabla (f_0+f^{(1)}) +\frac{q_r}{m_r}\Big[\bE^{(1)}+\frac{1}{c}\bV\times(\bb_0+\bb^{(1)}) \Big]\cdot\nabla_v (f_0+f^{(1)}) =0,
\end{equation}
or equivalently by using the r-species cyclotron frequency $\Omega_r=q_r B_0 / (m_r c)$
\begin{equation}
  \frac{\pr (f_0+f^{(1)})}{\pr t}+ \bV\cdot\nabla (f_0+f^{(1)}) +\frac{q_r}{m_r} \bE^{(1)}\cdot\nabla_v (f_0+f^{(1)})
  +\Omega_r \Big[ \bV\times(\boldsymbol{\hat{z}}+\frac{\bb^{(1)}}{B_0})\Big] \cdot\nabla_v (f_0+f^{(1)}) =0.
\end{equation}
The Vlasov equation is now expanded (i.e. linearized) by assuming that the ``(1)'' components are small, and that terms containing 2-small ``(1)'' quantities can
be neglected. At the leading order, the situation is similar as many times before, i.e.,
at very low frequencies $(\omega\ll\Omega_r)$ and very long spatial scales, the term proportional to $\Omega_r$ dominates and must be by itself equal to zero
\begin{equation} \label{eq:GyroRit}
\frac{q_r}{m_rc} (\bV\times\bb_0 )\cdot\nabla_v f_0 =0;  \qquad => \qquad \Omega_r \frac{\pr}{\pr\phi} f_0 =0,
\end{equation}  
where in the last step we used already calculated identity (\ref{eq:KinGyro}). The obtained result implies that at the longest spatial scales, the
distribution function cannot depend on the azimuthal angle $\phi$, or in another words, the distribution function must be isotropic in the perpendicular velocity
components and can depend only on $v_x^2+v_y^2=v_\perp^2$, i.e., the distribution function must be \emph{gyrotropic}.
The second most important principle for doing the linear kinetic hierarchy is to realize, that the hierarchy is \emph{linear}, 
and all the quantities will have to be linearized. Additionally, we are interested only in a simplified case where the plasma is perturbed around a homogeneous equilibrium
state, and we can assume that the equilibrium $f_0$ does not depend on time and position, so that $\pr f_0/\pr t=0$ and $\nabla f_0=0$.
Therefore, the distribution $f_0$ contains only density $n_0$ that is not $(t,\bx)$ dependent, or in another words $f(\bx,\bV,t)=f_0(\bV)+f^{(1)}(\bx,\bV,t)$. 
Perhaps a different way of looking at it is that the $f_0$ must satisfy the leading-order Vlasov equation
\begin{equation}
\frac{\pr f_0}{\pr t}+ \bV\cdot\nabla f_0 +\frac{q_r}{m_r}\Big[\bE_0+\frac{1}{c}\bV\times\bb_0 \Big]\cdot\nabla_v f_0 =0,
\end{equation}
which at long spatial scales and low frequencies further implies gyrotropy (\ref{eq:GyroRit}) and $\bE_0=0$, together with $\pr f_0/\pr t+\bV\cdot\nabla f_0=0$.

Terms that contain 2-small $(1)$ quantities in (\ref{eq:VlasovExp}) can be neglected, and by
putting the $f^{(1)}$ contributions to the left hand side and the $f_0$ contributions to the right hand side yields
\begin{equation} \label{eq:KinBase}
  \frac{\pr f^{(1)}}{\pr t}+ \bV\cdot\nabla f^{(1)} +\frac{q_r}{m_rc} (\bV\times\bb_0) \cdot\nabla_v f^{(1)} 
  = - \frac{q_r}{m_r}\Big[\bE^{(1)}+\frac{1}{c}\bV\times\bb^{(1)}\Big]\cdot\nabla_v f_0.
\end{equation}
This is the starting equation that expresses $f^{(1)}$ with respect to $f_0$ and that is used in plasma physics books to derive
the kinetic dispersion relation for waves in hot magnetized plasmas. The second term on the left hand side $\bV\cdot\nabla f^{(1)}$,
introduces the simplest forms of Landau damping. The most complicated term, by-far, is the 3-rd term $\Omega_r (\bV\times\bb_0/B_0) \cdot\nabla_v f^{(1)}$,
since it introduces non-gyrotropic $f^{(1)}$ effects. This term introduces the complicated integration around the unperturbed orbit
with associated sums over expressions containing Bessel functions, that are found in the full kinetic dispersion relations. It is this 3-rd term that
makes the collisionless damping (and the kinetic theory) a very complicated process, even at the linear level. Without this 3rd term, life would much
easier, and Landau fluid models would be an excellent match for a full kinetic description, at least at the linear level.

The 3-rd term is obviously equal to zero if the $f^{(1)}$ distribution function is assumed to be strictly gyrotropic (see (\ref{eq:KinGyro})). Or,
we can just neglect the term by hand, assuming that we are at low-frequencies and that $\omega \ll \Omega$, meaning, if we perform an ``overly-strict'',
and a bit ad-hoc-done low-frequency limit.
However, as we will see later in the 3D geometry section, it turns out that even if a strictly gyrotropic $f^{(1)}$ is assumed,
the 3-rd term can not be just eliminated from the onset. To obtain the correct $f^{(1)}$ in the gyrotropic limit, the 3-rd term has to be retained, 
the integration around the unperturbed orbit performed, and only then the term can be eliminated in a limit. 
It is emphasized that sophisticated Landau fluid models of (\cite{PassotSulem2007}), that we do not address here, do not neglect this 3rd term 
and these models do not assume the $f^{(1)}$ to be gyrotropic. It is exactly the deviations from gyrotropy that introduces the Bessel functions found
in kinetic theory and sophisticated Landau fluid models.

Now, for a moment we do not perform any calculations, and just reformulate the important equation (\ref{eq:KinBase}).
The (1)-st order fields are typically transformed to Fourier space $(\sim e^{i\boldsymbol{k}\cdot\bx-i\omega t})$, but we will postpone that for now. 
By defining operator
\begin{equation}
\frac{D}{Dt} \equiv \frac{\pr}{\pr t}+ \bV\cdot\nabla +\frac{q_r}{m_rc} (\bV\times\bb_0) \cdot\nabla_v,  
\end{equation}
that represents a rate of change along an unperturbed orbit (zero-order trajectory), the equation is rewritten as
\begin{equation}
  \frac{D f^{(1)}}{D t} = - \frac{q_r}{m_r}\Big[\bE^{(1)}+\frac{1}{c}\bV\times\bb^{(1)}\Big]\cdot\nabla_v f_0.
\end{equation}
To obtain the $f^{(1)}$, one therefore has to calculate the integral of the above equation, where also the integration of the r.h.s. must be naturally done along the zero-order trajectory
(along the unperturbed orbit) in order to cancel the $d/dt$ on the l.h.s.
The integration is denoted with
prime quantities, and the integral is performed along $dt'$. If the integral is performed from time $t'=t_0$ to $t'=t$, the integration of the
l.h.s. yields $f^{(1)}(\bx,\bV,t)-f^{(1)}(\bx,\bV,t_0)$, i.e. the result depends on the initial condition at time $t_0$. To remove this dependence,
the integral is performed from $t_0=-\infty$ and it is typically stated that in this case the initial condition $f^{(1)}(\bx,\bV,-\infty)$ can be neglected.
(This is however not that obvious and for example Stix have a rather long discussion in this regard on page 249). 
The distribution function $f^{(1)}(\bx,\bV,t)$ is therefore obtained by performing integral
\begin{equation} \label{eq:f1Int}
  f^{(1)}(\bx,\bV,t) = - \frac{q_r}{m_r}\int_{-\infty}^t \Big[\bE^{(1)}(\bx',t') +\frac{1}{c}\bV'\times\bb^{(1)}(\bx',t') \Big]\cdot\nabla_{v'} f_0 (\bV') \;  dt'.
\end{equation}
The calculation of this integral is cumbersome because of the required change of coordinates. We want to get the final $f^{(1)}$ expression and
we will repeat the algebra how to obtain it, but before doing that, let's consider the simplest possible case.
\subsection{The simplest case: 1D geometry, Maxwellian $f_0$}
Let's consider a particular situation, when (for whatever reason)
the 3rd term on the l.h.s of equation (\ref{eq:KinBase}) disappears, i.e. let's briefly consider
\begin{equation} \label{eq:B0f1}
  (\bV\times\bb_0) \cdot\nabla_v f^{(1)}=0,
\end{equation}
which according to (\ref{eq:GyroRit}) implies that $f^{(1)}$ is gyrotropic (it does not depend on the angle $\phi$).
Let's also consider the even more special case in which $f_0$  is isotropic.
In such a case, that is a specific case of (\ref{eq:B0f1}), the direction of $\bb^{(1)}$ does not matter at all for $f_0$ and naturally
\begin{equation} \label{eq:B0f0}
  (\bV\times\bb^{(1)}) \cdot\nabla_v f_0=0.
\end{equation}
To quickly double-check the correctness of the above expression, for isotropic $f_0(v)$ the velocity gradient is given by
$\pr f_0 / \pr v_i = (\pr f_0 / \pr v) (\pr v/\pr v_i)=f_0' v_i/v$ and the velocity gradient $\nabla_v f_0=f_0'\boldsymbol{\hat{v}}$ is in the direction of velocity $\bV$.
The result (\ref{eq:B0f0}) then immediately follows since $\epsilon_{ijk} v_j B^{(1)}_k v_i =0$.
The equation (\ref{eq:KinBase}) therefore reduces to
\begin{equation}
  \frac{\pr f^{(1)}}{\pr t}+ \bV\cdot\nabla f^{(1)} = - \frac{q_r}{m_r} \bE^{(1)}\cdot\nabla_v f_0.
\end{equation}
Fourier transforming the first-order quantities and $\frac{\pr}{\pr t}\to -i\omega$, $\nabla\to i\boldsymbol{k}$ yields
\begin{equation}
  \big(-i\omega +i \bV\cdot\bk\big) f^{(1)} = - \frac{q_r}{m_r} \bE^{(1)}\cdot\nabla_v f_0,
\end{equation}
which allows us to obtain expression for $f^{(1)}$ in the form
\begin{equation}
  f^{(1)} =  -i \frac{q_r}{m_r} \; \frac{\bE^{(1)}\cdot\nabla_v f_0}{\omega-\bV\cdot\bk}.
\end{equation}
Even though not necessary, it is useful to express the (electrostatic) electric field through the scalar potential $\bE^{(1)}=-\nabla\Phi$, which in Fourier space reads
$\bE^{(1)}=-i\bk\Phi$, yielding
\begin{equation}
  f^{(1)} =  - \frac{q_r}{m_r} \Phi\; \frac{\bk\cdot\nabla_v f_0}{\omega-\bV\cdot\bk}.
\end{equation}
Now we want to integrate the $f^{(1)}$, and obtain the linear ``kinetic'' moments for density, velocity (current), pressure (temperature), heat flux, and the 4-th order moment $r$
(or the correction $\tilde{r}$). To continue, we have to prescribe some distribution function $f_0$.

The 3D (isotropic) Maxwellian distribution is 
\begin{equation}
f_{0r} = n_{0r} \Big(\frac{\alpha_r}{\pi}\Big)^{3/2} e^{-\alpha_r v^2},
\end{equation}
where the isotropic $v^2=v_x^2+v_y^2+v_z^2$ and  $\alpha_r=m_r /(2T^{(0)}_r)=1/v_{\textrm{th} r}^2$.
For simplicity, let's drop the species index $r$, except for the charge $q_r$. The velocity gradient
\begin{eqnarray}
  \frac{\pr f_0}{\pr v_i} &=& n_0 \Big(\frac{\alpha}{\pi}\Big)^{3/2} (-\alpha) 2v_i e^{-\alpha v^2} = -2\alpha v_i f_0 = -\frac{m}{T^{(0)}} v_i f_0,\nn\\
  \nabla_v f_0 &=& -\frac{m}{T^{(0)}} \bV f_0.
\end{eqnarray}
Therefore, for a Maxwellian
\begin{equation}
  f^{(1)} =  + \frac{q_r}{T^{(0)}} \Phi\; \frac{\bk\cdot\bV}{\omega-\bV\cdot\bk} f_0.
\end{equation}
Before continuing, let's slightly re-arrange the above expression for $f^{(1)}$ and add $0=\omega-\omega$ to the numerator,
otherwise we will have to do this each time, when calculating the higher order moments. The rearrangement yields
\begin{eqnarray}
  f^{(1)} &=&  +\frac{q_r}{T^{(0)}} \Phi\; \frac{\bk\cdot\bV-\omega+\omega}{\omega-\bV\cdot\bk} f_0 = -\frac{q_r}{T^{(0)}} \Phi\; \Big(1+\frac{\omega}{\bV\cdot\bk-\omega} \Big) f_0.
\end{eqnarray}

For clarity, let's simplify even further and discuss the simplest possible 1D case, for a 1D Maxwellian distribution
\begin{equation}
  f_{0} = n_{0} \sqrt{\frac{\alpha}{\pi}}e^{-\alpha v^2}; \qquad \textrm{where}\quad  \alpha\equiv\frac{m}{2T^{(0)}}=\frac{1}{v_{\textrm{th}}^2}.
\end{equation}
Here we consider fluctuations along the magnetic field $B_0$ and the wavenumber is therefore denoted as $\kpar$. Note that the case is strictly
1D, and the velocity fluctuations are along the $B_0$ as well. For example from MHD perspective, we are therefore considering the parallel propagating
ion-acoustic mode. The $f^{(1)}$ for a Maxwellian $f_0$ is expressed as (dropping all the species indices 'r' except for the charge $q_r$)
\begin{eqnarray}
  f^{(1)} &=&  i \frac{q_r}{T^{(0)}} E^{(1)} \frac{v}{\omega-v\kpar}f_0, \\
  f^{(1)} &=&  -\frac{q_r}{T^{(0)}} \Phi \Big(1+\frac{\frac{\omega}{\kpar}}{v-\frac{\omega}{\kpar}}\Big)f_0.
\end{eqnarray}
Now we are ready to calculate the velocity integrals.
Let's start with the density $n^{(1)}$, by integrating
\begin{equation}
  n^{(1)} = \int_{-\infty}^\infty f^{(1)} dv = -\frac{q_r}{T^{(0)}} \Phi \Big( \underbrace{\int_{-\infty}^\infty f_0 dv}_{=n_0} + \int_{-\infty}^\infty
  \frac{\frac{\omega}{\kpar}f_0}{v-\frac{\omega}{\kpar}} dv \Big).  \label{eq:densityN1} 
\end{equation}
By using the prescribed Maxwellian $f_0$, the second integral is rewritten as
\begin{eqnarray}
\int_{-\infty}^\infty
\frac{\frac{\omega}{\kpar}f_0}{v-\frac{\omega}{\kpar}} dv
&=& n_0 \sqrt{\frac{\alpha}{\pi}} \frac{\omega}{\kpar}\int_{-\infty}^\infty \frac{e^{-\alpha v^2}}{v-\frac{\omega}{\kpar}} dv
= \bigg[ \begin{array}{c}
   \sqrt{\alpha} v = x  \\
   \sqrt{\alpha} dv = dx \end{array}\bigg]
=  n_0 \sqrt{\frac{\alpha}{\pi}} \frac{\omega}{\kpar}\int_{-\infty}^\infty \frac{e^{-x^2}}{x-\frac{\omega}{\kpar}\sqrt{\alpha}} dx
= \bigg[ \frac{\omega}{\kpar}\sqrt{\alpha}\equiv x_0 \bigg] \nn\\
&=& \frac{n_0}{\sqrt{\pi}} x_0 \int_{-\infty}^\infty \frac{e^{-x^2}}{x-x_0} dx. \label{eq:Landau}
\end{eqnarray}  
The notation $[\cdots ]$ just indicates change of a variable. We purposely wrote the integral with 
\begin{equation}
  x_0\equiv \frac{\omega}{k_\parallel}\sqrt{\alpha}=\frac{\omega}{k_\parallel v_{\textrm{th}}},
\end{equation}  
instead of the usual $\zeta$, since we want to define $\zeta$ slightly differently.
The integral is related to the famous \emph{plasma dispersion function} $Z(\zeta)$, that is responsible for the famous Landau damping.
Each plasma physics book devotes many pages to the discussion of Landau damping, that was first correctly described by \cite{Landau1946},
by considering
an initial value problem and using Laplace transforms. It was later shown by \cite{vanKampen1955}, that the Landau damping can be indeed obtained
by using the Fourier analysis. We refer the reader for example to books by Swanson, Stix, Akhiezer, Gary,
Gurnett and Bhattacharjee, Fitzpatrick, etc. Let's call the integral (\ref{eq:Landau}) the ``Landau integral''. 
Nevertheless, the very-well-known secret is, that even if one is armed with all these excellent books,
the Landau damping effect can still be very confusing (even at the linear level).  
We did not find any secret recipe that explains the Landau damping in a simplified and different way, and the reader is referred to the thick plasma physics books.
Here we want to concentrate only how to express the integral (\ref{eq:Landau}) through the plasma dispersion function.

\emph{Since the Landau integral can be very confusing and boring to explain, to increase the ``pedagogical'' value of this text,
let us talk a bit more freely on the next few pages.} The plasma dispersion function can be defined with a short definition
\begin{equation} \label{eq:PDFbasic}
  Z(\zeta) \equiv \frac{1}{\sqrt{\pi}} \int_{-\infty}^\infty \frac{e^{-x^2}}{x-\zeta}dx, \qquad \textrm{for} \quad Im(\zeta)>0.
\end{equation}  
In the definition of $x_0$, the thermal speed $v_{\textrm{th}}$ is always a positive real number, and we do not have to worry about it. 
Now, considering the specific case $k_\parallel>0$ and $Im(\omega)>0$, where we indeed have $Im(x_0)>0$, we can directly use the plasma dispersion
function and the result of the Landau integral (\ref{eq:Landau}) is $n_0 x_0 Z(x_0)$.
For this case, we are done. Really ? Yes, there is nothing else we can do for this case, we calculated the Landau integral.
Reeeaallyy ?? Yes, because the Landau integral can not be analytically ``calculated'',
the integral can not be expressed through elementary functions, unless the $Z(\zeta)$ function is somehow simplified, for example by expansion
for cases $|\zeta|\ll 1$ or $|\zeta|\gg 1$, or by considering the weak damping limit when $Im(x_0)$ is small (see plasma physics books).
We are not interested in these limits and the $Z(\zeta)$ function
has to be calculated numerically or looked up in the table. We are really done here ! \footnote{In old times, a good barber would loudly shout: The next in line for shaving!}
So why is the Landau integral so confusing for the other cases ? It is exactly because of that - that basically nothing gets ``really calculated''.

\subsection{The dreadful Landau integral $\int \frac{e^{-x^2}}{x-x_0} dx$}
There are many reasons why the ``Landau integral'' (\ref{eq:Landau}) can be so confusing.  
The first reason is, 1) that the integral (\ref{eq:Landau}) can not be expressed by using only elementary functions. 
If we did not arrive at this integral in the middle of a thick plasma physics book, but instead, encounter it
during our undergraduate studies of complex analysis, we would perhaps not have such a respect to this integral, and immediately attempted
to calculate it, by using the residue theorem. The integral appears to be so simple.
Instead of calculating $\int_{-\infty}^\infty$, we would calculate a different integral over a closed
contour in complex plane $\oint_C$. That integral can be calculated by using the residue theorem, that states that
$\oint_C =2\pi i \sum \textrm{Res}$, if the big path that encircles all the poles is counter-clockwise.
\footnote{As noted in the footnote of Appendix A of the book by Swanson, page 363, the rumor has it that the famous Cauchy's residue theorem, is
  actually due to Cauchy's dog, that usually went around leaving residues at every existing pole.}
An equivalent statement is that the integral is equal to $\oint_C=-2\pi i \sum \textrm{Res}$, if the big path that encircles all the poles is clockwise. 
In our case, there is always just one pole, at $x=x_0$, and the residue of $\frac{e^{-x^2}}{x-x_0}$ evaluated at $x=x_0$ is actually very simple, it is always
\begin{equation}
  \underset{x=x_0}{\textrm{Res}} \; \frac{e^{-x^2}}{x-x_0} = e^{-x_0^2},
\end{equation}
regardless of the value of $x_0$, since for a general function $f(x)$, the residue $\underset{x=x_0}{\textrm{Res}} \; \frac{f(x)}{x-x_0} = f(x_0)$.

However, to make the result $\oint_C$ useful for the calculation of our integral on the real axis $\int_{-\infty}^\infty$, we need to separate the closed contour integral to
$\oint_C=\int_{-\infty}^\infty + \int_{\textrm{arc}}$, where the $\int_{\textrm{arc}}$ represents the big half-circle at infinitely large radius. To preserve the direction
of integration along the real axis $\int_{-\infty}^\infty$, if the pole is in the upper complex plane, i.e. if $Im(x_0)>0$, we need to close the big
arc contour in the upper half of complex plane counter-clockwise.
Similarly, if the pole is in the lower complex plane, i.e. if $Im(x_0)<0$, we need to close the 
big arc contour clockwise. Importantly, in contrast to typical examples presented in basic complex analysis classes, the arc
integral $\int_{\textrm{arc}}$ does not disappear. The problem is, that the function $f(x)=e^{-z^2}$ is a very strongly decaying function on the Real axis (for $z=x\to\pm \infty$),
however, this is not true at all in the complex plane. Considering the purely Imaginary axis $z=\pm iy$, the function $e^{-z^2}=e^{+y^2}$ is a very strongly
diverging function as $y$ increases, and the arc integral $\int_{\textrm{arc}}$ cannot be neglected ! This is a very sad news, since now we clearly see, that with
$\int_{\textrm{arc}}\neq 0$, we will not be able to use the complex analysis to actually ``calculate'' the Landau integral (\ref{eq:Landau}).

We note that the well-known Gaussian integral $I = \int_{-\infty}^\infty e^{-x^2} dx =\sqrt{\pi}$,
is typically calculated in the Real plane by means of a trick which consists in evaluating $I^2$ in polar coordinates,
$I^2=\int_{-\infty}^\infty e^{-x^2-y^2} dx dy= 2\pi \int_0^\infty e^{-r^2} r dr$. The Gaussian integral can still be calculated in Complex plane by
using the residue theorem, even though quite sophisticated tricks are required.
\footnote{For example, by considering $\oint e^{i\pi z^2}/\sin(\pi z)dz$, calculated along lines with $45^\circ$ angle with the real axis, that encircle the pole at $z=0$.
and where the residue $\underset{z=0}{\textrm{Res}}=1/\pi$. } \\

The second reason why Landau damping is confusing is, 2) the necessity of analytic continuation. The third reason is very closely related to the second and it is
3) The analytic continuation has to be done differently for $k_\parallel>0$ and for $k_\parallel<0$. 
The big result of \cite{Landau1946} can be summarized as follows: if $k_\parallel>0$,  the path of integration always has to pass \emph{below} the pole $x=x_0$.
Therefore, starting with the basic case in the upper complex plane $Im(x_0)>0$, nothing has to be done and the integration is just along the real axis.
Now, if the pole is moved to the real axis, so $Im(x_0)=0$, one needs to go around that pole with a tiny half-circle from below. 
This creates a contribution of $1/2$ times $2\pi i$ times the residue at that pole, so the contribution is $\pi i e^{-x_0^2}$. If the pole $x_0$ is moved further down to the lower complex
plane, a full circle around the pole is required to enclose it from below, which yields a contribution of $2\pi i e^{-x_0^2}$.
The situation is demonstrated in the left panel of Figure 1. 
The integral (\ref{eq:Landau}) for $k_\parallel>0$ is therefore ``calculated'' as
\begin{equation}
  \int_C \frac{e^{-x^2}}{x-x_0} dx \overset{k_\parallel>0}{=}
  \begin{dcases}
    \int_{-\infty}^\infty \frac{e^{-x^2}}{x-x_0}dx, &\text{$Im(\omega)>0$};\; \text{$Im(x_0)>0$};\\
    V.P. \int_{-\infty}^\infty \frac{e^{-x^2}}{x-x_0}dx + \pi i e^{-x_0^2}, &\text{$Im(\omega)=0$}; \; \text{$Im(x_0)=0$};\\
    \int_{-\infty}^\infty \frac{e^{-x^2}}{x-x_0}dx+ 2\pi i e^{-x_0^2}, &\text{$Im(\omega)<0$}; \; \text{$Im(x_0)<0$}. 
  \end{dcases} 
\end{equation}
For the Cauchy principal value, we prefer the original French pronunciation \emph{``Valeur Principale''}, abbreviated as V.P. 

The above result is completely consistent with the definition of the plasma dispersion function,
since the plasma dispersion function was developed exactly to describe this integral. One starts with the definition in the upper complex plane (\ref{eq:PDFbasic}),
and analytically continues this function to a lower complex plane, according to
\begin{equation}
  Z(\zeta) \equiv \frac{1}{\sqrt{\pi}}   \int_C \frac{e^{-x^2}}{x-\zeta}dx =
  \frac{1}{\sqrt{\pi}}
  \begin{dcases}
    \int_{-\infty}^\infty \frac{e^{-x^2}}{x-\zeta}dx, &\text{$Im(\zeta)>0$};\\
    V.P. \int_{-\infty}^\infty \frac{e^{-x^2}}{x-\zeta}dx + \pi i e^{-\zeta^2}, &\text{$Im(\zeta)=0$};\\
    \int_{-\infty}^\infty \frac{e^{-x^2}}{x-\zeta}dx+ 2\pi i e^{-\zeta^2}, &\text{$Im(\zeta)<0$}. 
  \end{dcases} \label{eq:PDF}
\end{equation}
To save space in scientific papers and plasma physics books, the definition of $Z(\zeta)$ is often abbreviated as (\ref{eq:PDFbasic}),
i.e. only as a first line of (\ref{eq:PDF}), with a powerful statement
that for $Im(\zeta)<0$ the function is analytically continued. That statement indeed completely defines $Z(\zeta)$, since the powerful complex analysis tells us
that an analytic continuations of a function, if it exists, is unique. Another abbreviated definition is by essentially writing down only the second (middle) line of (\ref{eq:PDF}).
This is the most useful 1-line abbreviation because one can immediately recognize, how the $\sign(\kpar)$ was treated (as we will see soon). However, such a definition
of $Z(\zeta)$, with only specifying it for $Im(\zeta)=0$, would not be a complete definition of that function, and no powerful statement how the function is
extended above \& below from the x-axis is available.
So plasma physicists found a very smart workaround, how \emph{not to} write the $Im(\zeta)=0$ restriction
in the second line of (\ref{eq:PDF}) and how to completely define the $Z(\zeta)$ with this 1-line statement. Let's still consider the case $\kpar>0$, where our $x_0$ and
$\zeta$ are equivalent. It is often stated (e.g. Stix, bottom of page 190),
that ``the principal value of an integral through an isolated singular point may be considered the average of the two integrals that pass just above and just below
the point''. For example, for a specific situation when $x_0$ lies on the real x-axis, integrating along horizontal line below the x-axis yields the first
line of (\ref{eq:PDF}), and integration along horizontal line above the x-axis yields the third line of (\ref{eq:PDF}) since when the pole is encountered we have to pass
it from below. An average of the first line and third line of (\ref{eq:PDF}) yields the second line. The idea can now be generalized to an entire complex plane, for all
values of $Im(x_0)$, where two integrals are done. One integral along horizontal line that passes below the $x_0$ point (where nothing has to be done) and one integral
along horizontal line that passes above the $x_0$ point (and where a deformation that passes below the point has to be performed, accounting for the full residuum). 
Average of these two integrals yields an abbreviated $Z(\zeta)$ definition for all values of $\zeta$ in the form
\begin{equation}
  Z(\zeta) = \frac{1}{\sqrt{\pi}} V.P. \int_{-\infty}^\infty \frac{e^{-x^2}}{x-\zeta}dx + i\sqrt{\pi} e^{-\zeta^2}, \qquad \textrm{for} \quad \forall \;Im(\zeta), \label{eq:VP}
\end{equation}  
where the integration is said to go \emph{through} the pole. Of course, no integration can be really done ``through'' a singular point,
and what the wording means is that the integration is done along the horizontal axis that goes through $x_0$, i.e. the integration is along horizontal axis $Im(x_0)$.

It is possible to look at it from another (perhaps more illuminating) perspective. Consider the situation in which $x_0$ is somewhere in the upper half of the complex plane.
One can perform the integral along the real axis, so that the first line of (\ref{eq:PDF}) applies. Let's call this result $c_1$.
Alternatively, one can perform the integral along the horizontal line that passes through $Im(x_0)$
(with the required tiny half-circle passing below $x_0$),  and (\ref{eq:VP}) applies. Let's call this result $c_2$.
This two different integrals must be equal. Why? Because one can plot two vertical lines
(passing through $Re(x_0)=\pm \infty$) that together with the two horizontal integration lines, enclose an area that does not contain any pole, and integration around all
four lines (in a circular direction, let's say counter-clockwise) must yield zero. The two integrals along the vertical lines cancel each other, yielding that $c_1-c_2=0$,
the minus sign in front of $c_2$ appears since the integration along $c_2$ was now done in the opposite direction.
Even though perhaps a bit confusing when seen at first, the definition (\ref{eq:VP}) is very useful, and when encountered, it should be just interpreted as an
abbreviated definition of (\ref{eq:PDF}). 

Unfortunately, the plasma dispersion function was obviously developed only with the case $k_\parallel>0$ in mind. The Landau result
requires that for $k_\parallel<0$, the path of integration always encircles the pole \emph{from above}, see the right panel of Figure 1.
For $k_\parallel<0$, the Landau integral is defined as 
\begin{equation}
  \int_C \frac{e^{-x^2}}{x-x_0} dx \overset{k_\parallel<0}{=}
  \begin{dcases}
    \int_{-\infty}^\infty \frac{e^{-x^2}}{x-x_0}dx, &\text{$Im(\omega)>0$};\; \text{$Im(x_0)<0$};\\
    V.P. \int_{-\infty}^\infty \frac{e^{-x^2}}{x-x_0}dx - \pi i e^{-x_0^2}, &\text{$Im(\omega)=0$}; \; \text{$Im(x_0)=0$};\\
    \int_{-\infty}^\infty \frac{e^{-x^2}}{x-x_0}dx- 2\pi i e^{-x_0^2}, &\text{$Im(\omega)<0$}; \; \text{$Im(x_0)>0$}. 
  \end{dcases} 
\end{equation}
The two different cases for $k_\parallel>0$ and $k_\parallel<0$ can be easily combined together by using the sign of the wavenumber $k_\parallel$ function, that
is equal to $+1$ for $k_\parallel>0$, and equal to $-1$ for $k_\parallel<0$. However, one needs to forget the sign of $Im(x_0)$, and arrange the results only with
respect to the sign of $Im(\omega)$. The Landau integral with $x_0=\omega/(k_\parallel v_{\textrm{th}})$ therefore reads
\begin{equation}
  \int_C \frac{e^{-x^2}}{x-\frac{\omega}{k_\parallel v_{\textrm{th}}}} dx \overset{\forall k_\parallel}{=}
  \begin{dcases}
    \int_{-\infty}^\infty \frac{e^{-x^2}}{x-x_0}dx, &\text{$Im(\omega)>0$};\\
    V.P. \int_{-\infty}^\infty \frac{e^{-x^2}}{x-x_0}dx + \sign(k_\parallel) \pi i e^{-x_0^2}, &\text{$Im(\omega)=0$};\\
    \int_{-\infty}^\infty \frac{e^{-x^2}}{x-x_0}dx+ \sign(k_\parallel) 2\pi i e^{-x_0^2}, &\text{$Im(\omega)<0$}. 
  \end{dcases} 
\end{equation}
Obviously, it is the sign of $Im(\omega)$, and not the sign of $Im(x_0)$, that is the ``natural language'' of the Landau integral.
However, the connection to the plasma dispersion function $Z(\zeta)$ is unnecessarily difficult. Sometimes, the definition of the plasma
dispersion function is then altered so that the above expression is satisfied. Stix for example uses in addition to the usual $Z(\zeta)$, also a
different function $Z_0(\zeta)$ that can be defined with respect to the sign of $Im(\omega)$ instead of the sign of $Im(\zeta)$,
where as noted on page 202, $Z_0(\zeta)=Z(\zeta)$ for $k_\parallel>0$, and, $Z_0(\zeta)=-Z(-\zeta)$ for $k_\parallel<0$.
With $\zeta=\omega/(k_\parallel v_{\textrm{th}})$, the function $Z_0(\zeta)$ is defined according to
\begin{equation}
  Z_0(\zeta) = \frac{1}{\sqrt{\pi}}
  \begin{dcases}
    \int_{-\infty}^\infty \frac{e^{-x^2}}{x-\zeta}dx, &\text{$Im(\omega)>0$};\\
    V.P. \int_{-\infty}^\infty \frac{e^{-x^2}}{x-\zeta}dx + \sign(k_\parallel) \pi i e^{-\zeta^2}, &\text{$Im(\omega)=0$};\\
    \int_{-\infty}^\infty \frac{e^{-x^2}}{x-\zeta}dx+ \sign(k_\parallel) 2\pi i e^{-\zeta^2}, &\text{$Im(\omega)<0$}. 
  \end{dcases} \label{eq:PDFZ0}
\end{equation}
Again, the function $Z_0(\zeta)$ can be defined in an abbreviated form as the first line of (\ref{eq:PDFZ0}), with analytic continuation
for $Im(\omega)\le 0$ (Stix, page 206, eq. 91). The second possible abbreviated definition of $Z_0(\zeta)$, valid for all values of $Im(\omega)$,
is the trick with the principal value (Stix, page 206, eq. 92)
\begin{equation}
  Z_0(\zeta) = \frac{1}{\sqrt{\pi}} V.P. \int_{-\infty}^\infty \frac{e^{-x^2}}{x-\zeta}dx +\sign(k_\parallel)i\sqrt{\pi} e^{-\zeta^2},
  \qquad \textrm{for} \quad \forall \;Im(\omega), \label{eq:VPZ0}
\end{equation}  
where the integration path goes ``through'' the pole, i.e. the integration is done along the horizontal line $Im(\zeta)$.  
With the use of this new function $Z_0(\zeta)$ of Stix, we can therefore express the dreadful Landau integral for all values of $k_\parallel$ as
\begin{equation}
  \frac{1}{\sqrt{\pi}}\int_C \frac{e^{-x^2}}{x-\frac{\omega}{k_\parallel v_{\textrm{th}}}} dx  \overset{\forall k_\parallel}{=} Z_0(\zeta);
  \qquad \textrm{where}\quad \zeta = \frac{\omega}{k_\parallel v_{\textrm{th}}}. \label{eq:Landau1}
\end{equation}

However, we do not like this formulation with $Z_0$. Here we insist on using the original plasma dispersion function $Z$.
In our opinion, the most elegant solution, is the one that is used for example in the book by Peter Gary and in some Landau fluid papers,
and that is to use $|k_\parallel|$ in the definition of $\zeta$, by defining
\begin{equation} \label{eq:Dzeta}
\zeta \equiv \frac{\omega}{|k_\parallel| v_{\textrm{th}}}.
\end{equation}  
This amazingly convenient definition simplifies the expressions and represents ``natural language'' of the plasma dispersion function.
We note that $|\kpar| = \sign(\kpar) \kpar$, and also $\kpar=\sign(\kpar) |\kpar|$. With the new definition of $\zeta$, for $k_\parallel>0$ obviously nothing is changed
since $|k_\parallel|=k_\parallel$. However, for $k_\parallel<0$,
\begin{eqnarray}
  \int_{-\infty}^\infty \frac{e^{-x^2}}{x-\frac{\omega}{k_\parallel v_{\textrm{th}}}}dx &\overset{k_\parallel<0}{=}&
  \int_{-\infty}^\infty \frac{e^{-x^2}}{x+\zeta}dx
= \bigg[ \begin{array}{c}
    x = -y  \\
    dx = -dy \end{array}\bigg]
= \int_{\infty}^{-\infty} \frac{e^{-y^2}}{-y+\zeta} (-dy)  
= \bigg[ \begin{array}{c}
   \textrm{rename} \\
   y \to x
   \end{array}\bigg]
= - \int_{-\infty}^\infty \frac{e^{-x^2}}{x-\zeta}dx \nn\\
&=& \sign(k_\parallel)  \int_{-\infty}^\infty \frac{e^{-x^2}}{x-\zeta}dx,
\end{eqnarray}
where in the last step we used $-1 = \sign(k_\parallel)$. By examining the first and the last expression, the result is also obviously valid for $k_\parallel>0$,
and therefore for all $k_\parallel$. Or alternatively, (perhaps more confusingly, but keeping an exact track of the $\sign(k_\parallel)$), for all values of $k_\parallel$
\begin{eqnarray}
  \int_{-\infty}^\infty \frac{e^{-x^2}}{x-\frac{\omega}{k_\parallel v_{\textrm{th}}}}dx &\overset{\forall k_\parallel}{=}& \int_{-\infty}^\infty \frac{e^{-x^2}}{x-\sign(k_\parallel)\zeta}dx
  = \sign(k_\parallel) \int_{-\infty}^\infty \frac{e^{-x^2}}{\sign(k_\parallel) x-\zeta}dx 
= \bigg[ \begin{array}{c}
    y = \sign(k_\parallel) x  \\
    dy = \sign(k_\parallel)dx \end{array}\bigg]  \nn\\
&& = \sign(k_\parallel) \int_{-\infty \sign(k_\parallel)}^{+\infty \sign(k_\parallel)} \frac{e^{-y^2}}{y-\zeta} \big(\frac{dy}{\sign(k_\parallel)} \big)
= \int_{-\infty \sign(k_\parallel)}^{+\infty \sign(k_\parallel)} \frac{e^{-y^2}}{y-\zeta} dy = \sign(k_\parallel) \int_{-\infty}^{+\infty} \frac{e^{-y^2}}{y-\zeta} dy \nn\\
&& = \sign(k_\parallel) \int_{-\infty}^{+\infty} \frac{e^{-x^2}}{x-\zeta} dx, \label{eq:sign}
\end{eqnarray}
which is the same result as the one obtained above. The definition of $\zeta$ (\ref{eq:Dzeta}) therefore yields
\begin{equation}
  \int_C \frac{e^{-x^2}}{x-\frac{\omega}{k_\parallel v_{\textrm{th}}}} dx \overset{\forall k_\parallel}{=} \sign(k_\parallel)
  \begin{dcases}
    \int_{-\infty}^\infty \frac{e^{-x^2}}{x-\zeta}dx, &\text{$Im(\zeta)>0$};\\
    V.P. \int_{-\infty}^\infty \frac{e^{-x^2}}{x-\zeta}dx + \pi i e^{-\zeta^2}, &\text{$Im(\zeta)=0$};\\
    \int_{-\infty}^\infty \frac{e^{-x^2}}{x-\zeta}dx+ 2\pi i e^{-\zeta^2}, &\text{$Im(\zeta)<0$}. 
  \end{dcases} 
\end{equation}
This result allows us to use the original plasma dispersion function definition (\ref{eq:PDF}), and express the \emph{dreadful Landau integral} for all $k_\parallel$ simply as
\begin{equation}
  \frac{1}{\sqrt{\pi}}\int_C \frac{e^{-x^2}}{x-\frac{\omega}{k_\parallel v_{\textrm{th}}}} dx \overset{\forall k_\parallel}{=} \sign(k_\parallel) Z(\zeta);
  \qquad \textrm{where}\quad \zeta \equiv \frac{\omega}{|k_\parallel| v_{\textrm{th}}}. \label{eq:Landau2}
\end{equation}
Now we calculated the Landau integral to our satisfaction, and we can continue with the calculation of the linear kinetic hierarchy.
Wait. We had basically the same result several pages back ! For the case $\kpar>0$ and $Im(\omega)>0$.
The Landau integral was just expressed through the plasma dispersion function,
basically the same result as is done now, there is just one $\sign(\kpar)$ in front of the integral and one in the definition of $\zeta$.
Are you suggesting, that all these calculations, contour drawings and discussions, we did all of these things just to get a sign right ?
Affirmative. The Landau integral is all about chasing minus signs, but to get the correct signs is very important.
This is exactly the reason why the Landau damping is so confusing, and why it needed
the genius of Landau to correctly figure it out. Nevertheless, that the Landau damping \citep{Landau1946} is indeed very confusing can be understood from the fact,
that the effect was questioned for almost 20 years before it was experimentally verified by \cite{MalmbergWharton1966}. 

To conclude, and to summarize the differences between plasma physics books of Stix and Peter Gary, we have two equivalent recipes to ``calculate''
the Landau integral, that can be written as
\begin{equation}
  \boxed{
  \frac{1}{\sqrt{\pi}}\int_{-\infty}^\infty \frac{e^{-x^2}}{x-\frac{\omega}{k_\parallel v_{\textrm{th}}}} dx \; \overset{A.C.}{=} \;
  \frac{1}{\sqrt{\pi}}\int_C \frac{e^{-x^2}}{x-\frac{\omega}{k_\parallel v_{\textrm{th}}}} dx =
  \begin{dcases}
    Z_0(\zeta), & \text{$\zeta =\frac{\omega}{\kpar v_{\textrm{th}}}$};\\
    \sign(\kpar) Z(\zeta), & \text{$\zeta =\frac{\omega}{|\kpar| v_{\textrm{th}}}$},
  \end{dcases} } \label{eq:MAIN}
\end{equation} 
where the $A.C.$ stands for analytic continuation. It is important to emphasize that some plasma books, as for example by Gurnett and Bhattacharjee,
take a different approach and call the function $Z_0(\zeta)$ simply as $Z(\zeta)$, as is obvious from their expressions for $Z(\zeta)$ (pages 347-348) that
contain the $\sign(\kpar)$. Of course, this approach is fully kosher, however, one needs to be extra careful when adopting a numerical routine for the plasma dispersion
function. The second choice in (\ref{eq:MAIN}) appears inconvenient, however, it is not, since the expression (\ref{eq:Landau}) contains   
\begin{equation}
  \frac{1}{\sqrt{\pi}} \frac{\omega}{k_\parallel v_{\textrm{th}}} \int_C \frac{e^{-x^2}}{x-\frac{\omega}{k_\parallel v_{\textrm{th}}}} dx
  =\begin{dcases}
    \zeta Z_0(\zeta), & \text{$\zeta =\frac{\omega}{\kpar v_{\textrm{th}}}$};\\
  \frac{\omega}{k_\parallel v_{\textrm{th}}} \sign(\kpar) Z(\zeta) = \zeta Z(\zeta)  , & \text{$\zeta =\frac{\omega}{|\kpar| v_{\textrm{th}}}$}.
  \end{dcases}
\end{equation}
The book by Peter Gary, and many Landau fluid papers prefer the second choice, since this small trick with redefining $\zeta$
allows the use of the original plasma dispersion function $Z(\zeta)$, that was tabulated by \cite{FriedConte}.
\footnote{Peter Gary's book indeed appears to be the only ``recent'' plasma book, where $|\kpar|$ is used for the definition of $\zeta$.
  The only caveat of the book, which could be confusing, is the exclusion of the $\sqrt{2}$ in the definition of the thermal speed $v_{\textrm{th}}$. However,
some Landau fluid papers (\cite{HammettPerkins1990,Snyder1997}) use the same definition without the $\sqrt{2}$.}
We prefer it too, and therefore,
the integral that we will use frequently in the kinetic hierarchy is
\begin{equation}\boxed{
  \frac{x_0}{\sqrt{\pi}} \int_{-\infty}^\infty \frac{e^{-x^2}}{x-x_0}dx = \zeta Z(\zeta), \qquad \textrm{where}\quad
  x_0=\frac{\omega}{\kpar v_{\textrm{th}}}; \quad \zeta =\frac{\omega}{|\kpar| v_{\textrm{th}}},} \label{eq:LandauX}
\end{equation}
and obviously $x_0=\sign(k_\parallel) \zeta$.
Now we are able to finish the calculation of the density $n^{(1)}$, eq. (\ref{eq:densityN1}), that yields
\begin{equation} \label{eq:N1F}
n^{(1)} = -\frac{q_r}{T^{(0)}} \Phi \Big( n_0 +n_0 \zeta Z(\zeta) \Big) = -\frac{q_r n_0}{T^{(0)}} \Phi \Big( 1+ \zeta Z(\zeta) \Big). 
\end{equation}
The result can also be expressed by using the derivative $Z'(\zeta)=-2(1+\zeta Z(\zeta))$. 
The quantity $1+\zeta Z(\zeta)$ appears very frequently in kinetic calculations with Maxwellian distribution and
it is called the \emph{plasma response function}
\begin{equation} \boxed{
R(\zeta)=1+ \zeta Z(\zeta).}
\end{equation}
For a different (general) distribution function $f_0$,
the plasma response function $R(\zeta)$ can be defined according to what is obtained after calculating the density $n^{(1)} = -\frac{q_r n_0}{T^{(0)}} \Phi R(\zeta)$. The name
is very appropriate, since the $R(\zeta)$ describes, how plasma with some distribution function ``responds'' to an applied electric field (or a scalar potential).

\subsection{Short afterthoughts, after the Landau integral}
Why some ``analytic continuation'' has
to be done ? Even though we did not manage to express the Landau integral (\ref{eq:Landau}) through elementary functions, the integral appears to be well-defined in both 
upper and lower halves of the complex planes, regardless where the $x_0$ is. And it indeed is. So why the analytic continuation ? 
The very-deep reason why the analytic continuation is necessary, is that the integral is
\emph{not continuous} when crossing the real axis $Im(x_0)=0$ in the complex plane. \footnote{What actually matters is not the $x_0$, but the frequency $\omega$, and the crossing of the real
  axis $Im(\omega)=0$. This unfortunately yields that two separate cases for $k_\parallel>0$ and for $k_\parallel<0$ have to be considered.}
If a function is not continuous, it is \emph{not analytic} (a fancy well-defined language that says  that the function is not infinitely differentiable,
basically meaning that it matters from what direction that point is approached in the complex plane, very similarly to a derivative of function $|x|$ on real axis).
And if a function is analytic in some area, and not analytic outside of that area, we can sometimes push/extend the area of where the function \emph{is} analytic,
to/through the area where the function \emph{is not} analytic, therefore the term ``analytic continuation''.  

Why is the analytic continuation so important, why is it a big problem that the integral is not continuous when crossing the real axis ?
Because it directly relates to the \emph{causality} principle, that is, if something happens, then the response to this incident must come after,
and not before, the time in which that incident happened.
This can be perhaps more intuitively addressed by performing the Laplace transforms in time (instead of the Fourier transforms),
and considering an initial value problem, as was done by \cite{Landau1946}. For more information, see plasma physics books, for example \cite{Stix1992}, Chapter 3 on causality etc.

The necesity of analytic continuation and the definition of the plasma dispersion function can be nicely clarified by a formula 
  from a higher complex analysis, known as the Plemelj formula \citep{Plemelj1908}, which can be written in the following convenient form 
  \begin{equation} \label{eq:Plemelj}
\lim_{\epsilon\to 0^+} \frac{1}{x-x_0 \pm i\epsilon} = V.P. \frac{1}{x-x_0} \mp i\pi \delta(x-x_0).
 \end{equation}   
  The formula (\ref{eq:Plemelj}) is meant to be applied on a function $f(x)$ and integrated ``through'' the pole,
  i.e. along the horizontal line $Im(x_0)$. The easiest is to consider $x_0=0$ (or $Im(x_0)=0$) with integration along the real axis. The Dirac delta function
  $\delta(x-x_0)$ in (\ref{eq:Plemelj}) represents contributions of the Landau residue. If the Landau residue is neglected, i.e. if only the $V.P.$ part in
  (\ref{eq:Plemelj}) is considered as done by \cite{Vlasov1945}, yields that there is no damping present.
  In Section \ref{sec:PadeSec}, we will construct Pad\'e approximants of $Z(\zeta)$ and $R(\zeta)$.
  One can easily check, that by neglecting the Landau residue in the power-series expansions (the residue will be neglected in the asymptotic-series expansions),
  yields no collisionless damping. Therefore, as shown by \cite{vanKampen1955}, it is indeed possible to derive Landau damping by using Fourier analysis (an approach
  adobted here), provided the Landau residue in (\ref{eq:Plemelj}) is retained. 
  The formula (\ref{eq:Plemelj}) is often attributed only to \cite{Plemelj1908}, for his rigorous proof.
  Sometimes it is called the Sokhotski-Plemelj formula, because it is argued
  the formula was derived in the doctoral thesis of Y. V. Sokhotski in 1873, with a proof that can
  be viewed as sufficiently rigorous for mathematical standards that existed at those times, i.e. 35 years before the rigorous proof of Plemelj.
  The Sokhotski-Plemelj formula is used in many areas of physics, from the theory of elasticity to the quantum field theory.

\clearpage
\subsection{Easy Landau integrals $\int \frac{x^n e^{-x^2}}{x-x_0}dx$}
We want to calculate moments in velocity space all the way up to the 4th-order moment $r$, and (including the 3D geometry) we will need integrals only up to $n=5$.
To this aim, we will use frequently eq. (\ref{eq:LandauX}), where we find convenient to use $x_0$ and $\zeta$ instead of chasing the $\sign(\kpar)$,
and in the end we will just use the definition $x_0=\sign(\kpar)\zeta$. We already saw that the 0-th order moment was
\begin{eqnarray}
\frac{1}{\sqrt{\pi}} \int_{-\infty}^\infty \frac{e^{-x^2}}{x-x_0} dx = \sign(\kpar) Z(\zeta).\nn
\end{eqnarray}
Let us now calculate the higher order moments. Since we talked so much on the last pages, we will remain silent for a moment and we will just enjoy the calculation: 
\begin{eqnarray}
  \frac{1}{\sqrt{\pi}} \int_{-\infty}^\infty \frac{x e^{-x^2}}{x-x_0} dx &=& \frac{1}{\sqrt{\pi}} \int_{-\infty}^\infty \frac{x - x_0 + x_0 }{x-x_0} e^{-x^2}dx
  = \underbrace{\frac{1}{\sqrt{\pi}} \int_{-\infty}^\infty e^{-x^2} dx}_{=1} + \underbrace{\frac{x_0}{\sqrt{\pi}} \int_{-\infty}^\infty \frac{e^{-x^2}}{x-x_0}dx}_{=\zeta Z(\zeta)} \nn\\
  &=& 1+\zeta Z(\zeta) =R(\zeta). \label{eq:Lan_1}
\end{eqnarray}
\begin{eqnarray}
  \frac{1}{\sqrt{\pi}} \int_{-\infty}^\infty \frac{x^2 e^{-x^2}}{x-x_0} dx &=& \frac{1}{\sqrt{\pi}} \int_{-\infty}^\infty \frac{x^2 - x_0^2 + x_0^2 }{x-x_0} e^{-x^2}dx
  = \frac{1}{\sqrt{\pi}} \int_{-\infty}^\infty (x+x_0) e^{-x^2} dx + \underbrace{\frac{x_0^2}{\sqrt{\pi}} \int_{-\infty}^\infty \frac{e^{-x^2}}{x-x_0}dx}_{x_0 \zeta Z(\zeta)} \nn\\
  &=& \underbrace{\frac{1}{\sqrt{\pi}} \int_{-\infty}^\infty x e^{-x^2} dx}_{=0} + \underbrace{\frac{x_0}{\sqrt{\pi}} \int_{-\infty}^\infty e^{-x^2} dx}_{=x_0} + x_0 \zeta Z(\zeta) = 
  x_0 \big(1+\zeta Z(\zeta) \big) \nn\\
  &=& \sign(\kpar) \zeta R(\zeta).
\end{eqnarray}
\begin{eqnarray}
  \frac{1}{\sqrt{\pi}} \int_{-\infty}^\infty \frac{x^3 e^{-x^2}}{x-x_0} dx &=& 
   \underbrace{\frac{1}{\sqrt{\pi}} \int_{-\infty}^\infty \frac{x^3 - x_0^3}{x-x_0} e^{-x^2}dx}_{=\frac{1}{2}+x_0^2} +  \underbrace{\frac{x_0^3}{\sqrt{\pi}} \int_{-\infty}^\infty \frac{ e^{-x^2}}{x-x_0}dx}_{=\zeta^3 Z(\zeta)} 
  = \frac{1}{2}+\zeta^2 R(\zeta).\\ 
  \frac{1}{\sqrt{\pi}} \int_{-\infty}^\infty \frac{x^4 e^{-x^2}}{x-x_0} dx &=& 
   \sign(\kpar) \Big( \frac{1}{2}\zeta +\zeta^3 R(\zeta) \Big).\\ 
  \frac{1}{\sqrt{\pi}} \int_{-\infty}^\infty \frac{x^5 e^{-x^2}}{x-x_0} dx &=& \frac{3}{4} + \frac{\zeta^2}{2} +\zeta^4 R(\zeta). \label{eq:Lan_5}
\end{eqnarray}
That was easy ! If we ever need a higher order, we will just blindly calculate
\begin{eqnarray}
&&  \frac{1}{\sqrt{\pi}} \int_{-\infty}^\infty \frac{x^n e^{-x^2}}{x-x_0} dx 
  = \frac{1}{\sqrt{\pi}} \int_{-\infty}^\infty \frac{x^n - x_0^n}{x-x_0} e^{-x^2}dx
  +  \underbrace{\frac{x_0^n}{\sqrt{\pi}} \int_{-\infty}^\infty \frac{ e^{-x^2}}{x-x_0}dx}_{=x_0^n \sign(\kpar) Z(\zeta)} \nn\\
  && \qquad = \frac{1}{\sqrt{\pi}} \int_{-\infty}^\infty \big( x^{n-1} + x^{n-2} x_0 + x^{n-3} x_0^2 +\cdots +xx_0^{n-2}+ x_0^{n-1}\big) e^{-x^2}dx + \sign(\kpar)^{n+1} \zeta^n Z(\zeta),
\end{eqnarray}
and we do not worry right now if this general case can be expressed in some smarter way. Now we know how to calculate the kinetic Landau integrals,
so let's use this knowledge, to calculate the first few integrals of the linear ``kinetic hierarchy''.
\newpage
\section{1D geometry (electrostatic)} \label{section:1D}
\subsection{Kinetic moments for Maxwellian $f_0$}
With the previous integrals already calculated, the calculation of the linear kinetic hierarchy is an easy process.
However, it is important to emphasize, that the hierarchy is \emph{linear}, and must be calculated as such. Again, as emphasized before, the kinetic velocity $v$ is an
independent quantity, and is not linearized. The total density $n=\int f dv$, and at the first order of course $n_0=\int f_0 dv$. The expansion
$n_0+n^{(1)}=\int (f_0+f^{(1)}) dv$ implies $n^{(1)}=\int f^{(1)}dv$. The density $n^{(1)}$, already calculated in (\ref{eq:N1F}), was therefore
calculated correctly, and using the plasma response function
\begin{equation}
\frac{n^{(1)}}{n_0} = -\frac{q_r}{T^{(0)}} \Phi R(\zeta). 
\end{equation}

The velocity moment is $nu=\int v f dv$ and at the first order $n_0 u_0 = \int v f_0 dv$. In our specific case, because
we do not consider any drifts in the distribution function, $u_0=0$. Expanding $(n_0+n^{(1)})(u_0+u^{(1)}) =\int v (f_0+f^{(1)}) dv$ and neglecting the nonlinear
quantity $n^{(1)} u^{(1)}$, yields $n_0 u^{(1)} = \int v f^{(1)} dv$. The velocity moment calculates
\begin{eqnarray}
  n_0 u^{(1)} &=& \int v f^{(1)} dv = -\frac{q_r}{T^{(0)}} \Phi \int v \Big(1+\frac{\frac{\omega}{\kpar}}{v-\frac{\omega}{\kpar}}\Big)f_0 dv
  = -\frac{q_r}{T^{(0)}} \Phi \Big( \underbrace{\int v f_0 dv}_{=n_0 u_0 =0} + \int \frac{v\frac{\omega}{\kpar}}{v-\frac{\omega}{\kpar}} f_0 dv \Big) \nn\\
  &=& -\frac{q_r}{T^{(0)}} \Phi n_0 \sqrt{\frac{\alpha}{\pi}} \frac{\omega}{\kpar} \int \frac{v e^{-\alpha v^2}}{v-\frac{\omega}{\kpar}} dv
  = \bigg[ \begin{array}{c}
   \sqrt{\alpha} v = x  \\
   \sqrt{\alpha} dv = dx \end{array}\bigg]
  = -\frac{q_r}{T^{(0)}} \Phi \frac{n_0}{\sqrt{\pi}} \frac{\sqrt{\alpha}}{\sqrt{\alpha}}\frac{\omega}{\kpar} \int \frac{x e^{-x^2}}{x-\frac{\omega}{\kpar}\sqrt{\alpha}} dx\nn\\
   &=& \bigg[ x_0 = \frac{\omega}{\kpar}\sqrt{\alpha} \bigg]
  = - \frac{q_r}{T^{(0)}} \Phi \frac{n_0}{\sqrt{\alpha}} \frac{x_0}{\sqrt{\pi}} \int_{-\infty}^\infty \frac{x e^{-x^2}}{x-x_0} dx
  = - \frac{q_r}{T^{(0)}} \Phi \frac{n_0}{\sqrt{\alpha}} \sign(\kpar) \zeta R(\zeta),
\end{eqnarray}
and canceling $n_0$ and using $1/\sqrt{\alpha}=v_{\textrm{th}}=\sqrt{2T^{(0)}/m}$ yields
\begin{equation}
u^{(1)} = - \frac{q_r}{T^{(0)}} \Phi \sqrt{\frac{2T^{(0)}}{m}} \sign(\kpar) \zeta R(\zeta).
\end{equation}  

The definition of the scalar pressure is $p=m\int (v-u)^2 f dv$ and at the first order $p_0=m\int v^2 f_0 dv$, because again $u_0=0$.
The quantity $(v-u)^2=v^2-2vu+u^2$ is linearized as $v^2-2vu^{(1)}$, and expanding $p_0+p^{(1)}=m\int (v^2-2vu^{(1)})(f_0+f^{(1)})dv$, further linearizing by neglecting
$u^{(1)}f^{(1)}$, and using $u_0=0$ yields $p^{(1)}=m\int v^2 f^{(1)}dv$.
The pressure calculates
\begin{eqnarray}
  p^{(1)} &=& m \int v^2 f^{(1)} dv = -\frac{q_r}{T^{(0)}} \Phi \Big( \underbrace{m\int v^2 f_0 dv}_{=p_0} + m\int \frac{v^2\frac{\omega}{\kpar}}{v-\frac{\omega}{\kpar}} f_0 dv \Big)
 = \bigg[ \begin{array}{c}
   \sqrt{\alpha} v = x  \\
   \sqrt{\alpha} dv = dx \end{array}\bigg]  \nn\\
&=& -\frac{q_r}{T^{(0)}} \Phi \Big( p_0 + m \frac{n_0}{\sqrt{\pi}} \frac{\sqrt{\alpha}}{\alpha}\frac{\omega}{\kpar}
 \int \frac{x^2 e^{-x^2}}{x-\frac{\omega}{\kpar}\sqrt{\alpha}} dx \Big)
 = \bigg[ x_0 = \frac{\omega}{\kpar}\sqrt{\alpha} \bigg] \nn\\
&=&   -\frac{q_r}{T^{(0)}} \Phi \Big( p_0 + m \frac{n_0}{\alpha} \frac{x_0}{\sqrt{\pi}}
 \int \frac{x^2 e^{-x^2}}{x-x_0} dx \Big) 
 = -\frac{q_r}{T^{(0)}} \Phi \Big( p_0 + m \frac{n_0}{\alpha} \zeta^2 R(\zeta) \Big),
\end{eqnarray}
and dividing by $p_0$ and using $p_0=n_0 T^{(0)}$ to calculate $m n_0/(p_0\alpha)=2$, the pressure moment reads
\begin{equation}
\frac{p^{(1)}}{p_0} = -\frac{q_r}{T^{(0)}} \Phi \Big( 1+2\zeta^2 R(\zeta) \Big).
\end{equation}

We will also need the temperature $T^{(1)}$. The general temperature is defined $T=p/n$, i.e. the definition is nonlinear. The process of linearization is essentially
like doing a derivative 
\begin{eqnarray}  
T=\frac{p}{n} \quad \overset{\textrm{lin.}}{\rightarrow}\quad T^{(1)} = \frac{p^{(1)}}{n_0} - \frac{p_0}{n_0^2} n^{(1)},
\end{eqnarray}
and dividing by $T^{(0)}=p_0/n_0$ yields
\begin{equation} \label{eq:TempLin}
\frac{T^{(1)}}{T^{(0)}} = \frac{p^{(1)}}{p_0} - \frac{n^{(1)}}{n_0}.
\end{equation}
If one does not like the ``derivative'', the same result is obtained by writing $p=Tn$ instead, and linearizing
$(p_0+p^{(1)})=(T^{(0)}+T^{(1)})(n_0+n^{(1)})$. Which after subtracting $p_0=T^{(0)}n_0$, neglecting $T^{(1)}n^{(1)}$, yields $p^{(1)}=T^{(1)} n_0+ T^{(0)} n^{(1)}$,
which after dividing by $p_0$ yields (\ref{eq:TempLin}).
The temperature is therefore easily calculated as
\begin{equation} 
\frac{T^{(1)}}{T^{(0)}} = -\frac{q_r}{T^{(0)}} \Phi \Big( 1+2\zeta^2 R(\zeta) -R(\zeta) \Big).
\end{equation}

The scalar heat flux is defined as $q=m\int(v-u)^3 f dv$ and at the first order  $q_0=m\int v^3 f_0 dv$,
and for our case $q_0=0$. The quantity $(v-u)^3=v^3-3v^2u+3v u^2 -u^3$ is linearized as $v^3-3v^2 u^{(1)}$. Expanding
$q_0+q^{(1)}=m\int (v^3-3v^2 u^{(1)})(f_0+f^{(1)})dv$, neglecting $u^{(1)}f^{(1)}$, yields one contribution that is very easy to
overlook, and that is of the same order as the expected $m\int v^3 f^{(1)}dv$, and that is proportional to $m\int v^2 f_0 dv=p_0$.  
Therefore, the linearized heat flux $q^{(1)}$ must be correctly calculated according to
\begin{equation} \label{eq:HFkin}
q^{(1)}= m\int v^3 f^{(1)}dv -3 p_0 u^{(1)}.
\end{equation}
The first term calculates
\begin{eqnarray}
m\int v^3  f^{(1)}dv &=& -\frac{q_r}{T^{(0)}} \Phi \Big( \underbrace{m\int v^3 f_0 dv}_{=q_0=0} + m\int \frac{v^3\frac{\omega}{\kpar}}{v-\frac{\omega}{\kpar}} f_0 dv \Big)
 = \bigg[ \begin{array}{c}
   \sqrt{\alpha} v = x  \\
   \sqrt{\alpha} dv = dx \end{array}\bigg]  \nn\\
 &=& -\frac{q_r}{T^{(0)}} \Phi m \frac{n_0}{\sqrt{\pi}} \frac{\sqrt{\alpha}}{\alpha^{3/2}}\frac{\omega}{\kpar}
 \int \frac{x^3 e^{-x^2}}{x-\frac{\omega}{\kpar}\sqrt{\alpha}} dx
 = \bigg[ x_0 = \frac{\omega}{\kpar}\sqrt{\alpha} \bigg] \nn\\
&=&   -\frac{q_r}{T^{(0)}} \Phi m \frac{n_0}{\alpha^{3/2}} \frac{x_0}{\sqrt{\pi}}
 \int \frac{x^3 e^{-x^2}}{x-x_0} dx 
 = -q_r n_0 \Phi \sqrt{\frac{2T^{(0)}}{m}} \sign(\kpar) \Big( \zeta + 2\zeta^3 R(\zeta) \Big),
\end{eqnarray}  
where we used $\alpha^{-3/2}=(2T^{(0)}/m) \sqrt{2T^{(0)}/m}$. And the entire heat flux (\ref{eq:HFkin}) then reads
\begin{eqnarray}
q^{(1)} = -q_r n_0 \Phi \sqrt{\frac{2T^{(0)}}{m}} \sign(\kpar) \Big( \zeta  + 2\zeta^3 R(\zeta) -3\zeta R(\zeta) \Big).
\end{eqnarray}

The scalar 4th order moment is defined as $r=m\int(v-u)^4 f dv$ and at the first order of course $r_0=m\int v^4 f_0 dv$, since again $u_0=0$. Also, $r_0=3p_0^2/\rho_0$,
where $\rho_0=m n_0$. The quantity $(v-u)^4$ is linearized as $v^4-4v^3 u^{(1)}$. Expanding $r_0+r^{(1)}=m\int (v^4-4v^3 u^{(1)})(f_0+f^{(1)})dv$, the
quantity $m\int v^3 f_0=q_0=0$, which yields a simple $r^{(1)}=m\int v^4 f^{(1)} dv$. The 4th order moment calculates
\begin{eqnarray}
r^{(1)} &=& m\int v^4  f^{(1)}dv = -\frac{q_r}{T^{(0)}} \Phi \Big( \underbrace{m\int v^4 f_0 dv}_{=r_0} + m\int \frac{v^4\frac{\omega}{\kpar}}{v-\frac{\omega}{\kpar}} f_0 dv \Big)
 = \bigg[ \begin{array}{c}
   \sqrt{\alpha} v = x  \\
   \sqrt{\alpha} dv = dx \end{array}\bigg]  \nn\\
 &=& -\frac{q_r}{T^{(0)}} \Phi \Big(r_0+ m \frac{n_0}{\sqrt{\pi}} \frac{\sqrt{\alpha}}{\alpha^2}\frac{\omega}{\kpar}
 \int \frac{x^4 e^{-x^2}}{x-\frac{\omega}{\kpar}\sqrt{\alpha}} dx \Big)
 = \bigg[ x_0 = \frac{\omega}{\kpar}\sqrt{\alpha} \bigg] \nn\\
&=&   -\frac{q_r}{T^{(0)}} \Phi \Big( r_0 + m \frac{n_0}{\alpha^{2}} \frac{x_0}{\sqrt{\pi}}
 \int \frac{x^4 e^{-x^2}}{x-x_0} dx \Big)
 = -\frac{q_r p_0}{m} \Phi \Big( 3 + 2\zeta^2 + 4\zeta^4 R(\zeta) \Big),
\end{eqnarray} 
where we have used $1/\alpha^2=4T^{(0)2}/m^2$, and $mn_0/\alpha^2=4p_0^2/\rho_0$.

The entire nonlinear $r$ is decomposed as $r=3p^2/\rho + \widetilde{r}$. The first term can be linearized in a number of ways,
and of course, all techniques must yield the same result, since linearization must be unique. 
For example by using the derivative
\begin{equation}
\left( \frac{p^2}{\rho} \right )' = \frac{1}{\rho}2p p' - \frac{p^2}{\rho^2} \rho' = \frac{p^2}{\rho}\Big( \frac{2p'}{p}-\frac{\rho'}{\rho} \Big),
\end{equation}  
the term is easily linearized as
\begin{equation}
  \frac{p^2}{\rho} \overset{\textrm{lin.}}{=} \frac{p_0^2}{\rho_0}\Big( \frac{2p^{(1)}}{p_0}-\frac{\rho^{(1)}}{\rho_0} \Big)
  = \frac{p_0^2}{\rho_0}\Big( \frac{2p^{(1)}}{p_0}-\frac{n^{(1)}}{n_0} \Big),
\end{equation}  
and by further using (\ref{eq:TempLin}), also alternatively as
\begin{equation}
  \frac{p^2}{\rho} \overset{\textrm{lin.}}{=} \frac{p_0^2}{\rho_0}\Big( \frac{2T^{(1)}}{T^{(0)}}+\frac{n^{(1)}}{n_0} \Big).
\end{equation}
By using $r_0=3p_0^2/\rho_0$, therefore yields useful relations (valid for Maxwellian)
\begin{equation} \label{eq:crazyyy}
\frac{r^{(1)}}{r_0} = 2\frac{p^{(1)}}{p_0}-\frac{n^{(1)}}{n_0} +\frac{\widetilde{r}^{(1)}}{r_0} = 2\frac{T^{(1)}}{T^{(0)}}+\frac{n^{(1)}}{n_0} +\frac{\widetilde{r}^{(1)}}{r_0}.
\end{equation}
Another possibility (to double check the linearization), is to rewrite $\frac{p^2}{\rho}=\frac{1}{m}pT$, so that $r=\frac{3}{m} pT + \widetilde{r}$, and to linearize that one instead.
Expanding that expression into $r_0+r^{(1)} = \frac{3}{m}( (p_0+p^{(1)})(T^{(0)}+T^{(1)}) ) + \widetilde{r}^{(1)}$ (where by a definition/construction $\widetilde{r}^{(0)}=0$),
after subtracting $r_0=\frac{3}{m}p_0 T^{(0)}$, and neglecting $p^{(1)} T^{(1)}$, yields $r^{(1)} = \frac{3}{m}( p^{(1)}T^{(0)} + p_0T^{(1)}) + \widetilde{r}^{(1)}$. Dividing this
expression by $r_0$ yields
\begin{equation}
\frac{r^{(1)}}{r_0} = \frac{p^{(1)}}{p_0}+\frac{T^{(1)}}{T^{(0)}} +\frac{\widetilde{r}^{(1)}}{r_0},
\end{equation}
which when used with (\ref{eq:TempLin}), is equivalent to (\ref{eq:crazyyy}). 
Now we can easily calculate the $\widetilde{r}^{(1)}$ component as
\begin{equation}
  \widetilde{r}^{(1)} = r^{(1)} -3\frac{p_0^2}{\rho_0}\Big( \frac{2p^{(1)}}{p_0}-\frac{n^{(1)}}{n_0} \Big),
\end{equation}  
that directly yields
\begin{equation}
  \widetilde{r}^{(1)} = -\frac{q_r p_0}{m} \Phi \Big( 2\zeta^2 + 4\zeta^4 R(\zeta) +3R(\zeta)-3 -12\zeta^2R(\zeta) \Big).
\end{equation}
Now we are ready to explore the possible closures.

\clearpage
\subsection{Exploring possibilities of a closure}
Let's summarize the obtained linear hierarchy so that we can directly see the similarities. Let's also for a moment introduce back the species
index $r$, so that we are completely clear 
\begin{eqnarray}
\frac{n^{(1)}_r}{n_{0r}} &=& -\frac{q_r}{T^{(0)}_r} \Phi R(\zeta_r); \label{eq:RefPica-n}\\
u^{(1)}_r &=& - \frac{q_r}{T^{(0)}_r} \Phi \sqrt{\frac{2T^{(0)}_r}{m_r}} \sign(\kpar) \zeta_r R(\zeta_r);\\
\frac{p^{(1)}_r}{p_{0r}} &=& -\frac{q_r}{T^{(0)}_r} \Phi \Big( 1+2\zeta^2_r R(\zeta_r) \Big);\\
\frac{T^{(1)}_r}{T^{(0)}_r} &=& -\frac{q_r}{T^{(0)}_r} \Phi \Big( 1+2\zeta^2_r R(\zeta_r) -R(\zeta_r) \Big);\\
q^{(1)}_r &=& -q_r n_{0r} \Phi \sqrt{\frac{2T^{(0)}_r}{m_r}} \sign(\kpar) \Big( \zeta_r  + 2\zeta_r^3 R(\zeta_r) -3\zeta_r R(\zeta_r) \Big);\\
r^{(1)}_r &=& -\frac{q_r p_{0r}}{m_r} \Phi \Big( 3 + 2\zeta_r^2 + 4\zeta_r^4 R(\zeta_r) \Big);\\
\widetilde{r}^{(1)}_r &=& -\frac{q_r p_{0r}}{m_r} \Phi \Big( 2\zeta_r^2 + 4\zeta_r^4 R(\zeta_r) +3R(\zeta_r)-3 -12\zeta_r^2R(\zeta_r) \Big), \label{eq:RefPica-rr}
\end{eqnarray}  
with an emphasis that the charge $q_r$ should not be confused with the heat flux $q_r^{(1)}$. The $\zeta_r=\omega/(|\kpar| v_{\textrm{th} r})$ and
the thermal speed $v_{\textrm{th} r} = \sqrt{2T^{(0)}_r/m_r}$.  Note the presence of $\sign(\kpar)$ in the expressions for $u^{(1)}_r$ and $q^{(1)}_r$.
  The presence of $\sign(\kpar)$ can be verified aposteriori, for example by considering the simplest situation when the Landau damping is neglected,
  and the $R(\zeta_r)$ function yields only real numbers for real valued $\zeta_r$
  (i.e. the $R(\zeta_r)$ function can be approximated with Pad\'e approximants that contain only powers of $\zeta_r^2$).
  Simultaneously changing signs of $\kpar$ and $\omega$ in a Fourier mode should give its complex conjugate, i.e., the real part
  of expressions (\ref{eq:RefPica-n})-(\ref{eq:RefPica-rr}) can not change its sign in that transformation. This is indeed true because the
  expressions for $u^{(1)}_r$ and $q^{(1)}_r$ contain $\sign(\kpar)\zeta_r = \frac{\omega}{\kpar v_{\textrm{th} r}}$.

To better understand what is meant by ``a closure'', let's first examine what is not a closure. Let's examine the density $n^{(1)}$ equation. 
Since in this specific example we used the electrostatic electric field $\bE^{(1)}=-\nabla\phi$, the only Maxwell equation left is the $\nabla\cdot\bE^{(1)}=4\pi\sum_r q_r n_r$,
where $q_r$ is the charge and $n_r$ is the total density. Linearization of this equation, and using the natural charge neutrality that must be satisfied
at the 0-th order $\sum_r q_r n_{0r}=0$, yields $\nabla\cdot\bE^{(1)}=4\pi\sum_r q_r n_r^{(1)}$, or written with the scalar potential $-\nabla^2 \Phi=4\pi\sum_r q_r n_r^{(1)}$,
and transformed to Fourier space $k^2 \Phi = 4\pi\sum_r q_r n_r^{(1)}$. We consider 1D propagation parallel to $\bb_0$
with wavenumber $\kpar$, and to be consistent, we therefore continue with $\kpar$ and
\begin{equation}
\kpar^2 \Phi = 4\pi\sum_r q_r n_r^{(1)} = 4\pi\sum_r q_r n_{0r} \frac{n_r^{(1)}}{n_{0r}} = - 4\pi\sum_r n_{0r} \frac{q_r^2}{T^{(0)}_r} \Phi R(\zeta_r),
\end{equation}  
which can be rewritten as
\begin{equation} \label{eq:disprelrit}
\left( \kpar^2 + 4\pi\sum_r n_{0r} \frac{q_r^2}{T^{(0)}_r} R(\zeta_r) \right) \Phi =0.
\end{equation}  
Even though the system is now ``closed'', the eq. (\ref{eq:disprelrit}) does \emph{not} represent a fluid closure, and should be viewed only as a kinetic ``dispersion relation''.
To have a non-trivial solution for the potential $\Phi$, the expression inside of the bracket must be equal to zero.
By declaring that $\kpar\neq 0$ (the case $\kpar=0$ is trivial since we need some wavenumber), we can divide by $\kpar^2$.
By using the definition of the Debye length of r-species $\lambda_{Dr}=1/k_{Dr}$, where $k_{Dr}^2=4\pi n_{0r}q_r^2/T^{(0)}_r$, one obtains a dispersion relation
\footnote{An interesting observation (that is perhaps obvious if one considers how the Debye length is derived), is that the Debye length of r-species $\lambda_{Dr}$ does
  not depend on the mass $m_r$.}
\begin{equation} \label{eq:Estatic}
1 + \sum_r \frac{1}{\kpar^2 \lambda_{Dr}^2} R(\zeta_r) =0.
\end{equation}
If one replaces here $\kpar\to k$, the expression is actually equivalent to a multi-species dispersion relation, usually found
in plasma physics books under the \emph{electrostatic} waves in hot \emph{unmagnetized} plasmas, with Maxwellian $f_{0r}$.
See for example Gurnett \& Bhattachrjee, page 353, eq. (9.4.18). We are not interested here in studying unmagnetized plasmas, and
instead, we will just remember (\ref{eq:Estatic}) as the dispersion relation of the parallel propagating (to $\bb_0$) electrostatic mode in \emph{magnetized} plasma,
since this mode indeed does not contain any magnetic field fluctuations. 

Let's consider only the proton and electron species, $r=p,e$, so that
\begin{equation} \label{eq:SoundPEls}
1+ \frac{1}{\kpar^2 \lambda_{De}^2} \bigg[ \frac{T_e^{(0)}}{T_p^{(0)}} R(\zeta_p) +R(\zeta_e) \bigg] =0,
\end{equation}
where the proton Debye length was rewritten with the electron Debye length $\lambda_{De}=\lambda_{Dp} \sqrt{T_e^{(0)}/T_p^{(0)}}$.
For a general case, the dispersion relation has to be solved numerically, and again, can not be much simplified, unless one wants to consider
long wavelength limit $\kpar \lambda_{De}\ll 1$, where only the expression inside of the big brackets can be used.
The solution contains the usual Langmuir waves, that are obtained by neglecting the ion term (by making the ions immobile) and by
expanding the $R(\zeta_e)$ in the limit $|\zeta_e|\gg 1$, i.e. in the limit when the wave phase speed $\omega/k$ is much larger than the electron thermal speed $v_{\textrm{th} e}$.
Langmuir waves propagate with speeds that are higher than the electron plasma frequency $\omega_{pe}=\sqrt{4\pi n_{e0} e^2/m_e}$, which for us are extremely high frequencies.
The solution also contains the ``ion-acoustic mode'', which in plasma books is obtained in the limit $|\zeta_p|\gg 1$ and $|\zeta_e|\ll 1$,
i.e. in the limit where the wave phase speed is much larger than the proton thermal speed, $\omega/k \gg v_{\textrm{th} p}$,
but also where the phase speed is much smaller than the electron thermal speed, $\omega/k \ll v_{\textrm{th} e}$, for the result see for example
Gurnett and Bhattacharjee, page 356, eq. (9.4.28-29). 

So what about the limit $|\zeta_p|\ll 1$, when the phase speed is much smaller than the proton thermal speed $\omega/k \ll v_{\textrm{th} p}$ ?
The ion-acoustic mode does not exist in this limit ? Unfortunately, in the classical long wavelength limit,
the phase speeds do not become smaller and smaller, the phase speeds $\omega/k$ just become non-dispersive and constant. In the CGL description (with cold electrons),
the parallel propagating ion-acoustic mode has a phase speed $\omega/\kpar=\pm C_\parallel$, where the parallel sound speed $C_\parallel^2=3p_{\parallel p}^{(0)}/\rho_0=3T_{\parallel p}^{(0)}/m_p$.
The limit $|\zeta_p|\ll 1$ is never satisfied, because $C_\parallel \ll v_{\textrm{th}\parallel p}$ means $\sqrt{3T_{\parallel p}^{(0)}/m_p} \ll \sqrt{2T_{\parallel p}^{(0)}/m_p}$,
 which is never true. One can estimate the lowest possible value
of $|\zeta_p|$ to be roughly in the neighborhood of
$|\zeta_p|_{\textrm{min}} \approx C_\parallel^{\textrm{CGL}}/v_{\textrm{th}\parallel p} = \sqrt{3/2}$, or in another words $|\zeta_p|_{\textrm{min}} \approx 1$.
There is no expansion of the $Z(\zeta)$ for $|\zeta|\approx 1$ and the result has to be found only numerically. 

So what constitutes a Landau fluid closure ? We will use the following definition: 
Express the last retained moment through lower-order moments in such a way, that the kinetic $R(\zeta)$ function is eliminated
 (for example by using Pad\'e approximation), so that the closure is expressed only through fluid variables and it is prescribed for all $\zeta$ values.

\subsubsection{Preliminary closures for $|\zeta|\ll 1$}
As explained above, the limit $|\zeta|\ll 1$ is actually a bit unphysical for the proton species in the electrostatic limit,
and is physically plausible only for the electron species.
Nevertheless, briefly exploring the linear kinetic hierarchy in this limit allows us to explore what kind of closures might be possible.
In this limit, the plasma dispersion function can be expanded as
\begin{eqnarray}
  Z(\zeta) &=& i\sqrt{\pi} e^{-\zeta^2} -2\zeta \Big[ 1-\frac{2}{3}\zeta^2+\frac{4}{15}\zeta^4 -\frac{8}{105}\zeta^6 +\cdots
    +\frac{(-2)^n \zeta^{2n}}{(2n+1)!!}+\cdots \Big]; \qquad |\zeta|\ll 1, \\
Z(\zeta) &=& i\sqrt{\pi} e^{-\zeta^2} -2\zeta + \frac{4}{3}\zeta^3-\frac{8}{15}\zeta^5+\frac{16}{105}\zeta^7 +\cdots; 
\end{eqnarray}  
and the plasma response function as
\begin{equation}
  R(\zeta) = 1 + i\zeta\sqrt{\pi} e^{-\zeta^2} +\Big[ -2\zeta^2 + \frac{4}{3}\zeta^4-\frac{8}{15}\zeta^6+\frac{16}{105}\zeta^8+\cdots
  + \frac{(-2)^{n+1} \zeta^{2n+2}}{(2n+1)!!}+\cdots\Big] ; \qquad |\zeta|\ll 1, \label{eq:RzetaSMA}
\end{equation}
and where for small $\zeta$, the $e^{-\zeta^2}$ is naturally expanded as
\begin{equation}
e^{-\zeta^2} = 1-\zeta^2+\frac{\zeta^4}{2!}-\frac{\zeta^6}{3!}+\cdots+\frac{(-1)^n \zeta^{2n}}{n!}+\cdots; \qquad |\zeta|\ll 1, 
\end{equation}
yielding
\begin{eqnarray}
\qquad |\zeta|\ll 1: \quad Z(\zeta) &=& i\sqrt{\pi}-2\zeta-i\sqrt{\pi}\zeta^2+\frac{4}{3}\zeta^3+i\frac{\sqrt{\pi}}{2}\zeta^4-\frac{8}{15}\zeta^5-i\frac{\sqrt{\pi}}{6}\zeta^6
+\frac{16}{105}\zeta^7+\cdots;\\
R(\zeta) &=& 1 + i\sqrt{\pi}\zeta -2\zeta^2 -i\sqrt{\pi}\zeta^3 + \frac{4}{3}\zeta^4+i\frac{\sqrt{\pi}}{2}\zeta^5-\frac{8}{15}\zeta^6 -i\frac{\sqrt{\pi}}{6}\zeta^7
+\frac{16}{105}\zeta^8+\cdots. \label{eq:RzetaSmall1}
\end{eqnarray}
For our purposes it is sufficient to keep the series only up to $\zeta^3$, i.e. to work with the precision $o(\zeta^3)$.
The expressions entering the kinetic hierarchy in equations (\ref{eq:RefPica-n})-(\ref{eq:RefPica-rr}) are
\begin{eqnarray}
  R(\zeta) &=& 1+i\sqrt{\pi}\zeta -2\zeta^2 -i\sqrt{\pi}\zeta^3 +\cdots o(\zeta^3);\\
  \zeta R(\zeta) &=& \zeta + i\sqrt{\pi}\zeta^2 -2\zeta^3;\\
  1+2\zeta^2 R(\zeta) &=& 1+2\zeta^2+2i\sqrt{\pi}\zeta^3;\\
  1-R(\zeta)+2\zeta^2 R(\zeta) &=& -i\sqrt{\pi}\zeta +4\zeta^2+3i\sqrt{\pi}\zeta^3;\\
  \zeta +2\zeta^3R(\zeta)-3\zeta R(\zeta) &=& -2\zeta -3i\sqrt{\pi}\zeta^2+8\zeta^3;\\
  3+2\zeta^2+4\zeta^4R(\zeta) &=& 3+2\zeta^2+4\zeta^4;\\
  2\zeta^2 + 4\zeta^4 R(\zeta) +3R(\zeta)-3 -12\zeta^2 R(\zeta) &=& 3i\sqrt{\pi}\zeta-16\zeta^2-15i\sqrt{\pi}\zeta^3. 
\end{eqnarray}  
An interesting observation is that for small $\zeta$, moments $n^{(1)}$, $p^{(1)}$ and $r^{(1)}$ are finite, and
moments $u^{(1)}$, $T^{(1)}$, $q^{(1)}$ and $\widetilde{r}^{(1)}$ are proportional to $\zeta$ and therefore small. We want to make a simple closure for the heat flux $q^{(1)}$ or
the 4th order correction $\widetilde{r}^{(1)}$, and thus, let's concentrate on the moments that are small. To clarify how the closure is performed, let's write them down
only up to the precision $o(\zeta_r^2)$, so
\begin{eqnarray}
u^{(1)}_r &=& - \frac{q_r}{T^{(0)}_r} \Phi \sqrt{\frac{2T^{(0)}_r}{m_r}} \sign(\kpar) \Big( \zeta_r + i\sqrt{\pi}\zeta_r^2 \Big); \label{eq:S1a}\\
T^{(1)}_r &=& -q_r \Phi \Big(-i\sqrt{\pi}\zeta_r +4\zeta_r^2 \Big);\\
q^{(1)}_r &=& -q_r n_{0r} \Phi \sqrt{\frac{2T^{(0)}_r}{m_r}} \sign(\kpar) \Big(-2\zeta_r -3i\sqrt{\pi}\zeta_r^2 \Big);\\
\widetilde{r}^{(1)}_r &=& -\frac{q_r p_{0r}}{m_r} \Phi \Big( 3i\sqrt{\pi}\zeta_r-16\zeta_r^2 \Big).\label{eq:S1b}
\end{eqnarray}
If we further restrict ourselves to only precision $o(\zeta_r)$ and neglect the $\zeta_r^2$ terms, we can find an amazing result that we can express the heat flux
$q_r^{(1)}$ with respect to temperature $T_r^{(1)}$ according to
\begin{equation} \boxed{
   o(\zeta_r):\qquad q^{(1)}_r = -i\frac{n_{0r}}{\sqrt{\pi}}\sqrt{\frac{8T^{(0)}_r}{m_r}}\sign(\kpar) T_r^{(1)}
  =  -i\frac{2n_{0r}}{\sqrt{\pi}} v_{\textrm{th} r} \sign(\kpar) T_r^{(1)}. } \label{eq:R32_1closure}
\end{equation}
The above result is of upmost importance, because it emphasizes the major difference between \emph{collisionless} and \emph{collisional} systems.
At this point, the result is derived only with the assumption $|\zeta|\ll 1$, even though we will see later that
the result is not restricted to this limit, and the result has a much wider applicability. 
The result is the famous expression for \emph{collisionless heat flux}, that here reads $q \sim -i \sign(\kpar) T$, which is in strong contrast
to the usual collisional heat flux $q\sim -\nabla_\parallel T$ that in Fourier space reads $q\sim -i k_\parallel T$. We will come to this expression later. 

With the precision $o(\zeta_r)$, other obvious possibilities are to express $q_r^{(1)}$ with respect to velocity $u_r^{(1)}$,
or to express $\widetilde{r}^{(1)}$ through $u_r^{(1)}$, $T_r^{(1)}$, $q_r^{(1)}$ according to
\begin{eqnarray}
  o(\zeta_r):\qquad \quad q^{(1)}_r &=& -2n_{0r} T_r^{(0)} u_r^{(1)} = -2 p_{0r} u_r^{(1)} \label{eq:R21closure};\\
  \widetilde{r}_r^{(1)} &=&  i\frac{3}{2}\sqrt{\pi}v_{\textrm{th} r} p_{0r} \sign(\kpar) u_r^{(1)};\\
  \widetilde{r}_r^{(1)} &=&  -\frac{3}{2}v_{\textrm{th} r}^2 n_{0r} T_r^{(1)};\\
  \widetilde{r}_r^{(1)} &=&  -i\frac{3}{4}\sqrt{\pi}v_{\textrm{th} r} \sign(\kpar) q_r^{(1)}. \label{eq:R44closure}
\end{eqnarray}

However, if we did so much work that we consider the 4th order moment, it would be a shame not to increase the precision to $o(\zeta_r^2)$.
Obviously, we need to use a combination of at least 2 different lower order moments. For example, by trying
\begin{equation}
 o(\zeta_r^2):\qquad \quad \widetilde{r}^{(1)}_r = \alpha_q q_r^{(1)} + \alpha_T T_r^{(1)}.
\end{equation}
The proportionality constants $\alpha_q$, $\alpha_T$ are easily obtained by separation to two equations for $\zeta_r$ and $\zeta_r^2$ that must be satisfied
\begin{eqnarray}
  -\frac{p_{0r}}{m_r}3i\sqrt{\pi} = +2\alpha_q n_{0r} v_{\textrm{th}r} \sign(\kpar) +i\alpha_T \sqrt{\pi};\\
  16\frac{p_{0r}}{m_r} = +\alpha_q n_{0r} v_{\textrm{th}r} \sign(\kpar) 3i\sqrt{\pi} - 4\alpha_T.
\end{eqnarray}  
Playing with the algebra little bit (for example $p_{0r}/m_r=T_r^{(0)} n_{0r}/m_r = v_{\textrm{th} r}^2 n_{0r} /2$), the two equations can be solved easily for the
unknown quantities $\alpha_q$, $\alpha_T$, and the final result is
\begin{equation} \boxed{
  o(\zeta_r^2):\qquad \quad \widetilde{r}^{(1)}_r = -i \frac{2\sqrt{\pi}}{3\pi-8}v_{\textrm{th} r} \sign(\kpar) q_r^{(1)}
  + \frac{32-9\pi}{2(3\pi-8)}v_{\textrm{th} r}^2 n_{0r} T_r^{(1)}. } \label{eq:R43closure}
\end{equation}
There are naturally other possibilities and with the precision $o(\zeta_r^2)$, one can search for closures
\begin{eqnarray}
  o(\zeta_r^2):\qquad \quad  q^{(1)}_r &=& \alpha_T T_r^{(1)} + \alpha_u u_r^{(1)};\\
  \widetilde{r}^{(1)}_r &=& \alpha_q q_r^{(1)} + \alpha_u u_r^{(1)};\\
  \widetilde{r}^{(1)}_r &=& \alpha_T T_r^{(1)} + \alpha_u u_r^{(1)},
\end{eqnarray}
where the first choice yields a closure
\begin{equation} \boxed{
  o(\zeta_r^2):\qquad \quad  q^{(1)}_r = - i\frac{\sqrt{\pi}}{4-\pi} n_{0r} v_{\textrm{th} r} \sign(\kpar)T_r^{(1)}
  +\frac{3\pi-8}{4-\pi} n_{0r} T_r^{(0)} u_r^{(1)}, } \label{eq:R31closure}
\end{equation}
and the other two choices yield
\begin{eqnarray}
  o(\zeta_r^2):\qquad \quad \widetilde{r}^{(1)}_r &=& -i\frac{16-3\pi}{2\sqrt{\pi}}v_{\textrm{th} r} \sign(\kpar) q_r^{(1)}
  -i\frac{32-9\pi}{2\sqrt{\pi}} v_{\textrm{th} r} n_{0r} T_r^{(0)} \sign(\kpar) u_r^{(1)}; \\
  \widetilde{r}^{(1)}_r &=& - \frac{16-3\pi}{8-2\pi} v_{\textrm{th} r}^2 n_{0r}T_r^{(1)}
  +i\frac{2\sqrt{\pi}}{\pi-4} v_{\textrm{th} r} n_{0r} T_r^{(0)} \sign(\kpar) u_r^{(1)}.
\end{eqnarray}
For completeness, one can easily find a closure for $\widetilde{r}^{(1)}_r$ with precision $o(\zeta_r^3)$ (after updating (\ref{eq:S1a})-(\ref{eq:S1b}) to
precision $o(\zeta_r^3)$) by searching for a solution
\begin{equation}
o(\zeta_r^3):\qquad \quad \widetilde{r}^{(1)}_r = \alpha_q q_r^{(1)} + \alpha_T T_r^{(1)} + \alpha_u u_r^{(1)},
\end{equation}
and the solution reads
\begin{equation} \label{eq:R42closure}
\boxed{  
  o(\zeta_r^3):\qquad \widetilde{r}^{(1)}_r = -i\sqrt{\pi} \frac{10-3\pi}{16-5\pi} v_{\textrm{th} r} \sign(\kpar) q_r^{(1)}
  + \frac{21\pi-64}{2(16-5\pi)}v_{\textrm{th} r}^2 n_{0r} T_r^{(1)} + i\sqrt{\pi} \frac{9\pi-28}{16-5\pi}v_{\textrm{th} r} p_{0r}\sign(\kpar) u_r^{(1)}.}
\end{equation}  

We purposely kept the species index $r$ in the calculations, to clearly show that the closures are performed for each species separately,
and no Maxwell equations or other physical principles are used. The equations would be perhaps easier to read without the index r.

\clearpage
\subsubsection{Exploring the case $|\zeta|\gg 1$}
For large value of $|\zeta|$, we need to use an asymptotic expansion of the plasma dispersion function that reads
\begin{equation}
  Z(\zeta) = i\sigma \sqrt{\pi}e^{-\zeta^2} - \frac{1}{\zeta}\Big[ 1+\frac{1}{2\zeta^2} +\frac{3}{4\zeta^4}
    +\frac{15}{8\zeta^6}+ \frac{105}{16\zeta^8}\cdots +\frac{(2n-1)!!}{(2\zeta^2)^n}+\cdots\Big]; \qquad |\zeta|\gg 1,
\end{equation}  
where
\begin{equation}
\sigma =
  \begin{dcases}
    0, & \text{$Im(\zeta)>0$};\\
    1, & \text{$Im(\zeta)=0$};\\
    2, & \text{$Im(\zeta)<0$}.
  \end{dcases}
\end{equation}
The term with $\sigma$ comes directly from the definition of $Z(\zeta)$ and there is not much one can further do about it,
since there is no further asymptotic expansion for $exp(-\zeta^2)$ when $\zeta$ is large. The term is zero in the upper half of complex plane ($\sigma=0$).
When very close to the real axis, i.e. when $\sigma=1$, the term mainly contributes to the imaginary part of $Z(\zeta)$ (even though only very weakly) and
for the real part of $Z(\zeta)$, it's contribution can be neglected. However, when deeply down in the lower half of complex plane, the term can become
very large (for example if $\zeta=-iy$, $\exp(-\zeta^2)=\exp(y^2)$ and if $y$ is large the term obviously explodes). Deeply down in the lower complex plane the
term is a real trouble, and even some kinetic solvers such as WHAMP \citep{Ronnmark1982} have trouble with calculations when the damping is too large.

We will see shortly, that for our purposes the term can be completely neglected, but let's keep it for a moment.
The expansion of the Maxwellian plasma response function therefore reads
\begin{equation} \label{eq:RzetaL}
R(\zeta) = i\sigma\sqrt{\pi}\zeta e^{-\zeta^2} - \frac{1}{2\zeta^2}-\frac{3}{4\zeta^4} -\frac{15}{8\zeta^6} -\frac{105}{16\zeta^8}-\frac{945}{32\zeta^{10}}\cdots; \qquad |\zeta|\gg 1.
\end{equation}
Let's calculate the kinetic hierarchy, at least up to $1/\zeta^4$. After a short inspection, one immediately sees that the hierarchy calculates a bit
differently than in the previous case, and to get the 4th order moments with the precision $o(1/\zeta^4)$,
it is important to keep all the terms up to $\sim 1/\zeta^8$ in the $R(\zeta)$ expression, since the 4th order moments contain $\zeta^4 R(\zeta)$ terms.
The expressions entering the kinetic hierarchy dully calculate
\begin{eqnarray}
n^{(1)}\sim  R(\zeta) &=& i\sigma\sqrt{\pi}\zeta e^{-\zeta^2} - \frac{1}{2\zeta^2}-\frac{3}{4\zeta^4} + o(\frac{1}{\zeta^4});\\
u^{(1)}\sim  \zeta R(\zeta) &=& i\sigma\sqrt{\pi}\zeta^2 e^{-\zeta^2} - \frac{1}{2\zeta}-\frac{3}{4\zeta^3} - \frac{15}{8\zeta^5};\\
p^{(1)}\sim  1+2\zeta^2 R(\zeta) &=& 2i\sigma\sqrt{\pi}\zeta^3 e^{-\zeta^2} -\frac{3}{2\zeta^2}-\frac{15}{4\zeta^4};\\
T^{(1)}\sim  1-R(\zeta)+2\zeta^2 R(\zeta) &=& i\sigma\sqrt{\pi}e^{-\zeta^2}(2\zeta^3-\zeta) - \frac{1}{\zeta^2}-\frac{3}{\zeta^4};\\
q^{(1)}\sim \zeta +2\zeta^3R(\zeta)-3\zeta R(\zeta) &=& i\sigma\sqrt{\pi}e^{-\zeta^2}(2\zeta^4-3\zeta^2)-\frac{3}{2\zeta^3}-\frac{15}{2\zeta^5} ;\\
r^{(1)}\sim  3+2\zeta^2+4\zeta^4R(\zeta) &=& 4i\sigma\sqrt{\pi}\zeta^5 e^{-\zeta^2} -\frac{15}{2\zeta^2}-\frac{105}{4\zeta^4};\\
\widetilde{r}^{(1)}\sim  2\zeta^2 + 4\zeta^4 R(\zeta) +3R(\zeta)-3 -12\zeta^2 R(\zeta) &=& i\sigma\sqrt{\pi}e^{-\zeta^2}(4\zeta^5-12\zeta^3+3\zeta) -\frac{6}{\zeta^4}, 
\end{eqnarray}
where for brevity we suppressed the proportionality constants, including the $\sign(\kpar)$. Interestingly, the velocity $u^{(1)}$ decreases the slowest,
only as $1/\zeta$. The $n^{(1)}$, $p^{(1)}$, $r^{(1)}$ and also the temperature $T^{(1)}$, decrease as $1/\zeta^2$. The heat flux $q^{(1)}$ decreases as $1/\zeta^3$ and
the cumulant $\widetilde{r}^{(1)}$ decreases the fastest, as $1/\zeta^4$. This is not good news, since it is obvious that the direct closures that were easily obtained
for the small $\zeta$ case, can not be easily done here.   

To understand how the terms contribute to the real frequency and damping, it is useful to separate $\zeta=x+iy$ and calculate expressions with $y$ being small,
i.e. the weak growth rate (actually weak damping) approximation. The exponential term entering (\ref{eq:RzetaL}) can be approximated as
\begin{eqnarray}
  \zeta^2 &=& (x+iy)^2 = (x^2-y^2) + 2ixy \approx x^2 + 2ixy;\nn\\
  e^{-\zeta^2} &\approx& e^{-x^2} e^{-2ixy};\nn\\
  i\zeta e^{-\zeta^2} &=& i(x+iy) e^{-(x+iy)^2} \approx (-y+ix) e^{-x^2} e^{-2ixy}, \label{eq:expTerm}
\end{eqnarray}  
and the fractions of $\zeta$ are approximately
\begin{eqnarray}
  \frac{1}{\zeta} &=& \frac{1}{x+iy} = \frac{1}{x(1+i\frac{y}{x})} \approx \frac{1}{x} (1-i\frac{y}{x}) = \frac{1}{x}-i\frac{y}{x^2}; \\
  \frac{1}{\zeta^2} &=& \frac{1}{x^2(1+i\frac{y}{x})^2} \approx \frac{1}{x^2} (1-2i\frac{y}{x}) = \frac{1}{x^2}-2i\frac{y}{x^3}; \\
  \frac{1}{\zeta^3} &\approx& \frac{1}{x^3}-3i\frac{y}{x^4};\\
  \frac{1}{\zeta^4} &\approx& \frac{1}{x^4}-4i\frac{y}{x^5},
\end{eqnarray}
etc. For large $x$, the exponential term (\ref{eq:expTerm}) is strongly suppressed as $e^{-x^2}$ (with oscillations $e^{i2xy}$). Additionally, the real part of (\ref{eq:expTerm})
is proportional to $y$, which is also small, and its contribution to the real part of $R(\zeta)$ can be therefore completely neglected.
The imaginary part of the exponential term (\ref{eq:expTerm}) has to be kept, if one wants to match the approximate kinetic dispersion relations from plasma books
(usually calculated in the weak growth rate/damping approximation), for example for the damping of the Langmuir mode or the ion-acoustic mode.
However, even smart plasma physics books have trouble to analytically reproduce
the full kinetic dispersion relations that have to be solved numerically, see for example figures in Gurnett \& Bhattacharjee on pages 341 \& 355, that compare
the analytic and full solutions for the Langmuir mode and the ion-acoustic mode. The trouble is that the damping can become large, and the entire approach with the
weak damping invalid. If kinetic plasma books have trouble to analytically reproduce the damping with full accuracy under these conditions,
we would be naive to think that we can do better with a fluid model and we know we cannot be analytically exact for $|\zeta|\gg 1$ if the damping is too large.
If the damping is way-too large, and the imaginary frequency starts to be comparable to real frequency, the mode will be damped away very quickly.

In fact, even the well known kinetic solver WHAMP, neglects this term in calculation of $Z(\zeta)$ for large $\zeta$ values,
as can be verified in the WHAMP full manual \citep{Ronnmark1982} from the asymptotic expansion of $Z(\zeta)$, eq. III-6 on page 10, and the discussion of numerical errors
on page 13. The WHAMP solver uses an 8-pole Pad\'e approximant of $Z(\zeta)$, which is a very precise approximant, and imprecision starts to show up only if the damping
become too large. For example in the very damped regime when the $Im(\zeta)=-Re(\zeta)/2$, the error in real and imaginary values of $Z(\zeta)$ is still less than 2-3 \%, where the
calculation should be stopped (in less damped regime, the precision is much higher). 

If a full kinetic solver can neglect the exponential term for large $\zeta$ values, we can surely neglect it as well. It should be emphasized that the term is neglected only for large
$\zeta$ values (i.e. in the asymptotic expansion), the exponential term is otherwise fully retained and enters the Pad\'e approximation through the power series expansion for small $\zeta$. 
To summarize, the ``ideal'' large $\zeta$ asymptotic behavior that we would like to obtain reads 
\begin{eqnarray}
\frac{n^{(1)}_r}{n_{0r}} &=& -\frac{q_r}{T^{(0)}_r} \Phi \Big[ - \frac{1}{2\zeta^2}-\frac{3}{4\zeta^4} -\cdots     \Big];\label{eq:NlargeZZ}\\
u^{(1)}_r &=& - \frac{q_r}{T^{(0)}_r} \Phi v_{\textrm{th} r} \sign(\kpar) \Big[ - \frac{1}{2\zeta}-\frac{3}{4\zeta^3} -\cdots \Big];\\
\frac{p^{(1)}_r}{p_{0r}} &=& -\frac{q_r}{T^{(0)}_r} \Phi \Big[ -\frac{3}{2\zeta^2}-\frac{15}{4\zeta^4}-\cdots \Big];\\
\frac{T^{(1)}_r}{T^{(0)}_r} &=& -\frac{q_r}{T^{(0)}_r} \Phi \Big[- \frac{1}{\zeta^2}-\frac{3}{\zeta^4}-\cdots \Big];\\
q^{(1)}_r &=& -q_r n_{0r} \Phi v_{\textrm{th} r} \sign(\kpar) \Big[ -\frac{3}{2\zeta^3}-\frac{15}{2\zeta^5}-\cdots \Big];\label{eq:QlargeZZ}\\
r^{(1)}_r &=& -\frac{q_r p_{0r}}{m_r} \Phi \Big[ -\frac{15}{2\zeta^2}-\frac{105}{4\zeta^4}-\cdots \Big];\\
\widetilde{r}^{(1)}_r &=& -\frac{q_r p_{0r}}{m_r} \Phi \Big[ -\frac{6}{\zeta^4} -\cdots \Big].\label{eq:rrlargeZZ}
\end{eqnarray}

\newpage
\subsection{A brief introduction to Pad\'e approximants} \label{sec:PadeSec}
Pad\'e approximants, i.e. Pad\'e series approximation/expansion, is a very powerful mathematical technique, comparable to the
usual Taylor series and the Laurent series. Nevertheless, for some unknown reason, Pad\'e series seems to somehow
disappear from the modern educational system that a typical physicist encounter. The lack of Pad\'e series in classes
is even more surprising, if one realizes that the technique is in fact very simple, and anybody can fully grasp it in very short time. 
We therefore make a quick introduction to the technique here.

Pad\'e series consist of approximating a function as a ratio of two polynomials. If a power series (e.g. Taylor series) of a function
f(x) is known around some point with coefficients $c_n$, the goal is to express it as a ratio of two polynomials
\begin{equation}
c_0+c_1 x + c_2 x^2 +c_3 x^3+c_4 x^4+\cdots = \frac{a_0 + a_1 x + a_2 x^2 +\cdots}{1+b_1 x + b_2 x^2 +\cdots}. \label{eq:PICA}
\end{equation}
The choice of $b_0=1$ is an ad-hoc choice and the entire decomposition can be done without it, leading to the same results at the end. 
Multiplying the left hand side by the denominator $1+b_1x + b_2 x^2+\cdots$, and grouping the $x^n$ contributions together, that must be satisfied independently,
leads to the system of equations
\begin{eqnarray}
  a_0 &=& c_0;\nn\\
  a_1 &=& c_1+c_0 b_1;\nn\\
  a_2 &=& c_2+c_1 b_1 +c_0 b_2;\nn\\
  a_3 &=& c_3+c_2 b_1 +c_1 b_2 + c_0 b_3;\nn\\
  a_4 &=& c_4+c_3 b_1 +c_2 b_2 + c_1 b_3 + c_0 b_4;\nn\\
  a_5 &=& c_5+c_4 b_1 +c_3 b_2 + c_2 b_3 + c_1 b_4 + c_0 b_5;\nn\\
  a_6 &=& c_6+c_5 b_1 +c_4 b_2 + c_3 b_3 + c_2 b_4 + c_1 b_5 + c_0 b_6,
\end{eqnarray}   
etc. The necessary condition for the system being solvable, is that the number of variables is equivalent to the number of equations. Therefore,
if we want to approximate function f(x) with a ratio of two polynomials $P_m/Q_n$, of degrees $m$ and $n$, we will need the Taylor series
on the left hand side of (\ref{eq:PICA}) up to the order $m+n$. The Pad\'e approximation is sometimes denoted as $R_{m,n}$ or using a function
$f(x)$ that is being approximated as $f(x)_{m,n}$ or $[f(x)]_{m,n}$. If the Pad\'e approximation exists, it is unique. 

For example, the function $e^x$ has a Taylor series around the point $x=0$
\begin{equation}
e^x = 1+x+\frac{x^2}{2!}+\frac{x^3}{3!}+\cdots.
\end{equation}
Let's say we want to approximate $e^x$ as a ratio of two polynomials of 0-th and 1-st order $e^{x} \approx a_0/(1+b_1 x)$, i.e. we want to find
the Pad\'e approximant $[e^x]_{0,1}$. Respecting the $n+m$ rule, the approximation therefore consist of equating
\begin{equation}
\underbrace{c_0}_{=1} + \underbrace{c_1}_{=1} x = \frac{a_0}{1+b_1 x},
\end{equation}
that leads to the system of equations
\begin{eqnarray}
  a_0 &=& c_0 =1;\nn\\
  a_1 = 0 &=& c_1+ b_1 \quad => \quad b_1 = -c_1 = -1, 
\end{eqnarray}
yielding the Pad\'e approximation
\begin{equation}
\big[e^{x}\big]_{0,1} = \frac{1}{1-x}.
\end{equation}
To feel confident with the Pad\'e approximations, let's find another approximant of $e^x$, for example $[e^x]_{1,2}$. The system is written as
\begin{equation}
  \underbrace{1}_{=c_0}+\underbrace{1}_{=c_1}x+\underbrace{\frac{1}{2}}_{=c_2}x^2+\underbrace{\frac{1}{6}}_{=c_3}x^3
  = \frac{a_0+a_1 x}{1+b_1 x+b_2 x^2},
\end{equation}
and yields a system of equations
\begin{eqnarray}
  a_0 &=& 1;\nn\\
  a_1 &=& 1+b_1;\nn\\
  a_2= 0   &=& \frac{1}{2}+b_1 +b_2;\nn\\
  a_3= 0   &=& \frac{1}{6}+\frac{1}{2}b_1 +b_2 +0,
\end{eqnarray}  
which have a solution $b_1=-2/3$, $b_2=1/6$, $a_1=1/3$, and the Pad\'e approximant
\begin{equation}
\big[e^{x}\big]_{1,2} = \frac{1+\frac{1}{3}x}{1-\frac{2}{3}x+\frac{1}{6}x^2} = \frac{6+2x}{6-4x+x^2}.
\end{equation}
It is just a straightforward algebraic exercise to find other Pad\'e approximations, for example
\begin{eqnarray}
  \big[e^{x}\big]_{1,1} &=& \frac{1+\frac{1}{2}x}{1-\frac{1}{2}x} = \frac{2+x}{2-x};\nn\\
  \big[e^{x}\big]_{2,1} &=& \frac{1+\frac{2}{3}x+\frac{1}{6}x^2}{1-\frac{1}{3}x} = \frac{6+4x+x^2}{6-2x};\nn\\
  \big[e^{x}\big]_{3,1} &=& \frac{1+\frac{3}{4}x + \frac{1}{4}x^2 +\frac{1}{24}x^3}{1-\frac{1}{4}x} = \frac{24+18x+6x^2+x^3}{24-6x},
\end{eqnarray}
etc. Similarly, it is easy to find Pad\'e approximations to a function $e^{-x}$, and for example (obviously) 
\begin{equation}
  \big[e^{-x}\big]_{1,2} = \frac{6-2x}{6+4x+x^2}; \qquad \big[e^{-x}\big]_{1,1} = \frac{2-x}{2+x}; \qquad
  \big[e^{-x}\big]_{2,1} = \frac{6-4x+x^2}{6+2x}. \label{eq:Pemx}
\end{equation}  
The approximations were derived from Taylor expansion of $e^{-x}$ around $x=0$, and all 3 choices naturally have the correct limit
$\lim_{x\to 0} e^{-x}=1$. However, we can see that by choosing the degree of the Pad\'e approximation, we can also control what the
function is doing for large values of $x$. For example, for large values of $x$ the Pad\'e approximations (\ref{eq:Pemx}) go to
$0$, $-1$ and $+\infty$. Obviously, the smart choice is $[e^{-x}]_{1,2}$ which approximately reproduces the behavior of $e^{-x}$ also for large $x$.
The usefulness of Pad\'e approximation becomes especially apparent when considering analytically difficult functions, for example the $e^{-x^2}$,
where the ``smart'' lowest Pad\'e approximants are
\begin{equation}
  \big[e^{-x^2}\big]_{0,2} = \frac{1}{1+x^2};\qquad  \big[e^{-x^2}\big]_{0,4} = \frac{1}{1+x^2+\frac{1}{2}x^4};
  \qquad \big[e^{-x^2}\big]_{2,4} = \frac{6-2x^2}{6+4x^2+x^4}. 
\end{equation}
Therefore, depending on the required precision of a physical problem, instead of working with $e^{-x^2}$ (which for example does not have an indefinite
integral that can be expressed in elementary functions), one can approximate the function $e^{-x^2}$ \emph{for all} x, as $1/(1+x^2)$,
that is much easier to work with. Curiously, the reader might recognize that the $1/(1+x^2)$ is the Cauchy distribution function, often used in
plasma physics books to get better understanding of the complicated Landau damping. The Cauchy distribution therefore can be thought of as the
simplest Pad\'e approximation of the Maxwellian distribution. 

Now we are ready to use the Pad\'e approximation for the plasma dispersion function $Z(\zeta)$ or the plasma response function $R(\zeta)$.
We do not have to explore all the possibilities, and we can immediately pick up only the smart choices.
For large $\zeta$ (by neglecting the exponential term as discussed in the previous section),
at the first order $Z(\zeta)\sim 1/\zeta$ and $R(\zeta) \sim 1/\zeta^2$, and both functions approach zero as $\zeta$ increases.
Obviously, a smart choice worth exploring will always be a Pad\'e approximant $[\;\;]_{m,n}$ where $n>m$. In fact, we can be even more specific.
We know the asymptotic behavior for large $\zeta$, and obviously, even smarter choice is to concentrate only on approximants
$[Z(\zeta)]_{n-1,n}$ and $[R(\zeta)]_{n-2,n}$, since such a choice will naturally lead to the correct asymptotic behavior
\begin{equation}
\zeta\gg 1:\qquad \quad \big[Z(\zeta)\big]_{n-1,n}\sim \frac{1}{\zeta}; \qquad \quad \big[R(\zeta)\big]_{n-2,n} \sim \frac{1}{\zeta^2}.
\end{equation}  
Any other choice is not really interesting and therefore, the usual 2-digit notation of the Pad\'e approximation becomes redundant.
We can just use 1-digit notation with ``n'', that represents the degree of a chosen polynomial in the denominator,
and we can omit writing the $(n-1)$ and $(n-2)$, since this will always be the case (except for the $R_1(\zeta)$). The ``n'' represents the number of poles, and
we therefore talk about an ``n-pole Pad\'e approximation'' of $Z(\zeta)$ or $R(\zeta)$, and
\begin{equation} \label{eq:Rittt}
  Z_n(\zeta)= \frac{a_0+a_1 \zeta + \cdots +a_{n-1}\zeta^{n-1}}{1+b_1\zeta+\cdots +b_n \zeta^n};
  \qquad \quad R_n(\zeta) = \frac{a_0+a_1 \zeta + \cdots +a_{n-2}\zeta^{n-2}}{1+b_1\zeta+\cdots +b_n \zeta^n}.
\end{equation}
Note that one can directly work with Pad\'e approximants for both $Z_n(\zeta)$ and $R_n(\zeta)$, and that in general according to definitions (\ref{eq:Rittt}),
the approximants are not automatically equivalent. The difference is as if one does approximations to a function $f(x)$ or its derivative $f'(x)$.
Usually in papers, the approximant $Z_n(\zeta)$ is calculated, and $R_n(\zeta)$ is just defined according to
$R_n(\zeta)= 1+\zeta Z_n(\zeta)$. One can choose another (and better approach in our opinion) and to calculate directly approximants $R_n(\zeta)$,
and if really required (which should not be the case), obtain $Z_n(\zeta)$ approximants as $Z_n(\zeta)=(R_n(\zeta)-1)/\zeta$.

Moreover, we can do even better than (\ref{eq:Rittt}). We shall not be satisfied just by approximating the asymptotic trend
$\sim 1/\zeta$ for $\zeta\gg 1$, and hope for the best. For large $\zeta$, the correct asymptotic expansions are $Z(\zeta)\to -1/\zeta$
and $R(\zeta)\to -1/(2\zeta^2)$. By prescribing $a_{n-1}/b_n=-1$ for $Z(\zeta)$, and $a_{n-2}/b_n=-1/2$ for $R(\zeta)$,
we will obtain correct asymptotic behavior of these functions, at least at the first order. By doing this, we are not ``destroying'' the
Pad\'e approximation, since it is easy to argue that if an n-pole approximation is determined to be sufficient for small $\zeta$ values,
we can just add one more pole and use that one to control the asymptotic behavior for large $\zeta$ values. 
Of course, we will always use at least the first
term in the expansion for $\zeta\ll 1$, that yields $a_0=i\sqrt{\pi}$ for $Z_n(\zeta)$ and $a_0=1$ for $R_n(\zeta)$, otherwise the functions will have incorrect
values at $\zeta=0$. The ``smart'' choices worth considering therefore can be summarized as
\begin{equation}
  Z_{n}(\zeta)= \frac{i\sqrt{\pi}+a_1 \zeta + \cdots +a_{n-1}\zeta^{n-1}}{1+b_1\zeta+\cdots+b_{n-1}\zeta^{n-1} -a_{n-1} \zeta^n};
  \qquad \quad R_{n}(\zeta) = \frac{1+a_1 \zeta + \cdots +a_{n-2}\zeta^{n-2}}{1+b_1\zeta+\cdots+b_{n-1}\zeta^{n-1} -2a_{n-2} \zeta^n}, \label{eq:ZR}
\end{equation}
and have a property to correctly match the $Z$ and $R$ functions at $\zeta=0$ and, have the correct first order asymptotic expansion at $\zeta\gg 1$.   
The 1-pole approximant $R_1(\zeta)$ is an exception, and can be defined only as $R_1(\zeta)=1/(1+b_1\zeta)$.
This function obviously cannot have correct asymptotic expansion $\sim 1/\zeta^2$ and the only possibility is
to use $\zeta\ll 1$ expansion $R_1(\zeta)=1/(1+b_1\zeta) = 1+i\sqrt{\pi}\zeta$, which yields $b_1=-i\sqrt{\pi}$ and
\begin{equation}
R_1(\zeta) = \frac{1}{1-i\sqrt{\pi}\zeta}.
\end{equation}  
The 1-pole approximant $Z_1(\zeta)$ can be obtained directly from the definition (\ref{eq:ZR}), that yields
\begin{equation}
Z_1(\zeta) = \frac{i\sqrt{\pi}}{1-i\sqrt{\pi}\zeta},
\end{equation}
and that has correct asymptotic behavior $Z_1(\zeta) \to-1/\zeta$ for large $\zeta$ values, even though it has only precision $o(\zeta^0)$ for small
$\zeta$ values. Perhaps curiously, in this case $R_1(\zeta)=1+\zeta Z_1(\zeta)$ exactly. Alternatively, if the precision for small $\zeta$ is more important than the
exact asymptotic expansion for large $\zeta$, it is possible to increase the $Z_1$ precision to $o(\zeta^1)$ and write $Z_1(\zeta)=i\sqrt{\pi}/(1-2i\zeta/\sqrt{\pi})$.
In this case $R_1(\zeta)\neq 1+\zeta Z_1(\zeta)$ exactly, and the functions are equal only for small $\zeta$ and only with precision $o(\zeta^1)$.

Right now, in the definition (\ref{eq:ZR}), we just used 1 pole for the asymptotic series of $Z(\zeta)$ and $R(\zeta)$, but one can naturally use more poles.
By opening the possibility to increase the number of matching asymptotic points in the n-pole Pad\'e approximation (\ref{eq:ZR}), the number of possible approximants 
for a given n naturally increases. To keep track of all the possibilities, we obviously need some kind of classification scheme. It is useful
to modify the usual 1-index Pad\'e series notation for $Z_n(\zeta)$ and $R_n(\zeta)$ functions (that only specify the number of poles), to a two index notation
$ Z_{n,n'}(\zeta)$, $R_{n,n'}(\zeta)$. Now we have a wide range of possibilities how to define $n,n'$ and there is no clear ``natural'' winner. 

There are two different existing notations (likely more), introduced by \cite{Martin1980} and by \cite{HedrickLeboeuf1992}, that consider $Z_{n,n'}(\zeta)$ Pad\'e approximants. 
The first reference defines $n=$ number of points (equations) used in the power series expansion, and $n'=$ number of points (equations) used in the asymptotic series expansion. 
Even though perhaps clear, for example 3-pole approximants in this notation are expressed as $Z_{5,1}, Z_{4,2}, Z_{3,3}$ etc, so to get the number of poles
(which is the most important information), one has to calculate $(n+n')/2$. When using a lot of different approximants, this notation is a bit confusing and is rarely used.   

The notation of \cite{HedrickLeboeuf1992} can be interpreted as defining $Z_{n,n'}$ with $n=$ number of poles (which we like),
and $n'=$ number of \emph{additional} poles in the asymptotic expansion that is used, compared to some ``minimally interesting'' or ``basic'' definition $Z_n$, that
can be denoted as $Z_{n,0}$ (which we like too). The problem with the notation of \cite{HedrickLeboeuf1992} is  with the definition of the ``basic'' $Z_{n,0}$,
since the number of asymptotic points used in the definition of $Z_{n,0}$ keeps changing with $n$ (and is actually equal to $n$).
The notation is physically motivated, but the motivation is difficult to follow.
The $Z_{2,0}$ is defined with 2 asymptotic points, $Z_{3,0}$ with 3 asymptotic points and so on.
This can be easily deduced from their definitions of $Z_2-Z_5$, as we will discuss later. We find this notation confusing.

Importantly, both mentioned notations consider the Pad\'e approximants to $Z(\zeta)$.
We do not really care about $Z(\zeta)$, since all the kinetic moments are formulated with $R(\zeta)$ at this stage.
We want to calculate direct Pad\'e approximants to $R(\zeta)$, which is actually slightly less analytically complicated for a given $n$. 
Here we define the 2-index Pad\'e approximation to the plasma response function $R(\zeta)$ simply as 
\begin{equation}
  \boxed{
 R_{n,0}(\zeta) = \frac{1+a_1 \zeta + a_2\zeta^2+\cdots +a_{n-2}\zeta^{n-2} }{ 1+b_1\zeta+b_2\zeta^2 + \cdots+b_{n-1}\zeta^{n-1} -2a_{n-2} \zeta^n }, } \label{eq:Rn0}
\end{equation}
i.e. as having asymptote $-1/(2\zeta^2)$ for large $\zeta$, and
notation $R_{n,n'}(\zeta)$ means that $n'$ \emph{additional} asymptotic points are used compared to the basic definition $R_{n,0}(\zeta)$.
The notation feels natural, and the $n'=0$ index helps us to orient in the hierarchy of many possible $R(\zeta)$ approximants. It is easy to remember
that this asymptotic profile is the minimum ``desired'' profile that correctly captures the 0-th order (density) moment,
and any profile with less asymptotic points should be avoided if possible.
The $R_{n,0}(\zeta)$ has power series precision $o(\zeta^{2n-3})$ and asymptotic series precision $o(\zeta^{-2})$, so $R_{n,n'}(\zeta)$ has precision
$o(\zeta^{2n-3-n'})$ and $o(\zeta^{-2-n'})$.

Of course, we want to make the $R_{n,n'}(\zeta)$ and $Z_{n,n'}(\zeta)$ definitions fully consistent, and $Z_{n,n'}(\zeta)$ is defined so that 
\begin{equation} \label{eq:rigid}
R_{n,n'}(\zeta)=1+\zeta Z_{n,n'}(\zeta),
\end{equation}  
is satisfied. This dictates that in comparison to $Z_n(\zeta)$ definition (\ref{eq:ZR}), two additional asymptotic points must be used to define the $Z_{n,0}(\zeta)$.
We have no other choice and when calculating the $Z_{n,n'}(\zeta)$, we have to start counting from $n'=-2$, and we define
\begin{equation}
  Z_{n,-2}(\zeta)= \frac{i\sqrt{\pi}+a_1 \zeta + \cdots +a_{n-1}\zeta^{n-1}}{1+b_1\zeta+\cdots+b_{n-1}\zeta^{n-1} -a_{n-1} \zeta^n}.
\end{equation}
When calculating the hierarchy of plasma dispersion functions $Z(\zeta)$, the $-2$ index is actually a nice reminder that we are two asymptotic points short
of the ``desired'' profile (\ref{eq:Rn0}) for the plasma response function $R(\zeta)$.
We want to feel fully confident that we understand both Pad\'e approximants $R(\zeta)$ and $Z(\zeta)$, and we will calculate 2-pole and 3-pole approximants for both functions.
For 4-pole approximants and above, we will only work with $R(\zeta)$.

Pad\'e approximants were also used for other interesting physical problems, such as developing analytic models for the Rayleigh-Taylor
and Richtmyer-Meshkov instability \cite{Zhou2017-1,Zhou2017-2}.

\subsubsection{2-pole approximants of $R(\zeta)$ and $Z(\zeta)$} \label{section:2poleC}
Let's be patient and go slowly. A general 2-pole Pad\'e approximant to $R(\zeta)$ is
\begin{equation}
R_2 (\zeta) = \frac{a_0}{1+b_1\zeta+b_2\zeta^2},
\end{equation}
where $a_0=1$. The asymptotic expansion for large $\zeta$ values calculates
\begin{eqnarray}
  \frac{a_0}{1+b_1\zeta+b_2\zeta^2} &=& \frac{a_0}{b_2 \zeta^2 \Big( \frac{1}{b_2\zeta^2} + \frac{b_1}{b_2\zeta} +1\Big)}
  = \frac {a_0}{b_2 \zeta^2} \Big[ 1-\Big(  \frac{b_1}{b_2\zeta}+\frac{1}{b_2\zeta^2}\Big) +\Big( \frac{b_1}{b_2\zeta}+\frac{1}{b_2\zeta^2}\Big)^2   +\cdots\Big]\nn\\
  &=& \frac{a_0}{b_2 \zeta^2} -a_0\frac{b_1}{b_2^2 \zeta^3} + a_0 \frac{b_1^2-b_2}{b_2^3 \zeta^4}+\cdots; \qquad \zeta\gg 1,
\end{eqnarray}
and must be matched with the asymptotic expansion (\ref{eq:RzetaL})
\begin{equation}
R(\zeta) = -\frac{1}{2\zeta^2}-\frac{0}{\zeta^3}-\frac{3}{4\zeta^4}+\cdots; \qquad \zeta\gg 1.
\end{equation}
Matching the first point implies $b_2=-2a_0$, and this is how $R_{2,0}(\zeta)$ is defined. Then matching with 2 equation for the small $\zeta$ expansion,
eq. (\ref{eq:RzetaSmall1}), the classical Pad\'e approach yields
\begin{equation}
  R_{2,0}(\zeta)=\frac{a_0}{1+b_1\zeta-2a_0\zeta^2} = \underbrace{1}_{=c_0}+\underbrace{i\sqrt{\pi}}_{=c_1}\zeta; \quad => \quad R_{2,0}(\zeta)=\frac{1}{1-i\sqrt{\pi}\zeta-2\zeta^2}.
\end{equation}
To match additional asymptotic point (and to potentially find $R_{2,1}(\zeta)$), dictates that $b_1=0$. However, the resulting function
$R_{2,1}(\zeta)=1/(1-2\zeta^2)$ does not have any imaginary part for real valued $\zeta$,  since it uses too many asymptotic points and
the Landau residue is not accounted for. Therefore, the $R_{2,1}(\zeta)$
does not represent a valuable approximation of $R(\zeta)$, and this approximant is eliminated. 

Let's now explore possible 2-pole approximations of $Z(\zeta)$. A general 2-pole approximant is defined as
\begin{eqnarray}
  Z_{2}(\zeta) = \frac{a_0+a_1 \zeta}{1+b_1\zeta+b_2\zeta^2},
\end{eqnarray}
and has the following asymptotic expansion for large $\zeta$ values 
\begin{eqnarray}
  \frac{a_0+a_1 \zeta}{1+b_1\zeta+b_2\zeta^2} &=& \Big(\frac{a_1}{b_2}\Big) \frac{1}{\zeta} +\Big( \frac{a_0}{b_2}-\frac{a_1b_1}{b_2^2}\Big)\frac{1}{\zeta^2}
  +\Big( -\frac{a_0b_1}{b_2^2} + \frac{a_1(b_1^2-b_2)}{b_2^3} \Big) \frac{1}{\zeta^3} +\cdots; \qquad \zeta\gg 1.
\end{eqnarray}
The $Z(\zeta)$ has asymptotic expansion
\begin{equation}
Z(\zeta) = -\frac{1}{\zeta}-\frac{0}{\zeta^2}-\frac{1}{2\zeta^3}+\cdots;\qquad \zeta\gg 1,
\end{equation}
so by matching with $1/\zeta$ implies $b_2=-a_1$ (as already used previously) that defines $Z_{2,-2}$ (remember, we are starting to count with $n'=-2$).
By further matching with $1/\zeta^2$ implies $b_1=-a_0$,
that defines $Z_{2,-1}$, and by further matching with $1/\zeta^3$ implies $a_1=2$, that defines $Z_{2,0}$.

The calculation is continued by matching with the power series for small $\zeta$ values, i.e. by using the classical Pad\'e approach, that is described as
\begin{eqnarray}
  Z_{2,-2}(\zeta) &=& \frac{a_0+a_1 \zeta}{1+b_1\zeta-a_1\zeta^2} = \underbrace{i\sqrt{\pi}}_{=c_0} \underbrace{-2}_{=c_1}\zeta \underbrace{-i\sqrt{\pi}}_{=c_2}\zeta^2,
\end{eqnarray}
and the solution is
\begin{eqnarray}
  Z_{2,-2}(\zeta) &=& \frac{i\sqrt{\pi}+\frac{4-\pi}{\pi-2} \zeta}{1-\frac{i\sqrt{\pi}}{\pi-2}\zeta-\frac{4-\pi}{\pi-2}\zeta^2}.
\end{eqnarray}
Continuing with $Z_{2,-1}(\zeta)$, i.e. by using one more additional asymptotic term that dictates $b_1=-a_0$, the matching with the power series yields
\begin{equation}
  Z_{2,-1}(\zeta) = \frac{a_0+a_1\zeta}{1-a_0\zeta-a_1\zeta^2} = i\sqrt{\pi} - 2\zeta; \quad => \quad
  Z_{2,-1}(\zeta) = \frac{i\sqrt{\pi}+(\pi-2)\zeta}{1-i\sqrt{\pi}\zeta-(\pi-2)\zeta^2}.
\end{equation}
Similarly, considering $Z_{2,0}(\zeta)$ yields
\begin{equation}
  Z_{2,0}(\zeta) = \frac{a_0+2\zeta}{1-a_0\zeta-2\zeta^2} = i\sqrt{\pi}; \quad => \quad
  Z_{2,0}(\zeta) = \frac{i\sqrt{\pi}+2\zeta}{1-i\sqrt{\pi}\zeta-2\zeta^2}.
\end{equation}
Obviously, $R_{2,0}(\zeta)=1+\zeta Z_{2,0}(\zeta)$ exactly.
\subsubsection{3-pole approximants of $R(\zeta)$ and $Z(\zeta)$} \label{section:3poleC}
A general 3-pole approximant of $R(\zeta)$ is
\begin{eqnarray}
R_3(\zeta) = \frac{a_0+a_1\zeta}{1+b_1\zeta+b_2\zeta^2+b_3\zeta^3}.
\end{eqnarray}  
The asymptotic expansion calculates
\begin{eqnarray}
  \frac{1}{1+b_1\zeta+b_2\zeta^2+b_3\zeta^3} &=& \frac{1}{b_3 \zeta^3 \Big( \frac{1}{b_3\zeta^3} + \frac{b_1}{b_3\zeta^2}+\frac{b_2}{b_3\zeta} +1\Big)}
  = \frac {1}{b_3 \zeta^3} \Big[ 1-\Big(  \frac{b_2}{b_3\zeta}+\frac{b_1}{b_3\zeta^2} + \frac{1}{b_3\zeta^3}\Big)
    +\Big(  \frac{b_2}{b_3\zeta}+\frac{b_1}{b_3\zeta^2} + \frac{1}{b_3\zeta^3}\Big)^2   +\cdots\Big]\nn\\
  &=& \frac {1}{b_3 \zeta^3} \Big[ 1- \frac{1}{\zeta}\frac{b_2}{b_3}+\frac{1}{\zeta^2}\Big(-\frac{b_1}{b_3}+\frac{b_2^2}{b_3^2}\Big)+\cdots \Big]
  = \frac{1}{b_3\zeta^3} -\frac{b_2}{b_3^2\zeta^4}+\frac{b_2^2-b_1 b_3}{b_3^3\zeta^5}+\cdots,
\end{eqnarray}
so that
\begin{eqnarray}
  \frac{a_0+a_1\zeta}{1+b_1\zeta+b_2\zeta^2+b_3\zeta^3} &=& \frac{a_1}{b_3\zeta^2}+\frac{1}{\zeta^3}\Big( \frac{a_0}{b_3}-\frac{a_1 b_2}{b_3^2}\Big)
  +\frac{1}{\zeta^4} \Big( -\frac{a_0 b_2}{b_3^2}+a_1\frac{b_2^2-b_1 b_3}{b_3^3} \Big) +\cdots. \label{eq:R3Padee}
\end{eqnarray}
For $R_{3,0}(\zeta)$ this implies $b_3=-2a_1$, for $R_{3,1}(\zeta)$ additionally $b_2=-2a_0$, and for $R_{3,2}(\zeta)$ also $b_1=3a_1$.
The asymptotic expansions (\ref{eq:R3Padee}) can become very long for higher orders of $\zeta$, especially when more poles are considered. It is beneficial to write
down the following scheme, where in each line, we advance the matching with one more asymptotic point:
\begin{eqnarray}
  \frac{a_0+a_1\zeta}{1+b_1\zeta+b_2\zeta^2+b_3\zeta^3} &=& \underbrace{\frac{a_1}{b_3}}_{=-1/2}\frac{1}{\zeta^2}+\cdots \quad => \quad b_3=-2a_1;\\
R_{3,0}(\zeta)= \frac{a_0+a_1\zeta}{1+b_1\zeta+b_2\zeta^2-2a_1\zeta^3} &=& -\frac{1}{2\zeta^2} -\underbrace{\frac{a_0+\frac{b_2}{2}}{2a_1}}_{=0} \frac{1}{\zeta^3}+\cdots
  \quad => \quad b_2=-2a_0;\\
R_{3,1}(\zeta)=  \frac{a_0+a_1\zeta}{1+b_1\zeta-2a_0\zeta^2-2a_1 \zeta^3} &=& -\frac{1}{2\zeta^2} -\underbrace{\frac{b_1}{4a_1}}_{=3/4}\frac{1}{\zeta^4}+\cdots
  \quad => \quad b_1=3a_1;\\
R_{3,2}(\zeta)=  \frac{a_0+a_1\zeta}{1+3a_1\zeta-2a_0\zeta^2-2a_1 \zeta^3} &=& -\frac{1}{2\zeta^2} -\frac{3}{4\zeta^4} -\underbrace{\frac{1-3a_0}{4a_1}}_{=0}\frac{1}{\zeta^5}+\cdots
  \quad => \quad a_1\to\infty.
\end{eqnarray}  
In the last expression the $a_1\to\infty$ since $a_0=1$, implying the $R_{3,3}(\zeta)$ does not make sense and it is not defined. The scheme can be very quickly verified
by using Maple (or Mathematica) software, by using command $asympt(expression(\zeta),\zeta,n)$, where $\zeta$ is the variable, and
$n$ prescribes the precision of the expansion that is calculated up to the $o(\zeta^{-n})$ order.
Now by matching with the power series for small $\zeta$ values
\begin{eqnarray}
  R_{3,0}(\zeta) &=& \frac{a_0+a_1\zeta}{1+b_1\zeta+b_2\zeta^2-2a_1 \zeta^3} = 1 + i\sqrt{\pi}\zeta -2\zeta^2 -i\sqrt{\pi}\zeta^3;\\
  R_{3,1}(\zeta) &=& \frac{a_0+a_1\zeta}{1+b_1\zeta-2a_0\zeta^2-2a_1 \zeta^3} = 1 + i\sqrt{\pi}\zeta -2\zeta^2;\\
  R_{3,2}(\zeta) &=& \frac{a_0+a_1\zeta}{1+3a_1\zeta-2a_0\zeta^2-2a_1 \zeta^3} = 1 + i\sqrt{\pi}\zeta;
\end{eqnarray}  
and the solutions are
\begin{eqnarray}
  R_{3,0}(\zeta) &=& \frac{1-i\sqrt{\pi}\frac{\pi-3}{4-\pi}\zeta}{1-i\frac{\sqrt{\pi}}{4-\pi}\zeta -\frac{3\pi-8}{4-\pi}\zeta^2 +2i\sqrt{\pi}\frac{\pi-3}{4-\pi}\zeta^3 };\\
  R_{3,1}(\zeta) &=& \frac{1-i\frac{4-\pi}{\sqrt{\pi}}\zeta}{1-\frac{4i}{\sqrt{\pi}}\zeta-2\zeta^2+2i\frac{4-\pi}{\sqrt{\pi}}\zeta^3};\\
  R_{3,2}(\zeta) &=& \frac{1-\frac{i\sqrt{\pi}}{2}\zeta}{1-\frac{3i\sqrt{\pi}}{2}\zeta-2\zeta^2+i\sqrt{\pi} \zeta^3}.
\end{eqnarray}

A general 3-pole approximant of $Z(\zeta)$ is 
\begin{equation}
Z_3(\zeta) = \frac{a_0+a_1\zeta+a_2\zeta^2}{1+b_1\zeta+b_2\zeta^2+b_3\zeta^3},
\end{equation}
and has the following asymptotic expansion 
\begin{eqnarray}
  \frac{a_0+a_1\zeta+a_2\zeta^2}{1+b_1\zeta+b_2\zeta^2+b_3\zeta^3} &=&
  \frac{a_2}{b_3\zeta}+\Big(\frac{a_1}{b_3}-\frac{a_2 b_2}{b_3^2} \Big)\frac{1}{\zeta^2}
  +\Big(\frac{a_0}{b_3}-\frac{a_1 b_2}{b_3^2}+\frac{a_2(b_2^2-b_1 b_3)}{b_3^3}\Big)\frac{1}{\zeta^3}+\cdots. \label{eq:Z3pade}
\end{eqnarray}
By matching the first asymptotic term implies $b_3=-a_2$, which defines $Z_{3,-2}(\zeta)$. For $Z_{3,-1}(\zeta)$ the second term is matched as well and $b_2=-a_1$.
For $Z_{3,0}(\zeta)$ the third term is also matched and $b_1=-a_0+a_2/2$. To go higher requires higher order expansion (\ref{eq:Z3pade}). It is again easier to
write down the asymptotic expansion scheme step by step 
\begin{eqnarray}
  \frac{a_0+a_1\zeta+a_2\zeta^2}{1+b_1\zeta+b_2\zeta^2+b_3\zeta^3} &=& \underbrace{\frac{a_2}{b_3}}_{=-1}\frac{1}{\zeta}+\cdots; \quad => \quad b_3=-a_2;\\
Z_{3,-2}(\zeta)=  \frac{a_0+a_1\zeta+a_2\zeta^2}{1+b_1\zeta+b_2\zeta^2-a_2\zeta^3} &=& -\frac{1}{\zeta}-\underbrace{\frac{a_1+b_2}{a_2}}_{=0}\frac{1}{\zeta^2}+\cdots \quad => \quad b_2=-a_1;\\
Z_{3,-1}(\zeta)=  \frac{a_0+a_1\zeta+a_2\zeta^2}{1+b_1\zeta-a_1\zeta^2-a_2\zeta^3} &=& -\frac{1}{\zeta}-\frac{0}{\zeta^2}
  -\underbrace{\frac{a_0+b_1}{a_2}}_{=1/2}\frac{1}{\zeta^3}+\cdots \quad => \quad b_1=\frac{a_2}{2}-a_0;\\
Z_{3,0}(\zeta)=  \frac{a_0+a_1\zeta+a_2\zeta^2}{1+(\frac{a_2}{2}-a_0)\zeta-a_1\zeta^2-a_2\zeta^3} &=& -\frac{1}{\zeta}-\frac{1}{2\zeta^3}
  -\underbrace{\frac{2-a_1}{2a_2}}_{=0}\frac{1}{\zeta^4} +\cdots \quad => \quad a_1=2;\\
Z_{3,1}(\zeta)=  \frac{a_0+2\zeta+a_2\zeta^2}{1+(\frac{a_2}{2}-a_0)\zeta-2\zeta^2-a_2\zeta^3} &=& -\frac{1}{\zeta}-\frac{1}{2\zeta^3}
-\underbrace{\frac{a_2-2a_0}{4a_2}}_{=3/4}\frac{1}{\zeta^5} +\cdots \quad => \quad a_2=-a_0;\\
Z_{3,2}(\zeta) = \frac{a_0+2\zeta-a_0\zeta^2}{1-\frac{3}{2}a_0\zeta-2\zeta^2+a_0\zeta^3}.&&
\end{eqnarray}
Matching these results with an expansion for small $\zeta$ values is done according to
\begin{eqnarray}
  Z_{3,-2}(\zeta)&=&\frac{a_0+a_1\zeta+a_2\zeta^2}{1+b_1\zeta+b_2\zeta^2-a_2\zeta^3} = i\sqrt{\pi}-2\zeta-i\sqrt{\pi}\zeta^2+\frac{4}{3}\zeta^3+i\frac{\sqrt{\pi}}{2}\zeta^4;\\
  Z_{3,-1}(\zeta)&=&\frac{a_0+a_1\zeta+a_2\zeta^2}{1+b_1\zeta-a_1\zeta^2-a_2\zeta^3} = i\sqrt{\pi}-2\zeta-i\sqrt{\pi}\zeta^2+\frac{4}{3}\zeta^3;\\
  Z_{3,0}(\zeta)&=&\frac{a_0+a_1\zeta+a_2\zeta^2}{1+(-a_0+\frac{a_2}{2})\zeta-a_1\zeta^2-a_2\zeta^3}
  =i\sqrt{\pi}-2\zeta-i\sqrt{\pi}\zeta^2;\\
  Z_{3,1}(\zeta) &=& \frac{a_0+2\zeta+a_2\zeta^2}{1+(\frac{a_2}{2}-a_0)\zeta-2\zeta^2-a_2\zeta^3} = i\sqrt{\pi}-2\zeta;\\
  Z_{3,2}(\zeta) &=& \frac{a_0+2\zeta-a_0\zeta^2}{1-\frac{3}{2}a_0\zeta-2\zeta^2+a_0\zeta^3} = i\sqrt{\pi},
\end{eqnarray}
and the solutions are
\begin{eqnarray}
  Z_{3,-2}(\zeta) &=& \frac{i\sqrt{\pi}+\frac{3\pi^2-30\pi+64}{2(5\pi-16)}\zeta+\frac{i\sqrt{\pi}(9\pi-28)}{6(5\pi-16)}\zeta^2}{1-\frac{i\sqrt{\pi}(3\pi-10)}{2(5\pi-16)}\zeta
    +\frac{21\pi-64}{6(5\pi-16)}\zeta^2-\frac{i\sqrt{\pi}(9\pi-28)}{6(5\pi-16)}\zeta^3};
\end{eqnarray}
\begin{eqnarray}
  Z_{3,-1}(\zeta) &=& \frac{i\sqrt{\pi}+\frac{10-3\pi}{3(\pi-3)}\zeta+\frac{i(5\pi-16)}{3\sqrt{\pi}(\pi-3)}\zeta^2}{1-\frac{i(3\pi-8)}{3\sqrt{\pi}(\pi-3)}\zeta
    -\frac{10-3\pi}{3(\pi-3)}\zeta^2-\frac{i(5\pi-16)}{3\sqrt{\pi}(\pi-3)}\zeta^3};
\end{eqnarray}
\begin{eqnarray}
  Z_{3,0}(\zeta) &=& \frac{i\sqrt{\pi}+\frac{3\pi-8}{4-\pi}\zeta-2i\sqrt{\pi}\frac{\pi-3}{4-\pi}\zeta^2}{1-\frac{i\sqrt{\pi}}{4-\pi}\zeta-\frac{3\pi-8}{4-\pi}
    \zeta^2+2i\sqrt{\pi}\frac{\pi-3}{4-\pi}\zeta^3}.
\end{eqnarray}
\begin{eqnarray}
  Z_{3,1}(\zeta) &=& \frac{i\sqrt{\pi}+2\zeta-2i\frac{4-\pi}{\sqrt{\pi}}\zeta^2}{1-i\frac{4}{\sqrt{\pi}}\zeta-2\zeta^2+2i\frac{4-\pi}{\sqrt{\pi}}\zeta^3};\\
  Z_{3,2}(\zeta) &=& \frac{i\sqrt{\pi}+2\zeta-i\sqrt{\pi}\zeta^2}{1-\frac{3}{2}i\sqrt{\pi}\zeta-2\zeta^2+i\sqrt{\pi}\zeta^3}.
\end{eqnarray} 
Of course, the following relations now hold exactly 
\begin{eqnarray}
  R_{3,0}(\zeta) = 1+\zeta Z_{3,0}(\zeta);\\
  R_{3,1}(\zeta) = 1+\zeta Z_{3,1}(\zeta);\\
  R_{3,2}(\zeta) = 1+\zeta Z_{3,2}(\zeta).
\end{eqnarray}
\subsubsection{4-pole approximants of $R(\zeta)$ and $Z(\zeta)$} \label{section:4poleC}
As before, the procedure of matching with asymptotic expansion yields (for simplicity already assuming $a_0=1$)
\begin{eqnarray}
  \frac{1+a_1\zeta+a_2\zeta^2}{1+b_1\zeta+b_2\zeta^2+b_3\zeta^3+b_4\zeta^4} &=& \underbrace{\frac{a_2}{b_4}}_{=-1/2}\frac{1}{\zeta^2}+\cdots \quad => \quad b_4=-2a_2;\\
R_{4,0}(\zeta)=  \frac{1+a_1\zeta+a_2\zeta^2}{1+b_1\zeta+b_2\zeta^2+b_3\zeta^3-2a_2\zeta^4} &=& -\frac{1}{2\zeta^2} -\underbrace{\frac{a_1+\frac{b_3}{2}}{2a_2}}_{=0} \frac{1}{\zeta^3}+\cdots
  \quad => \quad b_3=-2a_1;\\
R_{4,1}(\zeta)=  \frac{1+a_1\zeta+a_2\zeta^2}{1+b_1\zeta+b_2\zeta^2-2a_1\zeta^3-2a_2\zeta^4} &=& -\frac{1}{2\zeta^2} -\underbrace{\frac{1+\frac{b_2}{2}}{2a_2}}_{=3/4}\frac{1}{\zeta^4}+\cdots
  \quad => \quad b_2=3a_2-2;\\
R_{4,2}(\zeta)=  \frac{1+a_1\zeta+a_2\zeta^2}{1+b_1\zeta+(3a_2-2)\zeta^2-2a_1\zeta^3-2a_2\zeta^4}&=& -\frac{1}{2\zeta^2} -\frac{3}{4\zeta^4} -\underbrace{\frac{b_1-3a_1}{4a_2}}_{=0}\frac{1}{\zeta^5}+\cdots
  \quad => \quad b_1=3a_1;\\
R_{4,3}(\zeta)=  \frac{1+a_1\zeta+a_2\zeta^2}{1+3a_1\zeta+(3a_2-2)\zeta^2-2a_1\zeta^3-2a_2\zeta^4}&=& -\frac{1}{2\zeta^2} -\frac{3}{4\zeta^4}
  -\underbrace{\frac{\frac{9}{2}a_2-2}{4a_2}}_{=15/8}\frac{1}{\zeta^6}+\cdots
  \quad => \quad a_2=-\frac{2}{3};\\
R_{4,4}(\zeta)=  \frac{1+a_1\zeta-\frac{2}{3}\zeta^2}{1+3a_1\zeta-4\zeta^2-2a_1\zeta^3+\frac{4}{3}\zeta^4}&=& -\frac{1}{2\zeta^2} -\frac{3}{4\zeta^4}-\frac{15}{8\zeta^6}
-\underbrace{\frac{9a_1}{8}}_{=0}\frac{1}{\zeta^7}\quad => \quad a_1=0,
\end{eqnarray}
where the last relation imply a possible approximant $R_{4,5}(\zeta)=(1-\frac{2}{3}\zeta^2)/(1-4\zeta^2+\frac{4}{3}\zeta^4)$. However, such an approximant
is not well behaved (it has zero imaginary part for real valued $\zeta)$ and the $R_{4,5}(\zeta)$ is eliminated.
Matching with the power series is performed according to
\begin{eqnarray}
  R_{4,0}(\zeta) &=& \frac{1+a_1\zeta+a_2\zeta^2}{1+b_1\zeta+b_2\zeta^2+b_3\zeta^3-2a_2\zeta^4} = 1 + i\sqrt{\pi}\zeta -2\zeta^2 -i\sqrt{\pi}\zeta^3
  +\frac{4}{3}\zeta^4+i\frac{\sqrt{\pi}}{2}\zeta^5;\\
  R_{4,1}(\zeta) &=& \frac{1+a_1\zeta+a_2\zeta^2}{1+b_1\zeta+b_2\zeta^2-2a_1\zeta^3-2a_2\zeta^4} = 1 + i\sqrt{\pi}\zeta -2\zeta^2 -i\sqrt{\pi}\zeta^3
  +\frac{4}{3}\zeta^4;\\
  R_{4,2}(\zeta) &=&  \frac{1+a_1\zeta+a_2\zeta^2}{1+b_1\zeta+(3a_2-2)\zeta^2-2a_1\zeta^3-2a_2\zeta^4} =  1 + i\sqrt{\pi}\zeta -2\zeta^2 -i\sqrt{\pi}\zeta^3;\\
  R_{4,3}(\zeta) &=&  \frac{1+a_1\zeta+a_2\zeta^2}{1+3a_1\zeta+(3a_2-2)\zeta^2-2a_1\zeta^3-2a_2\zeta^4}=  1 + i\sqrt{\pi}\zeta -2\zeta^2;\\
  R_{4,4}(\zeta) &=&  \frac{1+a_1\zeta-\frac{2}{3}\zeta^2}{1+3a_1\zeta-4\zeta^2-2a_1\zeta^3+\frac{4}{3}\zeta^4} = 1 + i\sqrt{\pi}\zeta,
\end{eqnarray}
and the results are
\begin{eqnarray}
  R_{4,0}(\zeta) = \frac{1+i\frac{\sqrt{\pi}}{2}\frac{(12\pi^2-67\pi+92)}{(6\pi^2-29\pi+32)}\zeta-
    \frac{(9\pi^2-69\pi+128)}{6(6\pi^2-29\pi+32)}\zeta^2}{1-i\frac{\sqrt{\pi}}{2}\frac{(9\pi-28)}{(6\pi^2-29\pi+32)}\zeta
    +\frac{(36\pi^2-195\pi+256)}{6(6\pi^2-29\pi+32)}\zeta^2
    -i\frac{\sqrt{\pi}(33\pi-104)}{6(6\pi^2-29\pi+32)}\zeta^3+ \frac{(9\pi^2-69\pi+128)}{3(6\pi^2-29\pi+32)}\zeta^4};
\end{eqnarray}
\begin{eqnarray}
  R_{4,1}(\zeta) = \frac{1-i\frac{\sqrt{\pi}}{3}\frac{(9\pi-28)}{(16-5\pi)}\zeta-\frac{(6\pi^2-29\pi+32)}{3(16-5\pi)}\zeta^2}{1
    -i\frac{2\sqrt{\pi}}{3}\frac{(10-3\pi)}{(16-5\pi)}\zeta
    -\frac{(21\pi-64)}{3(16-5\pi)}\zeta^2
    +i\frac{2\sqrt{\pi}}{3}\frac{(9\pi-28)}{(16-5\pi)}\zeta^3
    +\frac{2(6\pi^2-29\pi+32)}{3(16-5\pi)}\zeta^4};
\end{eqnarray}
\begin{eqnarray}
  R_{4,2}(\zeta) = \frac{1-i\sqrt{\pi}\frac{(10-3\pi)}{(3\pi-8)}\zeta-\frac{(16-5\pi)}{(3\pi-8)}\zeta^2}{1
    -i\sqrt{\pi}\frac{2}{(3\pi-8)}\zeta
    -\frac{(32-9\pi)}{(3\pi-8)}\zeta^2
    +i\sqrt{\pi}\frac{2(10-3\pi)}{(3\pi-8)}\zeta^3+\frac{2(16-5\pi)}{(3\pi-8)}\zeta^4}
\end{eqnarray}
\begin{eqnarray}
  R_{4,3}(\zeta) = \frac{1-i\frac{\sqrt{\pi}}{2}\zeta-\frac{(3\pi-8)}{4}\zeta^2}{1-i\frac{3\sqrt{\pi}}{2}\zeta-\frac{(9\pi-16)}{4}\zeta^2
    +i\sqrt{\pi}\zeta^3+\frac{(3\pi-8)}{2}\zeta^4};
\end{eqnarray}
\begin{eqnarray}
 R_{4,4}(\zeta) &=&  \frac{1-i\frac{\sqrt{\pi}}{2}\zeta-\frac{2}{3}\zeta^2}{1-i\frac{3\sqrt{\pi}}{2}\zeta-4\zeta^2+i\sqrt{\pi}\zeta^3+\frac{4}{3}\zeta^4}.
\end{eqnarray}
From the 4-pole approximants, perhaps the most known one is $R_{4,3}(\zeta)$ used for example by \cite{HammettPerkins1990}, \cite{PassotSulem2007} etc.,
and which can be written in a convenient form
\begin{eqnarray}
\boxed{  
  R_{4,3}(\zeta) = \frac{4-2i\sqrt{\pi}\zeta-(3\pi-8)\zeta^2}{4-6i\sqrt{\pi}\zeta-(9\pi-16)\zeta^2
    +4i\sqrt{\pi}\zeta^3+2(3\pi-8)\zeta^4}.}
\end{eqnarray}
 
Here we do not double check the derivation of the $Z_4(\zeta)$ approximants ``from scratch'', and for a given $R_4$ coefficients, the $Z_4$ coefficients are
of course easily obtained by
\begin{equation}
  R_4(\zeta) = \frac{1+a_1\zeta+a_2\zeta^2}{1+b_1\zeta+b_2\zeta^2+b_3\zeta^3+b_4\zeta^4}\quad => \quad
  Z_4(\zeta) =  \frac{(a_1-b_1)+(a_2-b_2)\zeta-b_3\zeta^2-b_4\zeta^3}{1+b_1\zeta+b_2\zeta^2+b_3\zeta^3+b_4\zeta^4}.
\end{equation}  
For completeness, the corresponding results are
\begin{eqnarray}
  Z_{4,0}(\zeta) = \frac{i\sqrt{\pi}-\frac{(15\pi^2-88\pi+128)}{2(6\pi^2-29\pi+32)}\zeta
    +i\frac{\sqrt{\pi}(33\pi-104)}{6(6\pi^2-29\pi+32)}\zeta^2- \frac{(9\pi^2-69\pi+128)}{3(6\pi^2-29\pi+32)}\zeta^3}
  {1-i\frac{\sqrt{\pi}}{2}\frac{(9\pi-28)}{(6\pi^2-29\pi+32)}\zeta
    +\frac{(36\pi^2-195\pi+256)}{6(6\pi^2-29\pi+32)}\zeta^2
    -i\frac{\sqrt{\pi}(33\pi-104)}{6(6\pi^2-29\pi+32)}\zeta^3+ \frac{(9\pi^2-69\pi+128)}{3(6\pi^2-29\pi+32)}\zeta^4};
\end{eqnarray}
\begin{eqnarray}
  Z_{4,1}(\zeta) = \frac{i\sqrt{\pi}-\frac{2(3\pi^2-25\pi+48)}{3(16-5\pi)}\zeta-i\frac{2\sqrt{\pi}}{3}\frac{(9\pi-28)}{(16-5\pi)}\zeta^2
    -\frac{2(6\pi^2-29\pi+32)}{3(16-5\pi)}\zeta^3}{1
    -i\frac{2\sqrt{\pi}}{3}\frac{(10-3\pi)}{(16-5\pi)}\zeta
    -\frac{(21\pi-64)}{3(16-5\pi)}\zeta^2
    +i\frac{2\sqrt{\pi}}{3}\frac{(9\pi-28)}{(16-5\pi)}\zeta^3
    +\frac{2(6\pi^2-29\pi+32)}{3(16-5\pi)}\zeta^4};
\end{eqnarray}
\begin{eqnarray}
  Z_{4,2}(\zeta) = \frac{i\sqrt{\pi}+\frac{4(4-\pi)}{(3\pi-8)}\zeta -i\sqrt{\pi}\frac{2(10-3\pi)}{(3\pi-8)}\zeta^2-\frac{2(16-5\pi)}{(3\pi-8)}\zeta^3    }
  {1 -i\sqrt{\pi}\frac{2}{(3\pi-8)}\zeta
    -\frac{(32-9\pi)}{(3\pi-8)}\zeta^2
    +i\sqrt{\pi}\frac{2(10-3\pi)}{(3\pi-8)}\zeta^3+\frac{2(16-5\pi)}{(3\pi-8)}\zeta^4}
\end{eqnarray}
\begin{eqnarray}
  Z_{4,3}(\zeta) = \frac{i\sqrt{\pi}+\frac{3\pi-4}{2}\zeta-i\sqrt{\pi}\zeta^2-\frac{(3\pi-8)}{2}\zeta^3}
  {1-i\frac{3\sqrt{\pi}}{2}\zeta-\frac{(9\pi-16)}{4}\zeta^2
    +i\sqrt{\pi}\zeta^3+\frac{(3\pi-8)}{2}\zeta^4};
\end{eqnarray}
\begin{eqnarray}
 Z_{4,4}(\zeta) &=&  \frac{i\sqrt{\pi}+\frac{10}{3}\zeta -i\sqrt{\pi}\zeta^2-\frac{4}{3}\zeta^3 }{1-i\frac{3\sqrt{\pi}}{2}\zeta-4\zeta^2+i\sqrt{\pi}\zeta^3+\frac{4}{3}\zeta^4}.
\end{eqnarray}
\clearpage
\subsection{Conversion of our 2-index $R_{n,n'}(\zeta)$ notation to other notations}
For clarity, we provide conversion tables of Pad\'e approximants in the notation of \cite{Martin1980} and \cite{HedrickLeboeuf1992}
to our notation. Comparing our analytic results to those of \cite{Martin1980} (introducing superscript M), can be done easily according to
\begin{eqnarray}
  &&  Z_{3,1}^{M} = Z_{2,-2}; \quad Z_{2,2}^{M} = Z_{2,-1}; \quad Z_{1,3}^{M} = Z_{2,0};\\
  &&  Z_{5,1}^{M} = Z_{3,-2}; \quad Z_{4,2}^{M} = Z_{3,-1}; \quad Z_{3,3}^{M} = Z_{3,0};\\
  &&  Z_{5,3}^{M} = Z_{4,0},   
\end{eqnarray}
and the general conversion can be written as
\begin{equation}
Z_{n,n'}^M = Z_{\frac{n+n'}{2}, n'-3}.
\end{equation} 
The Table 1 of \cite{Martin1980} can be now easily verified, which reveals a small obvious typo in their $Z_{4,2}^M$, where the coefficient $p_2$ is missing
the imaginary i number. 

To compare our results to those of \cite{HedrickLeboeuf1992}, it is useful to calculate asymptotic expansions of their $Z_{n}$ definitions
(that is defined as $Z_{n,0}$), that calculate
\begin{eqnarray}
  Z_{2}^{{HL}} &=& -\frac{a_1-\zeta}{a_0+a_1\zeta-\zeta^2} = -\frac{1}{\zeta}-\frac{0}{\zeta^2}+o(\frac{1}{\zeta^2});\\
  Z_{3}^{{HL}} &=& -\frac{\zeta^2-a_2\zeta-a_1+\frac{1}{2}}{\zeta^3-a_2\zeta^2-a_1\zeta-a_0} = -\frac{1}{\zeta} -\frac{1}{2\zeta^3} +o(\frac{1}{\zeta^3}); \label{eq:HLZ3}\\
  Z_{4}^{{HL}} &=& -\frac{\zeta^3-a_3\zeta^2-(a_2-\frac{1}{2}\zeta)-(a_1+a_3/2)}{\zeta^4-a_3\zeta^3-a_2\zeta^2-a_1\zeta-a_0}
  = -\frac{1}{\zeta} -\frac{1}{2\zeta^3} -\frac{0}{\zeta^4} + o(\frac{1}{\zeta^4});\\
  Z_{5}^{{HL}} &=& -\frac{\zeta^4-a_4\zeta^3-(a_3-\frac{1}{2})\zeta^2-(a_2+a_4/2)\zeta-(a_1+a_3/2-\frac{3}{4})}{\zeta^5-a_4\zeta^4-a_3\zeta^3-a_2\zeta^2-a_1\zeta-a_0}
  = -\frac{1}{\zeta} -\frac{1}{2\zeta^3} -\frac{3}{4\zeta^5} + o(\frac{1}{\zeta^5}),
\end{eqnarray}
where ``HL'' stands for \cite{HedrickLeboeuf1992}. As one can see, the number of asymptotic points used in their basic definition of $Z_{n}$, 
increases with the number of poles $n$. Compared to our definition, their $Z_{2,0}$ is defined as having another asymptotic point (for a total of 2),
$Z_{3,0}$ has another asymptotic point (for a total of 3), $Z_{4,0}$ another one (for a total of 4), and so on. Essentially, in their notation the basic $Z_{n,0}$
is defined as having ``n'' asymptotic points, and asymptotic precision $o(1/\zeta^n)$. The conversion between their and our notation is easy, and
\begin{eqnarray}
&&  Z_{2,0}^{HL} = Z_{2,-1}; \quad Z_{2,1}^{HL} = Z_{2,0}; \label{eq:HL1992f}\\
&&  Z_{3,1}^{HL} = Z_{3,1}; \quad Z_{3,2}^{HL} = Z_{3,2}; \\
&&  Z_{4,1}^{HL} = Z_{4,2}; \quad Z_{4,2}^{HL} = Z_{4,3}; \quad Z_{4,3}^{HL} = Z_{4,4}; \\
&&  Z_{5,1}^{HL} = Z_{5,3}; \quad Z_{5,2}^{HL} = Z_{5,4}; \quad Z_{5,3}^{HL} = Z_{5,5}; \quad  Z_{5,4}^{HL} = Z_{5,6}, \label{eq:HL1992l}
\end{eqnarray}
or the general conversion can be written as
\begin{equation}
Z_{n,n'}^{HL} = Z_{n,n'+n-3}.
\end{equation}
We checked the Table 1 of \cite{HedrickLeboeuf1992} that provides coefficients for the Pad\'e approximants (\ref{eq:HL1992f})-(\ref{eq:HL1992l}) and we
can confirm that the table is essentially correct, except for one coefficient. \footnote{Compared to our exact analytic expressions, there are also some rounding errors
in the last 1-2 digits in $Z_{4,1}^{HL}$, $Z_{5,1}^{HL}$, $Z_{5,2}^{HL}$.}
The coefficient where a simple typo is suspected, is the coefficient $a_1$ in $Z_{3,1}$. 
Rewriting our 3-pole approximant $R_3(\zeta)$ to the form used by \cite{PassotSulem2007} and \cite{HedrickLeboeuf1992} (that corresponds to the $Z_3^{HL}$ as
written in (\ref{eq:HLZ3}) ) yields
\begin{equation}
R_3 (\zeta) = \frac{-\frac{1}{2}\zeta-a_0}{\zeta^3-a_2\zeta^2-a_1\zeta-a_0}, 
\end{equation}
which further yields
\begin{eqnarray}
  R_{3,1}(\zeta) &=& \frac{-\frac{1}{2}\zeta-\frac{i\sqrt{\pi}}{2(4-\pi)}}{\zeta^3+\frac{i\sqrt{\pi}}{(4-\pi)}\zeta^2-\frac{2}{(4-\pi)}\zeta-\frac{i\sqrt{\pi}}{2(4-\pi)}};\\
  R_{3,2}(\zeta) &=& \frac{-\frac{1}{2}\zeta-\frac{i}{\sqrt{\pi}}}{\zeta^3+\frac{2i}{\sqrt{\pi}}\zeta^2 -\frac{3}{2}\zeta-\frac{i}{\sqrt{\pi}}},
\end{eqnarray}  
and our approximants are
\begin{eqnarray}
  R_{3,1}(\zeta)&:& \qquad a_0=\frac{i\sqrt{\pi}}{2(4-\pi)}=1.03241i; \qquad a_1=\frac{2}{4-\pi}=2.32990; \qquad a_2 = -\frac{i\sqrt{\pi}}{4-\pi}=-2.06482i;\\
  R_{3,2}(\zeta)&:& \qquad a_0=\frac{i}{\sqrt{\pi}}=0.56419i; \qquad a_1=\frac{3}{2}; \qquad a_2 = -\frac{2i}{\sqrt{\pi}} =  -1.12838i.
\end{eqnarray}
For the $a_1$ coefficient in $R_{3,1}$, both \cite{HedrickLeboeuf1992} and \cite{PassotSulem2007} use $a_1=2.23990$ instead of the correct $a_1=2.32990$.
The differences are of course small. Nevertheless, the new correct value explains the observation made by \cite{PassotSulem2007}, in the paragraph below their Figure 1,
where they write: ``It is conspicuous that $R_{3,2}$ provides a fit that is slightly better for small $\zeta$, but turns out to be globally less accurate
than $R_{3,1}$.'' Authors obviously noticed that something is not right, since for small $\zeta$, the $R_{3,1}$ has precision $o(\zeta^2)$ and $R_{3,2}$ only $o(\zeta)$,  
so the $R_{3,1}$ should be more precise. And it indeed is, authors were just misguided by the wrong value of $a_1$ introduced by \cite{HedrickLeboeuf1992}.

\clearpage
\subsection{Precision of $R(\zeta)$ approximants} \label{sec:PrecissionR}
\begin{figure*}[!htpb]
  $$\includegraphics[width=0.48\linewidth]{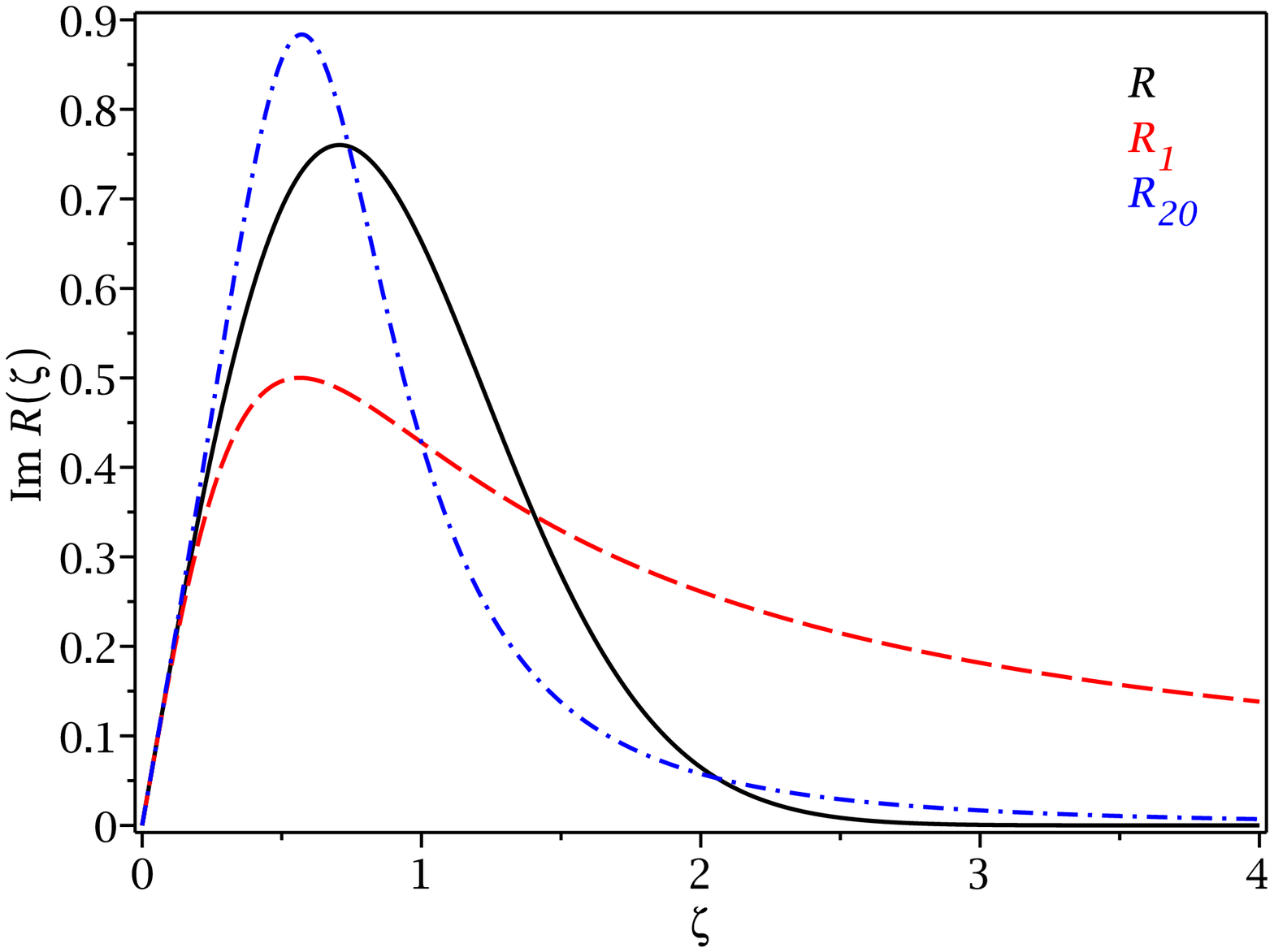}\hspace{0.03\textwidth}\includegraphics[width=0.48\linewidth]{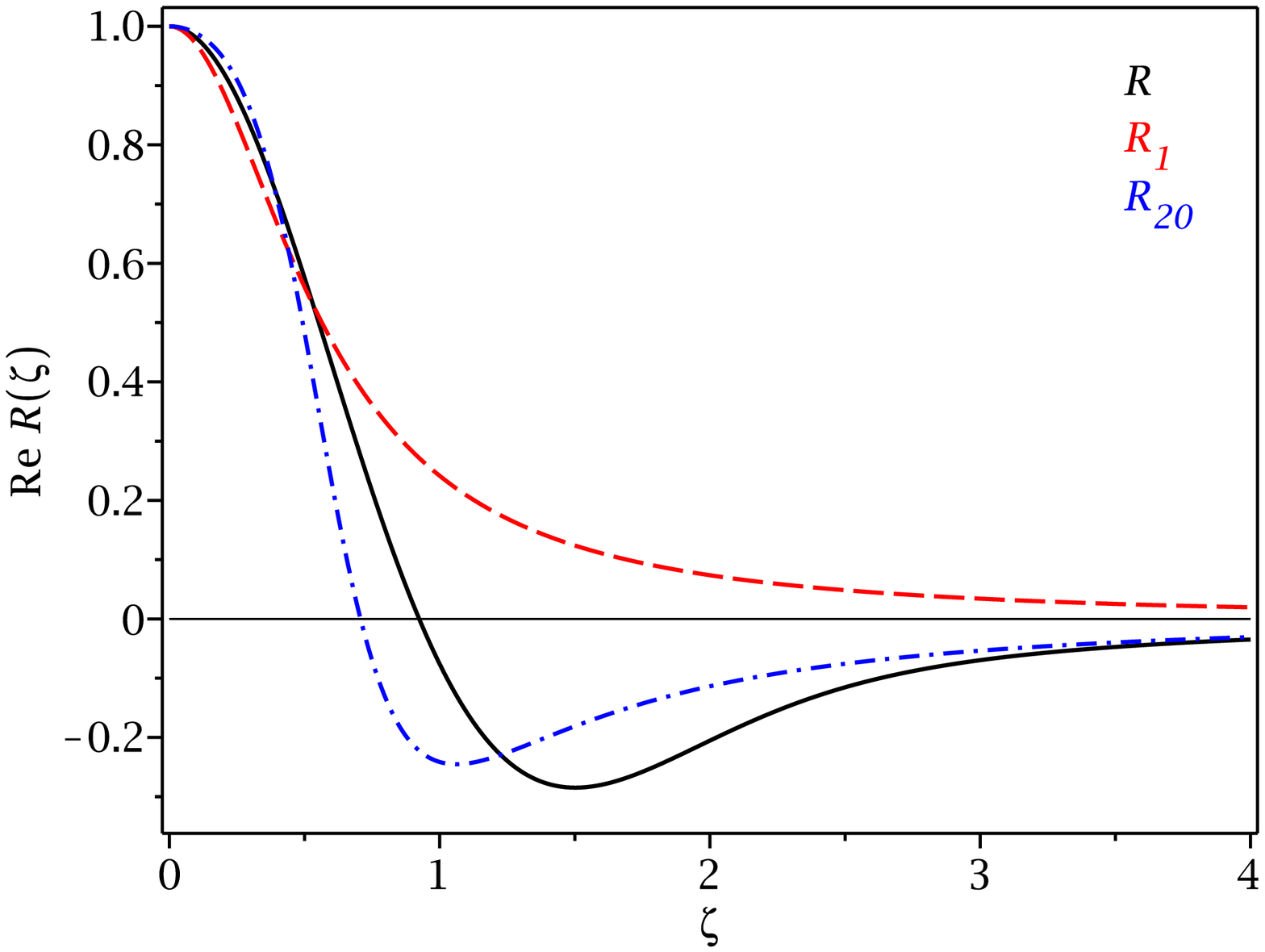}$$
  $$\includegraphics[width=0.48\linewidth]{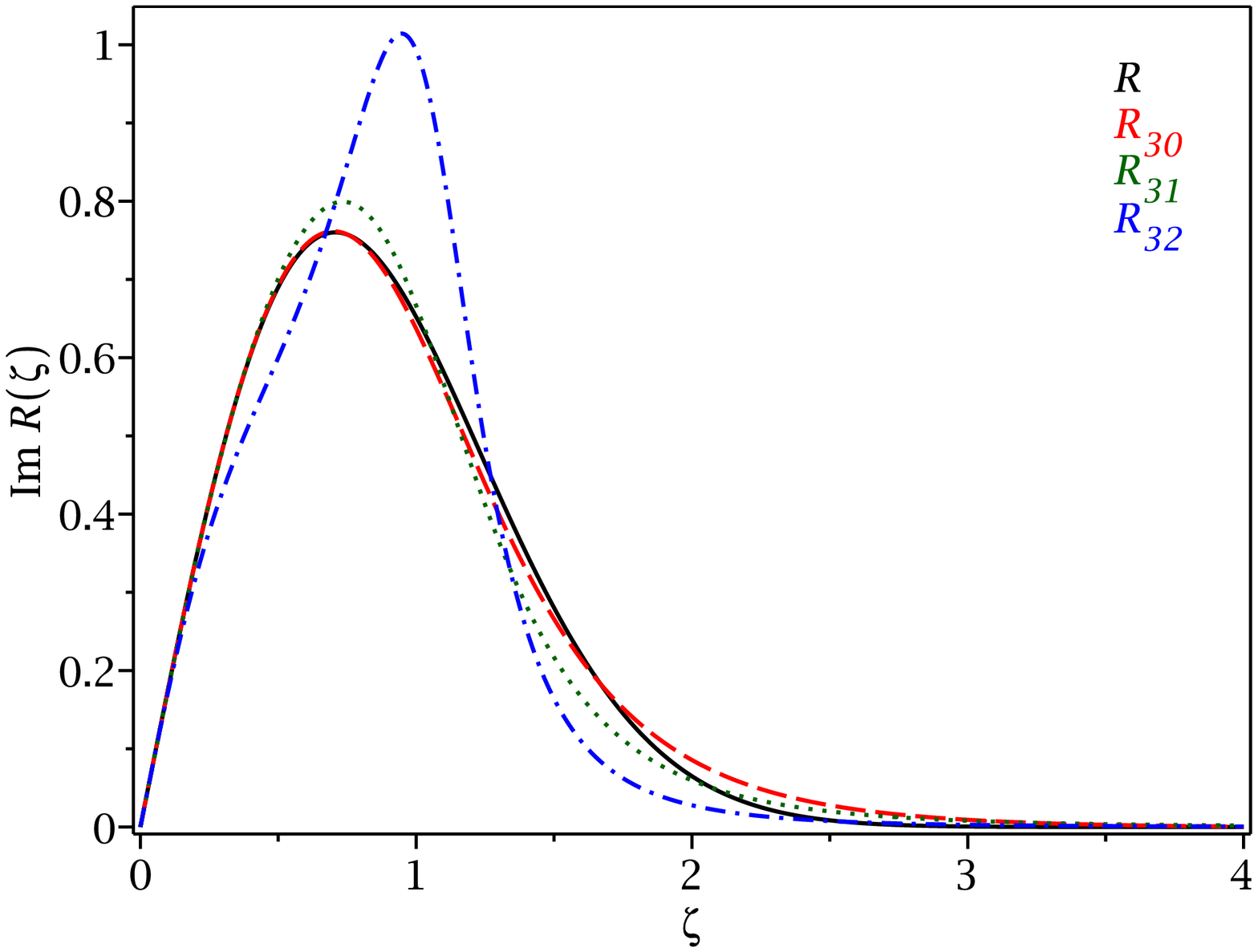}\hspace{0.03\textwidth}\includegraphics[width=0.48\linewidth]{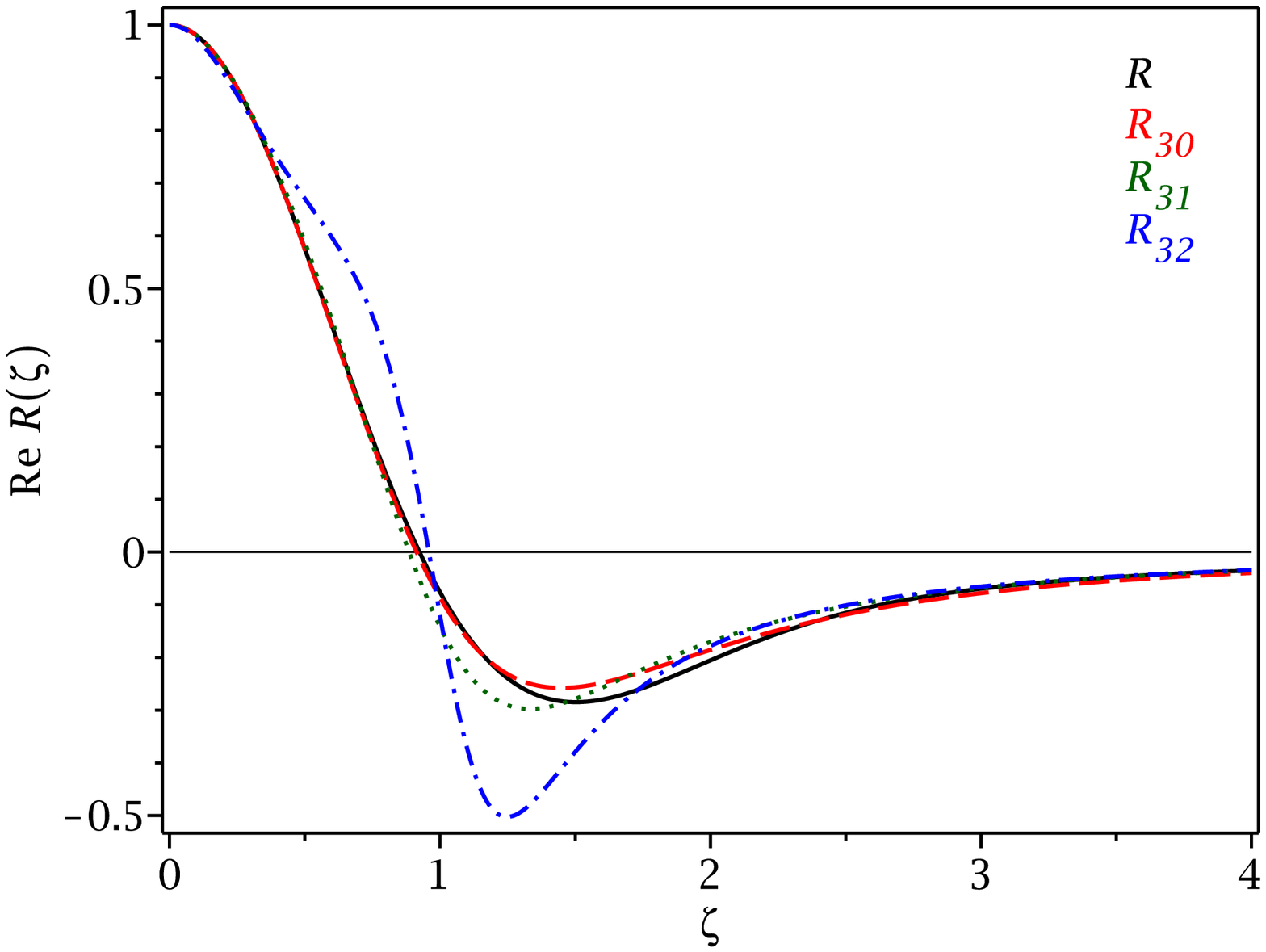}$$
  \caption{1-pole, 2-pole (top) and 3-pole (bottom) Pad\'e approximants of $R(\zeta)$. Left: $\textrm{Im} R(\zeta)$, Right: $\textrm{Re} R(\zeta)$, for $\zeta$ being real.} \label{fig:6}
\end{figure*}
\begin{figure*}[!htpb]
  $$\includegraphics[width=0.48\linewidth]{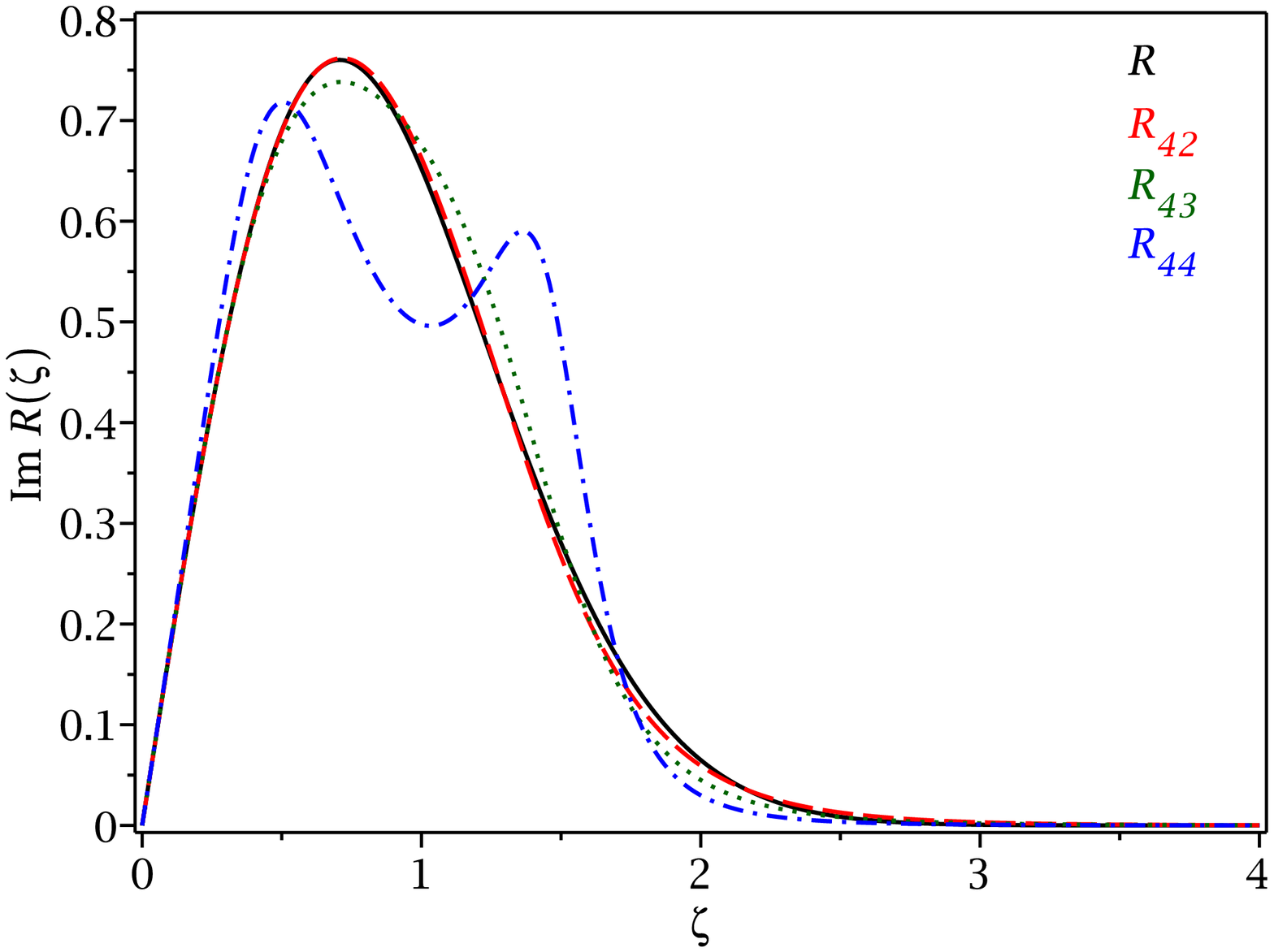}\hspace{0.03\textwidth}\includegraphics[width=0.48\linewidth]{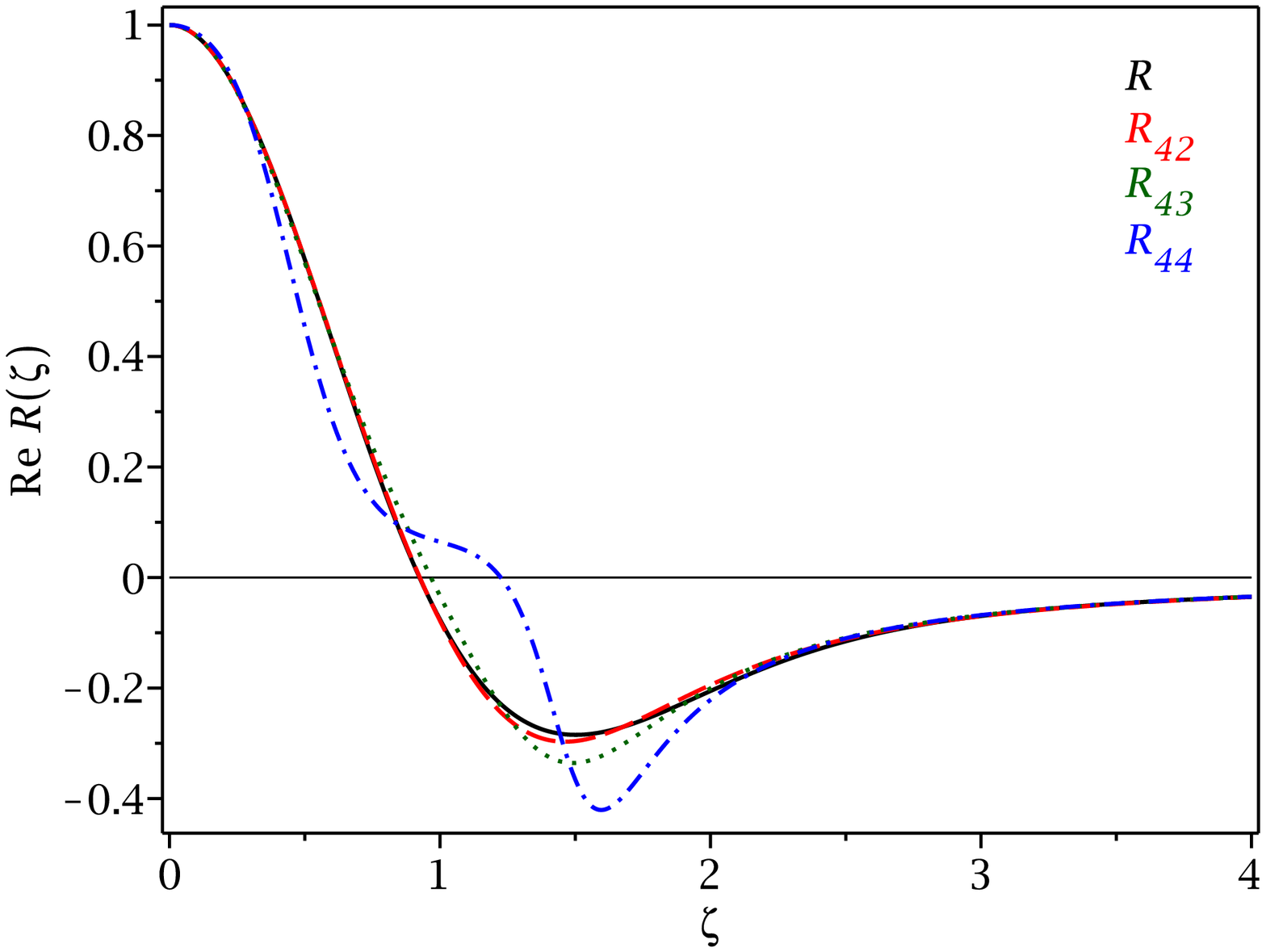}$$
  $$\includegraphics[width=0.48\linewidth]{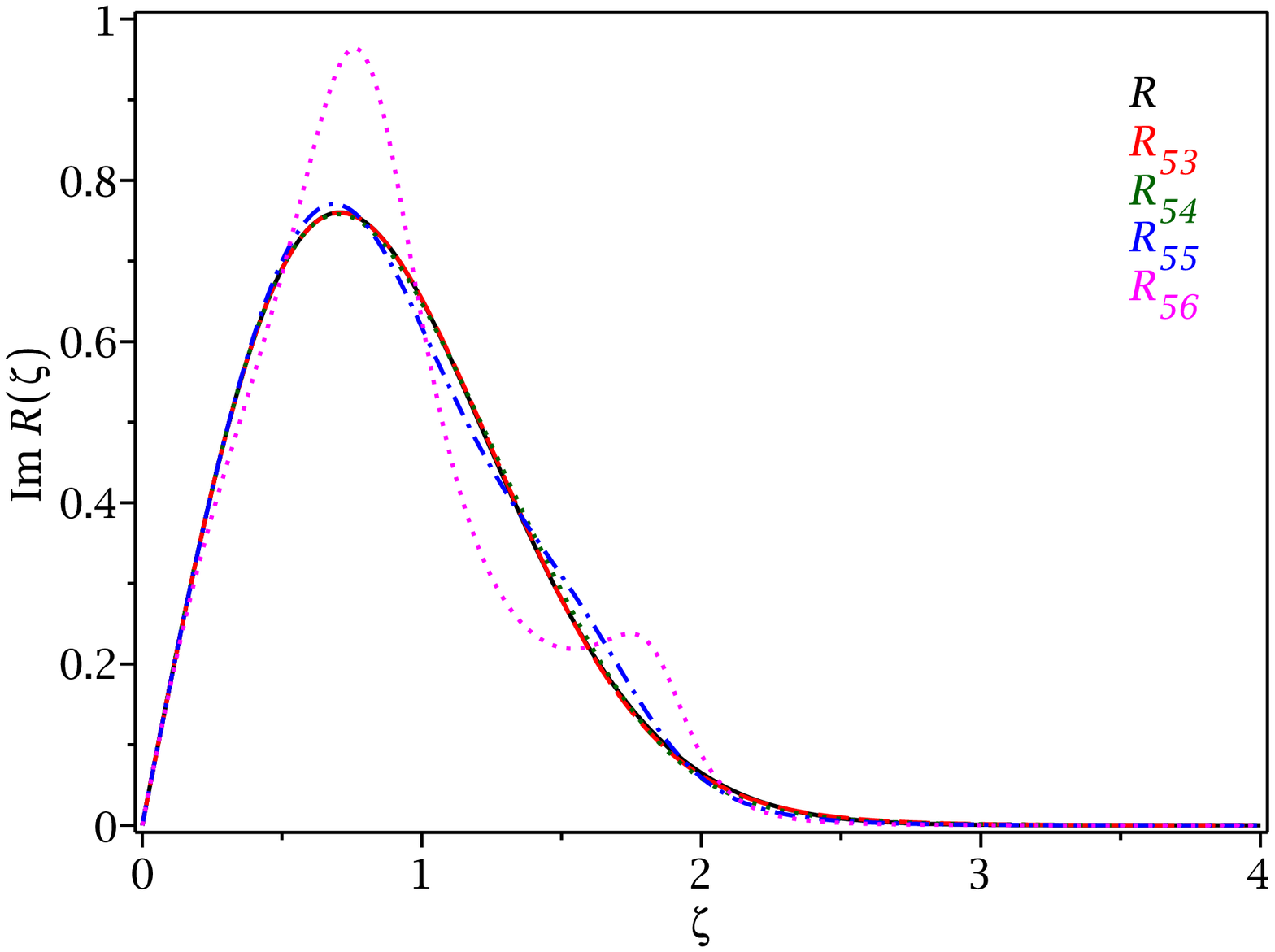}\hspace{0.03\textwidth}\includegraphics[width=0.48\linewidth]{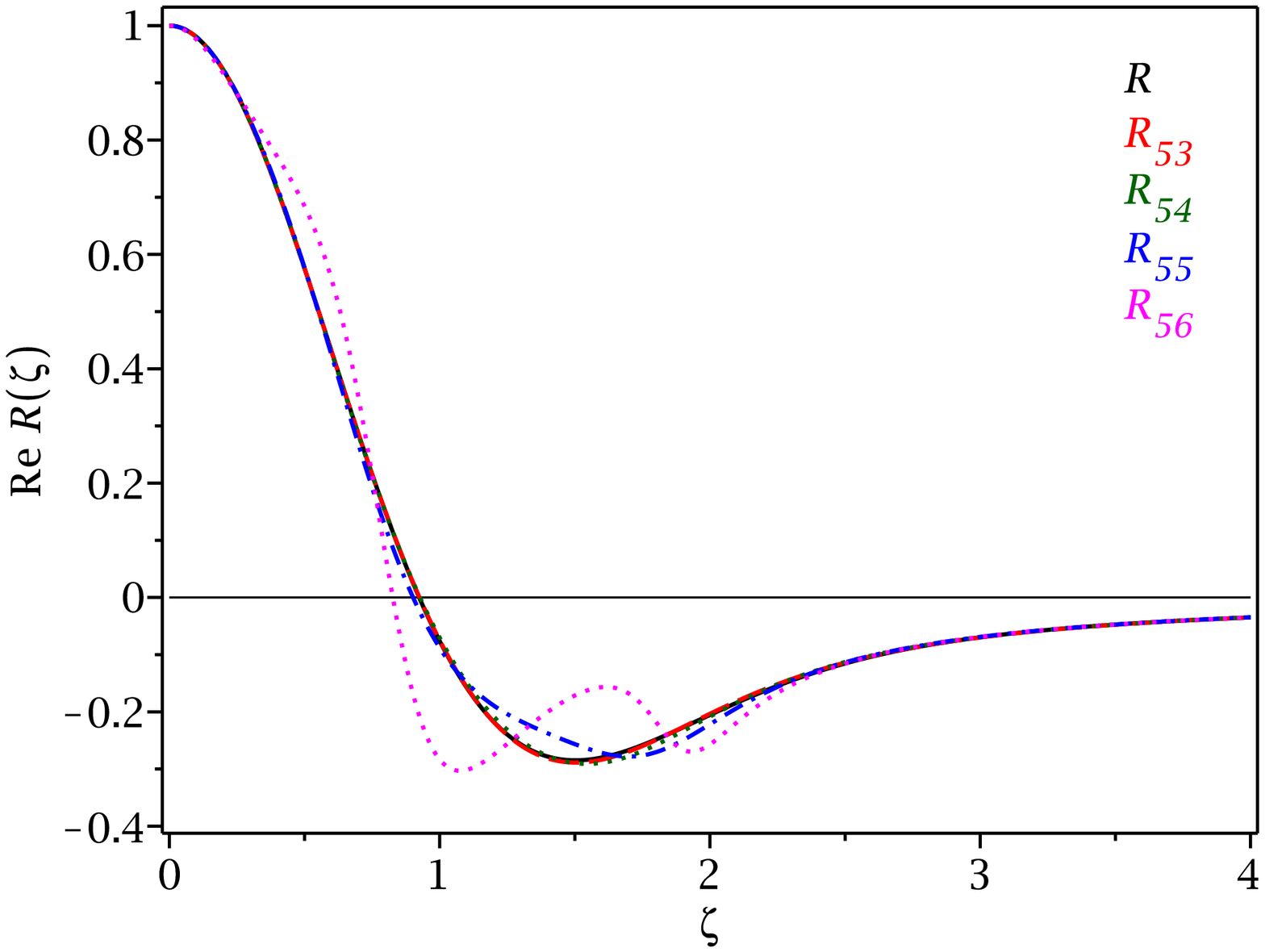}$$
  \caption{4-pole (top) and 5-pole (bottom) Pad\'e approximants of $R(\zeta)$.} \label{fig:7}
\end{figure*}

It is useful to compare the Pad\'e approximants to the exact $R(\zeta)=1+\zeta Z(\zeta)$, where the plasma dispersion function can be conveniently calculated
(for example in Maple) according to
\begin{eqnarray} \label{eq:ZzetaDefX}
  Z(\zeta) &=&  i\sqrt{\pi} e^{-\zeta^2} \Big( 1+\textrm{erf}(i\zeta) \Big),
\end{eqnarray}
where $\textrm{erf}(z)=\frac{2}{\sqrt{\pi}}\int_0^z e^{-t^2}dt$ is the well-known error function, defined for any complex z. We plot only approximants for
which we were able to obtain closures. The exact $R(\zeta)$ is plotted as a black solid line in all the Figures. Figure \ref{fig:6} top shows 1-pole and 2-pole approximants
$R_1(\zeta)$ (red dashed line) and $R_{2,0}(\zeta)$ (blue dot-dashed line). Figure \ref{fig:6} bottom shows 3-pole approximants $R_{3,0}(\zeta)$ (red dashed line), $R_{3,1}(\zeta)$
(green dotted line) and $R_{3,2}(\zeta)$ (blue dot-dashed line). Figures in the left column show imaginary part and figures in the right column show real part.
The input variable $\zeta$ plotted on the x-axis is prescribed to be real, i.e. states in the weak growth-rate/damping approximation are explored
(one might as well prescribe $Im(\zeta)=\pm 0.01 Re(\zeta)$ and plot essentially the same graphs, with only small differences in solutions). 

As expected, the very simple approximant $R_1(\zeta)$ is unprecise for larger values of $\zeta$, and above $\zeta>1$, the $\textrm{Re} R_1(\zeta)$ even has a wrong sign.
Nevertheless, the approximant is still a good approximant for small $\zeta\ll 1$ values, and it is also very valuable from a theoretical perspective,
since it is the only approximant that provides a quasi-static closure for the perpendicular heat flux $q_\perp$
 (see the 3D geometry Section \ref{section:3D}, closure (\ref{eq:QperpClosureX})).
This has one important implication : \newpage
\noindent
\emph{If one renders the approximant $R_1$ as not satisfactory (which is true unless $\zeta\ll 1$ or at least $\zeta <1$), 3D simulations with fluid models that contain Landau damping can be only
  performed with time-dependent heat flux equations.} All other approximants in Figure 1 perform reasonably well, and the most precise is $R_{3,0}(\zeta)$, followed by $R_{3,1}(\zeta)$.

Figure \ref{fig:7} shows selected 4-pole and 5-pole approximants for which we were able to obtain closures.
Unfortunately, approximants $R_{4,4}(\zeta)$ and $R_{5,6}(\zeta)$ show a bit unpleasant
behavior, and the associated closures obtained with these approximants are therefore difficult to recommend, unless the considered domain is $\zeta\ll 1$ or $\zeta\gg 1$,
or more specifically, at least $\zeta<0.5$ or $\zeta>2$. The behavior is not surprising, since approximants $R_{4,4}(\zeta)$ and $R_{5,6}(\zeta)$ have the maximum
available number of poles devoted to the asymptotic expansion $\zeta\gg 1$, without being ill-posed. The closures are therefore specifically suitable for $\zeta\gg 1$ regime,
for example in the \emph{low-temperature limit}, or, in the high-frequency (actually high phase speed) limit
(since $\zeta=\frac{\omega}{|\kpar|v_{\textrm{th}\parallel}}$). For $R_{4,4}(\zeta)$, the corresponding closures are the
quasi-static closure (\ref{eq:Static_R44}) and time-dependent closures (\ref{eq:Tclosure_R44}), (\ref{eq:T_R44_2}), (\ref{eq:T_R44_3}).
For $R_{5,6}(\zeta)$, the corresponding closure is time-dependent (\ref{eq:R56_closure}) and naturally, this is the most precise closure
in the $\zeta\gg 1$ regime, with precision $o(\zeta^{-8})$. Noticeably, the asymptotic precision is even better than the $R_{8,3}(\zeta)$ approximant used in the WHAMP code,
which has a precision $o(\zeta^{-5})$.

All other approximants in Figure \ref{fig:7} are very precise in the entire considered range of $\zeta$. To clearly see the precision, it is useful to calculate
 the maximum relative errors
\begin{equation}
\textrm{Im} \Big(\frac{R_{n,n'}(\zeta)-R(\zeta)}{R(\zeta)}\Big) 100\%;\qquad  \textrm{Re}\Big(\frac{R_{n,n'}(\zeta)-R(\zeta)}{R(\zeta)}\Big)  100\%,
\end{equation}
which we define this way instead of for example $\textrm{Re}(R_{n,n'}(\zeta)-R(\zeta))/ \textrm{Re} R(\zeta)$,
since the real part of $R(\zeta)$ is going through zero.  The maximum relative errors typically appear for $\zeta\in(0,4)$, even though some reported values are
 outside of this range. The $R_1(\zeta)$ approximant is excluded from the table since its relative error of the imaginary part
increases with $\zeta$. We omit if errors are positive or negative and the results are:\\

\noindent
2-pole and 3-pole approximants\\
\begin{tabular}{| c | c | c | c | c | c | c |}
  \hline			
                      & $R_{2,0}$ &   $R_{3,0}$    & $R_{3,1}$    & $R_{3,2}$\\
  \hline
  $\textrm{error}\; Im\;\%$ & $35$ &   $16.4$    & $\bf{13.3}$    & $44$\\
  \hline
   $ \textrm{error}\; Re\;\%$ & $44$ &   $\bf{14.7}$    & $16.6$    & $53$\\
  \hline
\end{tabular}\\
4-pole approximants\\
\begin{tabular}{| c | c | c | c | c | c | c | c |}
  \hline			
                      & $R_{4,0}$ &   $R_{4,1}$    & $R_{4,2}$    & $R_{4,3}$  & $R_{4,4}$\\
  \hline
  $\textrm{error}\; Im\;\%$ & $6.2$ &   $4.66$    & $\bf{4.57}$    & $11.6$  & $49$\\
  \hline
   $ \textrm{error}\; Re\;\%$ & $5.3$ &   $\bf{4.3}$    & $5.7$    & $12.3$  & $51$\\
  \hline
\end{tabular}\\
5-pole approximants\\
\begin{tabular}{| c | c | c | c | c | c | c | c |}
  \hline			
                      & $R_{5,0}$ &   $R_{5,1}$    & $R_{5,2}$    & $R_{5,3}$  & $R_{5,4}$  & $R_{5,5}$  & $R_{5,6}$\\
  \hline
  $\textrm{error}\; Im\;\%$ & $1.9$ &   $1.42$    & $\bf{1.34}$    & $1.46$  & $3.4$  & $10.3$  & $40$ \\
  \hline
   $ \textrm{error}\; Re\;\%$ & $2.0$ &   $\bf{1.26}$    & $\bf{1.26}$    & $1.81$  & $3.4$  & $10.0$  & $31$\\
  \hline
\end{tabular}\\
6-pole approximants\\
\begin{tabular}{| c | c | c | c | c | c | c | c | c | c |}
  \hline			
                      & $R_{6,0}$ &   $R_{6,1}$    & $R_{6,2}$    & $R_{6,3}$  & $R_{6,4}$  & $R_{6,5}$  & $R_{6,6}$  & $R_{6,7}$  & $R_{6,8}$\\
  \hline
  $\textrm{error}\; Im\;\%$ & $0.58$ &   $0.37$    & $\bf{0.35}$    & $0.38$  & $0.45$  & $1.0$  & $2.5$  & $7.6$  & $30$ \\
  \hline
   $ \textrm{error}\; Re\;\%$ & $0.64$ &   $0.40$    & $\bf{0.31}$    & $0.36$  & $0.54$  & $0.9$  & $2.5$  & $7.7$  & $39$\\
  \hline
\end{tabular}\\
7-pole approximants\\
\begin{tabular}{| c | c | c | c | c | c | c | c | c | c | c | c |}
  \hline			
                      & $R_{7,0}$ &   $R_{7,1}$    & $R_{7,2}$    & $R_{7,3}$  & $R_{7,4}$  & $R_{7,5}$  & $R_{7,6}$  & $R_{7,7}$  & $R_{7,8}$  & $R_{7,9}$ & $R_{7,10}$\\
  \hline
  $\textrm{error}\; Im\;\%$ & $0.18$ &   $0.10$    & $\bf{0.080}$    & $0.087$  & $0.10$  & $0.13$  & $0.29$  & $0.65$  & $1.8$  & $6.2$ & $35$\\
  \hline
   $ \textrm{error}\; Re\;\%$ & $0.2$ &   $0.11$    & $0.089$    & $\bf{0.080}$  & $0.10$  & $0.16$  & $0.26$  & $0.65$  & $1.9$  & $6.6$ & $33$\\
  \hline
\end{tabular}\\
8-pole approximants\\
\begin{tabular}{| c | c | c | c | c | c | c | c | c | c | c |}
  \hline			
                      & $R_{8,0}$ &   $R_{8,1}$    & $R_{8,2}$    & $R_{8,3}$  & $R_{8,4}$  & $R_{8,5}$  & $R_{8,6}$  & $R_{8,7}$  & $R_{8,8}$  & $R_{8,9}$\\
  \hline
  $\textrm{error}\; Im\;\%$ & $0.06$ &   $0.03$    & $0.021$    & $\bf{0.018}$  & $0.022$  & $0.028$  & $0.04$  & $0.08$  & $0.17$  & $0.43$\\
  \hline
   $ \textrm{error}\; Re\;\%$ & $0.06$ &   $0.03$    & $0.022$    & $\bf{0.020}$  & $\bf{0.020}$  & $0.027$  & $0.04$  & $0.07$  & $0.17$  & $0.46$\\
  \hline
\end{tabular}
\\\\
\begin{figure*}[!htpb]
  $$\includegraphics[width=0.3\linewidth]{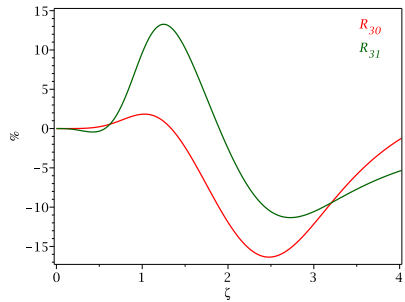}$$
  \caption{$\%$ error of imaginary part of $R_{3,0}(\zeta)$ (red line), and $R_{3,1}(\zeta)$ (green line).} \label{fig:8}
\end{figure*}
The numbers of course do not reveal the entire story, since the maximum error can occur for different $\zeta$ values. For example, from the plots of $\textrm{Im} R(\zeta)$ in
Figure \ref{fig:6}, the approximant $R_{3,0}(\zeta)$ captures the maximum (the peak around $\zeta\sim 1$) with much better accuracy than the approximant $R_{3,1}(\zeta)$. However,
according to the above table, the $R_{3,1}(\zeta)$ appears to be more precise globally. The discrepancy is easily understood from Figure \ref{fig:8} , where $\%$ errors of both approximants
are plotted with respect to $\zeta$. A similar table and figures can be created for the heavily damped regime, for example for $\zeta$ with the
  imaginary part $Im(\zeta)=-Re(\zeta)/2$, where the Pad\'e approximants are less precise.

\clearpage
\subsection{Landau fluid closures - fascinating closures for all $\zeta$}
Now, let's use various Pad\'e approximations of the plasma response function $R(\zeta)$, and calculate the kinetic moments.
Let's start with the simplest choice of replacing the exact $R(\zeta)$ with approximant $R_1(\zeta)=1/(1-i\sqrt{\pi}\zeta)$. Let's drop the index $r$.
The linear kinetic  moments (\ref{eq:RefPica-n})-(\ref{eq:RefPica-rr}) calculate
\begin{eqnarray}
R_1(\zeta):\qquad  
\frac{n^{(1)}}{n_{0}} &=& -\frac{q_r}{T^{(0)}} \Phi \frac{1}{1-i\sqrt{\pi}\zeta}\Big[ 1 \Big];\\
u^{(1)} &=& - \frac{q_r}{T^{(0)}} \Phi v_{\textrm{th}} \sign(\kpar)\frac{1}{1-i\sqrt{\pi}\zeta} \Big[ \zeta \Big];\\
\frac{p^{(1)}}{p_{0}} &=& -\frac{q_r}{T^{(0)}} \Phi \frac{1}{1-i\sqrt{\pi}\zeta}\Big[2\zeta^2-i\sqrt{\pi}\zeta+1 \Big];\\
T^{(1)} &=& -q_r \Phi \frac{1}{1-i\sqrt{\pi}\zeta}\Big[ 2\zeta^2-i\sqrt{\pi}\zeta\Big];\\
q^{(1)} &=& -q_r n_{0} \Phi v_{\textrm{th}} \sign(\kpar) \frac{1}{1-i\sqrt{\pi}\zeta}\Big[ 2\zeta^3-i\sqrt{\pi}\zeta^2-2\zeta \Big];\\
r^{(1)} &=& -q_r \frac{n_{0}}{2} v_{\textrm{th}}^2 \Phi \frac{1}{1-i\sqrt{\pi}\zeta}\Big[ 4\zeta^4-2i\sqrt{\pi}\zeta^3+2\zeta^2-3i\sqrt{\pi}\zeta+3 \Big];\\
\widetilde{r}^{(1)} &=& -q_r\frac{n_{0}}{2}v_{\textrm{th}}^2 \Phi \frac{1}{1-i\sqrt{\pi}\zeta}\Big[ 4\zeta^4-2i\sqrt{\pi}\zeta^3-10\zeta^2+3i\sqrt{\pi}\zeta \Big].
\end{eqnarray}
We are looking for a closure, and we want to express either $q^{(1)}$ or $\widetilde{r}^{(1)}$, as a linear combination of lower order moments.
To immediately see possible closures, it is always useful to pull out the denominator of the Pad\'e approximant out (as done above),
and concentrate only at the expressions inside the big brackets. Also, similarly to the closures explored for small $\zeta$, it is useful
to forget the $n^{(1)}$, $p^{(1)}$ and $r^{(1)}$ moments, and just concentrate at the $u^{(1)}$, $T^{(1)}, q^{(1)}$ and $\widetilde{r}^{(1)}$ moments.
Nevertheless, we will keep the $n^{(1)}$ moment, since it helps us to understand the expressions and to somehow ``maintain the touch with reality''.

By exploring the expressions inside of the brackets, it is obvious that it is impossible to express $q^{(1)}$ or $\widetilde{r}^{(1)}$ as
a linear combination of lower order moments that eliminate $\zeta$ dependence. Moreover, for large $\zeta$, the moments $q^{(1)}\sim \zeta^2$ and
$\widetilde{r}^{(1)}\sim \zeta^3$, which does not make physical sense, since these quantities should converge to zero with increasing $\zeta$,
as explored in the $\zeta\gg 1$ limit, see eqs. (\ref{eq:NlargeZZ})-(\ref{eq:rrlargeZZ}).
The $R_1(\zeta)$ approximant therefore does not yield a closure.
 The same conclusion is obtained by using the $R_{2,0}(\zeta)$ approximant, where no closure for $q^{(1)}$ or $\widetilde{r}^{(1)}$ is possible.
We note that the approximant $R_{2,1}(\zeta)$, that was eliminated because
it is not a good approximant of $R(\zeta)$ yields a closure $q^{(1)} = -2 p_0 u^{(1)}$, which is equivalent to the closure (\ref{eq:R21closure}),
that was obtained for small $\zeta$ with the precision $o(\zeta)$.
This closure is therefore disregarded. 

Let's try the 3-pole Pad\'e approximants. The moments with $R_{3,1}(\zeta)$ approximant are proportional to
\begin{eqnarray}
  R_{3,1}(\zeta):\qquad
  n^{(1)} &\sim& \frac{1}{1-\frac{4i}{\sqrt{\pi}}\zeta-2\zeta^2+2i\frac{4-\pi}{\sqrt{\pi}}\zeta^3}\Big[ 1+\frac{i}{\sqrt{\pi}}(\pi-4)\zeta \Big];\\
  u^{(1)} &\sim& \frac{1}{1-\frac{4i}{\sqrt{\pi}}\zeta-2\zeta^2+2i\frac{4-\pi}{\sqrt{\pi}}\zeta^3}\Big[ \frac{i}{\sqrt{\pi}}(\pi-4)\zeta^2+\zeta \Big];\\
  T^{(1)} &\sim& \frac{1}{1-\frac{4i}{\sqrt{\pi}}\zeta-2\zeta^2+2i\frac{4-\pi}{\sqrt{\pi}}\zeta^3}\Big[ -i\sqrt{\pi}\zeta \Big];\\
  q^{(1)} &\sim& \frac{1}{1-\frac{4i}{\sqrt{\pi}}\zeta-2\zeta^2+2i\frac{4-\pi}{\sqrt{\pi}}\zeta^3}\Big[ -\frac{i}{\sqrt{\pi}}(3\pi-8)\zeta^2-2\zeta \Big];\\
 \widetilde{r}^{(1)} &\sim& \frac{1}{1-\frac{4i}{\sqrt{\pi}}\zeta-2\zeta^2+2i\frac{4-\pi}{\sqrt{\pi}}\zeta^3}\Big[ -\frac{2i}{\sqrt{\pi}}(3\pi-8)\zeta^3-4\zeta^2+3i\sqrt{\pi}\zeta \Big],
\end{eqnarray}
 where we have suppressed writing all the multiplicative factors including the minus signs (it does not mean that these were neglected, full expressions
are considered, we are just not writing down the full expressions, which helps in spotting the possible closures).
 There is a possibility to express $q^{(1)}$ through the combination of the lower moments $T^{(1)}$ and $u^{(1)}$. The full expressions of these moments are
\begin{eqnarray}
  R_{3,1}(\zeta):\qquad D &=& \Big(1-\frac{4i}{\sqrt{\pi}}\zeta-2\zeta^2+2i\frac{4-\pi}{\sqrt{\pi}}\zeta^3 \Big);\\
  u^{(1)} &=& - \frac{q_r}{T^{(0)}} \Phi v_{\textrm{th}} \sign(\kpar)
  \frac{1}{D}\Big[ \frac{i}{\sqrt{\pi}}(\pi-4)\zeta^2+\zeta \Big];\\
  T^{(1)} &=& -q_r \Phi \frac{1}{D}\Big[ -i\sqrt{\pi}\zeta \Big];\\
  q^{(1)} &=& -q_r n_{0} \Phi v_{\textrm{th}} \sign(\kpar)
  \frac{1}{D}\Big[ -\frac{i}{\sqrt{\pi}}(3\pi-8)\zeta^2-2\zeta \Big],
\end{eqnarray}
where we have used a convenient notation $D$ for the denominator of the plasma response function, and the closure is
\begin{eqnarray} \label{eq:Static_R31_F} \boxed{
R_{3,1}(\zeta):\qquad q^{(1)} = \frac{3\pi-8}{4-\pi}n_0 T^{(0)} u^{(1)}-i\frac{\sqrt{\pi}}{4-\pi}n_0 v_{\textrm{th}}\sign(\kpar) T^{(1)},}
\end{eqnarray}  
which is equivalent to the (\ref{eq:R31closure}) closure (which was obtained for small $\zeta$ with the precision $o(\zeta^2)$).
Continuing with the next approximant $R_{3,2}(\zeta)$, the moments calculate
\begin{eqnarray}
  R_{3,2}(\zeta):\qquad D &=& \Big(1-\frac{3i\sqrt{\pi}}{2}\zeta-2\zeta^2+i\sqrt{\pi} \zeta^3\Big); \label{eq:R32_n}\\
 n^{(1)} &\sim& \frac{1}{D}\Big[ 1-\frac{i\sqrt{\pi}}{2}\zeta \Big];\\
 u^{(1)} &\sim& \frac{1}{D}\Big[-\frac{i\sqrt{\pi}}{2}\zeta^2+\zeta \Big];\\
 T^{(1)} &\sim& \frac{1}{D}\Big[-i\sqrt{\pi}\zeta \Big];\\
 q^{(1)} &\sim& \frac{1}{D}\Big[-2\zeta \Big]; \label{eq:R32_q}\\
 \widetilde{r}^{(1)} &\sim& \frac{1}{D}\Big[-4\zeta^2+3i\sqrt{\pi}\zeta \Big].
\end{eqnarray}
It is possible to express 1) $q^{(1)}$ through $T^{(1)}$; 2) $\widetilde{r}^{(1)}$ through the combination of $u^{(1)}$ and $q^{(1)}$;
3) $\widetilde{r}^{(1)}$ through the combination of $u^{(1)}$ and $T^{(1)}$. The first choice yields a closure
\begin{equation}
  \boxed{ R_{3,2}(\zeta):\qquad
    q^{(1)} = -i \frac{2}{\sqrt{\pi}} n_0 v_{\textrm{th}} \sign(\kpar)T^{(1)},
  }  \label{eq:Static_R32_main}  
\end{equation}  
that is equivalent to the (\ref{eq:R32_1closure}) closure obtained for small $\zeta$ with the precision $o(\zeta)$. This is indeed
the famous simplest possible Landau fluid closure that expresses the collisionless heat flux with respect to temperature, and it equivalent
to eq. (7) of \cite{HammettPerkins1990}.
\footnote{With their later found constant $\chi_1=2/\sqrt{\pi}$, and remembering that their thermal speeds are defined
  as $v_{\textrm{th}}=\sqrt{T^{(0)}/m}$, whereas ours are  $v_{\textrm{th}}=\sqrt{2 T^{(0)}/m}$.}
The closure is written here in Fourier space.  The important part is the $i\sign(\kpar)$ that typically written as $i\kpar/|\kpar|$,
and that in Real space rewrites as a Hilbert transform, which we will address later. The $R_{3,2}(\zeta)$ was obtained with $o(\zeta)$ power series expansion,
and $o(1/\zeta^4)$ asymptotic series expansion. How good is this closure ? By exploring expressions (\ref{eq:R32_n})-(\ref{eq:R32_q}),
the quantities $n^{(1)}$, $u^{(1)}$, $T^{(1)}$ have all correct asymptotic expansion for large $\zeta$ (including the proportionality constants),
however, the heat flux decreases only as $q^{(1)}\sim 1/\zeta^2$ instead of the correct $\sim 1/\zeta^3$, see eq. (\ref{eq:QlargeZZ}). For large $\zeta$, the heat flux is therefore overestimated
by this simple closure, which typically leads to an overestimation of the Landau damping in fluid models that use this simplest closure. Nevertheless,
the closure is very beneficial because it clarifies the distinction between the collisional and collisionless heat flux. 

The other two possible closures with $R_{3,2}(\zeta)$ are
\begin{eqnarray}
  R_{3,2}(\zeta): \qquad
  \widetilde{r}^{(1)} &=& -\frac{4i}{\sqrt{\pi}}v_{\textrm{th}} n_0 T^{(0)}\sign(\kpar) u^{(1)} -i\frac{(3\pi+8)}{4\sqrt{\pi}}v_{\textrm{th}}\sign(\kpar) q^{(1)};\label{eq:Static_R32_F2}\\
  \widetilde{r}^{(1)} &=& -\frac{4i}{\sqrt{\pi}}v_{\textrm{th}} n_0 T^{(0)}\sign(\kpar) u^{(1)} -\frac{(3\pi+8)}{2\pi}v_{\textrm{th}}^2 n_0 T^{(1)},\label{eq:Static_R32_F3}
\end{eqnarray}
and one can go from (\ref{eq:Static_R32_F2}) to (\ref{eq:Static_R32_F3}) by using (\ref{eq:Static_R32_main}). Obviously, it would be also possible to construct a closure
$ \widetilde{r}^{(1)}=\alpha_u u^{(1)}+\alpha_q q^{(1)}+\alpha_T T^{(1)}$, where $\alpha_u=-\frac{4i}{\sqrt{\pi}}v_{\textrm{th}} n_0 T^{(0)}\sign(\kpar)$, and where
$\alpha_q$ and $\alpha_T$ are related by satisfying $2n_0v_{\textrm{th}}\sign(\kpar)\alpha_q+i\sqrt{\pi}\alpha_T = -iv_{\textrm{th}}^2n_0 \frac{3\pi+8}{2\sqrt{\pi}}$,
i.e. one could consider a closure with a free parameter, which we will not consider. Additionally, all constructed closures should be checked
with respect to obtained dispersion relations, and closures (\ref{eq:Static_R32_F2}), (\ref{eq:Static_R32_F3}) will be
later disregarded as not well behaved  (see the discussion below eq. (\ref{eq:Table1T}), (\ref{eq:sound_rigid}) and (\ref{eq:Langmuir_R32})). 

For $R_{4,2}(\zeta)$, the kinetic moments calculate
\begin{eqnarray}
  R_{4,2}(\zeta):\qquad D &=& \Big(1
    -i\sqrt{\pi}\frac{2}{(3\pi-8)}\zeta
    -\frac{(32-9\pi)}{(3\pi-8)}\zeta^2
    +i\sqrt{\pi}\frac{2(10-3\pi)}{(3\pi-8)}\zeta^3+\frac{2(16-5\pi)}{(3\pi-8)}\zeta^4 \Big);\\
 n^{(1)} &\sim& \frac{1}{D}\Big[ \frac{(5\pi-16)}{(3\pi-8)}\zeta^2 +i\sqrt{\pi}\frac{(3\pi-10)}{(3\pi-8)}\zeta + 1 \Big];\\
 u^{(1)} &\sim& \frac{1}{D}\Big[\frac{(5\pi-16)}{(3\pi-8)}\zeta^3 +i\sqrt{\pi}\frac{(3\pi-10)}{(3\pi-8)}\zeta^2 + \zeta \Big];\\
 T^{(1)} &\sim& \frac{1}{D}\Big[\frac{2(5\pi-16)}{(3\pi-8)}\zeta^2-i\sqrt{\pi}\zeta  \Big];\\
 q^{(1)} &\sim& \frac{1}{D}\Big[-i\sqrt{\pi} \frac{(9\pi-28)}{(3\pi-8)}\zeta^2-2\zeta  \Big];\\
 \widetilde{r}^{(1)} &\sim& \frac{1}{D}\Big[-i\sqrt{\pi} \frac{2(9\pi-28)}{(3\pi-8)}\zeta^3-\frac{2(21\pi-64)}{(3\pi-8)}\zeta^2+3i\sqrt{\pi}\zeta   \Big].
\end{eqnarray}
The only possibility is to express $\widetilde{r}^{(1)}$ through a combination of $u^{(1)}$, $T^{(1)}$ and $q^{(1)}$, and the solution is
\begin{equation} \label{eq:Static_R42_F}
  \boxed{
  R_{4,2}(\zeta):\quad \widetilde{r}^{(1)} = -i\sqrt{\pi} \frac{(10-3\pi)}{(16-5\pi)} v_{\textrm{th}} \sign(\kpar) q^{(1)}
  + \frac{(21\pi-64)}{2(16-5\pi)}v_{\textrm{th}}^2 n_{0} T^{(1)} + i\sqrt{\pi} \frac{(9\pi-28)}{(16-5\pi)}v_{\textrm{th}} T^{(0)} n_{0}\sign(\kpar) u^{(1)},}
\end{equation}
which is equivalent to the closure (\ref{eq:R42closure}), that was obtained for small $\zeta$ with precision $o(\zeta^3)$. Obviously such a closure
is precise for small values of $\zeta$, however for large values of $\zeta$, the asymptotic behavior of $q^{(1)}\sim \zeta^{-2}$
and $\widetilde{r}^{(1)}\sim \zeta^{-1}$, instead of the correct $\zeta^{-3},\zeta^{-4}$ profiles (see eqs. (\ref{eq:QlargeZZ}), (\ref{eq:rrlargeZZ})),
and these quantities will be therefore overestimated.
Nevertheless, the solution is interesting and we are not aware of it being reporting in any literature. 

Continuing with $R_{4,3}(\zeta)$, the kinetic moments calculate
\begin{eqnarray}
  R_{4,3}(\zeta): \qquad D &=& \Big( 1-i\frac{3\sqrt{\pi}}{2}\zeta-\frac{(9\pi-16)}{4}\zeta^2 +i\sqrt{\pi}\zeta^3+\frac{(3\pi-8)}{2}\zeta^4 \Big); \label{eq:R43_D}\\
  n^{(1)} &\sim& \frac{1}{D} \Big[ \frac{(8-3\pi)}{4}\zeta^2-i\frac{\sqrt{\pi}}{2}\zeta+1 \Big] ;\\
  u^{(1)} &\sim& \frac{1}{D} \Big[ \frac{(8-3\pi)}{4}\zeta^3-i\frac{\sqrt{\pi}}{2}\zeta^2+\zeta \Big];\\
  T^{(1)} &\sim& \frac{1}{D}\Big[ \frac{(8-3\pi)}{2}\zeta^2-i\sqrt{\pi}\zeta \Big];\\
  q^{(1)} &\sim& \frac{1}{D} \Big[ -2\zeta \Big];\\
   \widetilde{r}^{(1)} &\sim& \frac{1}{D}\Big[ \frac{(9\pi-32)}{2}\zeta^2 +3i\sqrt{\pi}\zeta \Big].
\end{eqnarray}
It is possible to express $\widetilde{r}^{(1)}$ through the combination of $q^{(1)}$ and $T^{(1)}$ and the result is
\begin{equation} \label{eq:Static_R43_F} \boxed{
  R_{4,3}(\zeta):\qquad \widetilde{r}^{(1)} = -i \frac{2\sqrt{\pi}}{(3\pi-8)}v_{\textrm{th}} \sign(\kpar) q^{(1)}
  + \frac{(32-9\pi)}{2(3\pi-8)}v_{\textrm{th}}^2 n_{0} T^{(1)},}
\end{equation}
which is equivalent to the closure (\ref{eq:R43closure}), that was obtained for small $\zeta$ with the precision $o(\zeta^2)$.
The heat flux has a correct asymptotic behavior $q^{(1)}\sim \zeta^{-3}$ (even though with incorrect proportionality constant), and the
quantity $\widetilde{r}^{(1)}\sim \zeta^{-2}$ instead of the correct $\sim \zeta^{-4}$. The closure was first reported by \cite{HammettPerkins1990}, and
is equivalent to the (non-numbered) expression between their eq. (10) and (11).

Continuing with $R_{4,4}(\zeta)$ approximant, the kinetic moments are (let's stop writing down $n^{(1)}$ from now on since we know we can get it from $u^{(1)}$)
\begin{eqnarray}
  R_{4,4}(\zeta): \qquad D &=& \Big(1-i\frac{3\sqrt{\pi}}{2}\zeta-4\zeta^2+i\sqrt{\pi}\zeta^3+\frac{4}{3}\zeta^4 \Big);\\
  u^{(1)} &\sim& \frac{1}{D} \Big[ -\frac{2}{3}\zeta^3-\frac{i}{2}\sqrt{\pi}\zeta^2+\zeta \Big];\\
  T^{(1)} &\sim& \frac{1}{D}\Big[ -\frac{4}{3}\zeta^2-i\sqrt{\pi}\zeta \Big];\\
  q^{(1)} &\sim& \frac{1}{D} \Big[ -2\zeta \Big];\\
   \widetilde{r}^{(1)} &\sim& \frac{1}{D}\Big[ 3i\sqrt{\pi}\zeta \Big].
\end{eqnarray}
It is possible to express $\widetilde{r}^{(1)}$ through $q^{(1)}$ and the closure is
\begin{eqnarray} \label{eq:Static_R44_F}
\boxed{
R_{4,4}(\zeta):\qquad \widetilde{r}^{(1)} = -i\frac{3}{4}\sqrt{\pi} v_{\textrm{th}} \sign(\kpar) q^{(1)}.}
\end{eqnarray}  
The result is equivalent to the (\ref{eq:R44closure}) closure, that obtained for small $\zeta$ with precision $o(\zeta)$.
This very simple closure has only precision $o(\zeta)$, however, it does have the correct asymptotic behavior of the heat
flux $q^{(1)}\sim -3/(2\zeta^{3})$ (including the proportionality constant), and $\widetilde{r}^{(1)}\sim \zeta^{-3}$ that
is closer to the correct $\zeta^{-4}$ than the previous closure.  

\newpage
\subsection{Table of moments $(u_\parallel,T_\parallel,q_\parallel,\widetilde{r}_{\parallel\parallel})$ for various Pad\'e approximants}
To clearly see possibilities of a closure, it is useful to create the following summarizing table, that is self-explanatory
after reading the previous section, i.e. all the proportionality constants (including the minus signs) are suppressed. Even though the table here is created for 1D geometry,
we will see that exactly the same table is constructed for 3D geometry, where all the quantities are given a ``parallel'' sub-index, i.e. $u^{(1)}\to u_\parallel^{(1)}$,
$T^{(1)}\to T_\parallel^{(1)}$, $q^{(1)}\to q_\parallel^{(1)}$ and $\widetilde{r}^{(1)}\to \widetilde{r}_{\parallel\parallel}^{(1)}$.
The table is therefore useful to spot all the possible closures that can be constructed in 1D geometry for quantities $q^{(1)}, \widetilde{r}^{(1)}$, as well
as in 3D geometry for quantities $q_{\parallel}^{(1)}$ and $r_{\parallel\parallel}^{(1)}$.
The approximants $R_{2,1}, R_{4,5}, R_{6,9}$ and $R_{8,13}$ are marked with an asterisk ``*''. These approximants are not well-behaved (because the Landau residue is not
  accounted for) and are provided only for completeness,
these approximants should be disregarded.\\\\
1-pole and 2-pole approximants\\
\begin{tabular}{| c | c | c | c |}
  \hline
  & $R_1$   & $R_{2,0}$     &  $R_{2,1}^*$ \\
  \hline
  $u^{(1)}$ & $\zeta$  & $\zeta$      &  $\zeta$  \\
  $T^{(1)}$ & $\zeta^2,\zeta$  & $\zeta$      &  $0$ \\
  $q^{(1)}$ & $\zeta^3,\zeta^2,\zeta$      &$\zeta^2,\zeta$ & $\zeta$ \\
  $\widetilde{r}^{(1)}$ & $\zeta^4\cdots\zeta$   &$\zeta^3,\zeta^2,\zeta$ & $\zeta^2$ \\
  \hline
\end{tabular}
\\
3-pole approximants\\
\begin{tabular}{| c | c | c | c |}
  \hline
  & $R_{3,0}$     &  $R_{3,1}$ &  $R_{3,2}$ \\
  \hline
  $u^{(1)}$ & $\zeta^2,\zeta$    & $\zeta^2,\zeta$ & $\zeta^2,\zeta$   \\
  $T^{(1)}$ & $\zeta^2,\zeta$    &  $\zeta$   &  $\zeta$ \\
  $q^{(1)}$ & $\zeta^3,\zeta^2,\zeta$ & $\zeta^2,\zeta$  & $\zeta$\\
  $\widetilde{r}^{(1)}$ & $\zeta^4\cdots\zeta$ & $\zeta^3,\zeta^2,\zeta$ & $\zeta^2,\zeta$\\
  \hline
\end{tabular}
\\
4-pole approximants\\
\begin{tabular}{| c | c | c | c | c | c | c |}
  \hline			
          & $R_{4,0}$                                            & $R_{4,1}$               & $R_{4,2}$                & $R_{4,3}$                & $R_{4,4}$                & $R_{4,5}^*$ \\
  \hline
$u^{(1)}$ & $\zeta^3,\zeta^2,\zeta$                             & $\zeta^3,\zeta^2,\zeta$ & $\zeta^3,\zeta^2,\zeta$ &  $\zeta^3,\zeta^2,\zeta$ & $\zeta^3,\zeta^2,\zeta$ & $\zeta^3,\zeta$ \\
$T^{(1)}$ & $\zeta^3,\zeta^2,\zeta$                             & $\zeta^2,\zeta$         & $\zeta^2,\zeta$         &  $\zeta^2,\zeta$        & $\zeta^2,\zeta$         & $\zeta^2$ \\
$q^{(1)}$ & $\zeta^4\cdots\zeta$                     & $\zeta^3,\zeta^2,\zeta$  & $\zeta^2,\zeta$         &  $\zeta$               & $\zeta$                 & $\zeta$ \\
$\widetilde{r}^{(1)}$ & $\zeta^5\cdots\zeta$ & $\zeta^4\cdots\zeta$ & $\zeta^3,\zeta^2,\zeta$ & $\zeta^2, \zeta$ & $\zeta$                 & $0$ \\
  \hline  
\end{tabular}
\\
5-pole approximants\\
\begin{tabular}{| c | c | c | c | c | c | c | c |}
  \hline			
        & $R_{5,0}$ &   $R_{5,1}$    & $R_{5,2}$      & $R_{5,3}$       & $R_{5,4}$          & $R_{5,5}$        & $R_{5,6}$ \\
  \hline
$u^{(1)}$ & $\zeta^4\cdots\zeta$ & $\zeta^4\cdots\zeta$     & $\zeta^4\cdots\zeta$     & $\zeta^4\cdots\zeta$      &  $\zeta^4\cdots\zeta$     & $\zeta^4\cdots\zeta$   & $\zeta^4\cdots\zeta$ \\
$T^{(1)}$ & $\zeta^4\cdots\zeta$ & $\zeta^3,\zeta^2,\zeta$  & $\zeta^3,\zeta^2,\zeta$  & $\zeta^3,\zeta^2,\zeta$   &  $\zeta^3,\zeta^2,\zeta$   &$\zeta^3,\zeta^2,\zeta$ & $\zeta^3,\zeta^2,\zeta$ \\
$q^{(1)}$ & $\zeta^5\cdots\zeta$ & $\zeta^4\cdots\zeta$  & $\zeta^3,\zeta^2,\zeta$  & $\zeta^2,\zeta$           &  $\zeta^2, \zeta$         & $\zeta^2, \zeta$       & $\zeta^2, \zeta$\\
$\widetilde{r}^{(1)}$ & $\zeta^6\cdots\zeta$ & $\zeta^5\cdots\zeta$ & $\zeta^4\cdots\zeta$     & $\zeta^3,\zeta^2,\zeta$   & $\zeta^2, \zeta$          & $\zeta$                & $\zeta$ \\
  \hline  
\end{tabular}
\\
6-pole approximants\\
\begin{tabular}{| c | c | c | c | c | c | c | c | c | c | c |}
  \hline			
  & $R_{6,0}$ &   $R_{6,1}$    & $R_{6,2}$      & $R_{6,3}$       & $R_{6,4}$          & $R_{6,5}$        & $R_{6,6}$
  & $R_{6,7}$          & $R_{6,8}$        & $R_{6,9}^*$\\
  \hline
  $u^{(1)}$ & $\zeta^5\cdots\zeta$ & $\zeta^5\cdots\zeta$     & $\zeta^5\cdots\zeta$     & $\zeta^5\cdots\zeta$      &  $\zeta^5\cdots\zeta$     & $\zeta^5\cdots\zeta$   & $\zeta^5\cdots\zeta$
  &  $\zeta^5\cdots\zeta$     & $\zeta^5\cdots\zeta$   & $\zeta^5,\zeta^3,\zeta$\\
  $T^{(1)}$ & $\zeta^5\cdots\zeta$ & $\zeta^4\cdots\zeta$  & $\zeta^4\cdots\zeta$  & $\zeta^4\cdots\zeta$   &  $\zeta^4\cdots\zeta$   &$\zeta^4\cdots\zeta$ & $\zeta^4\cdots\zeta$
  &  $\zeta^4\cdots\zeta$   &$\zeta^4\cdots\zeta$ & $\zeta^4,\zeta^2$\\
  $q^{(1)}$ & $\zeta^6\cdots\zeta$ & $\zeta^5\cdots\zeta$  & $\zeta^4\cdots\zeta$  & $\zeta^3\cdots\zeta$           &  $\zeta^3\cdots \zeta$         & $\zeta^3\cdots\zeta$       & $\zeta^3\cdots \zeta$
  &  $\zeta^3\cdots \zeta$         & $\zeta^3\cdots\zeta$       & $\zeta^3, \zeta$\\
  $\widetilde{r}^{(1)}$ & $\zeta^7\cdots\zeta$& $\zeta^6\cdots\zeta$ & $\zeta^5\cdots\zeta$ & $\zeta^4\cdots\zeta$   & $\zeta^3\cdots\zeta$   & $\zeta^2,\zeta$     & $\zeta^2,\zeta$  &
  $\zeta^2,\zeta$ & $\zeta^2,\zeta$ & $\zeta^2$\\
  \hline  
\end{tabular}\\\\

It is obvious by now that any higher-order Pad\'e approximants will not help to achieve a closure. Or is it ? One might still hope for ``a miracle'' thinking that perhaps the
7-pole and 8-pole approximants with the maximum-possible number of poles devoted to the asymptotic series - the $R_{7,10}$ and $R_{8,12}$ - might yield a closure. However, this is
unfortunately not the
case, and the table for 7-pole and 8-pole approximants reads\\
\begin{tabular}{| c | c | c | c | c || c | c | c | c | c |}
  \hline			
  & $R_{7,0}$ &   $R_{7,1}$    & $\cdots$      & $R_{7,10}$       & $R_{8,0}$          & $R_{8,1}$        & $\cdots$
  & $R_{8,12}$          & $R_{8,13}^*$\\
  \hline
  $u^{(1)}$ & $\zeta^6\cdots\zeta$ & $\zeta^6\cdots\zeta$  & $\cdots$  & $\zeta^6\cdots\zeta$   &  $\zeta^7\cdots\zeta$   &$\zeta^7\cdots\zeta$ & $\cdots$
  &  $\zeta^7\cdots\zeta$   &$\zeta^7,\zeta^5,\zeta^3,\zeta$ \\
  $T^{(1)}$ & $\zeta^6\cdots\zeta$ & $\zeta^5\cdots\zeta$  & $\cdots$  & $\zeta^5\cdots\zeta$   &  $\zeta^7\cdots\zeta$   &$\zeta^6\cdots\zeta$ & $\cdots$
  &  $\zeta^6\cdots\zeta$   &$\zeta^6,\zeta^4,\zeta^2$ \\
  $q^{(1)}$ & $\zeta^7\cdots\zeta$ & $\zeta^6\cdots\zeta$  & $\cdots$  & $\zeta^4\cdots\zeta$           &  $\zeta^8\cdots \zeta$         & $\zeta^7\cdots\zeta$       & $\cdots$
  &  $\zeta^5\cdots \zeta$         & $\zeta^5,\zeta^3,\zeta$\\
  $\widetilde{r}^{(1)}$ & $\zeta^8\cdots\zeta$& $\zeta^7\cdots\zeta$ & $\cdots$ & $\zeta^3\cdots\zeta$   & $\zeta^9\cdots\zeta$   & $\zeta^8\cdots\zeta$     & $\cdots$  &
  $\zeta^4\cdots\zeta$ & $\zeta^4,\zeta^2$\\
  \hline  
\end{tabular}\\\\

By observing the entire table, there are 7 possible quasi-static closures (that were already addressed): 
\begin{eqnarray}
  R_{3,1}: \quad q^{(1)} &\overset{\textrm{\checkmark}}{=}& \alpha_u u^{(1)} + \alpha_T T^{(1)};\nn\\
  R_{3,2}: \quad q^{(1)} &\overset{\textrm{\checkmark}}{=}& \alpha_T T^{(1)}; \qquad
  \cancel{\widetilde{r}^{(1)}\overset{\textrm{!}}{=}\alpha_u u^{(1)} + \alpha_q q^{(1)}}; \qquad
  \cancel{\widetilde{r}^{(1)} \overset{\textrm{x}}{=} \alpha_u u^{(1)} + \alpha_T T^{(1)}}; \nn\\
  R_{4,2}: \quad \widetilde{r}^{(1)} &\overset{\textrm{\checkmark}}{=}& \alpha_u u^{(1)} + \alpha_T T^{(1)} + \alpha_q q^{(1)};\nn\\
  R_{4,3}: \quad \widetilde{r}^{(1)} &\overset{\textrm{\checkmark}}{=}& \alpha_T T^{(1)} + \alpha_q q^{(1)};\nn\\
  R_{4,4}: \quad \widetilde{r}^{(1)} &\overset{\textrm{\checkmark}}{=}& \alpha_q q^{(1)}. \label{eq:Table1S}
\end{eqnarray}
There are also 13 time-dependent closures (that are addressed in the next section):
\begin{eqnarray}
  R_{3,2}: \quad \zeta q^{(1)} + \alpha_q q^{(1)} &\overset{\textrm{\checkmark}}{=}& \alpha_u u^{(1)};\qquad  \cancel{\zeta q^{(1)} \overset{\textrm{x}}{=} \alpha_u u^{(1)}+ \alpha_T T^{(1)}};\nn\\
  R_{4,2}: \quad \zeta q^{(1)} + \alpha_q q^{(1)} &\overset{\textrm{\checkmark}}{=}& \alpha_u u^{(1)}+ \alpha_T T^{(1)};\nn\\
  R_{4,3}: \quad \zeta q^{(1)}+\alpha_q q^{(1)} &\overset{\textrm{\checkmark}}{=}& \alpha_T T^{(1)}; \qquad
  \cancel{\zeta \widetilde{r}^{(1)}+\alpha_r\widetilde{r}^{(1)} \overset{\textrm{!}}{=} \alpha_u u^{(1)} + \alpha_q q^{(1)}}; \qquad
  \cancel{\zeta \widetilde{r}^{(1)} \overset{\textrm{x}}{=} \alpha_u u^{(1)} + \alpha_T T^{(1)}+ \alpha_q q^{(1)}};\nn\\
  R_{4,4}: \quad \zeta q^{(1)}+\alpha_q q^{(1)} &\overset{\textrm{\checkmark}}{=}& \alpha_T T^{(1)}; \qquad
  \zeta \widetilde{r}^{(1)}+\alpha_r\widetilde{r}^{(1)} \overset{\textrm{\checkmark}}{=} \alpha_T T^{(1)}; \qquad
  \cancel{\zeta \widetilde{r}^{(1)} \overset{\textrm{x}}{=} \alpha_T T^{(1)}+ \alpha_q q^{(1)}}; \nn\\
  R_{5,3}: \quad \zeta \widetilde{r}^{(1)}+\alpha_r\widetilde{r}^{(1)} &\overset{\textrm{\checkmark}}{=}& \alpha_u u^{(1)}+\alpha_T T^{(1)} +\alpha_q q^{(1)};\nn\\
  R_{5,4}: \quad \zeta \widetilde{r}^{(1)}+\alpha_r\widetilde{r}^{(1)} &\overset{\textrm{\checkmark}}{=}& \alpha_T T^{(1)} +\alpha_q q^{(1)};\nn\\
  R_{5,5}: \quad \zeta \widetilde{r}^{(1)}+\alpha_r\widetilde{r}^{(1)} &\overset{\textrm{\checkmark}}{=}& \alpha_q q^{(1)};\nn\\
  R_{5,6}: \quad \zeta \widetilde{r}^{(1)}+\alpha_r\widetilde{r}^{(1)} &\overset{\textrm{\checkmark}}{=}& \alpha_q q^{(1)}. \label{eq:Table1T}
\end{eqnarray}
New closures should be always checked. Later on, we will consider propagation of the ion-acoustic mode, satisfying kinetic dispersion
relation (\ref{eq:soundKin}). We believe that a good ``reliable'' closure of a fluid model obtained with $R_{n,n'}(\zeta)$ approximant, should
yield a fluid dispersion relation that is equivalent to (\ref{eq:soundKin}), after $R(\zeta)$ is replaced with $R_{n,n'}(\zeta)$
(equivalent to the numerator of (\ref{eq:soundKin}) once both terms are written with the common denominator).
Closures that satisfy this requirement are marked with ``\checkmark'' in the above table. Closures that do not satisfy this requirement were
eliminated, and can be further split to two categories. Both eliminated categories actually appear to describe the ion-acoustic mode with the 
same accuracy as a corresponding ``reliable'' closure satisfying (\ref{eq:soundKin}), however, the difference is in the higher-order modes.
The first category of eliminated closures, marked with ``x'', produces higher-order modes with positive growth rate, and these closures can not
be used for numerical simulations. The second category, marked with ``!'', produces higher-order modes that are damped, and these
closures can still be useful. However, there is no guarantee that these closures will behave well in different circumstances
(for example when used in 3D geometry) and these closures were therefore eliminated.
\clearpage
\subsection{Going back from Fourier space to Real space - the Hilbert transform}
The quasi-static Landau fluid closures explored in the previous section, were constructed in Fourier space. For direct numerical simulations that can use
Fourier transforms (that are usually restricted to periodic boundaries), or for solving dispersion relations $\omega(\boldsymbol{k})$, this is the easiest and natural way
how to implement these closures. Nevertheless, it is very beneficial to see how these advanced fluid closures translate to Real space. 

Provided all equations are linear (and homogeneous), transformation between Real and Fourier space is usually very easy and so far we just needed
\begin{equation} \label{eq:InverseF}
  \frac{\pr}{\pr t} \to -i\omega; \qquad \nabla\to i\boldsymbol{k}; \qquad f(\boldsymbol{x},t)\to \hat{f}(\boldsymbol{k},\omega),
\end{equation}
where we did not even bother to write the hat symbol on the quantities in Fourier space, since it was obvious and not necessary.

With equations encountered in simple fluid models, transformation back to Real space is easy and one can usually just flip the direction of
the arrow in relations (\ref{eq:InverseF}). However, the constructed Landau fluid closures contain an unusual operator $i\sign(\kpar)=i\kpar/|\kpar|$.
How does this operator transforms to Real space ?
Considering just spatial 1D transformation between coordinates $z \leftrightarrow \kpar$, a general function Fourier transforms according to
\begin{eqnarray}
  f(z) &=& \frac{1}{2\pi}\int_{-\infty}^\infty \hat{f}(\kpar) e^{i\kpar z} d\kpar \equiv \mathcal{F}^{-1} \hat{f}(\kpar); \label{eq:Fback}\\
  \hat{f}(\kpar) &=& \int_{-\infty}^\infty f(z) e^{-i\kpar z} dz \equiv \mathcal{F} f(z), \label{eq:Fforw}
\end{eqnarray}
where the first equation is the inverse/backward Fourier transform and the second equation is the forward Fourier transform. As usual, we often do not bother
to write the hat symbols on quantities in Fourier space. The location of the normalization
factor $1/(2\pi)$ is an ad-hoc choice, but one has to be consistent, especially when calculating convolutions. 
As a first step, we need to calculate $\mathcal{F}^{-1}$ of a function $\sign(\kpar)$. However, if such an integral is calculated directly,
one will find out that the result is not clearly defined.

It is beneficial to use a small trick, where instead of a function $\sign(\kpar)$, one considers function $\sign(\kpar)e^{-\alpha|\kpar|}$,
where $\alpha$ is some small positive constant $\alpha>0$. And after the calculation, one performs the limit $\alpha\to 0$. The considered function is 
\begin{equation}
 \sign(\kpar)e^{-\alpha|\kpar|} = 
  \begin{dcases}
    -e^{+\alpha\kpar}; &\text{$\kpar<0$};\\
    +e^{-\alpha\kpar}; & \text{$\kpar>0$},
  \end{dcases} 
\end{equation}
and the integral calculates
\begin{eqnarray}
  \int_{-\infty}^\infty \sign(\kpar) e^{-\alpha|\kpar|} e^{i\kpar z} d\kpar &=& \int_{-\infty}^0 (-1)  e^{+\alpha\kpar} e^{i\kpar z} d\kpar + \int_{0}^\infty (+1)  e^{-\alpha\kpar} e^{i\kpar z} d\kpar \nn\\
  &=& - \int_{-\infty}^0 e^{(+\alpha+iz)\kpar} d\kpar + \int_{0}^\infty e^{(-\alpha+iz)\kpar}d\kpar \nn\\
  &=& -\frac{1}{(\alpha+iz)} e^{(+\alpha+iz)\kpar} \Big|_{\kpar=-\infty}^0 + \frac{1}{(-\alpha+iz)} e^{(-\alpha+iz)\kpar} \Big|_{\kpar=0}^\infty \nn\\
  &=& -\frac{1}{(\alpha+iz)} -  \frac{1}{(-\alpha+iz)} = \frac{-(\alpha-iz)+(\alpha+iz)}{(\alpha+iz)(\alpha-iz)}
  = \frac{2iz}{\alpha^2+z^2},
\end{eqnarray}
further yielding
\begin{equation}
\mathcal{F}^{-1} \Big[ i \sign(\kpar) e^{-\alpha|\kpar|} \Big] = \frac{1}{2\pi}\int_{-\infty}^\infty i \sign(\kpar) e^{-\alpha|\kpar|} e^{i\kpar z} d\kpar = -\frac{z}{\pi(\alpha^2+z^2)}.
\end{equation}  
By taking the limit $\alpha\to 0$,
\begin{eqnarray} \label{eq:weird}
 \mathcal{F}^{-1} \Big[ i \sign(\kpar) \Big] = \frac{1}{2\pi}\int_{-\infty}^\infty i \sign(\kpar) e^{i\kpar z} d\kpar = -\frac{1}{\pi z}.
\end{eqnarray}

In Landau fluid closures, the operator $i\sign(\kpar)$ acts on a variable $\hat{f}(\kpar)$, and to transform this to Real space, we need to use a convolution theorem
for Fourier transforms. To make sure that we get the normalization factors right, let's calculate it in detail. The convolution between two real functions is defined as
\begin{equation} \label{eq:Convolution}
f(z)*g(z) \equiv  \int_{-\infty}^\infty f(z-z') g(z') dz'. 
\end{equation}
For brevity, let's temporarily suppress the parallel subscript on $\kpar$ and use only $k$.
By decomposing the function $f(z-z')$ to waves (using the inverse Fourier transform), $f(z-z')=\frac{1}{2\pi} \int_{-\infty}^\infty \hat{f}(\kpar) e^{ik(z-z')} dk$,
splitting the $e^{ik(z-z')}=e^{ikz}e^{-ikz'}$, and changing the order of integrals
\begin{eqnarray}
\int_{-\infty}^\infty f(z-z') g(z') dz' &=& \int_{-\infty}^\infty \Big[\frac{1}{2\pi} \int_{-\infty}^\infty \hat{f}(k) e^{ik(z-z')} dk \Big] g(z') dz'
  = \frac{1}{2\pi} \int_{-\infty}^\infty \Big[ \underbrace{\int_{-\infty}^\infty g(z') e^{-i k z'}dz'}_{=\hat{g}(k)} \Big]  \hat{f}(k) e^{ikz} dk \nn\\
  &=& \frac{1}{2\pi} \int_{-\infty}^\infty \hat{f}(k) \hat{g}(k) e^{ikz} dk = \mathcal{F}^{-1} \big[\hat{f}(k) \hat{g}(k)\big]. 
\end{eqnarray}
For normalizations (\ref{eq:Fback}), (\ref{eq:Fforw}), the required convolution theorem therefore reads
\begin{equation}
\mathcal{F}^{-1} \big[\hat{f}(\kpar) \hat{g}(\kpar)\big] = \big[\mathcal{F}^{-1} \hat{f}(\kpar)\big] * \big[\mathcal{F}^{-1} \hat{g}(\kpar)\big] = f(z) * g(z),
\end{equation}
and of course, $f(z) * g(z) = g(z) * f(z)$. Now it is straightforward to calculate how the $i\sign(\kpar)\hat{f}(\kpar)$ transforms to Real space
\begin{eqnarray}
  \mathcal{F}^{-1} \big[ i\sign(\kpar) \hat{f}(\kpar) \big] &=& \big[\mathcal{F}^{-1} i\sign(\kpar) \big] * \big[\mathcal{F}^{-1} \hat{f}(\kpar)\big]
  = -\frac{1}{\pi z} * f(z)
\end{eqnarray}
The convolution of $\frac{1}{\pi z}$ with a function $f(z)$ is a famous transformation, called the \emph{Hilbert transform}.
According to the
definition (\ref{eq:Convolution}), the convolution $\frac{1}{z}*f(z)$ should be defined as $\int_{-\infty}^\infty \frac{1}{z-z'} f(z') dz'$. However, because of
the singularity $\frac{1}{z-z'}$, such an integral will likely not exist, and the convolution integral is defined with a principal value.
The definition of the Hilbert transform ``$H$'' that is acting on a function $f(z)$ reads
\begin{equation}
H f(z) \equiv \frac{1}{\pi z} * f(z) \equiv \frac{1}{\pi} V.P. \int_{-\infty}^\infty \frac{f(z')}{z-z'} dz'.
\end{equation}
The use of the Hilbert transform allows a very elegant notation, how the $i\sign(\kpar) \hat{f}(\kpar)$ transforms to Real space, it is according to
\begin{equation}
\mathcal{F}^{-1} \big[ i\sign(\kpar) \hat{f}(\kpar) \big] = -\frac{1}{\pi z} * f(z) = - H f(z).
\end{equation}
Performing a lot of calculations, we like shortcuts, and the quantity $i\sign(\kpar)$ can be viewed as an operator, that is acting on many possible
$\hat{f}(\kpar)$ variables, such as the velocity $u^{(1)}$, the heat flux $q^{(1)}$, etc. (see the Landau fluid closures). Therefore, in addition
to the usual shortcuts (\ref{eq:InverseF}), we can write an elegant shortcut for the operator $i\sign(\kpar)$, that is very useful for advanced fluid models when transforming
from Fourier to Real space, and that reads
\begin{equation}
  \boxed{
 i\sign(\kpar) \to -H.}
\end{equation}
I.e., the operator $i\sign(\kpar)$ in Fourier space, is the \emph{negative} Hilbert transform operator in Real space.
Curiously, doesn't the Hilbert transform integral $\int_{-\infty}^\infty \frac{f(z')}{z-z'} dz'$ reminds us something ?
What about, if we prescribe the quantity $f(z')$ to be a Maxwellian $f(z')=e^{-z'^2}$ ?
Oh yes, this is the dreadful Landau integral ! This is how the plasma dispersion function was essentially defined.
This is indeed the reason, why the paper by Fried and Conte 1961, that is well-known for tabulating the properties of the plasma dispersion function,
has a full title: \emph{``The Plasma Dispersion Function. The Hilbert Transform of the Gaussian.''} 

Now we are ready to reformulate
the Landau fluid closures in Real space. Purely for convenience, often in modern Landau fluid papers
another operator $\mathcal{H}$ is defined that is equivalent to the negative Hilbert transform, and that absorbs the minus sign, i.e.
\begin{equation}
\boxed{  
  \mathcal{H}=-H.}
\end{equation}
This ``$\mathcal{H}$'' operator is therefore defined as
\begin{eqnarray}
  \mathcal{H} f(z) \equiv -\frac{1}{\pi z} * f(z) &\equiv& - \frac{1}{\pi} V.P. \int_{-\infty}^\infty \frac{f(z')}{z-z'} dz'
  = \frac{1}{\pi} V.P. \int_{-\infty}^\infty \frac{f(z')}{z'-z} dz' \nn\\
  &=& - \frac{1}{\pi} V.P. \int_{-\infty}^\infty \frac{f(z-z')}{z'} dz',
\end{eqnarray}
and allows us to write
\begin{equation}
\mathcal{F}^{-1} \big[ i\sign(\kpar) \hat{f}(\kpar) \big] = \mathcal{H} f(z),
\end{equation}
or in the operator shortcut
\begin{equation}
  \boxed{
 i\sign(\kpar) \to \mathcal{H}.} \label{eq:GoldH}
\end{equation}
This new definition is of course not necessary. However, it is often used in Landau fluid papers, and there is indeed some logic behind it.
First of all, we do not have to remember another minus sign, and we will make less typos, perhaps.
Second, the definition is consistent with another ``spatial'' operator Fourier shortcut $i\kpar\to \pr_z$. Third, the Landau integral and
the plasma dispersion function were defined with integrals $\int \frac{f(x)}{x-x_0}dx$, and not as $\int \frac{f(x)}{x_0-x}dx$,
and the $\mathcal{H}$ operator therefore can feel more natural than $H$. Whatever the choice, we now talked about it detail, and
all possible confusion between $H$ and $\mathcal{H}$ should be clarified. We will use the $\mathcal{H}$ operator henceforth.

\subsection{Quasi-static closures in Real space}
With our new shortcut (\ref{eq:GoldH}) as discussed above, the transformation of closures from Fourier space to Real space is very easy.
For example, the heat flux closure obtained for $R_{3,2}(\zeta)$ that in Fourier space reads $q^{(1)} = -i \frac{2}{\sqrt{\pi}} n_0 v_{\textrm{th}} \sign(\kpar)T^{(1)}$,
is transferred to Real space as
\begin{equation}
  \boxed{
R_{3,2}(\zeta):\qquad q^{(1)} (z) = -\frac{2}{\sqrt{\pi}} n_0 v_{\textrm{th}} \mathcal{H} T^{(1)} (z). } \label{eq:Qcollisionless}
\end{equation} 
Again, the $\mathcal{H}$ operator shows its slight advantage over $H$ operator, because it is easy to remember that for the usual
\emph{collisional heat flux} $q\sim-\pr_z T$, whereas for the \emph{collisionless heat flux} $q\sim -\mathcal{H}T$.

Let's rewrite the Hilbert transform a bit further, so that we can clearly see what this distinction means physically. Rewriting the principal value
\begin{eqnarray}
  \mathcal{H} f(z) = - \frac{1}{\pi} V.P. \int_{-\infty}^\infty \frac{f(z-z')}{z'} dz'
  = -\frac{1}{\pi} \lim_{\epsilon\to +0} \Big[   \int_{-\infty}^{-\epsilon} + \int_{\epsilon}^\infty \Big] \frac{f(z-z')}{z'} dz',
\end{eqnarray}
using the substitution $z'=-y$ in the first integral (so that $dz'=-dy$ and $-\infty\to \infty$, $-\epsilon \to\epsilon$),
\begin{eqnarray}
  \mathcal{H} f(z) = -\frac{1}{\pi} \lim_{\epsilon\to +0} \Big[ \int_{\infty}^\epsilon \frac{f(z+y)}{-y} (-dy) +\int_{\epsilon}^\infty  \frac{f(z-z')}{z'} dz' \Big], 
\end{eqnarray}
and renaming back $y\to z'$, the $\mathcal{H}$ operator reads
\begin{eqnarray}
  \mathcal{H} f(z) &=& -\frac{1}{\pi} \lim_{\epsilon\to +0} \int_{\epsilon}^\infty \Big[-\frac{f(z+z')}{z'} dz' +\frac{f(z-z')}{z'} dz' \Big]\nn\\
  &=& \frac{1}{\pi} \lim_{\epsilon\to +0} \int_{\epsilon}^\infty \frac{f(z+z')-f(z-z')}{z'}dz'.
\end{eqnarray}
Instead of remembering the limit, it is more elegant to write the final result as $V.P.\int_{0}^\infty$.
In the simplest closure (\ref{eq:Qcollisionless}), the collisionless heat flux is therefore expressed with respect to the temperature as
\begin{equation}
  \boxed{
R_{3,2}(\zeta): \qquad q^{(1)} (z) = -\frac{2}{\pi^{\frac{3}{2}}} n_0 v_{\textrm{th}} V.P.\int_{0}^\infty \frac{T^{(1)}(z+z')-T^{(1)}(z-z')}{z'}dz',} \label{eq:LandauT}
\end{equation}
which is equivalent to the eq. (8) of \cite{HammettPerkins1990}.
Writing the Hilbert transform and the collisionless heat flux in this form is very useful, because it reveals what the
Hilbert transform of the temperature means physically. The equation says that to obtain the heat flux in Real space,
one has to calculate integrals - and sum the differences between temperatures according to (\ref{eq:LandauT}) - along the entire considered coordinate z.
Here we calculated the expressions in the linear setting/approximation, and in reality, the integrals (\ref{eq:LandauT})
should be performed along the magnetic field lines.

What is perhaps the most non-intuitive and most surprising about the expression (\ref{eq:LandauT}),
that the expression is telling us that the \emph{entire} temperature profile along a magnetic field line is important,
since it will be encountered in the integral (\ref{eq:LandauT}).
Therefore, the collisionless heat flux $q(z)$ at some spatial point z, depends on the temperature difference between that point,
and the temperature along the entire magnetic field line.
This effect is summarized with an appropriate word of \emph{non-locality}
of the collisionless heat flux, since it is in strong contrast with the usual collisional heat flux, that depends only at the
local gradient of the temperature at that point. For time-evolving systems, this effect is also directly associated with the
``isotropization'' of temperature along the magnetic field lines.
 Physically, the effect of non-locality in collisionless plasma is caused by particles that can freely stream along the magnetic field lines.
  Locality in collisional transport is caused by collisions, which introduces a mean free path.

To rewrite the other quasi-static closures that were explored in the previous section to the Real space is trivial, and for example
the quasi-static closure (\ref{eq:Static_R43_F}) of Hammett and Perkins 1990 obtained with $R_{4,3}(\zeta)$ reads
\begin{equation} \boxed{
    R_{4,3}(\zeta):\qquad \widetilde{r}^{(1)}(z) = -\frac{2\sqrt{\pi}}{(3\pi-8)}v_{\textrm{th}} \mathcal{H} q^{(1)}(z)
    + \frac{(32-9\pi)}{2(3\pi-8)}v_{\textrm{th}}^2 n_{0} T^{(1)}(z). \label{eq:Static_R43} }
\end{equation}
The closure (\ref{eq:Static_R42_F}) obtained with 4-pole approximant $R_{4,2}(\zeta)$ is rewritten to Real space as
\begin{equation} \label{eq:Static_R42} \boxed{
R_{4,2}(\zeta):\quad  \widetilde{r}^{(1)}(z) = - \frac{\sqrt{\pi}(10-3\pi)}{(16-5\pi)} v_{\textrm{th}} \mathcal{H} q^{(1)}(z)
+ \frac{(21\pi-64)}{2(16-5\pi)}v_{\textrm{th}}^2 n_{0} T^{(1)}(z) +  \frac{\sqrt{\pi}(9\pi-28)}{(16-5\pi)}v_{\textrm{th}} T^{(0)} n_{0}\mathcal{H} u^{(1)}(z),}
\end{equation}    
the closure (\ref{eq:Static_R44_F}) obtained with $R_{4,4}(\zeta)$ is rewritten as
\begin{equation} \label{eq:Static_R44} \boxed{
    R_{4,4}(\zeta):\quad \widetilde{r}^{(1)}(z) = -\frac{3}{4}\sqrt{\pi} v_{\textrm{th}} \mathcal{H} q^{(1)}(z),}
\end{equation}  
and the closure (\ref{eq:Static_R31_F}) obtained with $R_{3,1}(\zeta)$ reads
\begin{equation} \boxed{
R_{3,1}(\zeta):\qquad q^{(1)} = \frac{(3\pi-8)}{(4-\pi)}n_0 T^{(0)} u^{(1)}-\frac{\sqrt{\pi}}{(4-\pi)}n_0 v_{\textrm{th}}\mathcal{H} T^{(1)}.}
\end{equation}
The closures (\ref{eq:Static_R32_F2}), (\ref{eq:Static_R32_F3}) obtained with $R_{3,2}(\zeta)$ read 
\begin{eqnarray}
R_{3,2}(\zeta): \qquad
  \widetilde{r}^{(1)} &=& -\frac{4}{\sqrt{\pi}}v_{\textrm{th}} n_0 T^{(0)}\mathcal{H} u^{(1)} -\frac{(3\pi+8)}{4\sqrt{\pi}}v_{\textrm{th}}\mathcal{H} q^{(1)};\\
R_{3,2}(\zeta): \qquad  \widetilde{r}^{(1)} &=& -\frac{4}{\sqrt{\pi}}v_{\textrm{th}} n_0 T^{(0)}\mathcal{H} u^{(1)} -\frac{(3\pi+8)}{2\pi}v_{\textrm{th}}^2 n_0 T^{(1)},
\end{eqnarray}
however, these closures are not ``reliable'' and will be eliminated, see the discussion below eq. (\ref{eq:Table1T}), (\ref{eq:sound_rigid}) and (\ref{eq:Langmuir_R32}).
To summarize, we obtained altogether 7 quasi-static closures.
Additionally,  one closure was disregarded since
it was obtained with approximant $R_{2,1}(\zeta)$ that is not a well-behaved approximant.  
\clearpage
\subsection{Time-dependent (dynamic) closures}
In addition to the ``quasi-static'' closures explored above (sometimes called simply ``static''), it is possible to construct a different class of closures that we can call
``time-dependent'' closures, or ``dynamic'' closures.
For example, for the approximant $R_{4,3}(\zeta)$, the temperature $T^{(1)}$ and the heat flux $q^{(1)}$ read
\begin{eqnarray}
R_{4,3}(\zeta):\quad  T^{(1)} &=& -q_r\frac{\Phi}{D}\Big[ \frac{8-3\pi}{2}\zeta^2-i\sqrt{\pi}\zeta \Big];\\
  q^{(1)} &=& -q_r\frac{\Phi}{D}n_0 v_{\textrm{th}} \sign(\kpar)\Big[ -2\zeta \Big],
\end{eqnarray}  
where $D$ is the denominator of $R_{4,3}(\zeta)$ defined in (\ref{eq:R43_D}). Calculating the ratio
\begin{equation}
\frac{ T^{(1)}}{ q^{(1)}} = \frac{1}{n_0 v_{\textrm{th}} \sign(\kpar)} \Big[ \frac{3\pi-8}{4}\zeta + \frac{i\sqrt{\pi}}{2} \Big],
\end{equation}
using the definition $\zeta=\frac{\omega}{|\kpar|v_{\textrm{th}}}$ and multiplying by $|\kpar|v_{\textrm{th}}$ and $n_0 v_{\textrm{th}} \sign(\kpar)$, allows us to formulate
a closure
\begin{equation}
R_{4,3}(\zeta):\qquad \Big[ \frac{3\pi-8}{4}\omega + i\frac{\sqrt{\pi}}{2}v_{\textrm{th}} |\kpar| \Big] q^{(1)} = n_0 v_{\textrm{th}}^2 \kpar T^{(1)},
\end{equation}
that is further rewritten as
\begin{equation} \label{eq:T_R43}  
R_{4,3}(\zeta):\qquad \Big[ -i\omega + \frac{2\sqrt{\pi}}{(3\pi-8)}v_{\textrm{th}} |\kpar| \Big] q^{(1)} = - \frac{4n_0 v_{\textrm{th}}^2}{(3\pi-8)} i \kpar T^{(1)}.
\end{equation}
To go back to Real space, we need a recipe for the inverse Fourier transform of operator $|\kpar|$, that acts on a general quantity $\hat{f}(\kpar)$.
The transform calculates easily by using $|\kpar|=-(i\kpar)i\sign(\kpar)$ and writing
\begin{eqnarray}
  \mathcal{F}^{-1} \Big[ |\kpar| \hat{f}(\kpar) \Big] &=& \frac{1}{2\pi}\int_{-\infty}^\infty (-1)(i\kpar)i\sign(\kpar) \hat{f}(\kpar) e^{i\kpar z}d\kpar
  = -\frac{\pr}{\pr z} \frac{1}{2\pi}\int_{-\infty}^\infty i\sign(\kpar) \hat{f}(\kpar) e^{i\kpar z}d\kpar \nn\\
  &=&  -\frac{\pr}{\pr z} \mathcal{F}^{-1} \Big[ i\sign(\kpar) \hat{f}(\kpar) \Big] = -\frac{\pr}{\pr z} \mathcal{H} f(z),
\end{eqnarray}
that allows us to write a useful shortcut
\begin{equation}
  \boxed{
|\kpar| \to -\frac{\pr}{\pr z} \mathcal{H}.  }    
\end{equation}
The closure (\ref{eq:T_R43}) therefore transforms to Real space as
\begin{eqnarray}
  \Big[ \frac{\pr}{\pr t} - \frac{2\sqrt{\pi}}{(3\pi-8)}v_{\textrm{th}} \frac{\pr}{\pr z} \mathcal{H} \Big] q^{(1)}(z)
  = -\frac{4n_0 v_{\textrm{th}}^2}{(3\pi-8)} \frac{\pr}{\pr z} T^{(1)}(z), \label{eq:Tclosure_R43s}
\end{eqnarray}
and represents the time-dependent evolution equation for the heat flux.
The last step in these type of Landau fluid closures is to recover Galilean invariance, that is achieved by substituting $\pr/\pr t$ with
the convective derivative $d/dt$, and the final closure reads
\begin{equation}
  \boxed{ R_{4,3}(\zeta):\qquad
 \Big[ \frac{d}{d t} - \frac{2\sqrt{\pi}}{(3\pi-8)} v_{\textrm{th}}\frac{\pr}{\pr z} \mathcal{H} \Big] q^{(1)}(z)
  = -\frac{4n_0 v_{\textrm{th}}^2}{(3\pi-8)} \frac{\pr}{\pr z} T^{(1)}(z).} \label{eq:Tclosure_R43}
\end{equation}
To easily compare this expression with the existing literature, a small rearrangement yields
\begin{equation}  
  \Big[ \frac{d}{dt} + \frac{\sqrt{\pi}}{4(1-\frac{3\pi}{8})}v_{\textrm{th}} \frac{\pr}{\pr z} \mathcal{H} \Big] q^{(1)}(z)
  = \frac{n_0 v_{\textrm{th}}^2}{2(1-\frac{3\pi}{8})} \frac{\pr}{\pr z} T^{(1)}(z).
\end{equation}
The expression is equal for example to equation (57) in \cite{PassotSulem2003b} for the parallel heat flux $q_\parallel$
(where in that paper $v_{\textrm{th}}=\sqrt{T^{(0)}/m}$ is used, whereas ours here is  $v_{\textrm{th}}=\sqrt{2 T^{(0)}/m}$). 

The time-dependent closure (\ref{eq:Tclosure_R43}) was obtained with the approximant $R_{4,3}(\zeta)$. Interestingly, if the derivative $d/dt$ is neglected,
the closure is equivalent to the quasi-static closure (\ref{eq:Qcollisionless}) obtained with $R_{3,2}(\zeta)$ (which can be easily seen in Fourier space, or
by using $\mathcal{H}\mathcal{H}=-1$).
Also, it is useful to compare the time-dependent (\ref{eq:Tclosure_R43}) with the quasi-static closure (\ref{eq:Static_R43}),
that was obtained for the same approximant $R_{4,3}(\zeta)$. To compare these closures, we need to use a time-dependent heat flux equation where the closure
for $\widetilde{r}$ will be applied. In Part 1 of this guide, we derived nonlinear ``fluid'' equation for the parallel heat flux $q_\parallel$
(see Part 1, Section ``Collisionless damping in fluid models - Landau fluid models''). 
Quickly rewriting it in the 1D parallel geometry that we use here yields 
(dropping the parallel subscript)
\begin{equation}
\frac{\pr q}{\pr t} +\pr_z (qu) +\pr_z \widetilde{r} +3p\pr_z \big( \frac{p}{\rho}\big) +3q\pr_z u =0,
\end{equation}
where for brevity $\pr/\pr z=\pr_z$. The equation can be of course obtained by direct integration of the 1D Vlasov equation
$\pr f/\pr t +v\pr_z f + (q_r/m_r)E \pr f/\pr v =0$, as done by \cite{HammettPerkins1990}, and prescribing Maxwellian by $r=3p^2/\rho+\widetilde{r}$.  
The equation is nonlinear and to compare closures that were done at the linear level, we need to linearize the heat flux equation. This eliminates the
2nd and the last term, the 4rd term is linearized as $3(p_0/m)\pr_z T^{(1)}$, and since $p_0/m=n_0 v_{\textrm{th}}^2/2$, the linearized equation reads
\begin{equation} \label{eq:1DQstatic}
\frac{\pr q^{(1)}}{\pr t} +\pr_z \widetilde{r}^{(1)} +\frac{3}{2}n_0 v_{\textrm{th}}^2 \pr_z T^{(1)} =0.
\end{equation}
This is just a 1D linear heat flux equation, where no closure was imposed yet. The quantities $q^{(1)},\widetilde{r}^{(1)}, T^{(1)}$ were not calculated
from kinetic theory by using approximants to $R(\zeta)$, etc. The equation was obtained by a general ``fluid approach'', that we heavily used before we started
to consider kinetic calculations (perturbations around Maxwellian are assumed here because of the prescribed $r$).
The equation (\ref{eq:1DQstatic}) greatly clarifies relations between the quasi-static and time-dependent Landau fluid closures. For example, by using 
the quasi-static closure (\ref{eq:Static_R43}) in the heat flux equation (\ref{eq:1DQstatic}),
the time-dependent closure (\ref{eq:Tclosure_R43s}) is immediately recovered.

Often, time-dependent closures can not be straightforwardly constructed by a simple division of two moments as done above. It is useful to learn a new technique
that will allow us to see and construct possible closures in a quicker way. Let's explore the closure (\ref{eq:T_R43}). It is apparent that
whenever we attempt to use $\pr/\pr t$ of some moment (in this case $\pr q^{(1)}/\pr t$), it is logical to also use the same moment without the time derivative ($q^{(1)}$)
in the construction of the considered closure, i.e., in this case we search for a closure
\begin{equation}
R_{4,3}(\zeta):\qquad \big( \zeta + \alpha_q \big) q^{(1)} = \alpha_T T^{(1)},
\end{equation}
where $\alpha_q,\alpha_T$ need to be determined. By using expressions for $q^{(1)}, T^{(1)}$, the above closure is separated to 2 equations for $\zeta$ and $\zeta^2$
that must be satisfied independently if the closure is valid for all $\zeta$, and solving these 2 equations yields
\begin{equation}
  \alpha_q = i\frac{2\sqrt{\pi}}{(3\pi-8)}; \quad \alpha_T = \frac{4 n_0 v_{\textrm{th}}}{(3\pi-8)} \sign(\kpar).
\end{equation}
The closure therefore reads
\begin{equation}
R_{4,3}(\zeta):\qquad \Big[ \zeta + i\frac{2\sqrt{\pi}}{(3\pi-8)} \Big] q^{(1)} = \frac{4 n_0 v_{\textrm{th}}}{(3\pi-8)} \sign(\kpar) T^{(1)},
\end{equation}
and is of course equivalent to (\ref{eq:T_R43}).

We are now ready to construct all other possible time-dependent closures. Still considering $R_{4,3}(\zeta)$ approximant, another possible closure is
\begin{equation} \label{eq:T_R43_2}
R_{4,3}(\zeta):\qquad (\zeta+\alpha_r) \widetilde{r}^{(1)} = \alpha_u u^{(1)} + \alpha_q q^{(1)},
\end{equation}
which when separated into 3 equations for $\zeta,\zeta^2$ and $\zeta^3$ that must be each satisfied yields
\begin{equation}
   \alpha_r = \frac{i16\sqrt{\pi}}{(32-9\pi)(3\pi-8)};\quad
   \alpha_u = \frac{(32-9\pi)}{(3\pi-8)}v_{\textrm{th}} n_0 T^{(0)} \sign(\kpar); \quad
   \alpha_q =  \frac{(81\pi^2-552\pi+1024)}{2(32-9\pi)(3\pi-8)}v_{\textrm{th}}\sign(\kpar),
\end{equation}
the closure reads
\begin{eqnarray}
  R_{4,3}(\zeta):\quad \Big[-i\omega+\frac{16\sqrt{\pi}}{(32-9\pi)(3\pi-8)}v_{\textrm{th}} |\kpar| \Big] \widetilde{r}^{(1)}
  &=& -\frac{(32-9\pi)}{(3\pi-8)}v_{\textrm{th}}^2 n_0 T^{(0)} i\kpar u^{(1)} - \frac{(81\pi^2-552\pi+1024)}{2(32-9\pi)(3\pi-8)}v_{\textrm{th}}^2 i\kpar q^{(1)}; \nn\\
  \Big[\frac{d}{dt}-\frac{16\sqrt{\pi}}{(32-9\pi)(3\pi-8)}v_{\textrm{th}} \pr_z \mathcal{H}\Big] \widetilde{r}^{(1)}
  &=& -\frac{(32-9\pi)}{(3\pi-8)}v_{\textrm{th}}^2 n_0 T^{(0)} \pr_z u^{(1)} - \frac{(81\pi^2-552\pi+1024)}{2(32-9\pi)(3\pi-8)}v_{\textrm{th}}^2 \pr_z q^{(1)},\nn\\
\end{eqnarray}
and this closure will be eliminated.

Another closure with $R_{4,3}(\zeta)$ can be constructed as
\begin{eqnarray} \label{eq:T_R43_3}
&& R_{4,3}(\zeta):\qquad \zeta \widetilde{r}^{(1)} = \alpha_u u^{(1)} + \alpha_T T^{(1)} + \alpha_q q^{(1)};\\
&&  \alpha_u = \frac{(32-9\pi)}{(3\pi-8)}v_{\textrm{th}} n_0 T^{(0)} \sign(\kpar); \quad
  \alpha_T = -i\frac{8\sqrt{\pi}}{(3\pi-8)^2}v_{\textrm{th}}^2 n_0; \quad
  \alpha_q = -\frac{(27\pi^2-160\pi+256)}{2(3\pi-8)^2}v_{\textrm{th}}\sign(\kpar), \nn
\end{eqnarray}
so the closure reads 
\begin{eqnarray}
  R_{4,3}(\zeta):\quad -i\omega \widetilde{r}^{(1)} &=& -\frac{(32-9\pi)}{(3\pi-8)}v_{\textrm{th}}^2 n_0 T^{(0)} i\kpar u^{(1)}
  -\frac{8\sqrt{\pi}}{(3\pi-8)^2}v_{\textrm{th}}^3 n_0 |\kpar| T^{(1)} +\frac{(27\pi^2-160\pi+256)}{2(3\pi-8)^2}v_{\textrm{th}}^2 i\kpar q^{(1)};\nn\\
  \frac{d}{dt}\widetilde{r}^{(1)} &=& -\frac{(32-9\pi)}{(3\pi-8)}v_{\textrm{th}}^2 n_0 T^{(0)} \pr_z u^{(1)}
  +\frac{8\sqrt{\pi}}{(3\pi-8)^2}v_{\textrm{th}}^3 n_0 \pr_z \mathcal{H} T^{(1)} +\frac{(27\pi^2-160\pi+256)}{2(3\pi-8)^2}v_{\textrm{th}}^2 \pr_z q^{(1)},\nn\\
\end{eqnarray}
and this closure will be eliminated as well.
The time-dependent closures (\ref{eq:T_R43_3}) and (\ref{eq:T_R43_2}) are of course closely related, and one can go from one to another by using the
quasi-static closure (\ref{eq:Static_R43}) that expresses $\widetilde{r}^{(1)}$ as a combination of $T^{(1)}$ and $q^{(1)}$. 

Continuing with $R_{4,2}(\zeta)$ approximant, it is possible to construct the following closure
\begin{eqnarray}
&& R_{4,2}(\zeta):\qquad (\zeta +\alpha_q) q^{(1)} = \alpha_T T^{(1)} + \alpha_u u^{(1)};\\
&&  \alpha_q = i\sqrt{\pi}\frac{10-3\pi}{16-5\pi}; \quad \alpha_T = n_0 v_{\textrm{th}} \sign(\kpar) \frac{3\pi-8}{16-5\pi};
  \quad \alpha_u = i n_0 T^{(0)}\sqrt{\pi}\frac{9\pi-28}{16-5\pi},\nn
\end{eqnarray}
that implies
\begin{equation}
 \Big[ -i\omega +\sqrt{\pi}\frac{10-3\pi}{16-5\pi} v_{\textrm{th}} |\kpar| \Big] q^{(1)}
  = -n_0 v_{\textrm{th}}^2 \frac{3\pi-8}{16-5\pi} i\kpar T^{(1)}
  +n_0 T^{(0)}v_{\textrm{th}} \sqrt{\pi}\frac{9\pi-28}{16-5\pi} |\kpar| u^{(1)}; \nn
\end{equation}
\begin{equation} \boxed{
    R_{4,2}(\zeta):\qquad
     \Big[ \frac{d}{d t} -\sqrt{\pi}\frac{10-3\pi}{16-5\pi} v_{\textrm{th}} \pr_z\mathcal{H} \Big] q^{(1)}
  = -n_0 v_{\textrm{th}}^2 \frac{3\pi-8}{16-5\pi} \pr_z T^{(1)}
  -n_0 T^{(0)}v_{\textrm{th}} \sqrt{\pi}\frac{9\pi-28}{16-5\pi} \pr_z\mathcal{H} u^{(1)},}  \label{eq:R42_closure}
\end{equation}
and the result is consistent with using the quasi-static closure (\ref{eq:Static_R42}) in the linearized heat flux equation (\ref{eq:1DQstatic}).

Continuing with $R_{4,4}(\zeta)$, it is possible to construct
\begin{eqnarray}
&&  R_{4,4}(\zeta): \qquad (\zeta +\alpha_q) q^{(1)} = \alpha_T T^{(1)};\\
&& \alpha_q = i\frac{3\sqrt{\pi}}{4}; \quad \alpha_T = \frac{3}{2}n_0 v_{\textrm{th}} \sign(\kpar),\nn
\end{eqnarray}
and the closure reads
\begin{equation}
\Big[ -i\omega + \frac{3\sqrt{\pi}}{4}v_{\textrm{th}}|\kpar| \Big] q^{(1)} = -\frac{3}{2}n_0 v_{\textrm{th}}^2 i\kpar T^{(1)};\nn
\end{equation}
\begin{equation} \label{eq:Tclosure_R44} \boxed{
 R_{4,4}(\zeta): \qquad \Big[ \frac{d}{dt} - \frac{3\sqrt{\pi}}{4}v_{\textrm{th}}\pr_z \mathcal{H} \Big] q^{(1)}
  = -\frac{3}{2}n_0 v_{\textrm{th}}^2 \pr_z T^{(1)}.} 
\end{equation}  
The obtained closure is related to the quasi-static closure  (\ref{eq:Static_R44}), since by using the quasi-static closure  (\ref{eq:Static_R44}) 
in the linear heat flux equation (\ref{eq:1DQstatic}), the time-dependent closure (\ref{eq:Tclosure_R44}) is recovered.

Another closure with $R_{4,4}(\zeta)$ is
\begin{eqnarray}
&&  R_{4,4}(\zeta): \qquad (\zeta +\alpha_r) \widetilde{r}^{(1)} = \alpha_T T^{(1)}; \\
&& \alpha_r = i\frac{3\sqrt{\pi}}{4} ; \quad \alpha_T = -i\frac{9\sqrt{\pi}}{8}v_{\textrm{th}}^2 n_0;\nn\\
&& \Big[ -i\omega + \frac{3\sqrt{\pi}}{4}v_{\textrm{th}}|\kpar| \Big] \widetilde{r}^{(1)} = -\frac{9\sqrt{\pi}}{8}v_{\textrm{th}}^3 n_0 |\kpar| T^{(1)};\nn
\end{eqnarray}
\begin{equation} \boxed{
R_{4,4}(\zeta): \qquad \Big[ \frac{d}{dt} - \frac{3\sqrt{\pi}}{4}v_{\textrm{th}}\pr_z\mathcal{H} \Big] \widetilde{r}^{(1)} = +\frac{9\sqrt{\pi}}{8}v_{\textrm{th}}^3 n_0\pr_z\mathcal{H}  T^{(1)}.} \label{eq:T_R44_2}
\end{equation}
And yet another closure with $R_{4,4}(\zeta)$
\begin{eqnarray}
  &&  R_{4,4}(\zeta): \qquad \zeta \widetilde{r}^{(1)} = \alpha_T T^{(1)} +\alpha_q q^{(1)}; \\
  && \alpha_T = -i\frac{9\sqrt{\pi}}{8}v_{\textrm{th}}^2 n_0; \quad \alpha_q = -\frac{9\pi}{16}v_{\textrm{th}} \sign(\kpar);\nn\\
  && -i\omega \widetilde{r}^{(1)} = -\frac{9\sqrt{\pi}}{8}v_{\textrm{th}}^3 n_0 |\kpar| T^{(1)} +\frac{9\pi}{16}v_{\textrm{th}}^2 i\kpar q^{(1)};\nn\\
  && \frac{d}{dt}\widetilde{r}^{(1)} = \frac{9\sqrt{\pi}}{8}v_{\textrm{th}}^3 n_0 \pr_z \mathcal{H} T^{(1)} +\frac{9\pi}{16}v_{\textrm{th}}^2 \pr_z q^{(1)}. \label{eq:T_R44_3}
\end{eqnarray}
The closure (\ref{eq:T_R44_3}) is related to the closure (\ref{eq:T_R44_2}), because one can use the quasi-static closure (\ref{eq:Static_R44}) to
express $\widetilde{r}^{(1)}$ through $q^{(1)}$, however, the closure (\ref{eq:T_R44_3}) will be eliminated. 

The $R_{4,5}(\zeta)$ was eliminated because it is not a well-behaved approximant (see discussion above), nevertheless, for completeness the following closure can be constructed
\begin{eqnarray}
  && R_{4,5}(\zeta):\qquad \zeta q^{(1)} = \alpha_T T^{(1)}; \qquad \alpha_T = \frac{3}{2}n_0 v_{\textrm{th}} \sign(\kpar);\\
  && -i\omega q^{(1)} =-\frac{3}{2}n_0 v_{\textrm{th}}^2 i\kpar T^{(1)}; \qquad \frac{d}{dt} q^{(1)} =-\frac{3}{2}n_0 v_{\textrm{th}}^2 \pr_z T^{(1)}.\nn
\end{eqnarray}

With $R_{3,2}(\zeta)$, the following time-dependent closure can be constructed
\begin{eqnarray}
  && R_{3,2}(\zeta):\qquad \big( \zeta +\alpha_q \big) q^{(1)} = \alpha_u u^{(1)}; \\
  && \alpha_q = \frac{2i}{\sqrt{\pi}}; \qquad \alpha_u = -\frac{4i}{\sqrt{\pi}}n_0 T^{(0)};\nn\\
  && \Big[ -i\omega +\frac{2}{\sqrt{\pi}}v_{\textrm{th}}|\kpar| \Big] q^{(1)} = -\frac{4}{\sqrt{\pi}}n_0 T^{(0)}v_{\textrm{th}}|\kpar| u^{(1)};\nn
\end{eqnarray}
\begin{equation} \boxed{
 R_{3,2}(\zeta):\qquad  \Big[ \frac{d}{dt} -\frac{2}{\sqrt{\pi}}v_{\textrm{th}}\pr_z\mathcal{H} \Big] q^{(1)} = +\frac{4}{\sqrt{\pi}}n_0 T^{(0)}v_{\textrm{th}}\pr_z \mathcal{H} u^{(1)},}
\end{equation}
and similarly, yet another one
\begin{eqnarray}
  && R_{3,2}(\zeta):\qquad  \zeta q^{(1)} = \alpha_u u^{(1)} +\alpha_T T^{(1)}; \\
  && \alpha_u = -\frac{4i}{\sqrt{\pi}}n_0 T^{(0)}; \qquad \alpha_T = -\frac{4}{\pi} n_0 v_{\textrm{th}} \sign(\kpar);\nn\\
  && -i\omega q^{(1)} = -\frac{4}{\sqrt{\pi}}n_0 T^{(0)}v_{\textrm{th}} |\kpar| u^{(1)} +\frac{4}{\pi} n_0 v_{\textrm{th}}^2 i\kpar T^{(1)};\nn\\
  && \frac{d}{dt} q^{(1)} = +\frac{4}{\sqrt{\pi}}n_0 T^{(0)}v_{\textrm{th}} \pr_z \mathcal{H} u^{(1)} +\frac{4}{\pi} n_0 v_{\textrm{th}}^2 \pr_z T^{(1)},
\end{eqnarray}
however, the last closure will be eliminated. 
\clearpage
\subsection{Time-dependent closures with 5-pole approximants}
Now we can use this technique to construct time-dependent closures with 5-pole approximants of $R(\zeta)$.
Starting with the $R_{5,4}(\zeta)$ approximant
\begin{equation}
 R_{5,4}(\zeta) = \frac{1+a_1\zeta+a_2\zeta^2+a_3\zeta^3}{1+3(a_1+a_3)\zeta+(3a_2-2)\zeta^2+(3a_3-2a_1)\zeta^3-2a_2\zeta^4-2a_3\zeta^5},
\end{equation}
where the constants $a_1,a_2,a_3$ are given in the  Appendix (\ref{eq:ApxR54}), the kinetic moments calculate
\begin{eqnarray}
  R_{5,4}(\zeta):\qquad D &=& \Big(1+3(a_1+a_3)\zeta+(3a_2-2)\zeta^2+(3a_3-2a_1)\zeta^3-2a_2\zeta^4-2a_3\zeta^5\Big);\\
  u^{(1)} &=& - \frac{q_r}{T^{(0)}} \Phi v_{\textrm{th}} \sign(\kpar)
  \frac{1}{D}\Big[ a_3\zeta^4+a_2\zeta^3+a_1\zeta^2+\zeta \Big];\\
  T^{(1)} &=& -q_r \Phi \frac{1}{D}\Big[ 2a_3\zeta^3+2a_2\zeta^2+(2a_1+3a_3)\zeta \Big];\\
  q^{(1)} &=& -q_r n_{0} \Phi v_{\textrm{th}} \sign(\kpar)
  \frac{1}{D}\Big[ 3a_3\zeta^2-2\zeta\Big];\\
  \widetilde{r}^{(1)} &=& -q_r\frac{n_{0}}{2}v_{\textrm{th}}^2 \Phi \frac{1}{D}\Big[-(6a_2+4)\zeta^2-(6a_1+9a_3)\zeta \Big].
\end{eqnarray}
It is possible to construct time-dependent closure for $\widetilde{r}^{(1)}$, by searching for a solution
\begin{equation} \label{eq:R54_pic1}
(\zeta+\alpha_r) \widetilde{r}^{(1)} = \alpha_T T^{(1)} + \alpha_q q^{(1)}.
\end{equation}
Separating the equation to 3 equations for $\zeta,\zeta^2,\zeta^3$, the solution is
\begin{equation} \label{eq:R54_pic2}
\alpha_r = \frac{a_2}{a_3};\quad \alpha_T = -n_0v_{\textrm{th}}^2 \frac{3a_2+2}{2a_3}; \quad \alpha_q = - v_{\textrm{th}} \sign(\kpar) \frac{2a_1+3a_3}{2a_3}, 
\end{equation}
that evaluates as
\begin{equation}
  \alpha_r = i\frac{21\pi-64}{\sqrt{\pi}(9\pi-28)};\quad
  \alpha_T = in_0v_{\textrm{th}}^2 \frac{256-81\pi}{2(9\pi-28)\sqrt{\pi}}; \quad
  \alpha_q = v_{\textrm{th}} \sign(\kpar) \frac{32-9\pi}{2(9\pi-28)}. 
\end{equation}
The $R_{5,4}(\zeta)$ closure therefore reads
\begin{equation}
  \Big[ -i\omega + \frac{21\pi-64}{\sqrt{\pi}(9\pi-28)} v_{\textrm{th}} |\kpar| \Big] \widetilde{r}^{(1)}
  = n_0v_{\textrm{th}}^3 \frac{256-81\pi}{2(9\pi-28)\sqrt{\pi}} |\kpar| T^{(1)}
  -v_{\textrm{th}}^2 \frac{32-9\pi}{2(9\pi-28)} i\kpar q^{(1)},
\end{equation}
and transformation to Real space yields
\begin{eqnarray}
  \boxed{
R_{5,4}(\zeta):\qquad  \Big[ \frac{d}{dt} - \frac{21\pi-64}{\sqrt{\pi}(9\pi-28)} v_{\textrm{th}} \pr_z\mathcal{H} \Big] \widetilde{r}^{(1)}
  = -n_0v_{\textrm{th}}^3 \frac{256-81\pi}{2(9\pi-28)\sqrt{\pi}} \pr_z\mathcal{H} T^{(1)}
  -v_{\textrm{th}}^2 \frac{32-9\pi}{2(9\pi-28)} \pr_z q^{(1)}. } \label{eq:R54_closurE}
\end{eqnarray}
The closure is interesting, since the $R_{5,4}(\zeta)$ is a very precise $o(\zeta^3)$, $o(1/\zeta^6)$ approximant, and it is therefore only one of two closures
that have precision $o(\zeta^3)$. For large $\zeta$, the moments have correct asymptotic behavior up to the heat flux $q^{(1)}\sim -3/(2\zeta^3)$ (including the
proportionality constant) and the $\widetilde{r}^{(1)}\sim 1/\zeta^3$, which is not bad either. 
Additionally, the closure does not contain $u^{(1)}$, which is advantageous. 

Constructing a closure with $R_{5,5}(\zeta)$ is done quickly, by using $a_2=-2/3$ in the kinetic moments for $R_{5,4}(\zeta)$, so
\begin{eqnarray}
  R_{5,5}(\zeta):\qquad D &=& \Big(1+3(a_1+a_3)\zeta-4\zeta^2+(3a_3-2a_1)\zeta^3+\frac{4}{3}\zeta^4-2a_3\zeta^5\Big);\\
  q^{(1)} &=& -q_r n_{0} \Phi v_{\textrm{th}} \sign(\kpar)
  \frac{1}{D}\Big[ i\frac{(32-9\pi)}{3\sqrt{\pi}}\zeta^2-2\zeta\Big];\\
  \widetilde{r}^{(1)} &=& -q_r\frac{n_{0}}{2}v_{\textrm{th}}^2 \Phi \frac{1}{D}\Big[3i\sqrt{\pi}\zeta \Big],
\end{eqnarray}
where $a_1,a_3$ are given in the  Appendix (\ref{eq:ApxR55}). Searching for a closure $(\zeta+\alpha_r)\widetilde{r}^{(1)}=\alpha_q q^{(1)}$ has a solution
\begin{eqnarray}
  \Big[ \zeta + i\frac{6\sqrt{\pi}}{(32-9\pi)}  \Big]\widetilde{r}^{(1)} &=& \frac{9\pi}{2(32-9\pi)}v_{\textrm{th}} \sign(\kpar) q^{(1)};\\
  \Big[ -i\omega + \frac{6\sqrt{\pi}}{(32-9\pi)}v_{\textrm{th}} |\kpar| \Big]\widetilde{r}^{(1)} &=& -\frac{9\pi}{2(32-9\pi)}v_{\textrm{th}}^2 i\kpar q^{(1)},\nn
\end{eqnarray}
and the closure in Real space reads
\begin{equation}
\boxed{  
  R_{5,5}(\zeta):\qquad  \Big[ \frac{d}{dt} - \frac{6\sqrt{\pi}}{(32-9\pi)}v_{\textrm{th}} \pr_z\mathcal{H} \Big]\widetilde{r}^{(1)}
  = -\frac{9\pi}{2(32-9\pi)}v_{\textrm{th}}^2 \pr_z q^{(1)}.} \label{eq:R55_closurE}
\end{equation}
The $R_{5,5}(\zeta)$ approximant has precision $o(\zeta^2)$, $o(1/\zeta^7)$. The increase of the asymptotic precision reproduces correct asymptote
$\widetilde{r}^{(1)}\sim 1/\zeta^4$, even though with proportionality constant
$\widetilde{r}^{(1)}\sim -\frac{27\pi}{2(32-9\pi)\zeta^4}=-11.38/\zeta^4$ instead of the correct $-6/\zeta^4$.

Continuing with the approximant $R_{5,6}(\zeta)$, the kinetic moments calculate 
\begin{eqnarray}
  R_{5,6}(\zeta):\qquad D &=& \Big(1-i\sqrt{\pi}\frac{15}{8}\zeta-4\zeta^2+i\sqrt{\pi}\frac{5}{2}\zeta^3+\frac{4}{3}\zeta^4-i\frac{\sqrt{\pi}}{2}\zeta^5 \Big) ;\\
  q^{(1)} &=& -q_r n_{0} \Phi v_{\textrm{th}} \sign(\kpar) \frac{1}{D}\Big[ \frac{3i\sqrt{\pi}}{4}\zeta^2-2\zeta \Big];\\
  \widetilde{r}^{(1)} &=& -q_r\frac{n_{0}}{2}v_{\textrm{th}}^2 \Phi \frac{1}{D}\Big[3i\sqrt{\pi}\zeta \Big],
\end{eqnarray}
which yields a closure
\begin{eqnarray}
  \Big[ \zeta + i\frac{8}{3\sqrt{\pi}}  \Big]\widetilde{r}^{(1)} &=& 2v_{\textrm{th}} \sign(\kpar) q^{(1)};\\
  \Big[ -i\omega + \frac{8}{3\sqrt{\pi}} v_{\textrm{th}} |\kpar| \Big]\widetilde{r}^{(1)} &=& -2v_{\textrm{th}}^2 i\kpar q^{(1)},\nn
\end{eqnarray}
that in Real space reads
\begin{equation}
\boxed{  \label{eq:R56_closure}
R_{5,6}(\zeta):\qquad \Big[ \frac{d}{dt} - \frac{8}{3\sqrt{\pi}} v_{\textrm{th}} \pr_z\mathcal{H} \Big]\widetilde{r}^{(1)} = -2v_{\textrm{th}}^2 \pr_z q^{(1)}. }
\end{equation}
The $R_{5,6}(\zeta)$ approximant has precision $o(\zeta)$, $o(1/\zeta^8)$. Even though the precision for small $\zeta$ is relatively low,
the closure correctly reproduces the asymptotic behavior $\widetilde{r}^{(1)}\sim -6/\zeta^4$ (including the proportionality constant).

Finally, it is indeed possible to construct a closure with precision $o(\zeta^4)$, by using $R_{5,3}(\zeta)$. The approximant $R_{5,3}(\zeta)$ is defined as
\begin{equation}
 R_{5,3}(\zeta) = \frac{1+a_1\zeta+a_2\zeta^2+a_3\zeta^3}{1+b_1\zeta+(3a_2-2)\zeta^2+(3a_3-2a_1)\zeta^3-2a_2\zeta^4-2a_3\zeta^5},
\end{equation}
where the constants $a_1,a_2,a_3,b_1$ are given in the  Appendix (\ref{eq:ApxR53}). Using this approximant, the kinetic moments calculate
\begin{eqnarray}
  R_{5,3}(\zeta):\qquad D &=& \Big(1+b_1\zeta+(3a_2-2)\zeta^2+(3a_3-2a_1)\zeta^3-2a_2\zeta^4-2a_3\zeta^5\Big);\\
  u^{(1)} &=& - \frac{q_r}{T^{(0)}} \Phi v_{\textrm{th}} \sign(\kpar)
  \frac{1}{D}\Big[ a_3\zeta^4+a_2\zeta^3+a_1\zeta^2+\zeta \Big];\\
  T^{(1)} &=& -q_r \Phi \frac{1}{D}\Big[ 2a_3\zeta^3+2a_2\zeta^2+(b_1-a_1)\zeta \Big];\\
  q^{(1)} &=& -q_r n_{0} \Phi v_{\textrm{th}} \sign(\kpar)
  \frac{1}{D}\Big[ (b_1-3a_1)\zeta^2-2\zeta\Big];\\
  \widetilde{r}^{(1)} &=& -q_r\frac{n_{0}}{2}v_{\textrm{th}}^2 \Phi \frac{1}{D}\Big[(2b_1-6a_1-6a_3)\zeta^3-(6a_2+4)\zeta^2+(3a_1-3b_1)\zeta \Big].
\end{eqnarray}
It is possible to search for a closure 
\begin{equation}
 R_{5,3}(\zeta):\qquad \big( \zeta + \alpha_r \big) \widetilde{r}^{(1)} = \alpha_u u^{(1)} + \alpha_T T^{(1)} + \alpha_q q^{(1)},
\end{equation}
and the solution is
\begin{eqnarray}
  \alpha_r &=& \frac{a_2}{a_3}; \qquad
  \alpha_u = -v_{\textrm{th}} n_0 T^{(0)}\frac{(3a_1+3a_3-b_1)}{a_3}\sign(\kpar); \nn\\
  \alpha_T &=& -n_0 v_{\textrm{th}}^2 \frac{(3a_2+2)}{2a_3}; \qquad
  \alpha_q = -v_{\textrm{th}} \frac{(2a_1+3a_3)}{2a_3}\sign(\kpar).
\end{eqnarray}
The correctness of the algebra can be quickly checked by prescribing $b_1=3a_1+3a_3$, which immediately recovers
the closure (\ref{eq:R54_pic1})-(\ref{eq:R54_pic2}) that was obtained for $R_{5,4}(\zeta)$ with only asymptotic expansion coefficients
(and the power series coefficients unspecified), which yields $\alpha_u=0$. The $R_{5,3}(\zeta)$ closure reads
\begin{eqnarray}
  R_{5,3}(\zeta):\qquad \Big[ -i\omega - i\frac{a_2}{a_3}v_{\textrm{th}}|\kpar| \Big] \widetilde{r}^{(1)}
  &=& v_{\textrm{th}}^2 n_0 T^{(0)}\frac{(3a_1+3a_3-b_1)}{a_3} i\kpar u^{(1)}
  +i n_0 v_{\textrm{th}}^3 \frac{(3a_2+2)}{2a_3} |\kpar| T^{(1)} \nn\\
  && +v_{\textrm{th}}^2 \frac{(2a_1+3a_3)}{2a_3} i\kpar q^{(1)}.
\end{eqnarray}
By using the calculated coefficients from the  Appendix (\ref{eq:ApxR53}),
\begin{eqnarray}
  \frac{a_2}{a_3} &=& i\frac{(104-33\pi)\sqrt{\pi}}{2(9\pi^2-69\pi+128)}\equiv i\widetilde{\alpha}_r; \qquad 
  \frac{3a_1+3a_3-b_1}{a_3} = \frac{(135\pi^2-750\pi+1024)}{2(9\pi^2-69\pi+128)}\equiv \widetilde{\alpha}_u;\nn\\
  \frac{3a_2+2}{2a_3} &=& i\frac{3(160-51\pi)\sqrt{\pi}}{4(9\pi^2-69\pi+128)}\equiv i\widetilde{\alpha}_T; \qquad
  \frac{2a_1+3a_3}{2a_3} = \frac{(54\pi^2-333\pi+512)}{2(9\pi^2-69\pi+128)}\equiv\widetilde{\alpha}_q , \label{eq:R53_coeff}
\end{eqnarray}
the closure in Fourier and Real space then reads
\begin{eqnarray}
  R_{5,3}(\zeta):\qquad \Big[ -i\omega +\widetilde{\alpha}_r v_{\textrm{th}}|\kpar| \Big] \widetilde{r}^{(1)}
  &=& v_{\textrm{th}}^2 n_0 T^{(0)} \widetilde{\alpha}_u i\kpar u^{(1)}
  - n_0 v_{\textrm{th}}^3 \widetilde{\alpha}_T |\kpar| T^{(1)} +v_{\textrm{th}}^2 \widetilde{\alpha}_q i\kpar q^{(1)}; \\
  \Big[ \frac{d}{dt}  -\widetilde{\alpha}_r v_{\textrm{th}}\pr_z\mathcal{H} \Big] \widetilde{r}^{(1)}
  &=& v_{\textrm{th}}^2 n_0 T^{(0)} \widetilde{\alpha}_u \pr_z u^{(1)}
  + n_0 v_{\textrm{th}}^3 \widetilde{\alpha}_T \pr_z\mathcal{H} T^{(1)} +v_{\textrm{th}}^2 \widetilde{\alpha}_q \pr_z q^{(1)},
\end{eqnarray}
where the perhaps complicated appearing proportionality constants (that come from the Pad\'e approximation), are just constants, that are numerically evaluated as
\begin{eqnarray}
\widetilde{\alpha}_r &=& 5.13185; \qquad  \widetilde{\alpha}_u = 1.78706; \qquad \widetilde{\alpha}_T = -5.20074; \qquad  \widetilde{\alpha}_q = -10.53748.
\end{eqnarray}
For numerical simulations, we of course recommend to re-calculate these constants from the above analytic expressions, to fully match the numerical precision
of the considered simulation. For complete clarity, the fully expressed closure in Real space reads
\begin{empheq}[box=\fbox]{align}
R_{5,3}(\zeta):\qquad & \Big[ \frac{d}{dt} -\frac{(104-33\pi)\sqrt{\pi}}{2(9\pi^2-69\pi+128)} v_{\textrm{th}}\pr_z\mathcal{H} \Big] \widetilde{r}^{(1)}
  = v_{\textrm{th}}^2 n_0 T^{(0)}\frac{(135\pi^2-750\pi+1024)}{2(9\pi^2-69\pi+128)} \pr_z u^{(1)} \nn\\
  & \qquad + n_0 v_{\textrm{th}}^3 \frac{3(160-51\pi)\sqrt{\pi}}{4(9\pi^2-69\pi+128)} \pr_z\mathcal{H} T^{(1)} 
   +v_{\textrm{th}}^2 \frac{(54\pi^2-333\pi+512)}{2(9\pi^2-69\pi+128)} \pr_z q^{(1)}. \label{eq:R53_closure}
\end{empheq}
This is the only closure with precision $o(\zeta^4)$, and the asymptotic precision is $o(1/\zeta^5)$.
To conclude, we altogether obtained 13 time-dependent closures. Additionally, we also obtained 1 time-dependent closure for $R_{4,5}(\zeta)$ that
was disregarded since the $R_{4,5}(\zeta)$ is not a well-behaved approximant. 
\clearpage
\subsection{Parallel ion-acoustic (sound) mode, cold electrons}
After all the calculations, it is advisable to verify if we obtained anything useful. Let's consider only the proton species, make the electrons cold and neglect
electron inertia, so we have only 1-fluid model. Let's continue to work in physical units and later
we will switch to normalized units. From Part 1 of this guide, the linearized fluid equations (obtained by direct integration of the Vlasov equation
for a general distribution function $f$) can be written in physical units as 
\begin{eqnarray}
&&  -\omega \frac{n^{(1)}}{n_0} +\kpar u_z^{(1)} =0; \label{eq:sound_N1}\\
&&  -\omega u_z^{(1)} +\frac{v_{\textrm{th}\parallel}^2}{2} \kpar \frac{p_\parallel^{(1)}}{p_\parallel^{(0)}} =0; \label{eq:sound_U1}\\
&&  -\omega \frac{p_\parallel^{(1)}}{p_\parallel^{(0)}} +3 \kpar u_z^{(1)} +\kpar \frac{q_\parallel^{(1)}}{p_\parallel^{(0)}} =0; \label{eq:sound_P1}\\
  &&  -\omega \frac{q_\parallel^{(1)}}{p_\parallel^{(0)}} +\frac{3}{2}v_{\textrm{th}\parallel}^2\kpar \Big(\frac{p_\parallel^{(1)}}{p_\parallel^{(0)}}-\frac{n^{(1)}}{n_0}\Big)
  + \kpar \frac{\widetilde{r}_{\parallel\parallel}^{(1)}}{p_\parallel^{(0)}}=0, \label{eq:sound_Q1}  
\end{eqnarray}
where the fluctuating parallel temperature $T_\parallel^{(1)}$ is linearized as
\begin{equation}
\frac{T_\parallel^{(1)}}{T_\parallel^{(0)}} = \frac{p_\parallel^{(1)}}{p_\parallel^{(0)}} - \frac{n^{(1)}}{n_0}. \nn
\end{equation}
The superscript (1) on quantities $n^{(1)}$, $u_z^{(1)}$, $p_\parallel^{(1)}$, $q_\parallel^{(1)}$ (and $T_\parallel^{(1)}$) signifies that these are just fluctuating quantities,
the superscript does not mean here, that these quantities are obtained by integration over the kinetic $f^{(1)}$. 
This fluid model is accompanied by a closure for $\widetilde{r}_{\parallel\parallel}^{(1)}$, and that one was obtained from linear kinetic theory by integrating over the
kinetic $f^{(1)}$. Let's choose the $R_{4,3}(\zeta)$ closure, eq. (\ref{eq:Static_R43_F})  
\begin{equation}
  \frac{\widetilde{r}_{\parallel\parallel}^{(1)}}{p_\parallel^{(0)}} = \frac{32-9\pi}{2(3\pi-8)}v_{\textrm{th}\parallel}^2
  \Big(\frac{p_\parallel^{(1)}}{p_\parallel^{(0)}}-\frac{n^{(1)}}{n_0}\Big)
  -\frac{2\sqrt{\pi}}{3\pi-8}v_{\textrm{th}\parallel}  i\sign(\kpar) \frac{q_\parallel^{(1)}}{p_\parallel^{(0)}}.
\end{equation}
Now the model is closed, and calculating the determinant yields the following dispersion relation
\begin{equation}
  \omega^4+\frac{2i\sqrt{\pi}}{3\pi-8}\kpar v_{\textrm{th}\parallel} \sign(\kpar)\omega^3-\frac{(9\pi-16)}{2(3\pi-8)}\kpar^2 v_{\textrm{th}\parallel}^2 \omega^2
  -\frac{3i\sqrt{\pi}}{3\pi-8}\kpar^3 v_{\textrm{th}\parallel}^3 \sign(\kpar)\omega+\frac{2}{3\pi-8}\kpar^4 v_{\textrm{th}\parallel}^4 =0. 
\end{equation}
By examining the expression, an obvious substitution offers itself
\begin{equation}
\zeta = \frac{\omega}{\sign(\kpar)\kpar v_{\textrm{th}\parallel}} = \frac{\omega}{|\kpar| v_{\textrm{th}\parallel}},
\end{equation}
that transforms the polynomial to a completely dimensionless form
\begin{equation} \label{eq:Polyy_1}
\zeta^4+\frac{2i\sqrt{\pi}}{3\pi-8}\zeta^3-\frac{9\pi-16}{2(3\pi-8)}\zeta^2-\frac{3i\sqrt{\pi}}{3\pi-8}\zeta+\frac{2}{3\pi-8}=0.
\end{equation}
The $\zeta$ is obviously a very useful quantity, and one could rewrite the fluid equations (\ref{eq:sound_N1})-(\ref{eq:sound_Q1}) directly with this quantity. 
The polynomial (\ref{eq:Polyy_1}) can be solved numerically, and the approximate solutions are (writing only 3 decimal digits)
\begin{equation}
\zeta=   \pm 1.359 - 0.534i; \qquad \zeta=\pm 0.392 - 0.710i,
\end{equation}
yielding solutions in physical units
\begin{eqnarray}
  \omega  = |\kpar|v_{\textrm{th}\parallel}  (\pm 1.359 - 0.534i); \qquad   \omega  = |\kpar|v_{\textrm{th}\parallel}  (\pm 0.392 - 0.710i).  \label{eq:sound_Sol}
\end{eqnarray}
The first solution is the ion-acoustic (sound) mode and the second solution is ``a higher-order mode''. Both solutions are highly damped, and the
higher-order mode has actually higher damping rate than its real frequency.
We can now also see how important was to keep track of the $\sign(\kpar)$, since the modes are damped for both $\kpar>0$ and $\kpar<0$.  
If we have ignored the $\sign(\kpar)$, we would obtain that for $\kpar<0$ the sound mode has a positive growth rate and is unstable, which would be unphysical. 

Of course, each closure will yield a different dispersion relation. Exploring the simplest closure with quasi-static heat flux $q_\parallel^{(1)}$
obtained with $R_{3,2}(\zeta)$, the equations (\ref{eq:sound_N1})-(\ref{eq:sound_P1}) are closed by
\begin{equation}
\frac{q_\parallel^{(1)}}{p_\parallel^{(0)}} = -\frac{2}{\sqrt{\pi}} v_{\textrm{th}\parallel} i\sign(\kpar) \Big(\frac{p_\parallel^{(1)}}{p_\parallel^{(0)}}-\frac{n^{(1)}}{n_0}\Big),
\end{equation}
which yields a polynomial
\begin{equation} \label{eq:Polyy_2}
\zeta^3+\frac{2i}{\sqrt{\pi}}\zeta^2-\frac{3}{2}\zeta-\frac{i}{\sqrt{\pi}}=0.
\end{equation}
Numerical solutions are $\zeta=   \pm 1.041 - 0.327i$; $\zeta= - 0.474i$, 
showing that in this case the higher-order mode does not propagate and is purely damped. The ion-acoustic mode  is also very damped and has a dispersion relation
\begin{eqnarray}
  \omega  &=& |\kpar| v_{\textrm{th}\parallel} (\pm 1.041 - 0.327i).
\end{eqnarray}

We examine two more closures. The most precise quasi-static closure (\ref{eq:Static_R42_F}) obtained with $R_{4,2}(\zeta)$ yields the analytic dispersion relation
\begin{equation} \label{eq:Polyy_3}
\zeta^4+i\sqrt{\pi}\frac{10-3\pi}{16-5\pi}\zeta^3-\frac{32-9\pi}{32-10\pi}\zeta^2-\frac{i\sqrt{\pi}}{16-5\pi}\zeta+\frac{3\pi-8}{32-10\pi}=0,
\end{equation}
and the solutions are
\begin{eqnarray}
  \omega  = |\kpar| v_{\textrm{th}\parallel}  (\pm 1.294 - 0.790i);\qquad   \omega  = |\kpar| v_{\textrm{th}\parallel} (\pm 0.385 - 0.956i),
\end{eqnarray}
the first one being the ion-acoustic mode. Finally, the only available $o(\zeta^4)$ closure (\ref{eq:R53_closure}) obtained with $R_{5,3}(\zeta)$ yields
analytic dispersion relation
\begin{equation} \label{eq:Polyy_4}
  \zeta^5+i\widetilde{\alpha}_r \zeta^4 +(\widetilde{\alpha}_q-3)\zeta^3-i(3\widetilde{\alpha}_r-\widetilde{\alpha}_T)\zeta^2+\frac{1}{2}(\widetilde{\alpha}_u
  -3\widetilde{\alpha}_q+\frac{3}{2})\zeta   +\frac{i}{4}(3\widetilde{\alpha}_r-2\widetilde{\alpha}_T)=0,
\end{equation}
where the coefficients are specified in (\ref{eq:R53_coeff}), and the numerical solutions are
\begin{eqnarray}
  \omega  = |\kpar| v_{\textrm{th}\parallel}  (\pm 1.589 - 0.908i);\qquad
  \omega  = |\kpar| v_{\textrm{th}\parallel} (\pm 0.710 - 1.084i);\qquad
  \omega  = |\kpar| v_{\textrm{th}\parallel} (-1.147i),
\end{eqnarray}
the first being the ion-acoustic mode.

Let's compare the obtained results. Perhaps curiously, it appears that as the precision of closures increases, 
so does increases the real frequency and the damping rate of the ion-acoustic mode, and the differences are quite significant.
So what is the correct kinetic result, i.e., how close did we get to the kinetic theory ? That is not as easy question as it appears to be. 
By opening kinetic books, there is no such a discussion as long-wavelength limit of the ion-acoustic mode, when the electrons are cold.
Even the exact numerical solutions are usually considered only for $T_e/T_p \ge 1$, see for example Figure 9.18 on page 355 in Gurnett and Bhattacharjee. 

Let's examine the analytic dispersion relations (\ref{eq:Polyy_1}), (\ref{eq:Polyy_2}), (\ref{eq:Polyy_3}) and (\ref{eq:Polyy_4}), that were obtained
with approximants $R_{4,3}(\zeta)$, $R_{3,2}(\zeta)$, $R_{4,2}(\zeta)$ and $R_{5,3}(\zeta)$.  One notices that the dispersion relations exactly
match the \emph{denominators} of the associated Pad\'e approximants ! Or in another words, without doing any calculations whatsoever, it appears that
if a closure of a 1D fluid model is available for a $R_{n,n'}(\zeta)$ approximant, the dispersion relation is equivalent to the denominator of that $R_{n,n'}(\zeta)$.  
How is this possible ? The explanation is simple, if one considers the electrostatic kinetic  dispersion relation for the proton and electron 
  species (\ref{eq:SoundPEls}), which at scales that are much longer than the Debye length simplifies to (\ref{eq:soundKin}).
  By prescribing massless electrons yields $R(\zeta_e)=1$, implying dispersion relation $R(\zeta_p)=-T_p^{(0)}/T_e^{(0)}$. For cold electrons,
  both real and imaginary parts of $R(\zeta_p)$ obviously diverge, so that
\begin{equation}
\frac{1}{R(\zeta_p)} =0.\nn
\end{equation}
The above expression can be considered electrostatic dispersion relation of proton-electron plasma, where the electrons are  massless and completely cold. 
The reason why such an expression cannot be found in any plasma book is, that from the kinetic perspective, such an expression cannot be solved and is ill-defined.
The function $R(\zeta)$ is directly related to the derivative of $Z(\zeta)$ according to $Z'(\zeta)=-2R(\zeta)$. Infinitely large $R(\zeta)$ means that
$Z(\zeta)$ has infinitely large derivatives, i.e. that $Z(\zeta)$ is not continuous and, \emph{not analytic}, which contradicts the entire definition of $Z(\zeta)$ and
how the function was constructed. However, when Pad\'e approximants of these functions are considered, and when $R(\zeta)$ is
replaced by $R_{n,n'}(\zeta)$, so that
\begin{equation} \label{eq:Curious} \boxed{
\frac{1}{R_{n,n'}(\zeta_p)} =0,}
\end{equation}
such an expression does make sense, and is equivalent to the denominator of $R_{n,n'}(\zeta)$ being zero, i.e., it directly yields the dispersion
relations of the considered fluid models.  We note that while the plasma dispersion function corresponding to a Maxwellian
distribution function does not display singularities at finite distance in the complex plane,
this is not the case when considering kappa distribution functions, see e.g. \cite{Podesta2004}.

\subsubsection{The proton Landau damping does not disappear at long-wavelengths}
There are several extremely interesting phenomena worth discussing. 1) In the dispersion relation for the ion-acoustic (sound) mode (\ref{eq:sound_Sol}),
the usual phase speed $\omega/\kpar$ is constant, implying that
the Landau damping (of the parallel propagating sound mode) does not disappear,
however long-wavelengths are considered. With cold electrons as considered here, the parallel sound mode is always heavily damped,
and disappears in a few wavelengths, even on large astrophysical scales.  A very good discussion can be found for example
  in \cite{Howes2009}, who concluded that in general (unless electrons are hot),
  the MHD sound mode represents an unphysical spurious wave that does not exist in collisionless plasma.
2) The equations (\ref{eq:sound_N1})-(\ref{eq:sound_Q1}) do not even contain the parallel electric field $E_\parallel$. 
This might sound surprising, but the parallel electric field completely disappears at long wavelengths, even though the Landau damping
(as expressed through the constant phase speed), does not disappear. The parallel electric field does not disappear, if electrons have finite temperature, it also
enters (very weakly), if the electron inertia is included. In the 1D linearized geometry considered here, the contributions will be
\begin{equation}
  E_\parallel = -\frac{1}{en_0}\pr_z p_{\parallel e}-\frac{m_e}{e} \frac{\pr u_{ze}}{\pr t}.
\end{equation}
3) The presence of Landau damping in the long-wavelength limit is exactly the reason
why usual fluid models such as MHD or even much more sophisticated CGL description, \emph{do not} converge to the
collisionless kinetic theory, whatever long-wavelengths and low-frequencies are considered. 
There is always a mismatch in dispersion relations when the phase speed
is plotted, that depending on plasma parameters, can be quite large. This does not concern only the damping rate (which in MHD and CGL is of course zero), the differences in
the real frequency, which is always coupled to the imaginary frequency (for example through the polynomial (\ref{eq:Polyy_1}) for that specific closure), can be large too.  
4) If the heat flux $q_\parallel^{(1)}$ is prescribed to be zero, i.e. if a CGL model is prescribed, the dispersion relation of the parallel propagating
sound mode is determined only by the parallel velocity eq. (\ref{eq:sound_U1}) and parallel pressure eq. (\ref{eq:sound_P1}), yielding the CGL result
$\omega^2 = \frac{3}{2}v_{\textrm{th}\parallel}^2 \kpar^2$, so that
\begin{equation}
\omega^{\textrm{CGL}} = \pm |\kpar| v_{\textrm{th}\parallel} \sqrt{\frac{3}{2}} = \pm |\kpar| v_{\textrm{th}\parallel} 1.225.
\end{equation}   
For comparison, the MHD result can be written with the usual MHD sound speed $C_s^2=\gamma\frac{p_0}{\rho_0} = \frac{\gamma}{2}v_{\textrm{th}}^2$ where $\gamma=5/3$, so
\begin{equation}
\omega^{\textrm{MHD}} = \pm |\kpar| v_{\textrm{th}} \sqrt{\frac{5}{6}} = \pm |\kpar| v_{\textrm{th}} 0.913.
\end{equation}
It is important to examine the influence of isothermal electron species.
\clearpage
\subsection{Proton Landau damping, influence of isothermal electrons}
Let's prescribe electrons to be isothermal, with some finite electron temperature, but let's neglect the electron inertia.
 The proton momentum equation is changed to (\ref{eq:sound_U2}),
the electron pressure equation reads
\begin{equation}
  -\omega \frac{p_{\parallel e}^{(1)}}{p_{\parallel p}^{(0)}} + \kpar \tau u_z^{(1)} =0; \qquad \tau \equiv \frac{T_{\parallel e}^{(0)}}{T_{\parallel p}^{(0)}},
\end{equation}
where for brevity, we define the ratio of electron and proton temperature as $\tau$.
Using the $R_{4,3}(\zeta)$ closure as before, the coupled dispersion relation reads 
\begin{equation} \label{eq:R43_isothermal}
\zeta^4+\frac{2i\sqrt{\pi}}{3\pi-8}\zeta^3-\frac{9\pi-16+(3\pi-8)\tau}{2(3\pi-8)}\zeta^2-\frac{i\sqrt{\pi}(3+\tau)}{3\pi-8}\zeta+\frac{2(1+\tau)}{3\pi-8}=0.
\end{equation}
The above expression is equivalent to eq. (A6) in \cite{Hunana2011}.\footnote{Dispersion relations in the Appendix of that paper assumed $k>0$.
 We later noticed that (\ref{eq:R43_isothermal}), (\ref{eq:R42_isothermal}) are equivalent to the dispersion relation
  $\tau R_{n,n'}(\zeta_p) +1 =0$. We also noticed that for isothermal electrons the $R_{4,2}(\zeta)$ closure with eq. (\ref{eq:R42_isothermal}),
  can produce positive growth rate for high electron temperatures. The $R_{4,2}(\zeta)$ behaves correctly when the electron Landau damping is introduced, see
the next section \ref{section:PELD}.}
The $R_{4,2}(\zeta)$ closure yields dispersion relation
\begin{equation} \label{eq:R42_isothermal}
  \zeta^4+\frac{i\sqrt{\pi}(10-3\pi)}{16-5\pi}\zeta^3-\frac{32-9\pi+(16-5\pi)\tau}{32-10\pi}\zeta^2-\frac{i\sqrt{\pi}\big(2+(10-3\pi)\tau\big)}{32-10\pi}\zeta
  +\frac{(3\pi-8)(1+\tau)}{32-10\pi}=0.
\end{equation}

Let's use (\ref{eq:R43_isothermal}) and focus on the ion-acoustic mode, since the higher-order mode is always highly damped.
  Solutions for few different $\tau$ values are
\begin{eqnarray}
  \tau=0:\qquad \zeta &=& \pm 1.359-0.534i;\nn\\
  \tau=0.1:\qquad \zeta &=& \pm 1.367-0.514i;\nn\\
  \tau=0.5:\qquad \zeta &=& \pm 1.409-0.439i;\nn\\
  \tau=1.0:\qquad \zeta &=& \pm 1.481-0.361i;\nn\\
  \tau=2.0:\qquad \zeta &=& \pm 1.640-0.260i;\nn\\
  \tau=5.0:\qquad \zeta &=& \pm 2.054-0.131i;\nn\\
  \tau=10.0:\qquad \zeta &=& \pm 2.591-0.061i;\nn\\
  \tau=100.0:\qquad \zeta &=& \pm 7.180-0.001i.
\end{eqnarray}  
This is excellent, as in kinetic books, with increasing electron temperature, the Landau damping of the ion-acoustic mode  decreases.
Compared to kinetic calculations  (see the last column in (\ref{eq:Table1})),
the total Landau damping is here of course underestimated, especially for high electron temperatures,  
since here in the fluid model, only the proton Landau damping is contributing, and the electron Landau damping is turned off. Let's turn it on.
\subsection{Proton and Electron Landau damping} \label{section:PELD}
Considering wavelengths much longer than the Debye length, the exact kinetic dispersion relation reads
\begin{equation} \label{eq:soundKin} \boxed{
\frac{T_{\parallel e}^{(0)}}{T_{\parallel p}^{(0)}} R(\zeta_p) +R(\zeta_e) =0,}
\end{equation}
where the electron thermal velocity
\begin{equation}
v_{\textrm{th}\parallel e} = v_{\textrm{th}\parallel p} \sqrt{\frac{m_p}{m_e}} \sqrt{\frac{T_{\parallel e}^{(0)}}{T_{\parallel p}^{(0)}}},
\end{equation}  
is of course much higher than the proton thermal velocity (unless the electrons are cold), and by using the abbreviated
\begin{equation}
\tau \equiv \frac{T_{\parallel e}^{(0)}}{T_{\parallel p}^{(0)}} ;\qquad \mu\equiv \frac{m_e}{m_p},
\end{equation}  
so that
\begin{equation}
  \zeta_p = \frac{\omega}{|\kpar|v_{\textrm{th}\parallel p}}; \qquad
  \zeta_e = \frac{\omega}{|\kpar|v_{\textrm{th}\parallel e}} = \zeta_p \sqrt{\frac{\mu}{\tau}},
\end{equation}
and the exact kinetic dispersion relation reads
\begin{equation} \label{eq:soundKin_norm}
\tau R(\zeta_p) +R\Big(\zeta_p \sqrt{\mu/\tau}\Big) =0.
\end{equation}
Let's see how close did we get. One of the greatest advantages of Landau fluid models is that we do not have to resolve electron
motion to obtain the correct form of electron Landau damping at long wavelengths, and the electron inertia in the electron momentum equation
can be neglected. The correct electron-proton mass ratio can enters equations for
the electron heat flux $q_{\parallel e}$ and the 4th-order moment $\widetilde{r}_{\parallel\parallel e}$, and the electron inertia influence the solutions only
insignificantly. However, let's keep the electron inertia for a moment. The equations for the proton species read 
\begin{eqnarray}
&&  -\omega \frac{n^{(1)}}{n_0} +\kpar u_z^{(1)} =0; \label{eq:sound_N2}\\
  &&  -\omega u_z^{(1)} +\frac{v_{\textrm{th}\parallel p}^2}{2} \kpar
  \bigg( \frac{p_{\parallel p}^{(1)}}{p_{\parallel p}^{(0)}} + \frac{p_{\parallel e}^{(1)}}{p_{\parallel p}^{(0)}}      \bigg) -\omega \frac{m_e}{m_p} u_z^{(1)}=0; \label{eq:sound_U2}\\
&&  -\omega \frac{p_{\parallel p}^{(1)}}{p_{\parallel p}^{(0)}} +3 \kpar u_z^{(1)} +\kpar \frac{q_{\parallel p}^{(1)}}{p_{\parallel p}^{(0)}} =0; \label{eq:sound_P2}\\
  &&  -\omega \frac{q_{\parallel p}^{(1)}}{p_{\parallel p}^{(0)}} +\frac{3}{2}v_{\textrm{th}\parallel p}^2\kpar \Big(\frac{p_{\parallel p}^{(1)}}{p_{\parallel p}^{(0)}}-\frac{n^{(1)}}{n_0}\Big)
  + \kpar \frac{\widetilde{r}_{\parallel\parallel p}^{(1)}}{p_{\parallel p}^{(0)}}=0, \label{eq:sound_Q2}  
\end{eqnarray}
and the electron inertia represents the last term in (\ref{eq:sound_U2}). 
The electron equations are written in a form so that they are normalized with respect to the proton pressure
\begin{eqnarray}
&&  -\omega \frac{p_{\parallel e}^{(1)}}{p_{\parallel p}^{(0)}} +3 \kpar \tau u_z^{(1)} +\kpar \frac{q_{\parallel e}^{(1)}}{p_{\parallel p}^{(0)}} =0; \label{eq:sound_P3}\\
  &&  -\omega \frac{q_{\parallel e}^{(1)}}{p_{\parallel p}^{(0)}} +\frac{3}{2} v_{\textrm{th}\parallel p}^2 \frac{m_p}{m_e}\kpar
  \tau \Big(\frac{p_{\parallel e}^{(1)}}{p_{\parallel p}^{(0)}}-\tau \frac{n^{(1)}}{n_0}\Big)
  + \kpar \frac{\widetilde{r}_{\parallel\parallel e}^{(1)}}{p_{\parallel p}^{(0)}}=0. \label{eq:sound_Q3}  
\end{eqnarray}
Note that the electron fluid speed $u_{ze}^{(1)}=u_{zp}^{(1)}$ (so we omitted the index p). The fluid equations are accompanied by a closure from kinetic theory,
for example the $R_{4,3}(\zeta)$ closure
\begin{eqnarray}
  \frac{\widetilde{r}_{\parallel\parallel p}^{(1)}}{p_{\parallel p}^{(0)}} &=& \frac{32-9\pi}{2(3\pi-8)}v_{\textrm{th}\parallel p}^2
  \Big(\frac{p_{\parallel p}^{(1)}}{p_{\parallel p}^{(0)}}-\frac{n^{(1)}}{n_0}\Big)
  -\frac{2\sqrt{\pi}}{3\pi-8}v_{\textrm{th}\parallel p}  i\sign(\kpar) \frac{q_{\parallel p}^{(1)}}{p_{\parallel p}^{(0)}},\\
   \frac{\widetilde{r}_{\parallel\parallel e}^{(1)}}{p_{\parallel p}^{(0)}} &=& \frac{32-9\pi}{2(3\pi-8)}v_{\textrm{th}\parallel p}^2
  \frac{m_p}{m_e} \tau \Big(\frac{p_{\parallel e}^{(1)}}{p_{\parallel p}^{(0)}}-\tau \frac{n^{(1)}}{n_0}\Big)
  -\frac{2\sqrt{\pi}}{3\pi-8}v_{\textrm{th}\parallel p}\sqrt{\frac{m_p}{m_e}} \sqrt{\tau} i\sign(\kpar) \frac{q_{\parallel e}^{(1)}}{p_{\parallel p}^{(0)}}. \label{eq:sound_RparparQ4}
\end{eqnarray}
The equations (\ref{eq:sound_N2})-(\ref{eq:sound_RparparQ4}) now represents a fluid description of the ion-acoustic mode, and contain both proton and
electron Landau damping. It is rather mesmerizing, that the relatively complicated dispersion relation of this fluid model, can be shown to be
equivalent to the ``simple looking'' kinetic dispersion relation
\begin{equation} \label{eq:sound_R43}
\tau R_{4,3}(\zeta_p) +R_{4,3}\Big(\zeta_p \sqrt{\mu/\tau}\Big) =0,
\end{equation}
i.e. equivalent to the full kinetic dispersion relation (\ref{eq:soundKin_norm}), where the exact $R(\zeta)$ is replaced with the $R_{4,3}(\zeta)$ approximant
(by transferring the proton and electron terms of $R_{4,3}(\zeta)$ in the expression (\ref{eq:sound_R43}) to the common denominator
and making the resulting numerator of that expression equal to zero, Maple is great in this regard). 

Nevertheless, here we want clearly demonstrate that the electron inertia can be neglected and the electron Landau damping still nicely captured,
and we use fluid dispersion relations obtained from the system (\ref{eq:sound_N2})-(\ref{eq:sound_RparparQ4}), where the last term in (\ref{eq:sound_U2}) is neglected. 
It is important to normalize properly and for example the $R_{4,2}(\zeta)$ closure for electrons reads
\begin{eqnarray}
  \frac{\widetilde{r}_{\parallel\parallel e}^{(1)}}{p_{\parallel p}^{(0)}} &=&
  -i\sqrt{\pi} \frac{(10-3\pi)}{(16-5\pi)} v_{\textrm{th}\parallel p} \sqrt{\frac{m_p}{m_e}} \sqrt{\tau}\sign(\kpar) \frac{q_{\parallel e}^{(1)}}{p_{\parallel p}^{(0)}}
  + \frac{(21\pi-64)}{2(16-5\pi)}v_{\textrm{th}\parallel p}^2 \frac{m_p}{m_e} \tau \Big( \frac{p_{\parallel e}^{(1)}}{p_{\parallel p}^{(0)}}-\tau \frac{n^{(1)}}{n_0}\Big)\nn\\
  && + i\sqrt{\pi} \frac{(9\pi-28)}{(16-5\pi)}v_{\textrm{th}\parallel p} \sqrt{\frac{m_p}{m_e}} \tau^{3/2} \sign(\kpar) u_{z}^{(1)}.
\end{eqnarray}
In the table below, we compare these fluid solutions of the quasi-static $R_{4,3}(\zeta)$ and the $R_{4,2}(\zeta)$ closures, and the time-dependent
$R_{5,3}(\zeta)$ closure to the exact kinetic solutions, calculated from (\ref{eq:soundKin}), for various electron temperatures.
\vspace{0.2cm}
\begin{equation}
\begin{tabular}{| c | c | c | c | c |}
  \hline
                & $R_{4,3}(\zeta)$ closure        & $R_{4,2}(\zeta)$ closure & $R_{5,3}(\zeta)$ closure & Exact  \\
  \hline
  $\tau=1.0$    & $\zeta  = 1.487-0.373i$   & $1.434-0.506i$    & $1.511-0.591i$   & $1.457-0.627i$\\
  $\tau=2.0$    & $\zeta  = 1.645-0.276i$   & $1.629-0.372i$    & $1.691-0.393i$   & $1.692-0.425i$\\
  $\tau=5.0$    & $\zeta  = 2.057-0.154i$   & $2.080-0.212i$    & $2.116-0.188i$   & $2.136-0.189i$\\
  $\tau=10.0$   & $\zeta  = 2.593-0.094i$   & $2.627-0.123i$    & $2.635-0.089i$   & $2.640-0.072i$\\
  $\tau=20.0$   & $\zeta  = 3.417-0.069i$   & $3.446-0.075i$    & $3.432-0.052i$   & $3.421-0.046i$\\
  $\tau=50.0$   & $\zeta  = 5.157-0.078i$   & $5.170-0.072i$    & $5.156-0.070i$   & $5.153-0.073i$\\
  $\tau=100.0$  & $\zeta  = 7.180-0.105i$   & $7.186-0.099i$    & $7.179-0.102i$   & $7.177-0.103i$\\
\hline
\end{tabular} \label{eq:Table1}
\end{equation}
\vspace{0.2cm}

\begin{figure*}[!htpb]
  $$\includegraphics[width=0.48\linewidth]{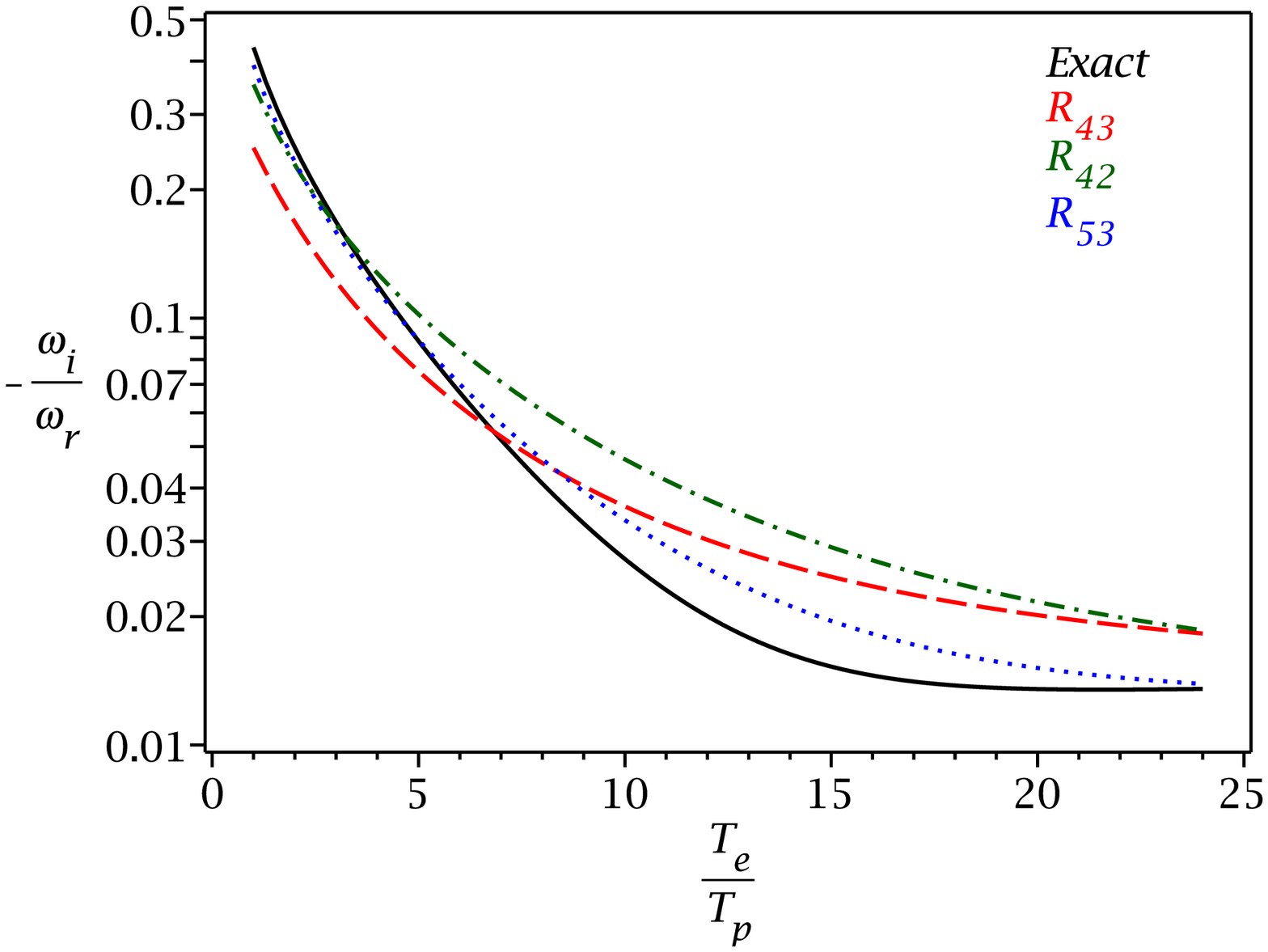}\hspace{0.03\textwidth}\includegraphics[width=0.48\linewidth]{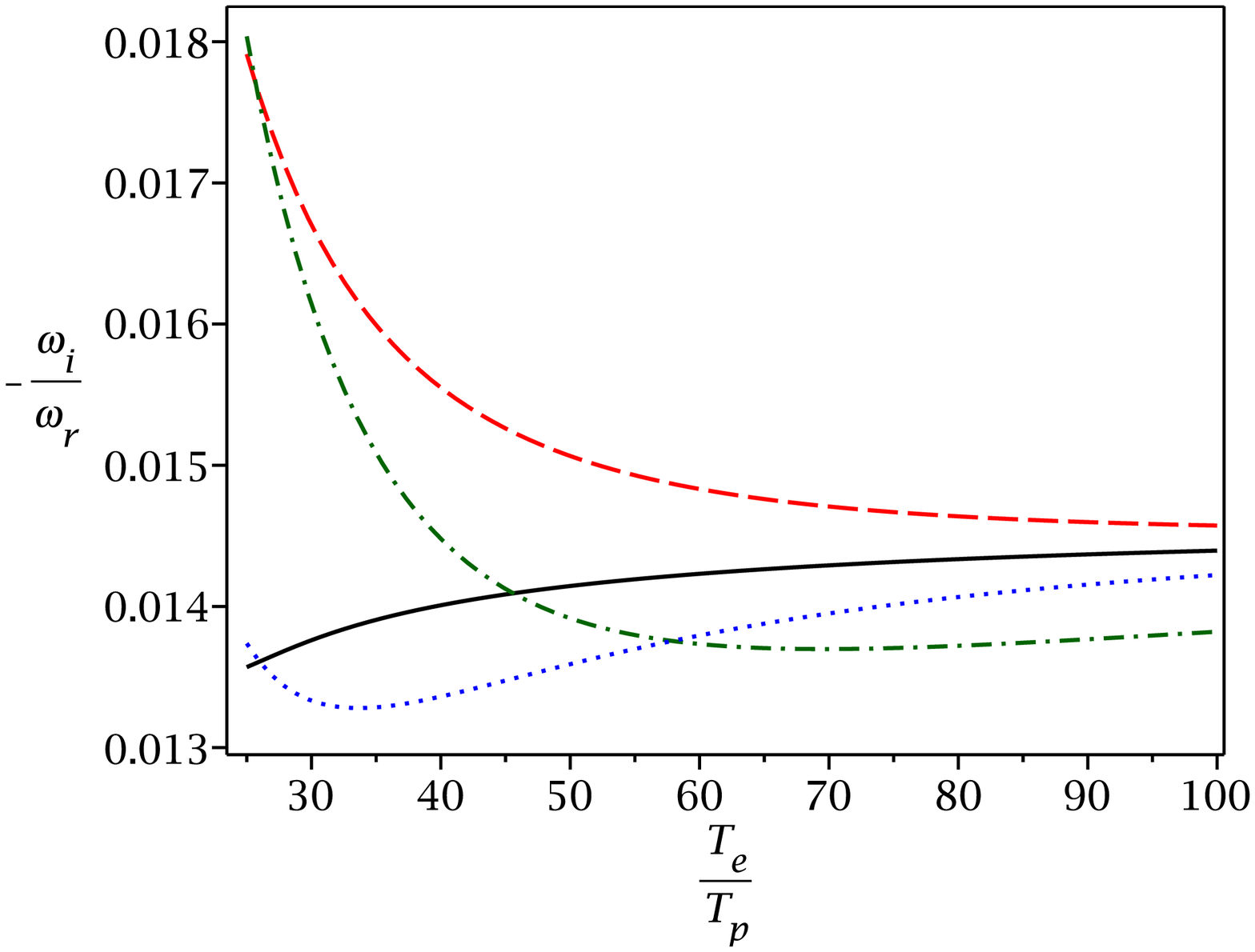}$$  
  \caption{Landau damping of the ion-acoustic (sound) mode. The black solid curve is the solution of exact kinetic dispersion relation (\ref{eq:soundKin}).
    The other curves are dispersion relations of a fluid model (\ref{eq:sound_N2})-(\ref{eq:sound_Q3}) where the electron inertia is neglected, supplemented by
    a closure for $r_{\parallel\parallel r}^{(1)}$. The red dashed line is the $R_{4,3}(\zeta)$ closure of Hammett and Perkins 1990, the green dash-dotted line is our new
    static closure $R_{4,2}(\zeta)$, and the blue dotted line is the new time-dependent closure $R_{5,3}(\zeta)$.} \label{fig:IAdamping1} 
\end{figure*}

\begin{figure*}[!htpb]
  $$\includegraphics[width=0.48\linewidth]{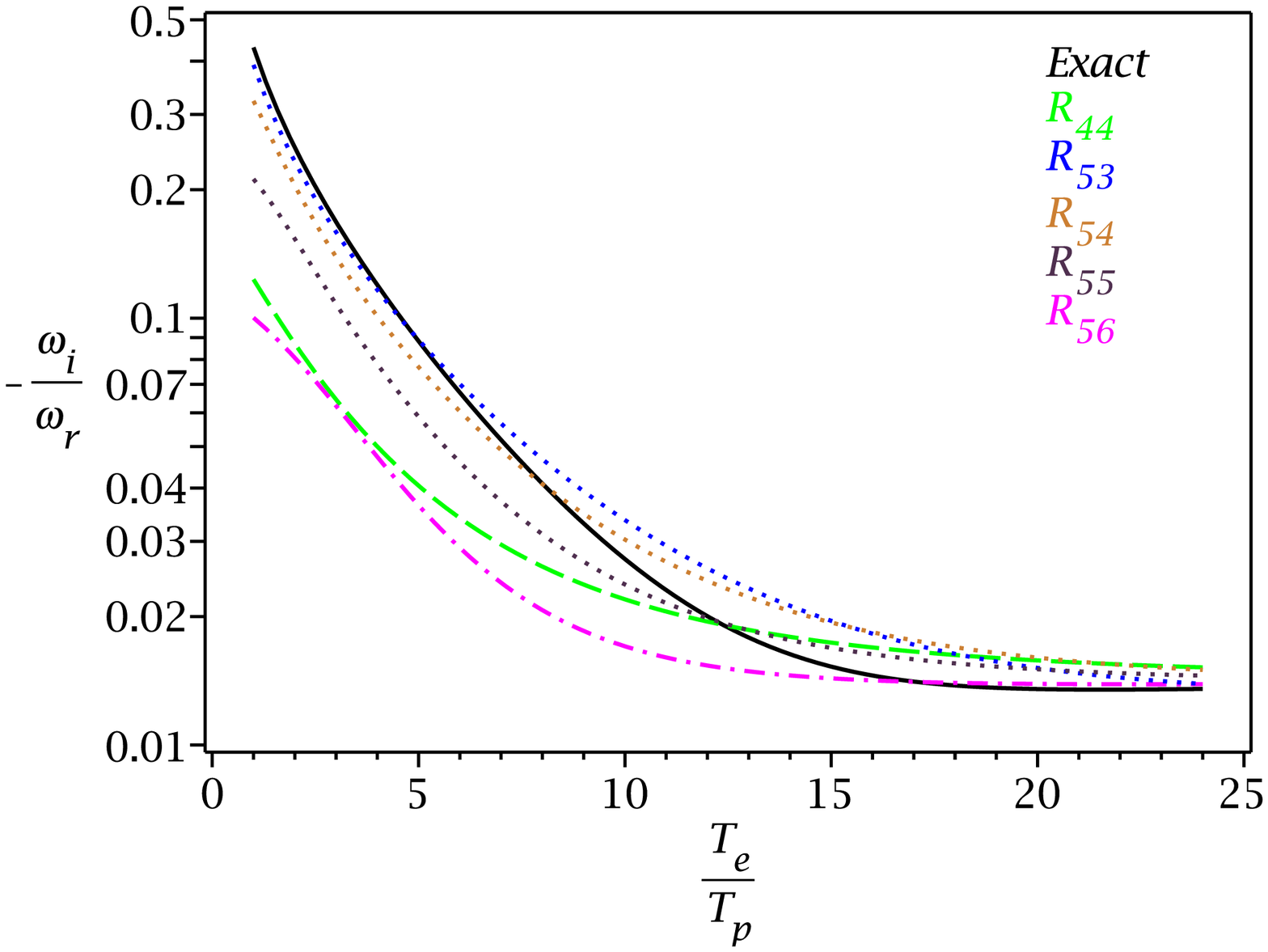}\hspace{0.03\textwidth}\includegraphics[width=0.48\linewidth]{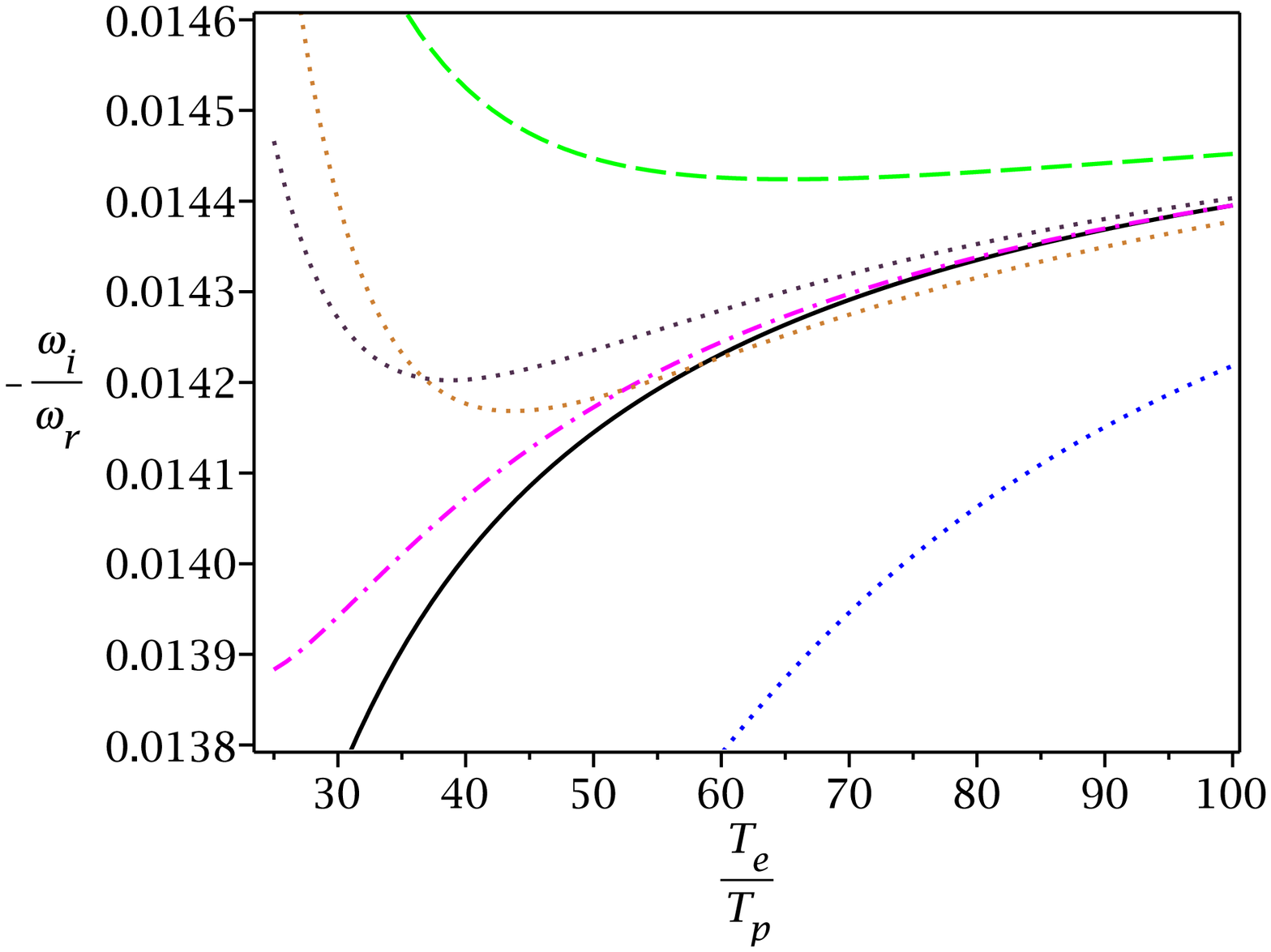}$$  
  \caption{Similar to Figure \ref{fig:IAdamping1}, but different closures are compared, and electron inertia is retained. The $R_{5,3}(\zeta)$ closure (blue dotted curve)
    is kept in the figure so that the comparison to other closures can be done easily. Also, by comparing the $R_{5,3}(\zeta)$ solution with Figure \ref{fig:IAdamping1},
    it is shown that the effect of electron inertia is negligible. 
    The $R_{4,4}(\zeta)$ is the only static closure, and all other closures are time-dependent.} \label{fig:IAdamping2}
\end{figure*}

Instead of a table, we can create a figure.
The Landau damping of the ion-acoustic sound mode, is nicely demonstrated for example in the plasma book of Gurnett and Bhattacharjee (Figure 9.18, page 355),
where on the x-axis is $\tau$, and on the y-axis (logarithmic), is the ratio of damping and real frequency. The same parameters are plotted in Figure \ref{fig:IAdamping1} left, and
in Figure \ref{fig:IAdamping1} right we extend the plot to higher electron temperatures. 
The figure shows that both new closures are very precise in the very important regime, where the electron temperature $\tau$ ranges between
$\tau=1$ and $\tau=5$. The $R_{5,3}(\zeta)$ closure is the most globally precise closure. If static closures are preferred, the comparison between $R_{4,2}(\zeta)$ and
$R_{4,3}(\zeta)$ is more difficult to summarize, the $R_{4,2}(\zeta)$ is definitely preferred in the regime $\tau\in[1,5]$ and perhaps also for $\tau\in[25,60]$, however,
the Hammett and Perkins closure $R_{4,3}(\zeta)$ is the better fit in the regime $\tau\in[5,25]$ and also for $\tau\in[60,100]$.
We checked that the inclusion of electron inertia is insignificant for all 3 fluid closures, and by eye inspection, it appears that the largest global difference
is seen for the $R_{5,3}(\zeta)$ closure, roughly for $\tau\in[30,60]$, making the closure (very slightly) more precise.
In Figure (\ref{fig:IAdamping2}),
we calculate the other selected obtained closures. We use the full dispersion relations with electron inertia included. The figure shows, that if static closures
are preferred, for value of roughly $\tau>15$, the best closure is actually the static closure $R_{4,4}(\zeta)$.
The most precise closure for $\tau>15$ is by far the time-dependent $R_{5,6}(\zeta)$ closure, which achieves an excellent accuracy for high values of tau. If a global
accuracy for all values of $\tau$ is required, our favorite closures are $R_{5,3}(\zeta)$ and $R_{5,4}(\zeta)$.

With the help of Maple software, we analytically investigated dispersion relations of all the obtained fluid closures, and we investigated if the resulting dispersion relation
(including the electron inertia) is equivalent to the kinetic result (\ref{eq:soundKin}), after replacing the $R(\zeta)$ with $R_{n,n'}(\zeta)$, i.e. if the fluid dispersion
relation is equivalent to the numerator of
\begin{equation} \label{eq:sound_rigid} \boxed{
\frac{T_{\parallel e}^{(0)}}{T_{\parallel p}^{(0)}} R_{n,n'}(\zeta_p) +R_{n,n'}(\zeta_e) =0.}
\end{equation}
All the closures considered in this subsection satisfied this requirement, however, some other previously obtained closures did not.
We concluded, that satisfying (\ref{eq:sound_rigid}) should be indeed considered as strong requirement for a physically meaningful closure,
and closures that did not satisfy this requirement were therefore eliminated. The results are summarized in the subsection
3.6 ``Table of moments $(u_\parallel,T_\parallel,q_\parallel,\widetilde{r}_{\parallel\parallel})$ for various Pad\'e approximants'', eqs.
(\ref{eq:Table1S}), (\ref{eq:Table1T}).

\clearpage
\subsection{Electron Landau damping of the Langmuir mode} \label{sec:Langmuir}
In addition to the ion-acoustic mode, let's calculate the Landau damping for the second (perhaps first) typical example, how Landau damping is addressed
in plasma physics book, the Langmuir mode.
Focusing on the electron species and making the proton species cold and ``very heavy'' with $m_p\to\infty$,
i.e. immobile with $u_{zp}^{(1)}=0$, the proton species completely decouple from the system, and their role is just to conserve the leading-order
charge neutrality $n_{0p}=n_{0e}=n_0$. Since we haven't dealt with such a system so far (not even in Part 1 of
the text), let's write down the basic equation nicely in Real space in physical units. Neglecting the electron heat flux, the basic system of linearized equations reads
\begin{eqnarray}
&&  \frac{\pr n_e^{(1)}}{\pr t} + n_0 \pr_z u_{ze}^{(1)} =0;\\
&&  \frac{\pr u_{ze}^{(1)}}{\pr t} +\frac{1}{m_e n_0} \pr_z p_{\parallel e}^{(1)} +\frac{e}{m_e} E_\parallel =0; \label{eq:Uz_lang}\\
&&  \frac{\pr p_{\parallel e}^{(1)}}{\pr t} +3p_{\parallel e}^{(0)} \pr_z u_{ze}^{(1)}=0.
\end{eqnarray}
By using the general electrostatic Maxwell's equation for the current (including the displacement current)
\begin{equation}
\boldsymbol{j}=\sum_r q_r n_r \bu_r = \cancel{\frac{c}{4\pi}\nabla\times\bb}-\frac{1}{4\pi}\frac{\pr \bE}{\pr t},
\end{equation}
that in our specific 1D linear case considered here reads 
\begin{equation}
j_\parallel = -e n_{0} u_{ze}^{(1)} = -\frac{1}{4\pi}\frac{\pr E_\parallel}{\pr t},
\end{equation}
prescribes the electric field time evolution, and the system of equations is closed. By applying $\pr/\pr t$ to the momentum equation
(\ref{eq:Uz_lang}), the equations can be combined, yielding a wave equation
\begin{equation}
\bigg[ \frac{\pr^2}{\pr t^2} -\frac{3p_{\parallel e}^{(0)}}{m_e n_0} \pr^2_z +\omega_{pe}^2 \bigg] u_{ze}^{(1)} =0,
\end{equation}
where the electron plasma frequency $\omega_{pe}^2=4\pi e^2n_0/m_e$. This wave equation describes the basic plasma physics mode, known as the Langmuir mode,
and the dispersion relation is
\begin{equation} \label{eq:Debye}
\omega^2 = \omega_{pe}^2+\frac{3T_{\parallel e}^{(0)}}{m_e} \kpar^2.
\end{equation}
If we ignored the displacement current $\pr E_\parallel /\pr t$, the $\omega_{pe}^2$ term would be absent.
By dividing with $\omega_{pe}^2$ and by using the Debye length $\lambda_{De}=1/k_{De}$ where
$k_{De}^2=4\pi e^2 n_{0e}/T_{\parallel e}^{(0)}$, so that the Debye length $\lambda_{De}^2 = T_{\parallel e}^{(0)}/ (m_e\omega_{pe}^2)$,
the dispersion relation (\ref{eq:Debye}) reads 
\begin{equation}
\frac{\omega^2}{\omega_{pe}^2} = 1+ 3 \lambda_{De}^2 \kpar^2.
\end{equation}
Obviously, the electron plasma frequency and the electron Debye length are the natural normalizing units of this system, and one should use normalized
quantities $\omega/\omega_{pe}$ and $\kpar \lambda_{De}$. A useful relation also is $\lambda_{De}^2=v_{\textrm{th}\parallel e}^2/(2\omega_{pe}^2).$

Often, in plasma physics books, the CGL adiabatic index $\gamma_\parallel=3$ in the above two equations, is substituted with a general adiabatic index $\gamma_e$,
so that a more ``general'' case can be considered. This is especially useful if the Langmuir waves, which are the basic waves of plasma physics,
are introduced early on (in an early chapter of a book), where the correct CGL value of $\gamma_\parallel=3$ is difficult to introduce. Again, we have an advantage
of not being a plasma book, and we are not describing the general electrostatic case, we are describing the fully electromagnetic case, but we are focusing only
on one mode - the electrostatic mode that propagates parallel to $B_0$. In the view presented here, and as elaborated in Part I of the text,
playing with adiabatic indices, does not make much sense. No adiabatic index can match the CGL and the MHD,
the CGL is always different from MHD, even for isotropic distribution function with $T_\parallel^{(0)}=T_\perp^{(0)}$. Therefore, we are not introducing any 
adiabatic index, and the correct CGL value is used, and fixed to 3. Instead, we introduce the electron heat flux and get closer to the kinetic theory
in a much more sophisticated way. 

The basic linearized fluid equations in Fourier space read
\begin{eqnarray}
&&  -\omega \frac{n_e^{(1)}}{n_0}+\kpar u_{ze}^{(1)}=0;\nn\\
  && -\omega u_{ze}^{(1)}+\frac{1}{2}v_{\textrm{th}\parallel e}^2 \kpar \frac{p_{\parallel e}^{(1)}}{p_{\parallel e}^{(0)}}
  +\frac{1}{2} \frac{v_{\textrm{th}\parallel e}^2}{\lambda_{De}^2} \frac{u_{ze}^{(1)}}{\omega}=0;\nn\\
 &&  -\omega \frac{p_{\parallel e}^{(1)}}{p_{\parallel e}^{(0)}} +3 \kpar u_{ze}^{(1)} +\kpar \frac{q_{\parallel e}^{(1)}}{p_{\parallel e}^{(0)}} =0;\\
  &&  -\omega \frac{q_{\parallel e}^{(1)}}{p_{\parallel e}^{(0)}} +\frac{3}{2} v_{\textrm{th}\parallel e}^2 \kpar
  \Big(\frac{p_{\parallel e}^{(1)}}{p_{\parallel e}^{(0)}}-\frac{n_e^{(1)}}{n_0}\Big)
  + \kpar \frac{\widetilde{r}_{\parallel\parallel e}^{(1)}}{p_{\parallel e}^{(0)}}=0, 
\end{eqnarray}
and are accompanied for example by the $R_{4,3}(\zeta)$ closure
\begin{equation}
   \frac{\widetilde{r}_{\parallel\parallel e}^{(1)}}{p_{\parallel e}^{(0)}} = \frac{32-9\pi}{2(3\pi-8)}v_{\textrm{th}\parallel e}^2
  \Big(\frac{p_{\parallel e}^{(1)}}{p_{\parallel e}^{(0)}}-\frac{n_e^{(1)}}{n_0}\Big)
  -\frac{2\sqrt{\pi}}{3\pi-8}v_{\textrm{th}\parallel e} i\sign(\kpar) \frac{q_{\parallel e}^{(1)}}{p_{\parallel e}^{(0)}}.
\end{equation}
The dispersion relation of this fluid model reads (suppressing $e$ in the electron Debye length $\lambda_{De}$)
\begin{equation}
  \zeta_e^4 +\frac{2i\sqrt{\pi}}{3\pi-8}\zeta_e^3 - \frac{3\pi-8+(9\pi-16)\kpar^2\lambda_{D}^2}{2(3\pi-8)\kpar^2\lambda_{D}^2}\zeta_e^2
  -\frac{i\sqrt{\pi}(1+3\kpar^2\lambda_{D}^2)}{(3\pi-8)\kpar^2\lambda_{D}^2}\zeta_e +\frac{2(1+\kpar^2\lambda_{D}^2)}{(3\pi-8)\kpar^2\lambda_{D}^2}=0, \label{eq:Lang_R43}
\end{equation}
where
\begin{equation} \label{eq:xydef}
\zeta_e=\frac{\omega}{|\kpar|v_{\textrm{th}\parallel e}}= \frac{\omega}{\omega_{pe}}\frac{1}{\sqrt{2}|\kpar|\lambda_D}.
\end{equation}
The exact kinetic dispersion relation reads 
\begin{equation} \label{eq:Langmuir_kin}\boxed{
1+\frac{1}{\kpar^2\lambda_D^2} R(\zeta_e)=0.}
\end{equation}
As can be verified, the fluid dispersion relation (\ref{eq:Lang_R43}) is equivalent to the kinetic one, if $R(\zeta_e)$ is replaced by $R_{4,3}(\zeta_e)$. 

Using the static $R_{4,2}(\zeta)$ closure, the dispersion relation reads
\begin{eqnarray}
  \zeta_e^4+i\sqrt{\pi}\frac{10-3\pi}{16-5\pi}\zeta_e^3-\frac{16-5\pi+(32-9\pi)\kpar^2\lambda_D^2}{2(16-5\pi)\kpar^2\lambda_D^2}\zeta_e^2
  -i\sqrt{\pi}\frac{10-3\pi+2\kpar^2\lambda_D^2}{2(16-5\pi)\kpar^2\lambda_D^2} \zeta_e +\frac{(3\pi-8)(1+\kpar^2\lambda_D^2)}{2(16-5\pi)\kpar^2\lambda_D^2}=0,
\end{eqnarray}
using the static $R_{4,4}(\zeta)$ closure yields
\begin{eqnarray}
  \zeta_e^4+i\sqrt{\pi}\frac{3}{4}\zeta_e^3-\frac{1+6\kpar^2\lambda_D^2}{2\kpar^2\lambda_D^2}\zeta_e^2
  -i\sqrt{\pi}\frac{3(1+3\kpar^2\lambda_D^2)}{8\kpar^2\lambda_D^2} \zeta_e +\frac{3(1+\kpar^2\lambda_D^2)}{4\kpar^2\lambda_D^2}=0,
\end{eqnarray}
and the simplest static $R_{3,2}(\zeta)$ closure yields
\begin{eqnarray} \label{eq:Langmuir_R32}
\zeta_e^3 +\frac{2i}{\sqrt{\pi}}\zeta_e^2-\frac{1+3\kpar^2\lambda_{D}^2}{2\kpar^2\lambda_D^2}\zeta_e -\frac{i(1+\kpar^2\lambda_D^2)}{\sqrt{\pi}\kpar^2\lambda_D^2}=0.
\end{eqnarray}
All dispersion relations are fully consistent with the kinetic dispersion relation (\ref{eq:Langmuir_kin}) when $R(\zeta_e)$ is replaced by
the corresponding $R_{4,2}(\zeta_e)$, $R_{4,4}(\zeta_e)$ and $R_{3,2}(\zeta_e)$ (equivalent to the numerator of the resulting expression). We verified that this
is also true for the static closure $R_{3,1}(\zeta_e)$ and actually all the ``reliable'' closures marked in (\ref{eq:Table1S}), (\ref{eq:Table1T})
with ``\checkmark'', including the time-dependent closures
$R_{3,2}(\zeta_e), R_{4,2}(\zeta_e), R_{4,3}(\zeta_e), R_{4,4}(\zeta_e), R_{5,3}(\zeta_e), R_{5,4}(\zeta_e), R_{5,5}(\zeta_e), R_{5,6}(\zeta_e)$. 
To clearly understand the obtained solutions, let's solve the simple $R_{3,2}(\zeta_e)$ dispersion relation (\ref{eq:Langmuir_R32}) for a few values of $\kpar\lambda_D$:
\begin{center}
\begin{tabular}{ l  l  l }
  $\kpar\lambda_D=0.001:$ & $\zeta_e = \pm 707.1-1.13i \times 10^{-6};$ & $\zeta_e=-1.13i;$\\
  $\kpar\lambda_D=0.01:$  & $\zeta_e = \pm 70.7-1.13i \times  10^{-4};$ & $\zeta_e=-1.13i;$\\
  $\kpar\lambda_D=0.1:$   & $\zeta_e = \pm 7.17-1.07i \times  10^{-2};$ & $\zeta_e=-1.11i;$\\
  $\kpar\lambda_D=0.2:$   & $\zeta_e = \pm 3.73-3.73i \times  10^{-2};$ & $\zeta_e=-1.05i;$\\
  $\kpar\lambda_D=0.3:$   & $\zeta_e = \pm 2.63-0.07i;$                & $\zeta_e=-0.99i;$\\
  $\kpar\lambda_D=1.0:$   & $\zeta_e = \pm 1.28-0.23i;$                & $\zeta_e=-0.67i.$
\end{tabular}
\end{center}
The first mode is the Langmuir mode, and the second mode is a purely damped higher-order mode. 
In the complete limit $\kpar\lambda_D\to 0$, the Langmuir mode becomes undamped with a solution $\zeta_e=1/(\sqrt{2}|\kpar|\lambda_D)$, which corresponds
to oscillations with electron plasma frequency $\omega=\omega_{pe}$; and the higher-order mode has a solution $\zeta_e=-2i/\sqrt{\pi}$. Considering the 
weak damping limit $\zeta_e=x+iy$, where $x\gg y$, at the leading order $\zeta_e^2=x^2+i2xy$ and $\zeta_e^3=x^3+i3x^2y$, which when used in the dispersion relation
(\ref{eq:Langmuir_R32}) that is separated to real and imaginary parts yields
\begin{eqnarray}
&&  x^3 -\frac{4}{\sqrt{\pi}}xy -\frac{1+3\kpar^2\lambda_D^2}{2\kpar^2\lambda_D^2}x=0;\nn\\
&&  3x^2y +\frac{2}{\sqrt{\pi}}x^2-\frac{1+3\kpar^2\lambda_D^2}{2\kpar^2\lambda_D^2}y -\frac{1}{\sqrt{\pi}} \frac{1+\kpar^2\lambda_D^2}{\kpar^2\lambda_D^2}=0,\nn
\end{eqnarray}
and for $x\gg y$ at the leading order
\begin{eqnarray} \label{eq:eqxy}
x^2=\frac{1+3\kpar^2\lambda_D^2}{2\kpar^2\lambda_D^2}; \qquad y=-\frac{1}{\sqrt{\pi}x^2} = -\frac{2}{\sqrt{\pi}}\frac{\kpar^2\lambda_D^2}{(1+3\kpar^2\lambda_D^2)}, 
\end{eqnarray}
which approximates the above numerical solutions reasonably well up to let's say $\kpar\lambda_D=0.3$, and from (\ref{eq:xydef}) the $x,y$ expressions are equivalent to
\begin{equation} \label{eq:R32damping}
\omega_r^2 = \omega_{pe}^2 (1+3\kpar^2\lambda_D^2); \qquad \omega_i= - \sqrt{\frac{8}{\pi}} \frac{ |\kpar|^3 \lambda_D^3}{1+3\kpar^2\lambda_D^2}\omega_{pe}.
\end{equation}
For $\kpar\lambda_D\ll 1$, the Landau damping of the Langmuir mode goes to zero, however, the damping rate is very overestimated. 
The approximate kinetic result found in  plasma books (see for example \cite{GurnettBhattacharjee2005}, page 349)
has of course the same real frequency, however, the damping rate reads
\begin{equation} \label{eq:LangKin}
  \omega_r^2 = \omega_{pe}^2 (1+3\kpar^2\lambda_D^2); \qquad
  \omega_i = - \sqrt{\frac{\pi}{8}} \frac{\omega_{pe}}{|\kpar|^3\lambda_D^3} e^{-\frac{1+3\kpar^2\lambda_D^2}{2\kpar^2\lambda_D^2}}.
\end{equation}
Since $\kpar\lambda_D\ll 1$, \cite{Landau1946} writes (see his eqs. 16 and 17)
\begin{equation} 
  \omega_r = \omega_{pe} (1+\frac{3}{2}\kpar^2\lambda_D^2); \qquad
  \omega_i = - \sqrt{\frac{\pi}{8}} \frac{\omega_{pe}}{|\kpar|^3\lambda_D^3} e^{-\frac{1}{2\kpar^2\lambda_D^2}}.
\end{equation}
\begin{figure*}[!htpb]
  $$\includegraphics[width=0.48\linewidth]{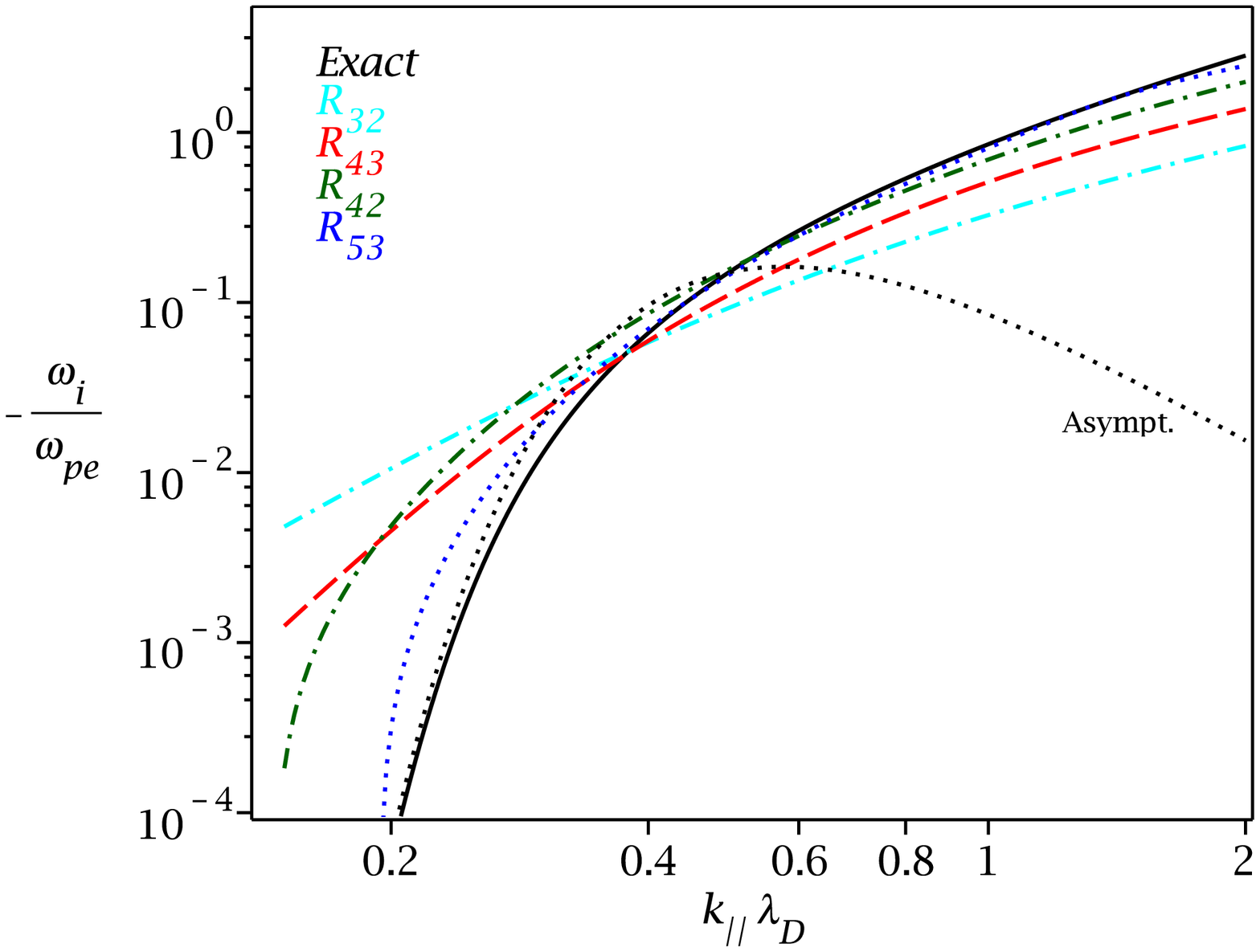}\hspace{0.03\textwidth}\includegraphics[width=0.48\linewidth]{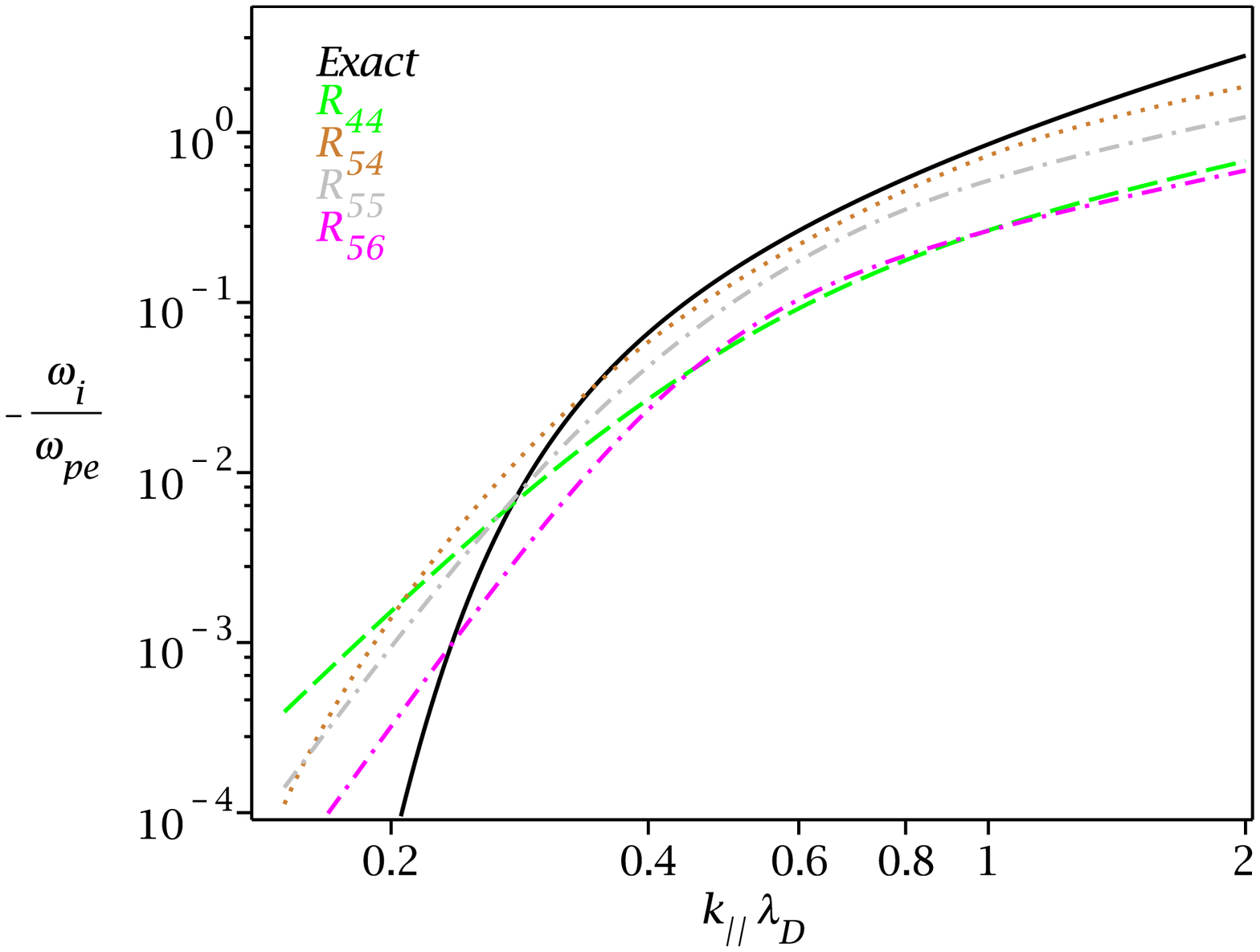}$$  
  \caption{Landau damping of the Langmuir mode. Numerical solution of the exact kinetic dispersion relation (\ref{eq:Langmuir_kin}) is the black solid line, and
  asymptotic kinetic solution (\ref{eq:LangKin}) is the black dotted line.} \label{fig:Langmuir1}
\end{figure*}
\begin{figure*}[!htpb]
  $$\includegraphics[width=0.48\linewidth]{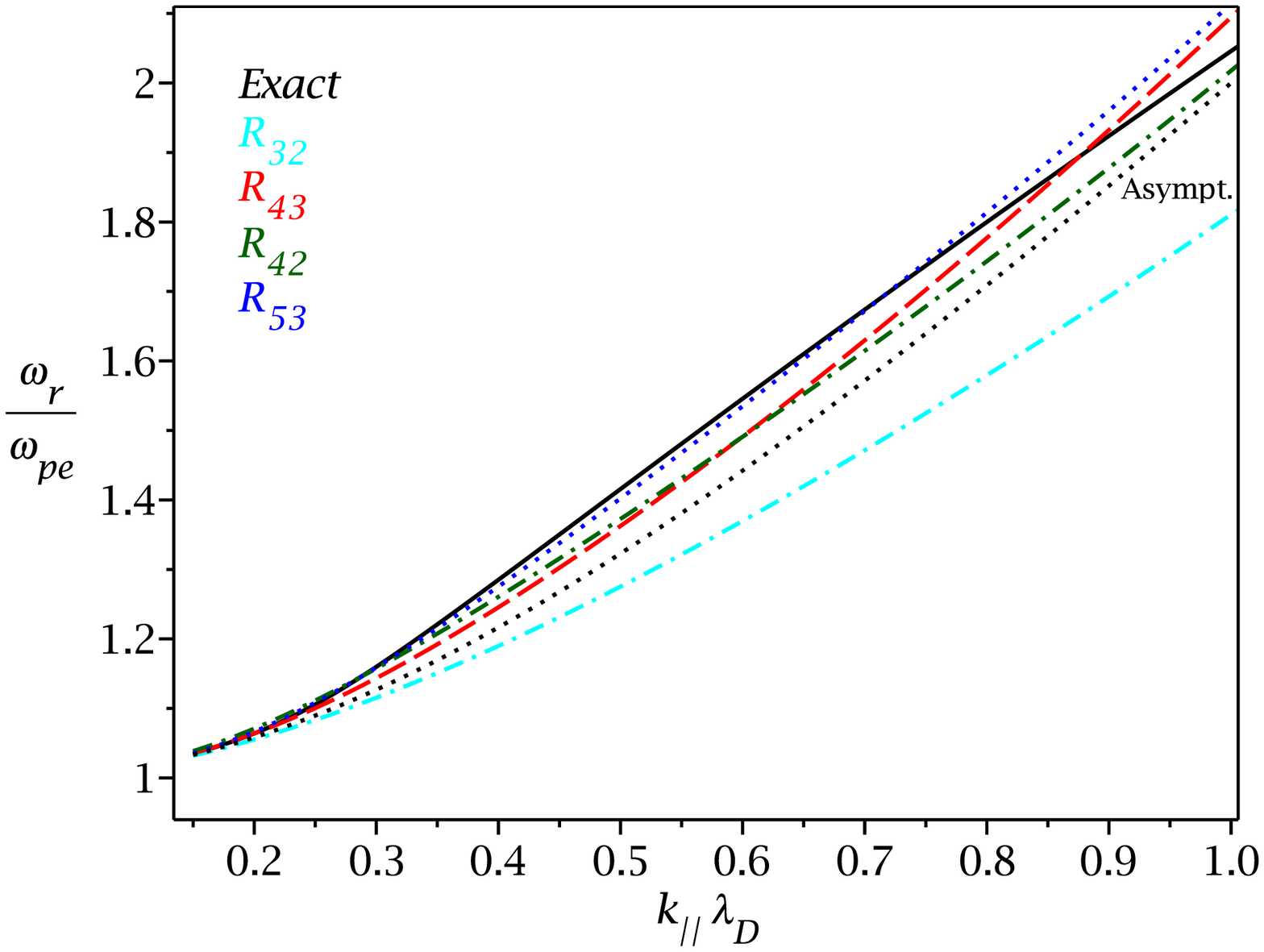}\hspace{0.03\textwidth}\includegraphics[width=0.48\linewidth]{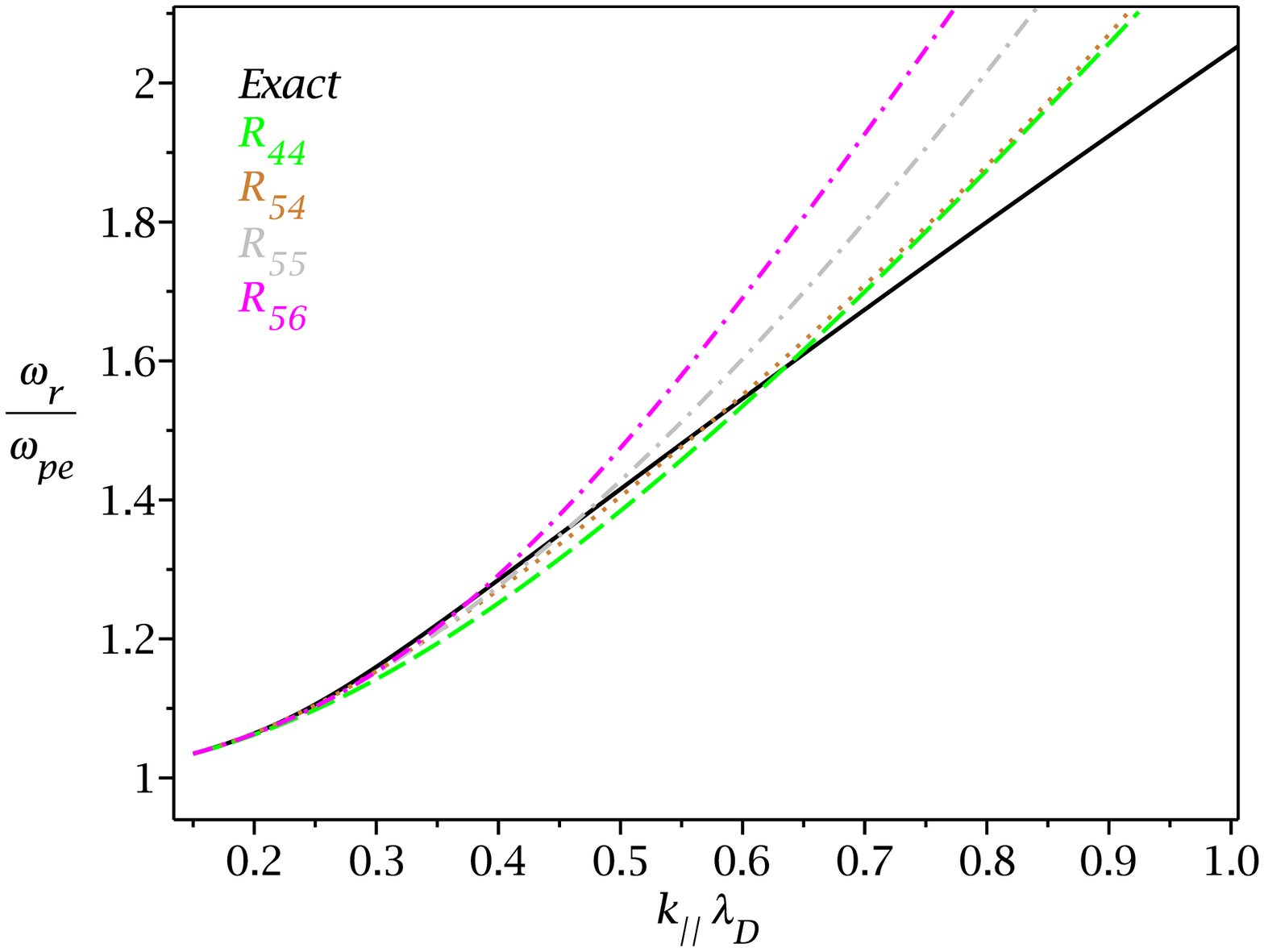}$$  
  \caption{Real frequency of the Langmuir mode.} \label{fig:Langmuir2}
\end{figure*}
For $\kpar \lambda_D\ll 1$, i.e. approaching long wavelengths, the exponential term suppresses the Landau damping much quicker than our result (\ref{eq:R32damping}).
To understand the discrepancy, let's quickly consider how the kinetic result (\ref{eq:LangKin}) was obtained.
The result is obtained by considering asymptotic expansion $|\zeta_e|\gg 1$ of the exact kinetic dispersion relation (\ref{eq:Langmuir_kin}), which in the weak
growth rate approximation (see eq. (\ref{eq:RzetaL}) with $\sigma=1$) reads
\begin{equation} \label{eq:eqxyy}
1+\frac{1}{\kpar^2\lambda_D^2} \bigg[-\frac{1}{2\zeta_e^2}-\frac{3}{4\zeta_e^4}+i\sqrt{\pi}\zeta_e e^{-\zeta_e^2} \bigg] =0.
\end{equation}
By using $\zeta_e=x+iy$ with $x\gg y$ and $\kpar \lambda_D\ll 1$ yields at the leading order $x^2=(1+3\kpar^2\lambda_D^2)/2\kpar^2\lambda_D^2$ which agrees with
(\ref{eq:eqxy}) and the damping rate is $y=-\sqrt{\pi} x^4 e^{-x^2}$, which recovers (\ref{eq:LangKin}). The $e^{-x^2}$ term in the damping rate comes from the last
term in (\ref{eq:eqxyy}), and as discussed previously, this term is neglected in the asymptotic expansion when constructing the Pad\'e approximants of $R(\zeta)$
(it is however included in the power series expansion), explaining the discrepancy. 

The damping rate of the Langmuir mode is plotted in Figure \ref{fig:Langmuir1} and the real frequency in Figure  \ref{fig:Langmuir2}, where solutions of various fluid models are compared
with exact kinetic dispersion relation (\ref{eq:Langmuir_kin}), depicted as the black solid line. Additionally, the 
asymptotic kinetic solution (\ref{eq:LangKin}) from plasma physics books is plotted as the black dotted line. Figure \ref{fig:Langmuir1} is plotted in log-log scale
and Figure \ref{fig:Langmuir2} uses linear scales. It is shown that for $\kpar\lambda_D>0.2$, fluid models can reproduce the damping of the Langmuir mode quite
accurately, and the most accurate closure is $R_{5,3}(\zeta)$. This closure also reproduces the real frequency of the Langmuir mode very accurately and
actually better than the asymptotic kinetic solution (\ref{eq:LangKin}).

Nevertheless, as discussed above, because of the missing exponential factor in fluid models, the Landau damping becomes very overestimated at
scales $\kpar\lambda_D<0.2$, i.e. it is the long-wavelength limit (and not the short-wavelength limit) that represents trouble. This is because
in the long-wavelength limit, the frequency of Langmuir mode does not go to zero but approaches electron plasma frequency $\omega_{pe}$, and
so the phase speed $\omega_r/\kpar$ (and the variable $\zeta_e$) becomes large and for $\kpar\lambda_D\to 0$ goes to infinity, where the fluid closures
become imprecise. Landau fluid simulations of the Langmuir mode should be therefore restricted to scales $\kpar\lambda_D>0.2$. 
At longer wavelengths, some closures can actually become ill-posed and instead of Landau damping, can produce a small positive growth rate.
For example, if one insists on numerical simulations in the domain below $\kpar\lambda_D<0.2$, the closures that have to be eliminated are closures
$R_{4,2}(\zeta), R_{5,3}(\zeta), R_{5,4}(\zeta)$, since they produce a small positive growth rate. We briefly checked, and all other closures seems to be
well-behaved all the way up to $\kpar\lambda_D = 10^{-4}$. At even longer scales, such as $\kpar\lambda_D = 10^{-5}$, two other closures become ill-posed,
the $R_{5,5}(\zeta)$ and $R_{5,6}(\zeta)$, and the remaining closures $R_{3,1}(\zeta),R_{3,2}(\zeta),R_{4,3}(\zeta),R_{4,4}(\zeta)$ do not appear to have
a length-scale restriction. It is useful to note that this is not only a problem of Landau fluid closures, but at long-wavelengths,
it is actually the kinetic theory itself that becomes very difficult to solve, and in the region $\kpar\lambda_D<0.1$, we were often not able to
obtain correct numerical solution when solving the exact dispersion relation (\ref{eq:Langmuir_kin}).

\newpage
\subsection{Selected closures for 5th-order moment}
Let's work in the 1D geometry and continue with the hierarchy. In Part 1 of this text, we called the n-th order moment $X^{(n)}$.
However, when linearizing, we want to use our $(1)$ superscript as before. Therefore, here we move the $(n)$ index of the n-th order moment down,
and refer to the n-th moment simply as $X_n$. The fifth-order moment $X_5 = m\int (v-u)^5 f dv$ is linearized according to
\begin{equation}
X_5^{(1)} = m\int v^5 f^{(1)} dv -5 u^{(1)} \underbrace{m\int v^4 f_0 dv}_{r_0},
\end{equation}  
and direct calculation yields (dropping species index $r$ everywhere except on charge $q_r$)
\begin{equation}
X_5^{(1)} = -q_r \Phi \sqrt{\frac{2T^{(0)}}{m}} \frac{p_{0}}{m}\sign(\kpar)\Big( 3\zeta+2\zeta^3+4\zeta^5 R(\zeta)-15\zeta R(\zeta) \Big),
\end{equation}
and alternatively $p_0/m=n_0 v_\textrm{th}^2/2$.

\noindent
The most precise (power-series) \emph{static} closure can be constructed with $R_{5,3}(\zeta)$ approximant
\begin{eqnarray}
R_{5,3}(\zeta):\qquad  X_5^{(1)} &=& -\frac{(104-33\pi)\sqrt{\pi}}{2(9\pi^2-69\pi+128)}i\sign(\kpar)v_\textrm{th} \widetilde{r}^{(1)}
  -\frac{(81\pi-256)}{2(9\pi^2-69\pi+128)} v_\textrm{th}^2 q^{(1)} \nn\\
  &&  -\frac{3(160-51\pi)\sqrt{\pi}}{4(9\pi^2-69\pi+128)}i\sign(\kpar )v_\textrm{th}^3 n_0 T^{(1)}
  -\frac{(135\pi^2-750\pi+1024)}{2(9\pi^2-69\pi+128)} v_\textrm{th}^2 n_0 T^{(0)} u^{(1)},
\end{eqnarray}
and other static closures with $R_{5,4}(\zeta)$ approximant
\begin{eqnarray}
  R_{5,4}(\zeta):\qquad X_5^{(1)} &=& -\frac{(21\pi-64)}{(9\pi-28)\sqrt{\pi}} i\sign(\kpar) v_\textrm{th}  \widetilde{r}^{(1)}
  +\frac{(45\pi-136)}{2(9\pi-28)}v_\textrm{th}^2 q^{(1)} \nn\\
&&  + \frac{(256-81\pi)}{2(9\pi-28)\sqrt{\pi}} i\sign(\kpar) n_0 v_\textrm{th}^3 T^{(1)},
\end{eqnarray}
with $R_{5,5}(\zeta)$ approximant
\begin{eqnarray}
  R_{5,5}(\zeta):\qquad X_5^{(1)} &=& \frac{6\sqrt{\pi}}{(9\pi-32)}i\sign(\kpar)v_\textrm{th} \widetilde{r}^{(1)}
  +\frac{3(15\pi-64)}{2(9\pi-32)}v_\textrm{th}^2 q^{(1)}, 
\end{eqnarray}
and with $R_{5,6}(\zeta)$ approximant
\begin{eqnarray}
  R_{5,6}(\zeta):\qquad X_5^{(1)} &=& -\frac{8}{3\sqrt{\pi}} i\sign(\kpar)v_\textrm{th}\widetilde{r}^{(1)} +5 v_\textrm{th}^2 q^{(1)}. 
\end{eqnarray}
In Part 1 of this guide, we derived directly from fluid hierarchy that at the linear level
\begin{equation}
\frac{\pr}{\pr t} \widetilde{r}^{(1)} +\pr_z X_5^{(1)} -3 v_\textrm{th}^2 \pr_z q^{(1)}  =0.
\end{equation}
Now, importantly, by using this equation, it is directly shown that the above static closures with $X_5^{(1)}$,
are equivalent to time-dependent (dynamic) closures with $\widetilde{r}^{(1)}$ obtained for the same $R(\zeta)$ approximants, closures
(\ref{eq:R53_closure}), (\ref{eq:R54_closurE}), (\ref{eq:R55_closurE}), (\ref{eq:R56_closure}). The process can be viewed as a
verification procedure. Indeed, it should be always possible to double check a dynamic closure, by calculating a static closure
at the next moment with the same Pad\'e approximant.

\begin{figure*}[!htpb]
  $$\includegraphics[width=0.95\linewidth]{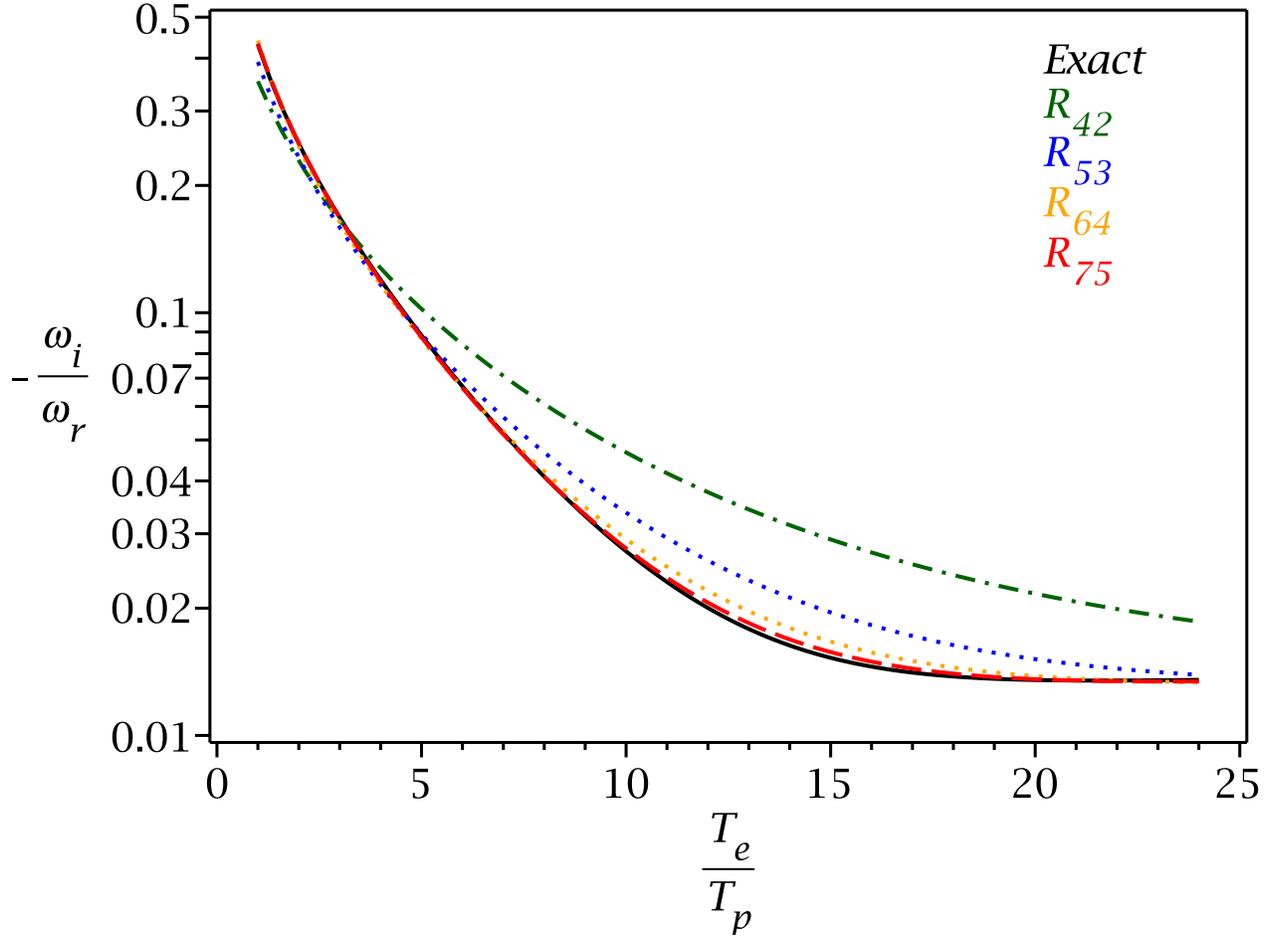}$$
  \caption{Landau damping of the ion-acoustic mode, calculated with exact $R(\zeta)$ - black line; $R_{4,2}(\zeta)$ - green line; $R_{5,3}(\zeta$) - blue line;
    $R_{6,4}(\zeta)$ - orange line; and $R_{7,5}(\zeta)$ - red line. The solutions represent the most precise dynamic closures that can be
    constructed for the 3rd-order moment (heat flux), 4th-order moment, 5th-order moment, and 6th-order moment.
    It was analytically verified that all closures are ``reliable'', i.e. equivalent to the kinetic dispersion relation once
    $R(\zeta)$ is replaced by the associated $R_{n,n'}(\zeta)$ approximant. The next most precise closure constructed for the 7th-order moment is $R_{8,6}(\zeta)$,
    which is not plotted, but we checked that the solution is basically not distinguishable (by eye) from the exact $R(\zeta)$ solution. 
    Figure shows that it is possible to reproduce Landau damping in the fluid framework to any desired precision.} \label{fig:IAmode}
\end{figure*}

The most precise (power series) \emph{dynamic} closure with $X_5$, is constructed with approximant $R_{6,4}(\zeta)$, by searching for a solution
\begin{equation}
\big[ \zeta  + \alpha_{x_5} \big] X_5^{(1)} = \alpha_r \widetilde{r}^{(1)} + \alpha_q q^{(1)} +  \alpha_t T^{(1)} +  \alpha_u u^{(1)}, 
\end{equation}
and the closure in real space reads
\begin{empheq}[box=\fbox]{align}
R_{6,4}(\zeta):\qquad &  \Big[ \frac{d}{dt}  - \frac{3(180\pi^2-1197\pi+1984)\sqrt{\pi}}{(801\pi^2-5124\pi+8192)} v_{\textrm{th}} \pr_z\mathcal{H} \Big] X_5^{(1)}
  =  - v_\textrm{th}^2 \frac{3(675\pi^2-4728\pi+8192)}{2(801\pi^2-5124\pi+8192)} \pr_z \widetilde{r}^{(1)} \nn\\
& \qquad  +  v_\textrm{th}^3 \frac{3(285\pi-896)\sqrt{\pi}}{2(801\pi^2-5124\pi+8192)} \pr_z\mathcal{H} q^{(1)}
  - v_\textrm{th}^4 n_0  \frac{3(945\pi^2-8184\pi+16384)}{4(801\pi^2-5124\pi+8192)} \pr_z T^{(1)} \nn\\
& \qquad  +  v_\textrm{th}^3 n_0 T_0 \frac{9(450\pi^2-2799\pi+4352)\sqrt{\pi}}{(801\pi^2-5124\pi+8192)}  \pr_z\mathcal{H} u^{(1)}. \label{eq:R64_closure}
\end{empheq}
The closure has precision $o(\zeta^5)$, $o(\zeta^{-6})$. It was verified that the closure is reliable, i.e. it satisfies (\ref{eq:soundKin})
once $R(\zeta)$ is replaced by $R_{6,4}(\zeta)$. The closure is plotted in Figure \ref{fig:IAmode} with orange line.
\clearpage
\subsection{Selected closures for 6th-order moment}
The sixth-order moment $X_6=m\int (v-u)^6 f dv$ is linearized simply as $X_6^{(1)}=m\int v^6 f^{(1)} dv$, and since
\begin{equation}
\frac{1}{\sqrt{\pi}} \int \frac{x^6 e^{-x^2}}{x-x_0} dx = \sign(\kpar) \Big( \frac{3}{4}\zeta +\frac{1}{2}\zeta^3 +\zeta^5 R(\zeta) \Big),
\end{equation}
and direct calculation yields
\begin{equation}
X_6^{(1)} = -q_r \Phi n_0 \frac{p_0^2}{\rho_0^2} \Big( 15+6\zeta^2+4\zeta^4+8\zeta^6 R(\zeta) \Big),
\end{equation}  
and alternatively $p_0/\rho_0=v_{\textrm{th}}^2/2$. Separating the deviation of this moment with tilde
(similarly to $\widetilde{r}$, see also Part 1 of this guide) is done according to
\begin{equation}
\widetilde{X}_6^{(1)} = X_6^{(1)} -15\frac{p_0^3}{\rho_0^2}\Big( 3\frac{p^{(1)}}{p_0} -2\frac{n^{(1)}}{n_0} \Big),
\end{equation}
which directly yields
\begin{equation}
\widetilde{X}_6^{(1)} = -q_r \Phi n_0 \frac{p_0^2}{\rho_0^2} \Big( 30R(\zeta)-30+6\zeta^2-90\zeta^2R(\zeta) +4\zeta^4+8\zeta^6 R(\zeta) \Big),
\end{equation}
Considering static closures, the most precise power series closure is constructed with $R_{6,4}(\zeta)$ and the closure reads
\begin{eqnarray}
  R_{6,4}(\zeta):\qquad \widetilde{X}_6^{(1)} &=& -\frac{3(180\pi^2-1197\pi+1984)\sqrt{\pi}}{(801\pi^2-5124\pi+8192)} v_{\textrm{th}} \mathcal{H} X_5^{(1)}
  +\frac{3(675\pi^2-4728\pi+8192)}{2(801\pi^2-5124\pi+8192)} v_{\textrm{th}}^2 \widetilde{r}^{(1)} \nn\\
  &&  -\frac{3(285\pi-896)\sqrt{\pi}}{2(801\pi^2-5124\pi+8192)} v_{\textrm{th}}^3 \mathcal{H} q^{(1)}
  -\frac{3(7065\pi^2-43056\pi+65536)}{4(801\pi^2-5124\pi+8192)}  n_0 v_{\textrm{th}}^4 T^{(1)} \nn\\
  && -\frac{9(450\pi^2-2799\pi+4352)\sqrt{\pi}}{(801\pi^2-5124\pi+8192)} n_0 T_0 v_{\textrm{th}}^3 \mathcal{H} u^{(1)}. \label{eq:R64_closureS}
\end{eqnarray}
This verifies that the dynamic $R_{6,4}(\zeta)$ closure (\ref{eq:R64_closure}) was calculated correctly,
since from the simple fluid approach (Part 1), the static and dynamics closures (\ref{eq:R64_closureS}), (\ref{eq:R64_closure}) must be related by
\begin{equation}
\frac{\pr}{\pr t} X_5^{(1)} +\pr_z \widetilde{X}_6^{(1)} +\frac{15}{2} n_0 v_\textrm{th}^4 \pr_z T^{(1)}  =0.
\end{equation}

The most precise (power series) dynamic closure for $\widetilde{X}_6^{(1)}$ can be constructed with approximant $R_{7,5}(\zeta)$, by searching for a solution   
\begin{equation}
\big[ \zeta  + \alpha_{x_6} \big] \widetilde{X}_6^{(1)} = \alpha_{x_5}X_5^{(1)} + \alpha_r \widetilde{r}^{(1)} + \alpha_q q^{(1)} +  \alpha_t T^{(1)} +  \alpha_u u^{(1)}, 
\end{equation}
and the closure in real space reads
\begin{empheq}[box=\fbox]{align}
  R_{7,5}(\zeta):\qquad &  \bigg[ \frac{d}{dt}  + \frac{18(1545\pi^2-9743\pi+15360)\sqrt{\pi} }{(10800\pi^3-120915\pi^2+440160\pi-524288)} v_{\textrm{th}}\pr_z\mathcal{H} \bigg]
  \widetilde{X}_6^{(1)} = \nn\\
&\qquad   + \frac{3(52425\pi^2-331584\pi+524288)}{2(10800\pi^3-120915\pi^2+440160\pi-524288)} v_{\textrm{th}}^2  \pr_z X_5^{(1)} \nn\\
  &\qquad   +  \frac{3(7875\pi^2-50490\pi+80896)\sqrt{\pi}}{(10800\pi^3-120915\pi^2+440160\pi-524288)} v_{\textrm{th}}^3 \pr_z\mathcal{H} \widetilde{r}^{(1)} \nn\\
  &\qquad + \frac{3(162000\pi^3-1758825\pi^2+6263040\pi-7340032)}{4(10800\pi^3-120915\pi^2+440160\pi-524288)} v_{\textrm{th}}^4 \pr_z  q^{(1)} \nn\\
  &\qquad  -    \frac{27(15825\pi^2-99260\pi+155648)\sqrt{\pi}}{2(10800\pi^3-120915\pi^2+440160\pi-524288)} v_{\textrm{th}}^5 n_0 \pr_z\mathcal{H} T^{(1)} \nn\\
  &\qquad + \frac{3(189000\pi^3-1612215\pi^2+4534656\pi-4194304)}{2(10800\pi^3-120915\pi^2+440160\pi-524288)}  v_{\textrm{th}}^4 n_0 T_0 \pr_z u^{(1)}. \label{eq:R75closure}
\end{empheq}
The closure has precision $o(\zeta^6)$, $o(\zeta^{-7})$, and it was verified that the closure is reliable. The closure is plotted in Figure \ref{fig:IAmode} with red line.


\clearpage
\subsection{Convergence of fluid and kinetic descriptions}
In general, for a given $X_n$, the most precise (power series) closures are of course \emph{dynamic} closures, and   
we have seen that for the 3rd-order moment it is $R_{4,2}(\zeta)$, for the 4th-order moment it is $R_{5,3}(\zeta)$, for the 5th-order moment it is $R_{6,4}(\zeta)$,
and for the 6th-order moment it is $R_{7,5}(\zeta)$. Therefore, it is reasonable to make a conjecture that for an nth-order moment $X_n$,
the most precise closure will be constructed with approximant $R_{n+1,n-1}(\zeta)$.

The dynamic closures above are directly related to the most precise (power series) \emph{static} closures that can be constructed, and we have seen that for the
3rd-order moment it is with approximant $R_{3,1}(\zeta)$, for the 4th-order moment it is $R_{4,2}(\zeta)$, for the 5th-order moment it is $R_{5,3}(\zeta)$, and
for the 6th-order moment it is $R_{6,4}(\zeta)$, and therefore for an nth-order moment, it will be with approximant $R_{n,n-2}(\zeta)$.  
Regardless if dynamic or static closures are used, this implies that one can reproduce the (linear) Landau damping phenomenon in the fluid framework, to any desired precision,     
which establishes convergence of fluid and kinetic descriptions.

\begin{figure*}[!htpb]
  $$\includegraphics[width=0.48\linewidth]{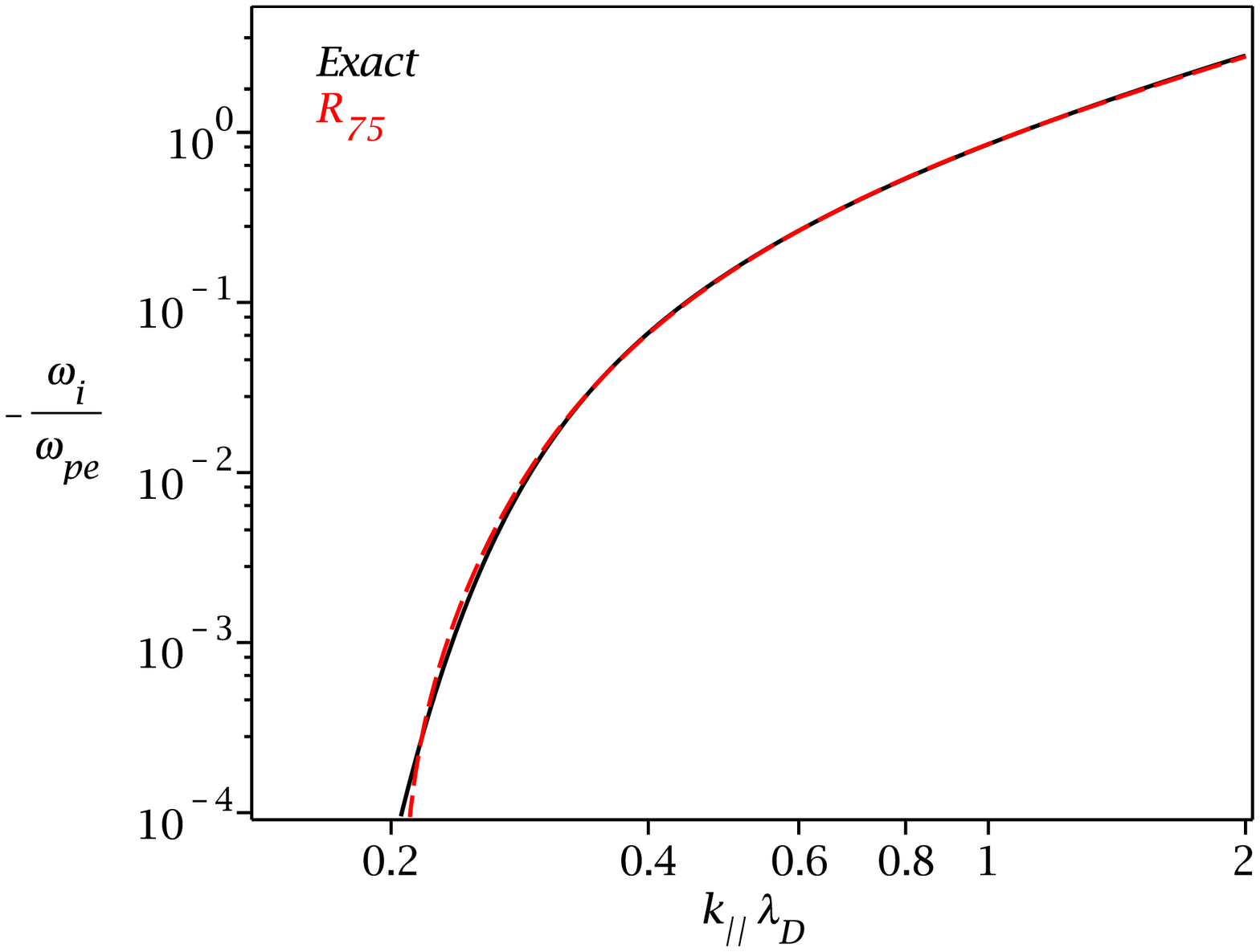}\hspace{0.03\textwidth}\includegraphics[width=0.48\linewidth]{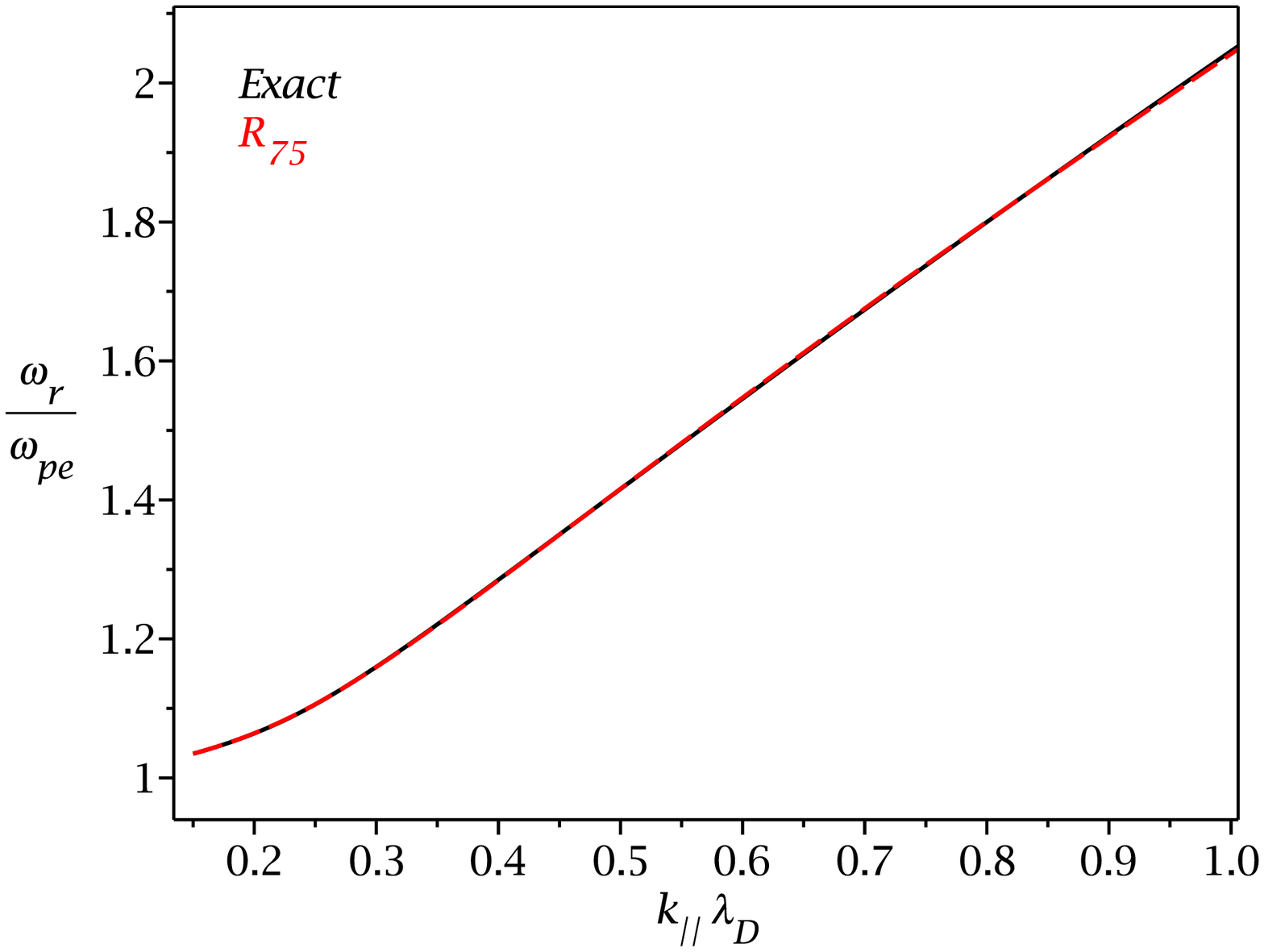}$$  
  \caption{Landau damping of the Langmuir mode (left), and real frequency (right), calculated with the exact $R(\zeta)$ - black line, and $R_{7,5}(\zeta)$ - red line.} \label{fig:Lmode}
\end{figure*}

The convergence was shown here in 1D (electrostatic) geometry, by considering the long-wavelength low-frequency ion-acoustic mode.
Nevertheless, the 1D closures have general validity, and are of course valid also for the Langmuir mode, that we considered in section 3.14.
However, see the discussion about limitations of the Langmuir mode modeling at the end of that section, since the closures can become unstable for $\kpar\lambda_D <0.2$,
i.e. in the long-wavelength limit. For a curious reader, the damping and real frequency of the Langmuir mode obtained with $R_{7,5}(\zeta)$, are plotted in Figure \ref{fig:Lmode}.    

If one wants to pursue a proof of our conjecture, the general Landau integral with $x^n$ can be calculated, for example by considering separate cases
for ``n'' being odd and even. The result can be expressed as
\begin{eqnarray}
  n=\textrm{odd}: \qquad \frac{1}{\sqrt{\pi}}\int^{\infty}_{-\infty} \frac{x^n e^{-x^2}}{x-x_0}dx &=&
  \zeta^{n-1} R(\zeta) +\sum_{l=0}^{(n-3)/2} \frac{(n-2l-2)!!}{2^{(n-2l-1)/2}} \zeta^{2l}; \nn\\
n=\textrm{even}: \qquad \frac{1}{\sqrt{\pi}}\int^{\infty}_{-\infty} \frac{x^n e^{-x^2}}{x-x_0}dx
  &=& \sign(\kpar)\bigg[ \zeta^{n-1} R(\zeta) + \sum_{l=0}^{(n-4)/2}  \frac{(n-2l-3)!!}{2^{(n-2l-2)/2}} \zeta^{2l+1} \bigg],
\end{eqnarray}
and it is valid for $n\ge 3$. Alternatively, one could say that the result is valid for $n\ge 1$ and that the sums are zero when
the upper index is negative. One can write expressions for the general n-th moment $X_n^{(1)}$, and the moment is proportional to $\zeta^n R(\zeta)$.
Therefore, considering static closures where the $X_n^{(1)}$ is expressed through all the lower-order moments $X_m^{(1)}$; $m=1\ldots n-1$ (for even moments
the deviations $\widetilde{X}^{(1)}$ have to be considered), it is obvious that the closure has to be achieved with n-th order Pad\'e approximant
of $R(\zeta)$. Similarly, considering dynamic closures where the $\zeta X_n^{(1)} \sim \zeta^{n+1} R(\zeta)$ is expressed through all the lower-order moments,
the closure has to be achieved with (n+1)-th order Pad\'e approximant of $R(\zeta)$. To finish the proof, one needs to show that the number of
required asymptotic points corresponds to $R_{n,n-2}(\zeta)$ and $R_{n+1,n-1}(\zeta)$, and that such a closure is ``reliable''.  

The next logical step would be to establish such analytic convergence of fluid and kinetic descriptions in 3D electromagnetic geometry
in the gyrotropic limit. However, in 3D, for a given n-th order tensor $\boldsymbol{X}_n$, the number of its gyrotropic moments is equal to $1+\textrm{int}[n/2]$
and increases with $n$. Therefore, it might be much more difficult to show the convergence in 3D, even though the convergence should still exist.

\clearpage
\section{3D geometry (electromagnetic)} \label{section:3D}
Considering gyrotropic $f_0$, let's remind ourselves the linearized Vlasov equation (\ref{eq:KinBase}), that reads  
\begin{equation}
  \frac{\pr f^{(1)}}{\pr t}+ \bV\cdot\nabla f^{(1)} -\frac{q_r B_0}{m_r c}\frac{\pr f^{(1)}}{\pr \phi} 
  = - \frac{q_r}{m_r}\Big(\bE^{(1)}+\frac{1}{c}\bV\times\bb^{(1)}\Big)\cdot\nabla_v f_0. \label{eq:Gyro_pic2}
\end{equation}
We want to describe the simplest kinetic effects and we demand that $f^{(1)}$ must be gyrotropic as well, so $\pr f^{(1)}/\pr\phi=0$.
This eliminates the third term on the left hand side of (\ref{eq:Gyro_pic2}) that is responsible for complicated non-gyrotropic effects with associated Bessel functions.
However, even without this term the equation still appears to be complicated. For gyrotropic $f_0$, the operator on the right hand side can be shown to be
(see Appendix, eq. (\ref{eq:Gyro_pic}), written in Fourier space)
\begin{eqnarray} 
  \Big( \bE + \frac{1}{c}\bV\times\bb \Big) \cdot \nabla_{v} f_0
  &=& ( E_x v_x + E_y v_y ) \bigg[ \Big( 1-\frac{v_\parallel \kpar}{\omega}\Big) \frac{1}{v_\perp} \frac{\pr f_0}{\pr v_\perp} + \frac{\kpar}{\omega}
    \frac{\pr f_0}{\pr v_\parallel} \bigg] \nn\\
  &+& E_z \bigg[ \frac{\pr f_0}{\pr v_\parallel}  -\frac{v_x k_x + v_y k_y}{\omega}\Big( \frac{\pr f_0}{\pr v_\parallel}
    - \frac{v_\parallel}{v_\perp} \frac{\pr f_0}{\pr v_\perp} \Big) \bigg].  
\end{eqnarray}
Written in the cylindrical co-ordinate system
\begin{eqnarray}
\bV =  \left( \begin{array}{c}
    v_\perp \cos \phi \\
    v_\perp \sin \phi \\
    v_\parallel
\end{array} \right),\qquad
\bk =  \left( \begin{array}{c}
    k_\perp \cos \psi \\
    k_\perp \sin \psi \\
    k_\parallel
  \end{array} \right),
\end{eqnarray}
so that
\begin{eqnarray}
  v_x k_x +v_y k_y &=& v_\perp k_\perp \cos\phi\cos\psi + v_\perp k_\perp \sin\phi\sin\psi = v_\perp k_\perp \cos( \phi-\psi);\\
  \bV\cdot\bk &=&   v_\parallel\kpar+ v_\perp k_\perp \cos( \phi-\psi),
\end{eqnarray}
which yields
\begin{eqnarray} 
  \Big( \bE + \frac{1}{c}\bV\times\bb \Big) \cdot \nabla_{v} f_0
  &=& ( E_x \cos\phi + E_y \sin\phi ) \bigg[ \Big( 1-\frac{v_\parallel \kpar}{\omega}\Big) \frac{\pr f_0}{\pr v_\perp} + \frac{v_\perp \kpar}{\omega}
    \frac{\pr f_0}{\pr v_\parallel} \bigg] \nn\\
  &+& E_z \bigg[ \frac{\pr f_0}{\pr v_\parallel}  -\frac{k_\perp \cos(\phi-\psi)}{\omega}\Big( v_\perp \frac{\pr f_0}{\pr v_\parallel}
    - v_\parallel \frac{\pr f_0}{\pr v_\perp} \Big) \bigg]. 
\end{eqnarray}
The Vlasov equation in Fourier space now reads
\begin{eqnarray}
  -i\Big(\omega- v_\parallel\kpar- v_\perp k_\perp \cos( \phi-\psi)\Big) f^{(1)} &=& -\frac{q_r}{m_r} \bigg\{
  ( E_x \cos\phi + E_y \sin\phi ) \bigg[ \Big( 1-\frac{v_\parallel \kpar}{\omega}\Big) \frac{\pr f_0}{\pr v_\perp} + \frac{v_\perp \kpar}{\omega}
    \frac{\pr f_0}{\pr v_\parallel} \bigg] \nn\\
  &+& E_z \bigg[ \frac{\pr f_0}{\pr v_\parallel}  -\frac{k_\perp \cos(\phi-\psi)}{\omega}\Big( v_\perp \frac{\pr f_0}{\pr v_\parallel}
    - v_\parallel \frac{\pr f_0}{\pr v_\perp} \Big) \bigg]\bigg\}. \label{eq:Strange_nonInt}
\end{eqnarray}
\emph{This equation is not very useful.} If the equation is divided by  the $(\omega-\bV\cdot\bk)$ to obtain $f^{(1)}$, and integration over $\int_0^{2\pi}d\phi$ is
attempted, leads to integrals that are not well defined. On the other hand, if (\ref{eq:Strange_nonInt}) is directly integrated over $d\phi$ (each
side separately), almost all the terms disappear since $\int_0^{2\pi} \cos(\phi-\psi)d\phi=0$ etc., except
\begin{equation} \label{eq:boring}
-i\big( \omega- v_\parallel\kpar\big) f^{(1)} = -\frac{q_r}{m_r} E_z \frac{\pr f_0}{\pr v_\parallel}, 
\end{equation}
and the system reduces to the simplest case of Landau damping that we have already described in detail (even though only in 1D geometry).
We could divide (\ref{eq:boring}) by $(\omega-v_\parallel\kpar)$, integrate the system in 3D geometry and consider Landau fluid closures,
but this would be a bit boring right now. We want to get a bit more kinetic effects out of the system. We need a different approach 
and we need to obtain a better gyrotropic limit for $f^{(1)}$.

It turns out that to obtain the correct gyrotropic limit for $f^{(1)}$, the 3rd term in the Vlasov equation (\ref{eq:Gyro_pic2}) cannot be just straightforwardly neglected.
The term has to be kept there, the relatively complicated integration around the unperturbed orbit has to be performed (see Appendix, Section \ref{sec:Orbit}),
and only then the term
can be removed in a limit. This is very similar to other mathematical techniques that were encountered earlier, for example when calculating the Fourier transform of $\sign(\kpar)$,
where instead of that function, one needs to consider $\sign(\kpar)e^{-\alpha|\kpar|}$, and only after the calculation the term is removed with the limit $\alpha\to 0$.
Without the additional term $e^{-\alpha|\kpar|}$ that was removed later, the calculations were not clearly defined, and a very similar situation is encountered now.
Nevertheless, it is indeed mind boggling that the complicated integration around the unperturbed orbit has to be performed to recover the gyrotropic limit.
This is exactly why the 3D case is so much more complicated than the previously studied 1D case, even though the Landau fluid closures will not be more complicated at all,
as we will see later. An alternative approach that we will discuss only very briefly, is to use the guiding center variables where the gyrotropic limit is recovered perhaps more
naturally. However, we will skip a huge amount of calculations that lead to do the guiding center approach, so the amount of complexity is probably similar at the end.
\subsection{Gyrotropic limit for $f^{(1)}$}
We need to consider the full kinetic $f^{(1)}$ with all non-gyrotropic effects, that is obtained in the Appendix, Section \ref{sec:Orbit}, eq. (\ref{eq:GrandF1}). 
 By using the z-component of the induction equation $\pr \bb/\pr t=-c\nabla\times\bE$ written in Fourier space (\ref{eq:Induct_Bz_end})
(that is an equation of general validity not introducing any simplifications), the general $f^{(1)}$ eq. (\ref{eq:GrandF1}) is slightly rewritten as 
\begin{eqnarray}
  f^{(1)}_r &=& - \frac{iq_r}{m_r} \sum_{n=-\infty}^\infty \sum_{m=-\infty}^\infty
  \frac{e^{+i (m-n)(\phi-\psi)}}{\omega-k_\parallel v_\parallel -n \Omega_r} J_m (\lambda_r) \bigg\{ \bigg[
    \frac{nJ_n(\lambda_r)}{\lambda_r}  \Big(E_x\cos\psi +E_y\sin\psi\Big)  \nn\\
&&\qquad    +i J_n'(\lambda_r) \frac{\omega}{ck_\perp}B_z  \bigg]
  \bigg[ \Big( 1- \frac{k_\parallel v_\parallel}{\omega}\Big) \frac{\pr f_{0r}}{\pr v_\perp}
    +\frac{k_\parallel v_\perp}{\omega}\frac{\pr f_{0r}}{\pr v_\parallel} \bigg]\nn\\
  && + E_z J_n(\lambda_r) \bigg[ \frac{\pr f_{0r}}{\pr v_\parallel}
    - \frac{n\Omega_r}{\omega}\Big(\frac{\pr f_{0r}}{\pr v_\parallel} -\frac{v_\parallel}{v_\perp} \frac{\pr f_{0r}}{\pr v_\perp}  \Big)
   \bigg] \bigg\}. \label{eq:GrandF1-a}
\end{eqnarray}
This $f^{(1)}$ contains all the information of linear kinetic theory, with associated Bessel functions $J_n(\lambda_r)$, where $\lambda_r=k_\perp v_\perp /\Omega_r$
and $\Omega_r=q_r B_0/(m_r c)$.  Two summations through integers ``n'' and ``m'' are present in (\ref{eq:GrandF1-a}),
  that originate in using identities (\ref{eq:StixB1}), (\ref{eq:StixB2}). The general (\ref{eq:GrandF1-a}) contains ``singularities'' where
  $\omega-k_\parallel v_\parallel -n \Omega_r$ becomes zero, that are called wave-particle resonances. For $n=0$ the resonance is called the Landau resonance, and
  resonances for $n\neq 0$ are called cyclotron resonances.
To get rid of the  summations and Bessel functions, we want to consider dynamics at spatial scales that are much larger than the particle gyroradius,
  which corresponds to limit $\lambda_r\ll 1$. Additionally, we will need to consider  low-frequency limit $\omega/\Omega_r\ll 1$.
   We find illuminating to first separate the $n=0$ resonance from all
the other expressions, without performing any approximations, i.e we want to separate
\begin{equation}
 f^{(1)}_r =  f^{(1)}_r \Big|_{n=0} + f^{(1)}_r \Big|_{n\neq 0}.
\end{equation}  
Separating the $n=0$ case directly yields
\begin{eqnarray}
f^{(1)}_r \Big|_{n=0} = - \frac{iq_r}{m_r} \sum_{m=-\infty}^\infty
  \frac{e^{+i m(\phi-\psi)}}{\omega-k_\parallel v_\parallel} J_m (\lambda_r) \bigg\{ && i J_0'(\lambda_r) \frac{B_z}{ck_\perp} 
  \bigg[ \Big( \omega-k_\parallel v_\parallel\Big) \frac{\pr f_{0r}}{\pr v_\perp}
    +k_\parallel v_\perp\frac{\pr f_{0r}}{\pr v_\parallel} \bigg]\nn\\
  && + E_z J_0(\lambda_r) \frac{\pr f_{0r}}{\pr v_\parallel} \bigg\}.
\end{eqnarray}
Note that $nJ_n(x)/x=(J_{n-1}(x)+J_{n+1}(x))/2$, which when evaluated for $n=0$ is zero exactly, since $J_{-1}(x)+J_1(x)=0$ exactly.
Since there is no dependence on angles $\phi,\psi$ inside of the big brackets, the sum can be summed (or put to its original form where it came from)
\begin{eqnarray}
f^{(1)}_r \Big|_{n=0} = - \frac{q_r}{m_r} \frac{e^{+i \lambda_r \sin(\phi-\psi)}}{\omega-k_\parallel v_\parallel} \bigg\{ && - J_0'(\lambda_r) \frac{B_z}{ck_\perp} 
  \bigg[ \Big( \omega-k_\parallel v_\parallel\Big) \frac{\pr f_{0r}}{\pr v_\perp}
    +k_\parallel v_\perp\frac{\pr f_{0r}}{\pr v_\parallel} \bigg] + i E_z J_0(\lambda_r) \frac{\pr f_{0r}}{\pr v_\parallel} \bigg\}.
\end{eqnarray}
Very interestingly, for one $B_z$ term, the complicated denominator $\omega-\kpar v_\parallel$ cancels out, yielding
\begin{eqnarray}
f^{(1)}_r \Big|_{n=0} = - \frac{q_r}{m_r} e^{+i \lambda_r \sin(\phi-\psi)} \bigg\{ && - J_0'(\lambda_r) \frac{B_z}{ck_\perp} 
  \bigg[ \frac{\pr f_{0r}}{\pr v_\perp}
    +\frac{k_\parallel v_\perp}{\omega-k_\parallel v_\parallel}\frac{\pr f_{0r}}{\pr v_\parallel} \bigg] +
  J_0(\lambda_r) \frac{iE_z}{\omega-k_\parallel v_\parallel} \frac{\pr f_{0r}}{\pr v_\parallel} \bigg\}. \label{eq:n0_last}
\end{eqnarray}
This is an exact kinetic expression for $f^{(1)}$ corresponding to $n=0$ resonances, that is accompanied by an expression for all the other resonances $f^{(1)} |_{n\neq 0}$
(that is equivalent to (\ref{eq:GrandF1-a}) where $n\neq 0$ is added below the sum with $n$).
Now considering the limit $\lambda_r\ll 1$, the Bessel functions $J_0(\lambda_r)=1$, $J_0'(\lambda_r)=-\lambda_r/2$, the exponential term disappears, which yields 
the final $f^{(1)}$ in the gyrotropic limit that reads
\begin{equation}  
\boxed{  f^{(1)}_r  = -  
   \frac{v_\perp}{2}\frac{B_z}{B_0} 
  \bigg[ \frac{\pr f_{0r}}{\pr v_\perp}
    +\frac{k_\parallel v_\perp }{(\omega-k_\parallel v_\parallel)}\frac{\pr f_{0r}}{\pr v_\parallel} \bigg]
  -  \frac{q_r}{m_r} \frac{i E_z}{(\omega-k_\parallel v_\parallel)}  \frac{\pr f_{0r}}{\pr v_\parallel}, } \label{eq:f1_gyrotropic}
\end{equation}
or alternatively 
\begin{equation}  \label{eq:f1_gyrotropic2}
f^{(1)}_r  = \underbrace{- \frac{B_z}{2B_0} v_\perp\frac{\pr f_{0r}}{\pr v_\perp}}_{\mu=\textrm{const.}}   
  \underbrace{- \frac{B_z}{2B_0}\frac{k_\parallel v_\perp^2 }{(\omega-k_\parallel v_\parallel)}\frac{\pr f_{0r}}{\pr v_\parallel}}_{\textrm{magnetic mirror force}}
  \underbrace{-  \frac{q_r}{m_r} \frac{i E_z}{(\omega-k_\parallel v_\parallel)}  \frac{\pr f_{0r}}{\pr v_\parallel}}_{\textrm{Coulomb force}}.
\end{equation}
As we will see shortly in Section \ref{sec:Forces}, the expression has a very nice physical interpretation, where the first term comes from the conservation of the magnetic moment $\mu$,
the second term comes from the magnetic mirror force and the third term comes from the Coulomb force.
The same expression is obtained by directly picking up the $m=0,n=0$ contributions from the general (\ref{eq:GrandF1-a}).
Up to replacing $B_z$ with $|\bb|$, the expression agrees for example with eq. (19) of \cite{FerriereAndre2002}, and is of course equivalent to expressions of \cite{Snyder1997}
(formulated in the gyrofluid formalism). In those works, the expression is derived perhaps more elegantly, in the so-called
\emph{guiding-center limit of the Vlasov equation} (see \cite{Kulsrud1983}). The difference between $B_z$ and $|\bb|$ arises, because the fully kinetic $f^{(1)}$ in (\ref{eq:GrandF1-a})
is linearized completely. 

Note that to obtain the gyrotropic limit (\ref{eq:f1_gyrotropic}), we did not have to  explicitly perform the low-frequency limit $\omega/\Omega_r\ll 1$.
However, it is important to emphasize that by only picking up the $n=0$ resonances,  we have performed the low-frequency limit implicitly.
The power series expansion of the Bessel functions for $n\ge 0$ reads (with integer $n$)  
\begin{equation}
  J_n(x) = \sum_{s=0}^{\infty} \frac{(-1)^s}{s!(n+s)!} \Big(\frac{x}{2}\Big)^{n+2s}; \qquad
  J_{-n}(x) = \sum_{s=0}^{\infty} \frac{(-1)^{n+s}}{s!(n+s)!} \Big(\frac{x}{2}\Big)^{n+2s},
\end{equation}
where the second expression can be easily replaced by $J_{-n}(x)=(-1)^n J_n(x)$. The first few terms are
\begin{eqnarray}
  J_0(x) = 1-\frac{x^2}{4}+\frac{x^4}{64}+\cdots; \qquad J_1(x) &=& \frac{x}{2}-\frac{x^3}{16}+\cdots; \qquad J_2(x) = \frac{x^2}{8}-\frac{x^4}{96}+\cdots;\nn\\
  J_{-1}(x) &=& -\frac{x}{2}+\frac{x^3}{16}+\cdots; \qquad J_{-2}(x) = \frac{x^2}{8}-\frac{x^4}{96}+\cdots,
\end{eqnarray}
and the derivatives of these functions read 
\begin{eqnarray}
  J_0'(x) = -\frac{x}{2}+\frac{x^3}{16}+\cdots; \qquad J_1'(x) &=& \frac{1}{2}-\frac{3x^2}{16}+\cdots; \qquad J_2'(x) = \frac{x}{4}-\frac{x^3}{24}+\cdots;\nn\\
  J_{-1}'(x) &=& -\frac{1}{2}+\frac{3x^2}{16}+\cdots; \qquad J_{-2}'(x) = \frac{x}{4}-\frac{x^3}{24}+\cdots,
\end{eqnarray}
and the derivatives can be also calculated by using identity $J_n'(x)=(J_{n-1}(x)-J_{n+1}(x))/2$. In the full equation (\ref{eq:GrandF1-a}),
the term with $E_x,E_y$ components contains $J_m(x)(J_{n-1}(x)+J_{n+1}(x))$, so for $m=0$ terms with resonances $n=\pm 1$ do not disappear in the limit $\lambda_r\ll 1$.
Similar situation is for the $B_z$ components (which for $n\neq 0$ is actually easier to reformulate to the original formulation without the $B_z$ induction equation to
recover the correct limit). If we like it or not, to get rid of these terms and to obtain the gyrotropic limit (\ref{eq:f1_gyrotropic}),
one has to do the low frequency limit $\omega\ll \Omega_r$ as well.

A few notes are in order. 1) If we now calculate the kinetic moments with $f^{(1)}$ described by (\ref{eq:f1_gyrotropic}), which was obviously obtained in the low-frequency limit,
and find possible fluid closures for the heat fluxes $q_\parallel,q_\perp$ or the 4th-order moments
$\widetilde{r}_{\parallel \parallel},\widetilde{r}_{\parallel\perp},\widetilde{r}_{\perp\perp}$,
such a fluid model will not become necessarily restricted only to a low frequency regime $\omega\ll \Omega$.
At the linear level, the parallel propagating ion-cyclotron and whistler modes are completely independent from the Landau fluid closures, and
these modes remain undamped.\footnote{The situation is different in nonlinear numerical simulations, where the modes are damped by nonlinear
coupling with the strongly Landau damped ion-acoustic (sound) mode, see for example Landau fluid simulations of \cite{Hunana2011}.} 
For example Figure 6 in Part 1 remains unchanged, and the simplest ion-cyclotron resonance where $\omega\to\Omega$ for high wavenumbers (neglecting FLRs),
will not be suddenly ``removed'' by using a low-frequency Landau fluid closure. All figures for the (strictly) parallel firehose instability remain unchanged,
and the same applies to the perpendicular fast mode.  

2) There is nothing ``esoteric'' about ion-cyclotron resonances. Similarly to the kinetic effect of Landau damping, the ion-cyclotron
resonances just represent some integral, which indeed has some wave-particle ``resonance'', i.e. the integral has some singularity in the denominator. 
Similarly to Landau damping, in the case of bi-Maxwellian $f_0$ this singularity can be expressed through the plasma dispersion function $Z(\zeta)$
(similar generalizations exist for a bi-Kappa distribution etc.).
The variable $\zeta$ is only modified to include the resonances, and for $n=\pm 1$ one can work with
\begin{equation}
  \zeta_{+1} = \frac{(\omega+\Omega)\sqrt{\alpha_\parallel}}{|\kpar|}=\frac{(\omega+\Omega)}{|\kpar| v_{\textrm{th}\parallel}};
  \qquad \zeta_{-1} = \frac{(\omega-\Omega)\sqrt{\alpha_\parallel}}{|\kpar|}=\frac{(\omega-\Omega)}{|\kpar| v_{\textrm{th}\parallel} },
\end{equation}
or for general $n$ with $\zeta_n=(\omega+n\Omega)/(|\kpar| v_{\textrm{th}\parallel})$. 
No new discussion how to treat this singularity is required. The singular point $x_0$ in the complex plane is only moved to some other location,
and all the previous discussion about the Landau integral fully applies. We could potentially integrate over all the ion-cyclotron resonances
and obtain expressions for the heat flux or the 4th-order moments (with the same techniques as plasma physics books do,
even though they usually stop at the 1st-order velocity moment, since it is enough to obtain the kinetic dispersion relation). Even though complicated in
detail, these would be just standard kinetic calculations. The difference between advanced fluid and kinetic description is, that 
we need to find a closure after all of these kinetic calculations. I.e., we need to find a way to express the last considered moment through lower
order moments, that the closure is valid for all the $\zeta$ values, for example, by using the Pad\'e approximation for $R(\zeta)$. Such a closure remains elusive 
for the ion-cyclotron resonances.  

3) Advanced fluid models are not restricted to work with $f^{(1)}$ in the gyrotropic limit (\ref{eq:f1_gyrotropic}). 
In Landau fluid models of \cite{PassotSulem2007}, no assumption about the size of the gyroradius is made, and only the low-frequency condition is used 
and therefore, the $f^{(1)}$ of these fluid models contain Bessel functions $J_n(\lambda_r)$.
The integrals over $d v_\perp$ are slightly more difficult, and for example
if a term proportional to $J_0(\lambda)J_0(\lambda) f_0$ is encountered, the integration over $d v_\perp$ ($d^3v=v_\perp dv_\perp dv_\parallel d\phi$)
is calculated as
\begin{equation}
\int_0^\infty x J_n^2 (ax) e^{-x^2}dx = \frac{1}{2}e^{-a^2/2}I_n(a^2/2),
\end{equation}
implying
\begin{equation}
  \int_0^\infty J_0^2\big( \frac{k_\perp}{\Omega} v_\perp \big) e^{-\alpha_\perp v_\perp^2} v_\perp dv_\perp = \Big[x=\sqrt{\alpha_\perp}v_\perp \Big]
  =  \frac{1}{\alpha_\perp} \int_0^\infty J_0^2\big( \frac{k_\perp}{\Omega\sqrt{\alpha_\perp}} x \big) e^{-x^2} x dx
  = \frac{1}{2\alpha_\perp} e^{-b} I_0 (b), 
\end{equation}
where the new parameter $b$ (which should not be confused with the magnetic field unit vector $\hat{\boldsymbol{b}}$) is
\begin{equation}
b = \frac{k_\perp^2}{2\Omega^2\alpha_\perp} = \frac{k_\perp^2 v_{\textrm{th}\perp}^2}{2\Omega^2} = \frac{k_\perp^2 T_\perp^{(0)}}{m\Omega^2}=\frac{1}{2} k_\perp^2 r_L^2.
\end{equation}
Calculations like this lead to the functions $\Gamma_0(b)= e^{-b} I_0(b)$ and $\Gamma_1(b)= e^{-b} I_1(b)$. We note that the 
limit $b\to 0$ yields $\Gamma_0(b)\to 1$ and $\Gamma_1(b)\to 0$.  

\clearpage
\subsection{Coulomb force \& mirror force (Landau damping \& transit-time damping)} \label{sec:Forces}
The gyrotropic limit (\ref{eq:f1_gyrotropic}) has a very meaningful physical interpretation.
To clearly understand what kind of forces are present in such a system, one needs to consider that a particle quickly gyrates
around its slower moving center, called the ``guiding center'', 
and express the full velocity of a particle $\bV$ as being composed of the quick gyration $\bV_{\textrm{gyro}}$,
and a motion of the guiding center, that is further decomposed to its free motion parallel to the magnetic field line $v_\parallel\bhat$, and all the possible drifts of the guiding center :  
the ExB drift $\bu_E$, the grad-B drift, the curvature drift, the polarization drift etc.  
The plasma physics books by Fitzpatrick and \cite{GurnettBhattacharjee2005} have detailed introductions about single-particle motions in the presence
of Lorentz force, where the drifts of the guiding center are calculated. Then one should follow the \emph{gyrofluid} approach, and by performing integrals over $d\phi$ (gyro-averaging) and
by expanding for example with respect to Larmor radius, one should get the ``guiding center limit'' of the Vlasov equation and the expression for $f^{(1)}$.
One should follow \cite{Kulsrud1983,Snyder1997} etc. Very useful paper is also by \cite{FerriereAndre2002},
that explores the discrepancy between the usual CGL and the long-wavelength low-frequency kinetic theory in great detail and that we follow here.

Without going through the lengthy derivation, it can be shown that at the leading order (for low frequencies $\omega/\Omega$ and long wavelengths $kr_L$),
it is sufficient to consider the motion of the guiding center with velocity
\begin{equation}
  \bV = v_\parallel \bhat +\bu_E; \qquad \bu_E = c\frac{\bE\times\bb}{|\bb|^2},
\end{equation}
where the perpendicular equation of motion satisfies the conservation of the magnetic moment
\begin{equation}
  \mu=\frac{m v_\perp^2}{2|\bb|}=const.;\quad => \quad\frac{d}{d t}\Big(\frac{v_\perp^2}{|\bb|}\Big)=0; \quad => \quad
  \frac {dv_\perp}{dt} = \frac{v_\perp}{2|\bb|} \frac{d|\bb|}{dt}, \label{eq:vperp_gyro}
\end{equation}
and the parallel equation of motion satisfies
\begin{equation}
m\frac{dv_\parallel}{dt} = q E_\parallel -\mu \bhat\cdot\nabla|\bb|-m\bhat \cdot \frac{d\bu_E}{dt},\label{eq:vpar_gyro}
\end{equation}
where $E_\parallel=\bhat\cdot\bE$ and $d/dt=\pr/\pr t+\bV\cdot\nabla = \pr/\pr t+(v_\parallel \bhat +\bu_E)\cdot\nabla$.
The first term on the right hand side of the above equation is the \emph{Coulomb force},
responsible for acceleration of particles along the magnetic field lines. The second term is
the \emph{magnetic mirror force}, responsible for trapping of particles in the magnetic bottle. The third term is
 a non-inertial force associated with the time dependence of the ExB drift of the gyrocenter.
The similarity of the Coulomb force and the magnetic mirror force can be emphasized by using the scalar potential $\Phi$ and rewriting $E_\parallel=-\nabla \Phi$, which yields
\begin{equation}
  m\frac{dv_\parallel}{dt} = \underbrace{-q \bhat\cdot\nabla \Phi}_{\textrm{Coulomb}}
  \underbrace{-\mu \bhat\cdot\nabla|\bb|}_{\textrm{mirror}}-m\bhat \cdot \frac{d\bu_E}{dt}. 
\end{equation}
The similarity is immediately apparent, one just needs to replace the charge of the particle with its magnetic moment $q\to\mu$ and replace $\Phi\to |\bb|$.
Therefore, in a similar way as a charged particle reacts to electric field, a gyrating particle has a magnetic moment 
that reacts with the gradient of the strength (absolute value) of the magnetic field. The damping effects associated with the Coulomb force are called
\emph{Landau damping}. The damping effects associated with the mirror force are called \emph{transit-time damping}  or \emph{Barnes damping} \citep{Barnes1966}.
Therefore, it is often stated that
\emph{the transit-time damping is a ``magnetic analogue'' of Landau damping}. Often, the two effects are not separated since both represent the $n=0$ particle
resonance and one talks only about Landau damping. Nevertheless, it is emphasized that Landau fluid models in 3D geometry contain both damping mechanisms,
and these models contain both the Coulomb force and the mirror force.\footnote{In the 1D geometry where only $v_\parallel$ is considered,
  the gyration of particles, the magnetic mirror force and the transit-time damping of course disappear, since these effects naturally require $v_\perp$ as well.} 

The equations of motion (\ref{eq:vperp_gyro}), (\ref{eq:vpar_gyro}) should be used in gyro-averaged Vlasov equation
\begin{equation}
\frac{\pr f}{\pr t} +(v_\parallel \bhat +\bu_E)\cdot \nabla f + \Big[ \frac{dv_\perp}{dt} \frac{\pr}{\pr v_\perp} +\frac{dv_\parallel}{dt} \frac{\pr}{\pr v_\parallel} \Big] f =0, 
\end{equation}
and the equation should be expanded $f=f_0+f^{(1)}$. We are interested only in linear solutions, and we can simplify.
To avoid discussing compatibility conditions for $f_0$ (see \cite{Kulsrud1983}),
we can just simply claim, that $f_0$ does not have any time or spatial dependence.  
By further noticing that linearization of $\bu_E \overset{\textrm{\tiny lin}}{=} \bu_E^{(0)}+\bu_E^{(1)}$ yields $\bu_E^{(0)}=c\bE_0\times \bb_0/B_0^2=0$ since $E_0=0$,
we can immediately write that at the linear level
\begin{equation}
\frac{\pr f^{(1)}}{\pr t} +v_\parallel \pr_z f^{(1)} =- \Big[ \frac{dv_\perp}{dt} \frac{\pr}{\pr v_\perp} +\frac{dv_\parallel}{dt} \frac{\pr}{\pr v_\parallel} \Big] f_0. 
\end{equation}
Noticing that the ExB drift $\bu_E$ is always perpendicular to the direction of $\bb$ (and also $\bE$) implies $\bhat\cdot\bu_E=0$, and the last term in the $d\vpar/dt$ equation
(\ref{eq:vpar_gyro}) rewrites
\begin{equation}
-\bhat \cdot \frac{d\bu_E}{dt} =-\frac{d}{dt}(\underbrace{\bhat\cdot\bu_E}_{=0}) +\bu_E\cdot \frac{d\bhat}{dt},
\end{equation}
which at the linear level disappears, since
\begin{equation}
  \bu_E\cdot \frac{d\bhat}{dt} \overset{\textrm{\tiny lin}}{=}
  \underbrace{\bu_E^{(0)}}_{=0}\cdot \frac{d\bhat^{(1)}}{dt} + \bu_E^{(1)}\cdot \underbrace{\frac{d\bhat^{(0)}}{dt}}_{=0} =0.
\end{equation}
The magnetic mirror force contains $\pr_z|\bb|=\pr_z \sqrt{B_x^2+B_y^2+B_z^2} = \bhat\cdot \pr_z \bb$, where linearization yields
$\pr_z|\bb| \overset{\textrm{\tiny lin}}{=} \pr_z B_z$. Similarly, the $dv_\perp/dt$ equation (\ref{eq:vperp_gyro}) contains
$d|\bb|/dt=\bhat\cdot d\bb/dt$ that linearizes as
$d|\bb|/dt\overset{\textrm{\tiny lin}}{=} (\pr/\pr t+ \vpar \pr_z)B_z$. The linearized equations of motion therefore read
\begin{eqnarray}
  \frac{dv_\parallel}{dt} &\overset{\textrm{\tiny lin}}{=}& \frac{q}{m} E_\parallel -\frac{v_\perp^2}{2B_0} \pr_z B_z;\\
   \frac {dv_\perp}{dt} &\overset{\textrm{\tiny lin}}{=}& \frac{v_\perp}{2B_0} \Big( \frac{\pr}{\pr t} +\vpar \pr_z \Big) B_z,
\end{eqnarray}
yielding the final expression for $f^{(1)}$ in real space 
\begin{equation}
\Big( \frac{\pr}{\pr t} +\vpar \pr_z \Big) f^{(1)} = - \frac{v_\perp}{2B_0} \Big( \frac{\pr}{\pr t} +\vpar \pr_z \Big) B_z \frac{\pr f_0}{\pr v_\perp}
  -\Big( \frac{q}{m} E_\parallel -\frac{v_\perp^2}{2B_0} \pr_z B_z  \Big) \frac{\pr f_0}{\pr v_\parallel}, \label{eq:f1_gyrot_real}
\end{equation}
which when Fourier transformed recovers the $f^{(1)}$ in the gyrotropic limit (\ref{eq:f1_gyrotropic}). Instead of fully linearized equations with $B_z$,
one can also work with $|\bb|$, i.e. one can write the leading order equations of motion as
\begin{eqnarray}
  \frac{dv_\parallel}{dt} &=& \frac{q}{m} E_\parallel -\frac{v_\perp^2}{2B_0} \pr_z |\bb|;\\
   \frac {dv_\perp}{dt} &=& \frac{v_\perp}{2B_0} \Big( \frac{\pr}{\pr t} +\vpar \pr_z \Big) |\bb|,
\end{eqnarray}
which yields analogous equations (\ref{eq:f1_gyrot_real}), (\ref{eq:f1_gyrotropic}) where $B_z$ is just replaced by $|\bb|$.
\clearpage
\subsection{Kinetic moments for Bi-Maxwellian $f_0$}
Since in the Vlasov expansion the gyrotropic $f_0$ was assumed to dependent only on $\bV$, i.e. $f_0(v_\perp^2,v_\parallel^2)$ and be $\bx,t$
independent, the fluid velocity $\bu$ is removed from the distribution function and the ``pure'' bi-Maxwellian is
\begin{eqnarray}
f_0 = n_{0} \sqrt{\frac{\alpha_\parallel}{\pi}} \frac{\alpha_\perp}{\pi} e^{-\alpha_\parallel v_\parallel^2 - \alpha_\perp v_\perp^2},
\end{eqnarray}  
where
\begin{equation}
  \alpha_\parallel = \frac{m}{2T_\parallel^{(0)}};\qquad \alpha_\perp = \frac{m}{2T_\perp^{(0)}},
\end{equation}
or in the language of thermal speeds,
\begin{equation}
  v_{\textrm{th}\parallel}^2=\frac{ 2T^{(0)}_{\parallel} }{m} = \alpha_\parallel^{-1};\qquad  v_{\textrm{th}\perp}^2=\frac{2T^{(0)}_{\perp}}{m}=\alpha_\perp^{-1}.
\end{equation}
We prefer the $\alpha$ notation instead of the thermal speed $v_{\textrm{th}}$, since in long analytic calculations, there is a less chance of an error.
We work without the species index $r$ except for charge $q_r$ and mass $m_r$. 
It is straightforward to calculate that  
\begin{eqnarray}
  \frac{\pr f_0}{\pr v_\parallel} &=& -2 \alpha_\parallel v_\parallel f_0 = -\frac{m_r}{T_{\parallel}^{(0)}} v_\parallel f_0;\\ 
  \frac{\pr f_0}{\pr v_\perp} &=& -2 \alpha_\perp v_\perp f_0 = -\frac{m_r}{T_{\perp}^{(0)}} v_\perp f_0,
\end{eqnarray}
Instead of $E_z$, we will work with the scalar potential $\Phi$ as before
\begin{equation}
E_z = -\nabla_\parallel \Phi; \qquad => \qquad iE_z = \kpar \Phi.
\end{equation}  
The $f^{(1)}$ that we want to integrate reads
\begin{equation}
  f^{(1)} = \frac{B_z}{B_0} \alpha_\perp \bigg[ v_\perp^2 f_0 +\frac{\alphapa}{\alphape} \frac{\kpar \vpar \vperp^2}{(\omega-\kpar\vpar)}f_0 \bigg]
  +\Phi \frac{q_r}{m_r} 2\alphapa \frac{\kpar\vpar}{(\omega-\kpar\vpar)}f_0,
\end{equation}
or alternatively expressed with temperatures
\begin{equation}
  f^{(1)} = \frac{B_z}{B_0} \frac{m_r}{2 T_\perp^{(0)}}\bigg[ v_\perp^2 f_0 + \frac{T_\perp^{(0)}}{T_\parallel^{(0)}}\frac{\kpar \vpar \vperp^2}{(\omega-\kpar\vpar)}f_0 \bigg]
  +\Phi \frac{q_r}{T_\parallel^{(0)}} \frac{\kpar\vpar}{(\omega-\kpar\vpar)}f_0.
\end{equation}
Now we want to calculate the linear ``kinetic'' moments over this distribution function. 
The kinetic moments are
\begin{eqnarray}
  n^{(1)} &=& \int f^{(1)} d^3v; \qquad u_\parallel^{(1)} = \frac{1}{n_0} \int \vpar f^{(1)} d^3v;\\
  p_\parallel^{(1)} &=& m_r \int \vpar^2 f^{(1)} d^3v; \qquad p_\perp^{(1)} = \frac{m_r}{2} \int v_\perp^2 f^{(1)} d^3v;\\
  q_\parallel^{(1)} &=& m_r \int \vpar^3 f^{(1)} d^3v - 3p_\parallel^{(0)} u_\parallel^{(1)};\qquad
  q_\perp^{(1)} = \frac{m_r}{2} \int \vpar \vperp^2 f^{(1)} d^3v - p_\perp^{(0)} u_\parallel^{(1)};\\
  r_{\parallel\parallel}^{(1)} &=& m_r \int \vpar^4 f^{(1)} d^3 v; \qquad r_{\parallel\perp}^{(1)} = \frac{m_r}{2} \int \vpar^2\vperp^2 f^{(1)} d^3 v; \qquad
  r_{\perp\perp}^{(1)} = \frac{m_r}{4} \int \vperp^4 f^{(1)} d^3 v.
\end{eqnarray}
We have so far avoided integration in the cylindrical co-ordinate system, and all the previous integral were done in Cartesian co-coordinate system.
In the cylindrical system, $d^3 v=v_\perp dv_\perp dv_\parallel d\phi$ and the integral with respect to $v_\perp$ is from 0 to $\infty$. The Gaussian integrals are
\begin{eqnarray}
&&  \int_0^\infty e^{-ax^2} dx = \frac{1}{2}\sqrt{\frac{\pi}{a}}; \quad \int_0^\infty x e^{-ax^2} dx = \frac{1}{2a}; \quad
  \int_0^\infty x^2 e^{-ax^2} dx = \frac{1}{4a}\sqrt{\frac{\pi}{a}}; \quad \int_0^\infty x^3 e^{-ax^2} dx = \frac{1}{2a^2}; \nn\\
  &&  \int_0^\infty x^4 e^{-ax^2} dx = \frac{3}{8a^2}\sqrt{\frac{\pi}{a}}; \quad \int_0^\infty x^5 e^{-ax^2} dx = \frac{1}{a^3}; \quad 
  \int_0^\infty x^6 e^{-ax^2} dx = \frac{15}{16a^3}\sqrt{\frac{\pi}{a}}; \quad \int_0^\infty x^7 e^{-ax^2} dx = \frac{3}{a^4}.\nn
\end{eqnarray}
Therefore, integrating over $dv_\perp d\phi$ is straightforward and 
\begin{equation}
  \int f_0 v_\perp dv_\perp d\phi = 2\pi \int_0^\infty f_0 v_\perp dv_\perp
  = 2\pi n_{0} \sqrt{\frac{\alpha_{\parallel}}{\pi}} e^{-\alpha_{\parallel} v_\parallel^2} \frac{\alpha_{\perp}}{\pi}
  \underbrace{\int_0^\infty v_\perp e^{- \alpha_{\perp} v_\perp^2} dv_\perp}_{=1/(2\alpha_\perp)} = n_0  \sqrt{\frac{\alpha_{\parallel}}{\pi}} e^{-\alpha_{\parallel} v_\parallel^2},
\end{equation}
and similarly
\begin{eqnarray}
  \int f_0 v_\perp^3 dv_\perp d\phi &=& n_0  \sqrt{\frac{\alpha_{\parallel}}{\pi}} e^{-\alpha_{\parallel} v_\parallel^2} \frac{1}{\alphape};\qquad 
  \int f_0 v_\perp^5 dv_\perp d\phi = n_0  \sqrt{\frac{\alpha_{\parallel}}{\pi}} e^{-\alpha_{\parallel} v_\parallel^2} \frac{2}{\alphape^2};\nn\\
  \int f_0 v_\perp^7 dv_\perp d\phi &=& n_0  \sqrt{\frac{\alpha_{\parallel}}{\pi}} e^{-\alpha_{\parallel} v_\parallel^2} \frac{6}{\alphape^3}, \label{eq:Smart_split}
\end{eqnarray}
and these are all of integrals over $dv_\perp$ that are needed right now. The basic integrals (without singularity) calculate
\begin{eqnarray}
  \int f_0 d^3v &=& n_0;\qquad
  \int \vpar f_0 d^3v = 0;\qquad
  \int \vpar^2 f_0 d^3v = n_0 \frac{1}{2\alphapa};\nn\\
  \int \vpar^3 f_0 d^3v &=& 0;\qquad
  \int \vpar^4 f_0 d^3v = n_0 \frac{3}{4 \alphapa^2},
\end{eqnarray}  
and each integral yields further 3 cases from (\ref{eq:Smart_split}) just by multiplying, so
\begin{eqnarray}
  \int v_\perp^2 f_0 d^3v &=& n_0 \frac{1}{\alphape};\qquad
  \int v_\perp^4 f_0 d^3v = n_0 \frac{2}{\alphape^2};\qquad
  \int v_\perp^6 f_0 d^3v = n_0 \frac{6}{\alphape^3};\nn
\end{eqnarray}
\begin{eqnarray}
  \int \vpar^2 \vperp^2 f_0 d^3v &=& n_0 \frac{1}{2\alphapa}\frac{1}{\alphape};\qquad
  \int \vpar^2 \vperp^4 f_0 d^3v = n_0 \frac{1}{2\alphapa}\frac{2}{\alphape^2};\qquad
  \int \vpar^2 \vperp^6 f_0 d^3v = n_0 \frac{1}{2\alphapa}\frac{6}{\alphape^3};\nn
\end{eqnarray}
\begin{eqnarray}
  \int \vpar^4 \vperp^2 f_0 d^3v &=& n_0 \frac{3}{4 \alphapa^2}\frac{1}{\alphape};\qquad
  \int \vpar^4 \vperp^4 f_0 d^3v = n_0 \frac{3}{4 \alphapa^2}\frac{2}{\alphape^2};\qquad
  \int \vpar^4 \vperp^6 f_0 d^3v = n_0 \frac{3}{4 \alphapa^2}\frac{6}{\alphape^3}.
\end{eqnarray}  
 By using Landau integrals (\ref{eq:Lan_1})-(\ref{eq:Lan_5}), the following integrals can be calculated  
\begin{eqnarray}
  \int \frac{\kpar v_\parallel f_0}{\omega-\kpar v_\parallel} d^3 v
  &=& -n_0 R(\zeta);\qquad
  \int \frac{\kpar v_\parallel^2 f_0}{\omega-\kpar v_\parallel} d^3 v
  =  -\frac{n_0}{\sqrt{\alpha_\parallel}} \sign(\kpar)\zeta R(\zeta);\nn\\
  \int \frac{\kpar v_\parallel^3 f_0}{\omega-\kpar v_\parallel} d^3 v
  &=& -\frac{n_0}{\alpha_\parallel} \Big( \frac{1}{2}+\zeta^2 R(\zeta) \Big);\qquad
  \int \frac{\kpar v_\parallel^4 f_0}{\omega-\kpar v_\parallel} d^3 v
  =  -\frac{n_0}{\alpha_\parallel^{3/2}} \sign(\kpar) \Big( \frac{1}{2}\zeta+\zeta^3 R(\zeta) \Big);\nn\\
  \int \frac{\kpar v_\parallel^5 f_0}{\omega-\kpar v_\parallel} d^3 v
  &=&  -\frac{n_0}{\alpha_\parallel^{2}} \Big( \frac{3}{4}+\frac{\zeta^2}{2}+\zeta^4 R(\zeta) \Big),
\end{eqnarray}
and each of these integrals yields further 3 cases from (\ref{eq:Smart_split}) just by multiplying, so that
\begin{eqnarray}
  \int \frac{\kpar v_\parallel \vperp^2 f_0}{\omega-\kpar v_\parallel} d^3 v &=& -n_0 R(\zeta) \frac{1}{\alphape} ;\qquad
  \int \frac{\kpar v_\parallel \vperp^4 f_0}{\omega-\kpar v_\parallel} d^3 v = -n_0 R(\zeta) \frac{2}{\alphape^2} ;\nn\\
  \int \frac{\kpar v_\parallel \vperp^6 f_0}{\omega-\kpar v_\parallel} d^3 v &=& -n_0 R(\zeta) \frac{6}{\alphape^3};
\end{eqnarray}
\begin{eqnarray}
  \int \frac{\kpar v_\parallel^2 \vperp^2 f_0}{\omega-\kpar v_\parallel} d^3 v &=& -\frac{n_0}{\sqrt{\alpha_\parallel}} \sign(\kpar)\zeta R(\zeta) \frac{1}{\alphape};\qquad
  \int \frac{\kpar v_\parallel^2 \vperp^4 f_0}{\omega-\kpar v_\parallel} d^3 v = -\frac{n_0}{\sqrt{\alpha_\parallel}} \sign(\kpar)\zeta R(\zeta) \frac{2}{\alphape^2};\nn\\
  \int \frac{\kpar v_\parallel^2 \vperp^6 f_0}{\omega-\kpar v_\parallel} d^3 v &=& -\frac{n_0}{\sqrt{\alpha_\parallel}} \sign(\kpar)\zeta R(\zeta) \frac{6}{\alphape^3};
\end{eqnarray}
\begin{eqnarray}
  \int \frac{\kpar v_\parallel^3 \vperp^2 f_0}{\omega-\kpar v_\parallel} d^3 v &=&  -\frac{n_0}{\alpha_\parallel} \Big( \frac{1}{2}+\zeta^2 R(\zeta) \Big) \frac{1}{\alphape};\qquad
  \int \frac{\kpar v_\parallel^3 \vperp^4 f_0}{\omega-\kpar v_\parallel} d^3 v =  -\frac{n_0}{\alpha_\parallel} \Big( \frac{1}{2}+\zeta^2 R(\zeta) \Big) \frac{2}{\alphape^2};\nn\\
  \int \frac{\kpar v_\parallel^3 \vperp^6 f_0}{\omega-\kpar v_\parallel} d^3 v &=&  -\frac{n_0}{\alpha_\parallel} \Big( \frac{1}{2}+\zeta^2 R(\zeta) \Big) \frac{6}{\alphape^3};
\end{eqnarray}
\begin{eqnarray}  
  \int \frac{\kpar v_\parallel^4 \vperp^2 f_0}{\omega-\kpar v_\parallel} d^3 v &=& -\frac{n_0}{\alpha_\parallel^{3/2}} \sign(\kpar) \Big( \frac{1}{2}\zeta+\zeta^3 R(\zeta) \Big)\frac{1}{\alphape};\qquad
  \int \frac{\kpar v_\parallel^4 \vperp^4 f_0}{\omega-\kpar v_\parallel} d^3 v = -\frac{n_0}{\alpha_\parallel^{3/2}} \sign(\kpar) \Big( \frac{1}{2}\zeta+\zeta^3 R(\zeta) \Big)\frac{2}{\alphape^2};\nn\\
  \int \frac{\kpar v_\parallel^4 \vperp^6 f_0}{\omega-\kpar v_\parallel} d^3 v &=& -\frac{n_0}{\alpha_\parallel^{3/2}} \sign(\kpar) \Big( \frac{1}{2}\zeta+\zeta^3 R(\zeta) \Big)\frac{6}{\alphape^3};
\end{eqnarray}
\begin{eqnarray}
  \int \frac{\kpar v_\parallel^5 \vperp^2 f_0}{\omega-\kpar v_\parallel} d^3 v  &=& -\frac{n_0}{\alpha_\parallel^{2}} \Big( \frac{3}{4}+\frac{\zeta^2}{2}+\zeta^4 R(\zeta) \Big)\frac{1}{\alphape};\qquad
  \int \frac{\kpar v_\parallel^5 \vperp^4 f_0}{\omega-\kpar v_\parallel} d^3 v  = -\frac{n_0}{\alpha_\parallel^{2}} \Big( \frac{3}{4}+\frac{\zeta^2}{2}+\zeta^4 R(\zeta) \Big)\frac{2}{\alphape^2};\nn\\
  \int \frac{\kpar v_\parallel^5 \vperp^6 f_0}{\omega-\kpar v_\parallel} d^3 v  &=& -\frac{n_0}{\alpha_\parallel^{2}} \Big( \frac{3}{4}+\frac{\zeta^2}{2}+\zeta^4 R(\zeta) \Big)\frac{6}{\alphape^3}.
\end{eqnarray}  
Now it is easy to calculate the kinetic moments.
\subsubsection*{Density}
\noindent The density calculates
\begin{eqnarray}
  n^{(1)} &=& \frac{B_z}{B_0} \alpha_\perp \bigg[ \int v_\perp^2 f_0d^3v +\frac{\alphapa}{\alphape} \int \frac{\kpar \vpar \vperp^2}{(\omega-\kpar\vpar)}f_0 d^3v \bigg]
  +\Phi \frac{q_r}{m_r} 2\alphapa \int \frac{\kpar\vpar}{(\omega-\kpar\vpar)}f_0 d^3v\nn\\
  &=& \frac{B_z}{B_0} \alpha_\perp \bigg[ \frac{n_0}{\alphape} +\frac{\alphapa}{\alphape} (-n_0)R(\zeta)\frac{1}{\alphape} \bigg]
  +\Phi \frac{q_r}{m_r} 2\alphapa (-n_0)R(\zeta),\nn
\end{eqnarray}
so that the ratio
\begin{eqnarray}
 \frac{n^{(1)}}{n_0} &=& \frac{B_z}{B_0} \bigg[ 1 -\frac{\alphapa}{\alphape}R(\zeta) \bigg]
 -\Phi \frac{q_r}{m_r} 2\alphapa R(\zeta),\nn
\end{eqnarray}
and the final result reads
\begin{eqnarray}
 \frac{n^{(1)}}{n_0} &=& \frac{B_z}{B_0} \bigg[ 1 -\frac{T_\perp^{(0)}}{T_\parallel^{(0)}}R(\zeta) \bigg]
 -\Phi \frac{q_r}{T_\parallel^{(0)}} R(\zeta).
\end{eqnarray}
\subsubsection*{Parallel velocity}
\noindent
The parallel velocity calculates
\begin{eqnarray}
  n_0 u_\parallel ^{(1)} &=& \frac{B_z}{B_0} \alpha_\perp \bigg[ \underbrace{\int \vpar v_\perp^2 f_0d^3v}_{=0}
    +\frac{\alphapa}{\alphape} \int \frac{\kpar \vpar^2 \vperp^2}{(\omega-\kpar\vpar)}f_0 d^3v \bigg]
  +\Phi \frac{q_r}{m_r} 2\alphapa \int \frac{\kpar\vpar^2}{(\omega-\kpar\vpar)}f_0 d^3v\nn\\
  &=& -\frac{B_z}{B_0} \frac{n_0}{\sqrt{\alphapa}}\frac{\alphapa}{\alphape}\sign(\kpar)\zeta R(\zeta)  
  -\Phi \frac{q_r}{m_r} 2\alphapa \frac{n_0}{\sqrt{\alphapa}}\sign(\kpar)\zeta R(\zeta)\nn\\
  &=& - \frac{n_0}{\sqrt{\alphapa}}\sign(\kpar)\zeta R(\zeta) \bigg[ \frac{B_z}{B_0}\frac{\alphapa}{\alphape} +\Phi\frac{q_r}{m_r}2\alphapa\bigg],
\end{eqnarray}  
so that
\begin{eqnarray}
  u_\parallel ^{(1)} &=&
  -\sqrt{\frac{2T_\parallel^{(0)}}{m_r}}\sign(\kpar)\zeta R(\zeta) \bigg[ \frac{B_z}{B_0}\frac{T_\perp^{(0)}}{T_\parallel^{(0)}} + \Phi\frac{q_r}{T_\parallel^{(0)}} \bigg].
\end{eqnarray}
\subsubsection*{Parallel pressure}
\noindent
The parallel pressure calculates
\begin{eqnarray}
  p_\parallel^{(1)} &=& m_r\frac{B_z}{B_0} \alpha_\perp \bigg[ \int \vpar^2 v_\perp^2 f_0d^3v +\frac{\alphapa}{\alphape} \int \frac{\kpar \vpar^3 \vperp^2}{(\omega-\kpar\vpar)}f_0 d^3v \bigg]
  +m_r \Phi \frac{q_r}{m_r} 2\alphapa \int \frac{\kpar\vpar^3}{(\omega-\kpar\vpar)}f_0 d^3v\nn\\
  &=& m_r\frac{B_z}{B_0} \alpha_\perp \bigg[ \frac{n_0}{2\alphapa\alphape} -\frac{\alphapa}{\alphape} \frac{n_0}{\alphapa}\Big(\frac{1}{2}+\zeta^2 R(\zeta)\Big)\frac{1}{\alphape} \bigg]
  -m_r \Phi \frac{q_r}{m_r} 2\alphapa \frac{n_0}{\alphapa}\Big(\frac{1}{2}+\zeta^2 R(\zeta)\Big),\nn
\end{eqnarray}
so that
\begin{eqnarray}
  \frac{p_\parallel^{(1)}}{n_0} &=& m_r\frac{B_z}{B_0} \frac{1}{2\alphapa}\bigg[ 1
    -\frac{\alphapa}{\alphape} \Big(1+2\zeta^2 R(\zeta)\Big)\bigg]
  -\Phi q_r \Big(1+2\zeta^2 R(\zeta)\Big),\nn
\end{eqnarray}
and
\begin{eqnarray}
  \frac{p_\parallel^{(1)}}{p_\parallel^{(0)}} &=& \frac{B_z}{B_0}\bigg[ 1
    -\frac{T_\perp^{(0)}}{T_\parallel^{(0)}} \Big(1+2\zeta^2 R(\zeta)\Big)\bigg]
  -\Phi \frac{q_r}{T_\parallel^{(0)}} \Big(1+2\zeta^2 R(\zeta)\Big)\nn\\
   &=& \frac{B_z}{B_0}
  -  \Big(1+2\zeta^2 R(\zeta)\Big)\bigg[ \frac{B_z}{B_0} \frac{T_\perp^{(0)}}{T_\parallel^{(0)}} + \Phi \frac{q_r}{T_\parallel^{(0)}} \bigg].
\end{eqnarray}
\subsubsection*{Parallel temperature}
\noindent
The parallel temperature calculates (linearizing $p_\parallel=T_\parallel n$)
\begin{equation}
\frac{T_\parallel^{(1)}}{T_\parallel^{(0)}} = \frac{p_\parallel^{(1)}}{p_\parallel^{(0)}}-\frac{n^{(1)}}{n_0},
\end{equation}
that yields
\begin{eqnarray}
  \frac{T_\parallel^{(1)}}{T_\parallel^{(0)}} &=&  -\frac{B_z}{B_0} \frac{T_\perp^{(0)}}{T_\parallel^{(0)}} \bigg( 1+2\zeta^2 R(\zeta) -R(\zeta) \bigg)
  -\Phi \frac{q_r}{T_\parallel^{(0)}}\bigg( 1+2\zeta^2 R(\zeta) -R(\zeta) \bigg)\\
   &=& -\bigg( 1+2\zeta^2 R(\zeta) -R(\zeta) \bigg) \bigg[ \frac{B_z}{B_0} \frac{T_\perp^{(0)}}{T_\parallel^{(0)}} + \Phi \frac{q_r}{T_\parallel^{(0)}} \bigg].\nn
\end{eqnarray}  
\subsubsection*{Perpendicular pressure}
\noindent
The perpendicular pressure calculates
\begin{eqnarray}
  p_\perp^{(1)} &=& \frac{m_r}{2}\frac{B_z}{B_0} \alpha_\perp \bigg[ \int v_\perp^4 f_0d^3v +\frac{\alphapa}{\alphape} \int \frac{\kpar \vpar \vperp^4}{(\omega-\kpar\vpar)}f_0 d^3v \bigg]
  +\frac{m_r}{2}\Phi \frac{q_r}{m_r} 2\alphapa \int \frac{\kpar\vpar\vperp^2}{(\omega-\kpar\vpar)}f_0 d^3v\nn\\
   &=& \frac{m_r}{2}\frac{B_z}{B_0} \alpha_\perp \bigg[ n_0\frac{2}{\alphape^2} +\frac{\alphapa}{\alphape} (-n_0)R(\zeta)\frac{2}{\alphape^2} \bigg]
  +\frac{m_r}{2}\Phi \frac{q_r}{m_r} 2\alphapa (-n_0)R(\zeta)\frac{1}{\alphape},\nn
\end{eqnarray}
so that
\begin{equation}
\frac{p_\perp^{(1)}}{n_0} = m_r\frac{B_z}{B_0} \frac{1}{\alpha_\perp} \bigg[ 1 -\frac{\alphapa}{\alphape} R(\zeta) \bigg]
  -\Phi q_r \frac{\alphapa}{\alphape} R(\zeta),
\end{equation}
further yielding
\begin{equation}
\frac{p_\perp^{(1)}}{p_\perp^{(0)}} = \frac{B_z}{B_0} 2\bigg[ 1 -\frac{T_\perp^{(0)}}{T_\parallel^{(0)}} R(\zeta) \bigg]
  -\Phi \frac{q_r}{T_\parallel^{(0)}} R(\zeta).
\end{equation}
\subsubsection*{Perpendicular temperature}
\noindent
The perpendicular temperature calculates (linearizing $p_\perp=T_\perp n$)
\begin{equation}
\frac{T_\perp^{(1)}}{T_\perp^{(0)}} = \frac{p_\perp^{(1)}}{p_\perp^{(0)}}-\frac{n^{(1)}}{n_0},
\end{equation}
that yields
\begin{equation}
\frac{T_\perp^{(1)}}{T_\perp^{(0)}} =  \frac{B_z}{B_0} \bigg[ 1 -\frac{T_\perp^{(0)}}{T_\parallel^{(0)}} R(\zeta) \bigg].
\end{equation}
We might be tired of calculations at this stage, but, this result nicely shows that (at the linear level and at the long-scales and low-frequencies considered here),
the Landau damping ($\sim E_z$) does \emph{not} influence the perpendicular temperature, however, the transit-time damping still does.  
\subsubsection*{Parallel heat flux}
\noindent
The parallel heat flux calculates
\begin{eqnarray}
  q_\parallel^{(1)} &=& m_r \frac{B_z}{B_0} \alpha_\perp \bigg[ \underbrace{\int \vpar^3 v_\perp^2 f_0d^3v}_{=0}
    +\frac{\alphapa}{\alphape} \int \frac{\kpar \vpar^4 \vperp^2}{(\omega-\kpar\vpar)}f_0 d^3v \bigg]
  +m_r \Phi \frac{q_r}{m_r} 2\alphapa \int \frac{\kpar\vpar^4}{(\omega-\kpar\vpar)}f_0 d^3v-3p_\parallel^{(0)}u_\parallel^{(1)}\nn\\
  &=& -m_r \frac{B_z}{B_0} \alphapa \frac{n_0}{\alphapa^{3/2}}\sign(\kpar)\bigg( \frac{\zeta}{2}+\zeta^3 R(\zeta) \bigg) \frac{1}{\alphape} 
  -m_r \Phi \frac{q_r}{m_r} 2\alphapa\frac{n_0}{\alphapa^{3/2}}\sign(\kpar)\bigg( \frac{\zeta}{2}+\zeta^3 R(\zeta) \bigg) -3p_\parallel^{(0)}u_\parallel^{(1)}\nn\\
  &=& -\frac{n_0 m_r}{\alphapa^{3/2}}\sign(\kpar)\bigg( \frac{\zeta}{2}+\zeta^3 R(\zeta) \bigg)
  \bigg[ \frac{B_z}{B_0}\frac{\alphapa}{\alphape}+\Phi \frac{q_r}{m_r} 2\alphapa\bigg]  -3p_\parallel^{(0)}u_\parallel^{(1)}\nn,
\end{eqnarray}
so that
\begin{eqnarray}
\frac{q_\parallel^{(1)}}{p_\parallel^{(0)}}  &=& -\frac{1}{\sqrt{\alphapa}}\sign(\kpar)\bigg( \zeta+2\zeta^3 R(\zeta) \bigg)
\bigg[ \frac{B_z}{B_0}\frac{\alphapa}{\alphape}+\Phi \frac{q_r}{m_r} 2\alphapa\bigg]  -3u_\parallel^{(1)}\nn\\
  &=& -\frac{1}{\sqrt{\alphapa}}\sign(\kpar)\bigg( \zeta+2\zeta^3 R(\zeta) -3\zeta R(\zeta)\bigg)
\bigg[ \frac{B_z}{B_0}\frac{\alphapa}{\alphape}+\Phi \frac{q_r}{m_r} 2\alphapa\bigg],
\end{eqnarray}  
or alternatively
\begin{eqnarray}
q_\parallel^{(1)} &=& -\sqrt{\frac{2T_\parallel^{(0)}}{m_r}} n_0 T_\parallel^{(0)} \sign(\kpar)\bigg( \zeta+2\zeta^3 R(\zeta) -3\zeta R(\zeta)\bigg)
\bigg[ \frac{B_z}{B_0}\frac{T_\perp^{(0)}}{T_\parallel^{(0)}} + \Phi\frac{q_r}{T_\parallel^{(0)}} \bigg].
\end{eqnarray}
\subsubsection*{Perpendicular heat flux}
\noindent
The perpendicular heat flux calculates
\begin{eqnarray}
  q_\perp^{(1)} &=& \frac{m_r}{2}\frac{B_z}{B_0} \alpha_\perp \bigg[ \underbrace{\int \vpar v_\perp^4 f_0d^3v}_{=0}
    +\frac{\alphapa}{\alphape} \int \frac{\kpar \vpar^2 \vperp^4}{(\omega-\kpar\vpar)}f_0 d^3v \bigg]
  +\frac{m_r}{2}\Phi \frac{q_r}{m_r} 2\alphapa \int \frac{\kpar\vpar^2\vperp^2}{(\omega-\kpar\vpar)}f_0 d^3v -p_\perp^{(0)}u_\parallel^{(1)} \nn\\
   &=& -\frac{m_r}{2}\frac{B_z}{B_0} \alphapa n_0\frac{1}{\sqrt{\alphapa}}\sign(\kpar)\zeta R(\zeta)\frac{2}{\alphape^2} 
  -\frac{m_r}{2}\Phi \frac{q_r}{m_r} 2\alphapa n_0\frac{1}{\sqrt{\alphapa}}\sign(\kpar)\zeta R(\zeta)\frac{1}{\alphape}-p_\perp^{(0)}u_\parallel^{(1)} \nn\\
  &=& -\frac{m_r}{2}\frac{n_0}{\sqrt{\alphapa}}\sign(\kpar)\zeta R(\zeta)\frac{1}{\alphape}\bigg[ \frac{B_z}{B_0}\frac{2\alphapa}{\alphape}
    +\Phi\frac{q_r}{m_r}2\alphapa\bigg] -p_\perp^{(0)}u_\parallel^{(1)},\nn
\end{eqnarray}
so that
\begin{eqnarray}
\frac{q_\perp^{(1)}}{p_\perp^{(0)}}  &=& -\frac{1}{\sqrt{\alphapa}}\sign(\kpar)\zeta R(\zeta)\bigg[ \frac{B_z}{B_0}\frac{2\alphapa}{\alphape}
  +\Phi\frac{q_r}{m_r}2\alphapa\bigg] -u_\parallel^{(1)}\nn\\
&=& -\frac{1}{\sqrt{\alphapa}}\sign(\kpar)\zeta R(\zeta)\frac{B_z}{B_0}\frac{\alphapa}{\alphape},
\end{eqnarray}
or alternatively
\begin{equation}
q_\perp^{(1)}= -\sqrt{\frac{2T_\parallel^{(0)}}{m_r}} p_\perp^{(0)} \frac{T_\perp^{(0)}}{T_\parallel^{(0)}}\sign(\kpar)\zeta R(\zeta)\frac{B_z}{B_0}.
\end{equation}
The perpendicular heat flux $q_\perp^{(1)}$ (similarly to the perpendicular temperature $T_\perp^{(1)}$), is also not directly influenced by the
Landau damping $(\sim E_z)$, even though it is influenced by the transit-time damping $(\sim B_z)$.
\clearpage
\subsubsection*{4th-order moment $\lowercase{r}_{\parallel\parallel}$}
\noindent
The 4th-order moment $r_{\parallel\parallel}$ calculates
\begin{eqnarray}
  r_{\parallel\parallel}^{(1)} &=& m_r \frac{B_z}{B_0} \alpha_\perp \bigg[ \int \vpar^4 v_\perp^2 f_0d^3v
    +\frac{\alphapa}{\alphape} \int \frac{\kpar \vpar^5 \vperp^2}{(\omega-\kpar\vpar)}f_0 d^3v \bigg]
  +m_r \Phi \frac{q_r}{m_r} 2\alphapa \int \frac{\kpar\vpar^5}{(\omega-\kpar\vpar)}f_0 d^3v\nn\\
  &=& m_r \frac{B_z}{B_0} \alpha_\perp \bigg[ n_0 \frac{3}{4\alphapa^2\alphape}
    +\frac{\alphapa}{\alphape} \frac{(-n_0)}{\alphapa^2}\Big( \frac{3}{4}+\frac{\zeta^2}{2}+\zeta^4 R(\zeta) \Big)\frac{1}{\alphape} \bigg]
  +m_r \Phi \frac{q_r}{m_r} 2\alphapa \frac{(-n_0)}{\alphapa^2}\Big( \frac{3}{4}+\frac{\zeta^2}{2}+\zeta^4 R(\zeta) \Big)\nn\\
  &=& m_r \frac{B_z}{B_0} n_0 \frac{3}{4\alphapa^2}-m_r n_0 \Big( \frac{3}{4}+\frac{\zeta^2}{2}+\zeta^4 R(\zeta) \Big) \frac{1}{\alphapa^2}
    \bigg[ \frac{B_z}{B_0}\frac{\alphapa}{\alphape} +\Phi\frac{q_r}{m_r}2\alphapa\bigg],
\end{eqnarray}
so that
\begin{eqnarray}
  r_{\parallel\parallel}^{(1)} &=& \frac{p_\parallel^{(0)} T_\parallel^{(0)}}{m_r} \bigg\{ 3\frac{B_z}{B_0}
  -\Big( 3+2\zeta^2+4\zeta^4 R(\zeta) \Big) \bigg[ \frac{B_z}{B_0}\frac{T_\perp^{(0)}}{T_\parallel^{(0)}} + \Phi\frac{q_r}{T_\parallel^{(0)}} \bigg] \bigg\}.
\end{eqnarray}
\subsubsection*{4th-order moment deviation $\widetilde{\lowercase{r}}_{\parallel\parallel}$}
\noindent
The 4th-order moment ``deviation'' $\widetilde{r}_{\parallel\parallel}^{(1)}$ calculates
(linearizing $r_{\parallel\parallel}=\frac{3}{m_r}p_\parallel T_\parallel +\widetilde{r}_{\parallel\parallel}$ with definitions
$r_{\parallel\parallel}^{(0)}=\frac{3}{m_r} p_\parallel^{(0)}T_\parallel^{(0)}$ and $\widetilde{r}_{\parallel\parallel}^{(0)}=0$) 
\begin{equation}
  \frac{r_{\parallel\parallel}^{(1)}}{r_{\parallel\parallel}^{(0)}} = \frac{p_\parallel^{(1)}}{p_\parallel^{(0)}}+ \frac{T_\parallel^{(1)}}{T_\parallel^{(0)}}
  +  \frac{\widetilde{r}_{\parallel\parallel}^{(1)}}{r_{\parallel\parallel}^{(0)}},
\end{equation}
or equivalently
\begin{equation}
  \widetilde{r}_{\parallel\parallel}^{(1)} = r_{\parallel\parallel}^{(1)}-\frac{3}{m_r}p_\parallel^{(0)}T_\parallel^{(0)}
  \bigg( \frac{p_\parallel^{(1)}}{p_\parallel^{(0)}}+ \frac{T_\parallel^{(1)}}{T_\parallel^{(0)}} \bigg),
\end{equation}
which yields
\begin{equation}
  \widetilde{r}_{\parallel\parallel}^{(1)} = -\frac{p_\parallel^{(0)} T_\parallel^{(0)}}{m_r} \Big( 2\zeta^2+4\zeta^4R(\zeta)+3R(\zeta)-3-12\zeta^2 R(\zeta) \Big)
  \bigg[ \frac{B_z}{B_0}\frac{T_\perp^{(0)}}{T_\parallel^{(0)}} + \Phi\frac{q_r}{T_\parallel^{(0)}} \bigg].
\end{equation}
\subsubsection*{4th-order moment $\lowercase{r}_{\parallel\perp}$}
\noindent
The 4th-order moment $r_{\parallel\perp}$ calculates
\begin{eqnarray}
  r_{\parallel\perp}^{(1)} &=& \frac{m_r}{2} \frac{B_z}{B_0} \alpha_\perp \bigg[ \int \vpar^2 v_\perp^4 f_0d^3v
    +\frac{\alphapa}{\alphape} \int \frac{\kpar \vpar^3 \vperp^4}{(\omega-\kpar\vpar)}f_0 d^3v \bigg]
  +\frac{m_r}{2} \Phi \frac{q_r}{m_r} 2\alphapa \int \frac{\kpar\vpar^3\vperp^2}{(\omega-\kpar\vpar)}f_0 d^3v\nn\\
  &=& \frac{m_r}{2} \frac{B_z}{B_0} \alpha_\perp \bigg[ \frac{n_0}{\alphapa \alphape^2}
    -\frac{\alphapa}{\alphape} n_0\frac{2}{\alphapa\alphape^2}\Big(\frac{1}{2}+\zeta^2 R(\zeta) \Big) \bigg]
  -\frac{m_r}{2} \Phi \frac{q_r}{m_r} 2\alphapa \frac{n_0}{\alphapa\alphape}\Big(\frac{1}{2}+\zeta^2 R(\zeta) \Big)  \nn\\
 &=& \frac{m_r}{2} \frac{B_z}{B_0} \frac{n_0}{\alphapa \alphape}
    -m_rn_0\Big(\frac{1}{2}+\zeta^2 R(\zeta) \Big)\frac{1}{\alphapa\alphape} \bigg[ \frac{B_z}{B_0}\frac{\alphapa}{\alphape}
  +\frac{1}{2}\Phi \frac{q_r}{m_r}2\alphapa \bigg], \nn
\end{eqnarray}
so that
\begin{eqnarray}
  r_{\parallel\perp}^{(1)} &=& \frac{p_\parallel^{(0)}T_\perp^{(0)}}{m_r} \bigg\{ 2\frac{B_z}{B_0} -\Big( 2+4\zeta^2R(\zeta) \Big)
   \bigg[ \frac{B_z}{B_0}\frac{T_\perp^{(0)}}{T_\parallel^{(0)}} + \frac{1}{2}\Phi\frac{q_r}{T_\parallel^{(0)}} \bigg] \bigg\}.
\end{eqnarray}
\subsubsection*{4th-order moment deviation $\widetilde{\lowercase{r}}_{\parallel\perp}$}
\noindent
The 4th-order moment ``deviation'' $\widetilde{r}_{\parallel\perp}^{(1)}$ calculates
(for example linearizing $r_{\parallel\perp}=\frac{1}{m_r}p_\parallel T_\perp +\widetilde{r}_{\parallel\perp}$ with definitions
$r_{\parallel\perp}^{(0)}=\frac{1}{m_r} p_\parallel^{(0)} T_\perp^{(0)}$ and $\widetilde{r}_{\parallel\perp}^{(0)}=0$) 
\begin{equation}
  \frac{r_{\parallel\perp}^{(1)}}{r_{\parallel\perp}^{(0)}} = \frac{p_\parallel^{(1)}}{p_\parallel^{(0)}}+ \frac{T_\perp^{(1)}}{T_\perp^{(0)}}
  +  \frac{\widetilde{r}_{\parallel\perp}^{(1)}}{r_{\parallel\perp}^{(0)}},
\end{equation}
or equivalently
\begin{equation}
  \widetilde{r}_{\parallel\perp}^{(1)} = r_{\parallel\perp}^{(1)}-\frac{1}{m_r}p_\parallel^{(0)}T_\perp^{(0)}
  \bigg( \frac{p_\parallel^{(1)}}{p_\parallel^{(0)}}+ \frac{T_\perp^{(1)}}{T_\perp^{(0)}} \bigg),
\end{equation}
and the result is
\begin{equation}
\widetilde{r}_{\parallel\perp}^{(1)} = -\frac{p_\perp^{(0)}T_\perp^{(0)}}{m_r} \frac{B_z}{B_0}  \Big( 1+2\zeta^2 R(\zeta)-R(\zeta) \Big).
\end{equation}
\subsubsection*{4th-order moment $\lowercase{r}_{\perp\perp}$}
\noindent
The 4th-order moment $r_{\perp\perp}$ calculates
\begin{eqnarray}
  r_{\perp\perp}^{(1)} &=& \frac{m_r}{4}\frac{B_z}{B_0} \alpha_\perp \bigg[ \int v_\perp^6 f_0d^3v +\frac{\alphapa}{\alphape} \int \frac{\kpar \vpar \vperp^6}{(\omega-\kpar\vpar)}f_0 d^3v \bigg]
  +\frac{m_r}{4}\Phi \frac{q_r}{m_r} 2\alphapa \int \frac{\kpar\vpar\vperp^4}{(\omega-\kpar\vpar)}f_0 d^3v\nn\\
  &=& \frac{m_r}{4}\frac{B_z}{B_0} \alpha_\perp \bigg[ n_0\frac{6}{\alphape^3} -\frac{\alphapa}{\alphape}n_0 R(\zeta)\frac{6}{\alphape^3}  \bigg]
  -\frac{m_r}{4}\Phi \frac{q_r}{m_r} 2\alphapa n_0 R(\zeta)\frac{2}{\alphape^2}, \nn\\
\end{eqnarray}  
and the result is
\begin{equation}
  r_{\perp\perp}^{(1)} = \frac{2 p_\perp^{(0)}T_\perp^{(0)}}{m_r} \bigg\{ 3\frac{B_z}{B_0}\Big( 1-\frac{T_\perp^{(0)}}{T_\parallel^{(0)}} R(\zeta) \Big)
  -\Phi \frac{q_r}{T_\parallel^{(0)}} R(\zeta) \bigg\}.
\end{equation}
\subsubsection*{4th-order moment deviation $\widetilde{\lowercase{r}}_{\perp\perp}$}
\noindent
The 4th-order moment deviation $\widetilde{r}_{\perp\perp}^{(1)}$ calculates
(for example linearizing $r_{\perp\perp}=\frac{2}{m_r}p_\perp T_\perp +\widetilde{r}_{\perp\perp}$ with definitions
$r_{\perp\perp}^{(0)}=\frac{2}{m_r} p_\perp^{(0)} T_\perp^{(0)}$ and $\widetilde{r}_{\perp\perp}^{(0)}=0$)
\begin{equation}
  \frac{r_{\perp\perp}^{(1)}}{r_{\perp\perp}^{(0)}} = \frac{p_\perp^{(1)}}{p_\perp^{(0)}}+ \frac{T_\perp^{(1)}}{T_\perp^{(0)}}
  +  \frac{\widetilde{r}_{\perp\perp}^{(1)}}{r_{\perp\perp}^{(0)}},
\end{equation}
or equivalently
\begin{equation}
\widetilde{r}_{\perp\perp}^{(1)} = r_{\perp\perp}^{(1)}-\frac{2p_\perp^{(0)}T_\perp^{(0)}}{m_r} \bigg(\frac{p_\perp^{(1)}}{p_\perp^{(0)}} +  \frac{T_\perp^{(1)}}{T_\perp^{(0)}} \bigg), 
\end{equation}
which yields
\begin{equation}
\widetilde{r}_{\perp\perp}^{(1)} = 0.
\end{equation}
This is an excellent news, since we will not have to consider closures for $\widetilde{r}_{\perp\perp}^{(1)}$.

\clearpage
\subsection{Landau fluid closures in 3D} \label{sec:3DclosuresX}
Let's separate the kinetic moments to two groups. The first group: 
\begin{eqnarray}
 u_\parallel ^{(1)} &=&
 -\sqrt{\frac{2T_\parallel^{(0)}}{m_r}}\sign(\kpar)\zeta R(\zeta) \bigg[ \frac{B_z}{B_0}\frac{T_\perp^{(0)}}{T_\parallel^{(0)}} + \Phi\frac{q_r}{T_\parallel^{(0)}} \bigg];\\
 \frac{T_\parallel^{(1)}}{T_\parallel^{(0)}}
 &=& -\bigg( 1+2\zeta^2 R(\zeta) -R(\zeta) \bigg) \bigg[ \frac{B_z}{B_0} \frac{T_\perp^{(0)}}{T_\parallel^{(0)}} + \Phi \frac{q_r}{T_\parallel^{(0)}} \bigg];\\
 q_\parallel^{(1)} &=& -\sqrt{\frac{2T_\parallel^{(0)}}{m_r}} n_0 T_\parallel^{(0)} \sign(\kpar)\bigg( \zeta+2\zeta^3 R(\zeta) -3\zeta R(\zeta)\bigg)
 \bigg[ \frac{B_z}{B_0}\frac{T_\perp^{(0)}}{T_\parallel^{(0)}} + \Phi\frac{q_r}{T_\parallel^{(0)}} \bigg];\\
 \widetilde{r}_{\parallel\parallel}^{(1)} &=& -\frac{p_\parallel^{(0)} T_\parallel^{(0)}}{m_r} \Big( 2\zeta^2+4\zeta^4R(\zeta)+3R(\zeta)-3-12\zeta^2 R(\zeta) \Big)
  \bigg[ \frac{B_z}{B_0}\frac{T_\perp^{(0)}}{T_\parallel^{(0)}} + \Phi\frac{q_r}{T_\parallel^{(0)}} \bigg].
\end{eqnarray}
And the second group:
\begin{eqnarray}
\frac{T_\perp^{(1)}}{T_\perp^{(0)}} &=&   \bigg[ 1 -\frac{T_\perp^{(0)}}{T_\parallel^{(0)}} R(\zeta) \bigg]\frac{B_z}{B_0};\\
  q_\perp^{(1)} &=& -\sqrt{\frac{2T_\parallel^{(0)}}{m_r}} p_\perp^{(0)} \frac{T_\perp^{(0)}}{T_\parallel^{(0)}}\sign(\kpar)\zeta R(\zeta)\frac{B_z}{B_0};\label{eq:QperpPP}\\
\widetilde{r}_{\parallel\perp}^{(1)} &=& -\frac{p_\perp^{(0)}T_\perp^{(0)}}{m_r}   \Big( 1+2\zeta^2 R(\zeta)-R(\zeta) \Big)\frac{B_z}{B_0}.
\end{eqnarray}
One immediately notices that the moments in the first group, are extremely similar to the moments we obtained in the simplified case of 1D geometry,
where we neglected the transit-time damping, i.e. in the system (\ref{eq:RefPica-n})-(\ref{eq:RefPica-rr}). In fact, the system is completely the same,
if the variable $\frac{B_z}{B_0}\frac{T_\perp^{(0)}}{T_\parallel^{(0)}} + \Phi\frac{q_r}{T_\parallel^{(0)}}$ is replaced by $\Phi\frac{q_r}{T_\parallel^{(0)}}$
\emph{Therefore there is nothing more we can do here, and all the discussion and closures from 1D geometry, applies here in 3D geometry to closures for
  $q_\parallel^{(1)}$ and $\widetilde{r}_{\parallel\parallel}^{(1)}$ without any changes}. So for example,
\begin{eqnarray}
&&  R_{3,2}(\zeta):\qquad q_\parallel^{(1)} = -\frac{2}{\sqrt{\pi}} n_0 v_{\textrm{th}\parallel} i\sign(\kpar)T_\parallel^{(1)}; \label{eq:Qpar_Again}\\
&&  R_{4,3}(\zeta):\qquad \widetilde{r}_{\parallel\parallel}^{(1)} = - \frac{2\sqrt{\pi}}{(3\pi-8)}v_{\textrm{th}\parallel} i\sign(\kpar) q_\parallel^{(1)}
  + \frac{(32-9\pi)}{2(3\pi-8)}v_{\textrm{th}\parallel}^2 n_{0} T_\parallel^{(1)},
\end{eqnarray}
and similarly for all the other closures that we considered in the 1D geometry.
\clearpage
\subsubsection{Closures for $q_\perp$ and $\widetilde{r}_{\parallel\perp}$} 
For the second group, we do not have much choices and the calculations are quite simpler.
In comparison to the first group, the expressions for $q_\perp^{(1)}$ and $\widetilde{r}_{\parallel\perp}^{(1)}$ contain only powers $\zeta$ and $\zeta^2$. On one hand, this is good news since
the analytic calculations are simpler and we will explore all possible cases of closure very quickly. On the other hand, this means that we will be able to
use only relatively low-order Pad\'e approximants to $R(\zeta)$, implying that the closures will be less accurate.

To easily spot closures, it is perhaps beneficial to use 
\begin{equation}
v_{\textrm{th}\parallel}=\sqrt{\frac{2T_\parallel^{(0)}}{m_r}};\qquad a_p = \frac{T_\perp^{(0)}}{T_\parallel^{(0)}}; \qquad \widetilde{B}_z = \frac{B_z}{B_0},
\end{equation}
and the moments read
\begin{eqnarray}
\frac{T_\perp^{(1)}}{T_\perp^{(0)}} &=& \widetilde{B}_z \Big[ 1 -a_p R(\zeta) \Big] ;\\
 q_\perp^{(1)} &=& -v_{\textrm{th}\parallel} p_\perp^{(0)} a_p \sign(\kpar)\widetilde{B}_z \Big[ \zeta R(\zeta) \Big];\\
\widetilde{r}_{\parallel\perp}^{(1)} &=& -v_{\textrm{th}\parallel}^2 p_\perp^{(0)} \frac{a_p}{2} \widetilde{B}_z \Big[ 1+2\zeta^2 R(\zeta)-R(\zeta) \Big].
\end{eqnarray}
Before proceeding with Pad\'e approximants, it is very beneficial to briefly consider the limit $\zeta\ll 1$, where the $R(\zeta)\to 1$.
And a problem is immediately apparent. The quantities $q_\perp^{(1)}$ and $\widetilde{r}_{\parallel\perp}^{(1)}$ are small and converge to zero, however,
this is in general not true for the perpendicular temperature $T_\perp^{(1)}$, where the result depends on the temperature anisotropy ratio $a_p$.  
With anisotropic mean temperatures ($a_p\neq 1$), the quantity $T^{(1)}_\perp$ will remain finite and will not converge to zero  due to coupling
with magnetic field perturbations $B_z$, essentially because of conservation of magnetic moment. The quantity
$T^{(1)}_\perp$, at least as is written now, is therefore not suitable for construction of closures. Or in another words, the technique with Pad\'e approximants
of $R(\zeta)$ will not work, since the technique is based on matching the expressions for all $\zeta$ values. To consider closures, we have
to separate this finite contribution, so that the Pad\'e technique can be used, i.e. by writing
\begin{eqnarray}
\frac{T_\perp^{(1)}}{T_\perp^{(0)}} &=& \widetilde{B}_z \Big[ 1 -a_p+a_p -a_p R(\zeta) \Big] = 
\widetilde{B}_z (1 -a_p) + \widetilde{B}_z a_p \Big[ 1-R(\zeta) \Big],\nn
\end{eqnarray}
and by moving the finite contribution to the left hand side
\begin{eqnarray}
\underbrace{\frac{T_\perp^{(1)}}{T_\perp^{(0)}} +\widetilde{B}_z (a_p-1)}_{=\mathcal{T_\perp}} &=& \widetilde{B}_z a_p \Big[ 1-R(\zeta) \Big].
\end{eqnarray}
Therefore, instead of looking for closures with $T_\perp^{(1)}$, we have to look for closures with a quantity that is proportional to the left hand side of this equation,
that we call $\mathcal{T}_\perp$ (T written with ``mathcal'' command in latex), and for clarity written with the full notation
\begin{equation} \boxed{
    \mathcal{T}_\perp \equiv \frac{T_\perp^{(1)}}{T_\perp^{(0)}} + \Big(\frac{T_\perp^{(0)}}{T_\parallel^{(0)}}-1\Big)\frac{B_z}{B_0};}
  \qquad \qquad  \boxed{\mathcal{T}_\perp = \frac{B_z}{B_0} \frac{T_\perp^{(0)}}{T_\parallel^{(0)}} \Big[ 1-R(\zeta) \Big],}
\end{equation}
where on the left is the definition of the new quantity, and on the right is the kinetic moment that this new quantity satisfies. 
Only now we are ready to use the Pad\'e approximants of $R(\zeta)$ and construct closures.
\clearpage
\subsubsection*{1-pole closure}
By using approximant $R_1(\zeta)$, the moments calculate
\begin{eqnarray}
R_1(\zeta):\qquad  D &=& \Big( 1-i\sqrt{\pi}\zeta \Big);\\
  \mathcal{T_\perp} &=& a_p \widetilde{B}_z \frac{1}{D}\Big[ -i\sqrt{\pi}\zeta\Big] ;\\
  q_\perp^{(1)} &=& -v_{\textrm{th}\parallel} p_\perp^{(0)} a_p \sign(\kpar)\widetilde{B}_z \frac{1}{D}\Big[ \zeta \Big]; \label{eq:Qperp_R1}\\
\widetilde{r}_{\parallel\perp}^{(1)} &=& -v_{\textrm{th}\parallel}^2 p_\perp^{(0)} \frac{a_p}{2} \widetilde{B}_z \frac{1}{D}\Big[ 2\zeta^2 -i\sqrt{\pi}\zeta \Big],
\end{eqnarray}
and the heat flux $q_\perp^{(1)}$ can be directly expressed through $\mathcal{T}_\perp$ according to
\begin{equation}
  R_1(\zeta): \qquad q_\perp^{(1)} =-\frac{p_\perp^{(0)}}{\sqrt{\pi}}  v_{\textrm{th}\parallel}
  i\sign(\kpar)\mathcal{T}_\perp,
\end{equation}
and using full notation and transforming to real space
\begin{equation} \boxed{
  R_1(\zeta): \qquad q_\perp^{(1)} =-\frac{p_\perp^{(0)}}{\sqrt{\pi}}  v_{\textrm{th}\parallel}
  \mathcal{H}\bigg[ \frac{T_\perp^{(1)}}{T_\perp^{(0)}} + \Big(\frac{T_\perp^{(0)}}{T_\parallel^{(0)}}-1\Big)\frac{B_z}{B_0} \bigg].} \label{eq:QperpClosureX}
\end{equation}
Up to the replacement of $B_z$ with $|\bb|$, the closure is equivalent for example to eq. (40) of \cite{Snyder1997} (their thermal speeds do not contain the factors of 2). 
The closure is similar to the corresponding closure for the parallel heat flux (\ref{eq:Qpar_Again}), and for isotropic temperatures the term $\sim B_z$ disappears.
The closure is therefore very useful for understanding of the collisionless heat flux, however, the closure is not very accurate and for $\zeta\gg 1$,
the heat flux (\ref{eq:Qperp_R1}) does not disappear and instead, converges to an asymptotic value.
Alternatively, since later on, the normalization is always done with respect to parallel quantities
\begin{equation}
  R_1(\zeta): \qquad \frac{q_\perp^{(1)}}{p_\parallel^{(0)}} =-\frac{ v_{\textrm{th}\parallel}}{\sqrt{\pi}} 
  \mathcal{H}\bigg[ \frac{T_\perp^{(1)}}{T_\parallel^{(0)}} +a_p \Big( a_p-1\Big)\frac{B_z}{B_0} \bigg],
\end{equation}
and when the temperature $T_\perp^{(1)}$ is expressed through the pressure and density, it is useful to note the difference between
\begin{eqnarray}
\frac{T_\perp^{(1)}}{T_\perp^{(0)}} &=& \frac{p_\perp^{(1)}}{p_\perp^{(0)}}-\frac{n^{(1)}}{n_0}; \nn\\
\frac{T_\perp^{(1)}}{T_\parallel^{(0)}} &=& \frac{p_\perp^{(1)}}{p_\parallel^{(0)}}-a_p\frac{n^{(1)}}{n_0}.
\end{eqnarray}

\subsubsection*{2-pole closures}
Continuing with the $R_{2,0}(\zeta)$ approximant, the moments calculate
\begin{eqnarray}
R_{2,0}(\zeta):\qquad  D &=& \Big( 1-i\sqrt{\pi}\zeta -2\zeta^2 \Big);\\
  \mathcal{T}_\perp &=& a_p \widetilde{B}_z \frac{1}{D}\Big[ -2\zeta^2-i\sqrt{\pi}\zeta \Big] ;\\
  q_\perp^{(1)} &=& -v_{\textrm{th}\parallel} p_\perp^{(0)} a_p \sign(\kpar)\widetilde{B}_z \frac{1}{D}\Big[ \zeta \Big];\\
\widetilde{r}_{\parallel\perp}^{(1)} &=& -v_{\textrm{th}\parallel}^2 p_\perp^{(0)} \frac{a_p}{2} \widetilde{B}_z \frac{1}{D}\Big[ -i\sqrt{\pi}\zeta \Big].
\end{eqnarray}
The $\widetilde{r}_{\parallel\perp}^{(1)}$ can be expressed through $q_\perp^{(1)}$ and the closure reads
\begin{equation}
R_{2,0}(\zeta): \qquad \widetilde{r}_{\parallel\perp}^{(1)} = -\frac{\sqrt{\pi}}{2} v_{\textrm{th}\parallel} i\sign(\kpar) q_\perp^{(1)},
\end{equation}
or in real space
\begin{equation}
\boxed{
  R_{2,0}(\zeta): \qquad \widetilde{r}_{\parallel\perp}^{(1)} = -\frac{\sqrt{\pi}}{2} v_{\textrm{th}\parallel} \mathcal{H} q_\perp^{(1)}.} \label{eq:R20_boring1}
\end{equation}
The closure with $R_{2,0}(\zeta)$ is naturally more precise that the closure with $R_1(\zeta)$, and both $q_\perp^{(1)}$ and $\widetilde{r}_{\parallel\perp}^{(1)}$ at least
converge to zero for $\zeta\gg 1$. The closure is equivalent to eq. (35) of \cite{Snyder1997}.

There are 3 another closures that can be constructed with $R_{2,0}(\zeta)$, all of them time-dependent. The first one is obtained by searching for
$(\zeta+\alpha_q)q_{\perp}^{(1)} = \alpha_T \mathcal{T}_\perp$, and the solution is
\begin{eqnarray}
  R_{2,0}(\zeta):\qquad \Big[\zeta+ \frac{i\sqrt{\pi}}{2} \Big] q_{\perp}^{(1)} &=& \frac{p_\perp^{(0)}}{2}v_{\textrm{th}\parallel} \sign(\kpar) \mathcal{T}_\perp ;\nn\\
  \Big[-i\omega+ \frac{\sqrt{\pi}}{2}v_{\textrm{th}\parallel} |\kpar| \Big] q_{\perp}^{(1)} &=& -\frac{p_\perp^{(0)}}{2}v_{\textrm{th}\parallel}^2 i\kpar \mathcal{T}_\perp,
\end{eqnarray}
and in real space
\begin{equation} \boxed{
    R_{2,0}(\zeta):\qquad  \Big[\frac{d}{dt}- \frac{\sqrt{\pi}}{2}v_{\textrm{th}\parallel} \pr_z\mathcal{H} \Big] q_{\perp}^{(1)} =
    -\frac{p_\perp^{(0)}}{2} v_{\textrm{th}\parallel}^2 \pr_z \Big[ \frac{T_\perp^{(1)}}{T_\perp^{(0)}} + \Big(\frac{T_\perp^{(0)}}{T_\parallel^{(0)}}-1\Big)\frac{B_z}{B_0} \Big]. }
  \label{eq:R20_boring11}
\end{equation}
Alternatively, considering future normalization with parallel quantities
\begin{equation}
    R_{2,0}(\zeta):\qquad  \Big[\frac{d}{dt}- \frac{\sqrt{\pi}}{2}v_{\textrm{th}\parallel} \pr_z\mathcal{H} \Big] \frac{q_{\perp}^{(1)}}{p_\parallel^{(0)}} =
    -\frac{v_{\textrm{th}\parallel}^2}{2}  \pr_z \Big[ \frac{T_\perp^{(1)}}{T_\parallel^{(0)}} + a_p\Big(a_p-1\Big)\frac{B_z}{B_0} \Big].
\end{equation}
The closures (\ref{eq:R20_boring11}) and (\ref{eq:R20_boring1}) are related. In the companion paper (Part 1), we derived ``fluid'' nonlinear equation for
perpendicular heat flux $\pr q_\perp/\pr t$. Linearizing this equation yields
\begin{equation}
\frac{\pr q_\perp^{(1)}}{\pr t} +\pr_z \widetilde{r}_{\parallel\perp}^{(1)} +\frac{n_0}{2}v_{\textrm{th}\parallel}^2 \pr_z T_\perp^{(1)}
+\frac{p_\perp^{(0)}}{2}v_{\textrm{th}\parallel}^2 (a_p-1)\pr_z \frac{B_z}{B_0} =0, \label{eq:Qperp_lin_boring}
\end{equation} 
where since at the linear level $\pr_z\hat{b}_z\overset{\textrm{\tiny lin}}{=}0$, the
quantity $\nabla\cdot\bhat \overset{\textrm{\tiny lin}}{=} \frac{1}{B_0}(\pr_x B_x+\pr_y B_y)=-\frac{1}{B_0}\pr_z B_z$. Now by plugging
the quasi-static closure (\ref{eq:R20_boring1}) into the linearized heat flux equation (\ref{eq:Qperp_lin_boring}),
immediately recovers the time-dependent closure (\ref{eq:R20_boring11}). As discussed before, the difference between $B_z$ and $|\bb|$ again arises only from
how ``deeply'' the linearization is done. For example, exact calculation of $\nabla\cdot\bhat$ yields
\begin{equation}
  \nabla\cdot\bhat = \nabla\cdot \Big(\frac{\bb}{|\bb|}\Big) = \frac{1}{|\bb|} \underbrace{\nabla\cdot\bb}_{=0} +\bb\cdot\nabla \Big(\frac{1}{|\bb|} \Big)
  = -\frac{\bhat}{|\bb|} \cdot\nabla |\bb|,
\end{equation}
and instead of linearizing completely, it is possible to stop the linearization at the level $\nabla\cdot\bhat \overset{\textrm{\tiny lin}}{=} -\frac{1}{B_0}\pr_z |\bb|$. 

Another closure can be constructed by searching for $(\zeta+\alpha_r)r_{\parallel\perp}^{(1)}=\alpha_T \mathcal{T}_\perp$, and the solution is
\begin{eqnarray}
  R_{2,0}(\zeta):\qquad \Big[ \zeta+\frac{i\sqrt{\pi}}{2}  \Big] r_{\parallel\perp}^{(1)} &=& -\frac{i\sqrt{\pi}}{4} p_\perp^{(0)} v_{\textrm{th}\parallel}^2 \mathcal{T}_\perp;\\
  \Big[ -i\omega+\frac{\sqrt{\pi}}{2} v_{\textrm{th}\parallel} |\kpar| \Big] r_{\parallel\perp}^{(1)} &=&
  -\frac{\sqrt{\pi}}{4} p_\perp^{(0)} v_{\textrm{th}\parallel}^3 |\kpar| \mathcal{T}_\perp; \nn\\
   \Big[ \frac{d}{dt}-\frac{\sqrt{\pi}}{2} v_{\textrm{th}\parallel} \pr_z \mathcal{H} \Big] r_{\parallel\perp}^{(1)} &=&
   +\frac{\sqrt{\pi}}{4} p_\perp^{(0)} v_{\textrm{th}\parallel}^3 \pr_z \mathcal{H}
   \Big[ \frac{T_\perp^{(1)}}{T_\perp^{(0)}} + \Big(\frac{T_\perp^{(0)}}{T_\parallel^{(0)}}-1\Big)\frac{B_z}{B_0} \Big], \label{eq:R20_boring2}
\end{eqnarray}
and yet another related one by searching for  $\zeta r_{\parallel\perp}^{(1)}=\alpha_q q_\perp^{(1)}+\alpha_T \mathcal{T}_\perp$, with solution
\begin{eqnarray}
  R_{2,0}(\zeta):\qquad \zeta r_{\parallel\perp}^{(1)} &=&
  -\frac{\pi}{4}v_{\textrm{th}\parallel}\sign(\kpar) q_\perp^{(1)} -\frac{i\sqrt{\pi}}{4} p_\perp^{(0)} v_{\textrm{th}\parallel}^2 \mathcal{T}_\perp;\\
  -i\omega r_{\parallel\perp}^{(1)} &=&
  +\frac{\pi}{4}v_{\textrm{th}\parallel}^2 i\kpar q_\perp^{(1)} -\frac{\sqrt{\pi}}{4} p_\perp^{(0)} v_{\textrm{th}\parallel}^3 |\kpar| \mathcal{T}_\perp;\nn\\
  \frac{d}{dt} r_{\parallel\perp}^{(1)} &=&
  +\frac{\pi}{4}v_{\textrm{th}\parallel}^2 \pr_z q_\perp^{(1)} +\frac{\sqrt{\pi}}{4} p_\perp^{(0)} v_{\textrm{th}\parallel}^3 \pr_z\mathcal{H}
  \Big[ \frac{T_\perp^{(1)}}{T_\perp^{(0)}} + \Big(\frac{T_\perp^{(0)}}{T_\parallel^{(0)}}-1\Big)\frac{B_z}{B_0} \Big]. \label{eq:R20_boring3}
\end{eqnarray}  
The last closure (\ref{eq:R20_boring3}) can also be directly obtained from (\ref{eq:R20_boring2}) by using the quasi-static closure (\ref{eq:R20_boring1}) and
$\mathcal{H}\mathcal{H}=-1$. 
Both closures (\ref{eq:R20_boring3}), (\ref{eq:R20_boring2}) are not very interesting,
since the quasi-static closure (\ref{eq:R20_boring1}) for $r_{\parallel\perp}^{(1)}$  and the time-dependent closure (\ref{eq:R20_boring11}) for the heat flux $q_\perp^{(1)}$
are of the same precision and much simpler to implement. Importantly, after checking the dispersion relations, closure (\ref{eq:R20_boring3})
has to be disregarded since it can produce positive growth rate.  

For completeness, there is also 1 time-dependent closure with $R_{2,1}$ approximant $\zeta q_\perp^{(1)}=\alpha_T \mathcal{T}_\perp$ that is not considered and is disregarded,
since that approximant is not well-behaved.

\subsubsection*{3-pole closures}
As in the 1D case, we can suppress writing the proportionality constants (including the minus signs) and
concentrate only on expressions inside of the big brackets. Continuing with the $R_{3,1}(\zeta)$ approximant
\begin{eqnarray}
R_{3,1}(\zeta):\qquad  D &=& \Big( 1-\frac{4i}{\sqrt{\pi}}\zeta -2\zeta^2 +2i\frac{(4-\pi)}{\sqrt{\pi}}\zeta^3\Big);\\
  \mathcal{T}_\perp &\sim&  \frac{1}{D}\Big[  2i\frac{(4-\pi)}{\sqrt{\pi}}\zeta^3-2\zeta^2-i\sqrt{\pi}\zeta \Big] ;\\
  q_\perp^{(1)} &\sim&  \frac{1}{D}\Big[ -i\frac{(4-\pi)}{\sqrt{\pi}}\zeta^2+\zeta \Big];\\
\widetilde{r}_{\parallel\perp}^{(1)} &\sim&  \frac{1}{D}\Big[ -i\sqrt{\pi}\zeta \Big].
\end{eqnarray}
No quasi-static closures are possible. A time-dependent closure can be constructed
by searching for $(\zeta+\alpha_r)\widetilde{r}_{\parallel\perp}^{(1)} = \alpha_q q_\perp^{(1)}$, and
the solution reads
\begin{eqnarray}
  R_{3,1}(\zeta):\qquad \Big[\zeta+ \frac{i\sqrt{\pi}}{4-\pi} \Big] \widetilde{r}_{\parallel\perp}^{(1)} &=& v_{\textrm{th}\parallel} \frac{\pi}{2(4-\pi)} \sign(\kpar) q_\perp^{(1)};\nn\\
   \Big[-i \omega+ \frac{\sqrt{\pi}}{4-\pi} v_{\textrm{th}\parallel} |\kpar| \Big] \widetilde{r}_{\parallel\perp}^{(1)}
   &=& -v_{\textrm{th}\parallel}^2 \frac{\pi}{2(4-\pi)} i\kpar q_\perp^{(1)},
\end{eqnarray}
and in real space
\begin{equation}
\boxed{  
R_{3,1}(\zeta):\qquad   \Big[\frac{d}{dt}- \frac{\sqrt{\pi}}{4-\pi} v_{\textrm{th}\parallel} \pr_z \mathcal{H} \Big] \widetilde{r}_{\parallel\perp}^{(1)}
   = -v_{\textrm{th}\parallel}^2 \frac{\pi}{2(4-\pi)} \pr_z q_\perp^{(1)}.} \label{eq:PS1closure}
\end{equation}

Continuing with the $R_{3,2}(\zeta)$ approximant
\begin{eqnarray}
R_{3,2}(\zeta):\qquad  D &=& \Big( 1-\frac{3}{2}i\sqrt{\pi}\zeta -2\zeta^2 +i\sqrt{\pi}\zeta^3\Big);\\
  \mathcal{T}_\perp &\sim&  \frac{1}{D}\Big[ i\sqrt{\pi}\zeta^3-2\zeta^2 -i\sqrt{\pi}\zeta \Big] ;\\
  q_\perp^{(1)} &\sim&  \frac{1}{D}\Big[ -\frac{i}{2}\sqrt{\pi}\zeta^2+\zeta \Big];\\
\widetilde{r}_{\parallel\perp}^{(1)} &\sim&  \frac{1}{D}\Big[ -i\sqrt{\pi}\zeta \Big].
\end{eqnarray}
By searching for $(\zeta+\alpha_r)\widetilde{r}_{\parallel\perp}^{(1)} = \alpha_q q_\perp^{(1)}$, yields a closure
\begin{eqnarray}
  R_{3,2}(\zeta):\qquad \Big[\zeta+ \frac{2i}{\sqrt{\pi}} \Big] \widetilde{r}_{\parallel\perp}^{(1)} &=& v_{\textrm{th}\parallel} \sign(\kpar) q_\perp^{(1)};\nn\\
  \Big[-i\omega+ \frac{2}{\sqrt{\pi}}v_{\textrm{th}\parallel} |\kpar| \Big] \widetilde{r}_{\parallel\perp}^{(1)} &=& -v_{\textrm{th}\parallel}^2 i\kpar q_\perp^{(1)},
\end{eqnarray}
and in real space
\begin{equation}
  \boxed{
  R_{3,2}(\zeta):\qquad  \Big[\frac{d}{dt}- \frac{2}{\sqrt{\pi}}v_{\textrm{th}\parallel} \pr_z \mathcal{H}\Big] \widetilde{r}_{\parallel\perp}^{(1)}
  = -v_{\textrm{th}\parallel}^2 \pr_z q_\perp^{(1)}.} \label{eq:PS2closure}
\end{equation}
Closures (\ref{eq:PS1closure}), (\ref{eq:PS2closure}) are equivalent to closures of \cite{PassotSulem2007},
after one prescribes gyrotropic limit in that paper (and replaces the wrong coefficient in the $R_{3,1}(\zeta)$ closure introduced by \cite{HedrickLeboeuf1992}).

Finally, it is indeed possible to construct an $o(\zeta^3)$ closure for the perpendicular quantities considered, by using the $R_{3,0}(\zeta)$ approximant. 
The moments calculate
\begin{eqnarray}
R_{3,0}(\zeta):\qquad  D &=& \Big( 1-i\frac{\sqrt{\pi}}{4-\pi}\zeta -\frac{3\pi-8}{4-\pi}\zeta^2 +2i\sqrt{\pi}\frac{\pi-3}{4-\pi}\zeta^3\Big);\\
  \mathcal{T}_\perp &\sim&  \frac{1}{D}\Big[ 2i\sqrt{\pi}\frac{\pi-3}{4-\pi}\zeta^3-\frac{3\pi-8}{4-\pi}\zeta^2 -i\sqrt{\pi}\zeta \Big] ;\\
  q_\perp^{(1)} &\sim&  \frac{1}{D}\Big[ -i\sqrt{\pi}\frac{\pi-3}{4-\pi}\zeta^2+\zeta \Big];\\
\widetilde{r}_{\parallel\perp}^{(1)} &\sim&  \frac{1}{D}\Big[ \frac{16-5\pi}{4-\pi}\zeta^2-i\sqrt{\pi}\zeta \Big],
\end{eqnarray}
and by searching for $(\zeta+\alpha_r)\widetilde{r}_{\parallel\perp}^{(1)}=\alpha_q  q_\perp^{(1)} + \alpha_T  \mathcal{T}_\perp$ yields a closure
\begin{eqnarray}
  R_{3,0}(\zeta): \qquad \Big[\zeta+\frac{i}{2\sqrt{\pi}} \frac{(3\pi-8)}{(\pi-3)}  \Big] \widetilde{r}_{\parallel\perp}^{(1)} &=&
  v_{\textrm{th}\parallel} \frac{4-\pi}{2(\pi-3)} \sign(\kpar) q_\perp^{(1)} + p_{\perp}^{(0)}v_{\textrm{th}\parallel}^2 \frac{i}{4\sqrt{\pi}}
  \frac{(16-5\pi)}{(\pi-3)} \mathcal{T}_\perp;\nn\\
  \Big[-i\omega+\frac{(3\pi-8)}{2\sqrt{\pi}(\pi-3)} v_{\textrm{th}\parallel} |\kpar| \Big] \widetilde{r}_{\parallel\perp}^{(1)} &=&
  -v_{\textrm{th}\parallel}^2 \frac{4-\pi}{2(\pi-3)} i\kpar q_\perp^{(1)} + p_{\perp}^{(0)}v_{\textrm{th}\parallel}^3
  \frac{(16-5\pi)}{4\sqrt{\pi}(\pi-3)} |\kpar| \mathcal{T}_\perp,
\end{eqnarray}
and the full expression in real space reads \citep{HunanaPRL2018}
\begin{empheq}[box=\fbox]{align}
  R_{3,0}(\zeta): \qquad  \Big[\frac{d}{dt}-\frac{(3\pi-8)}{2\sqrt{\pi}(\pi-3)} v_{\textrm{th}\parallel} \pr_z \mathcal{H} \Big] \widetilde{r}_{\parallel\perp}^{(1)} = &
  -v_{\textrm{th}\parallel}^2 \frac{4-\pi}{2(\pi-3)} \pr_z q_\perp^{(1)} \nn\\ & - p_{\perp}^{(0)}v_{\textrm{th}\parallel}^3
  \frac{(16-5\pi)}{4\sqrt{\pi}(\pi-3)} \pr_z \mathcal{H} \Big[ \frac{T_\perp^{(1)}}{T_\perp^{(0)}} + \Big(\frac{T_\perp^{(0)}}{T_\parallel^{(0)}}-1\Big)\frac{B_z}{B_0} \Big],
  \label{eq:3D-R30closure}
\end{empheq}
or again considering normalization with respect to parallel quantities
\begin{eqnarray}
  R_{3,0}(\zeta): \qquad  \Big[\frac{d}{dt}-\frac{(3\pi-8)}{2\sqrt{\pi}(\pi-3)} v_{\textrm{th}\parallel} \pr_z \mathcal{H} \Big] \frac{\widetilde{r}_{\parallel\perp}^{(1)}}{p_{\parallel}^{(0)}} &=&
  -v_{\textrm{th}\parallel}^2 \frac{4-\pi}{2(\pi-3)} \pr_z \frac{q_\perp^{(1)}}{p_{\parallel}^{(0)}} \nn\\ && - v_{\textrm{th}\parallel}^3
  \frac{(16-5\pi)}{4\sqrt{\pi}(\pi-3)} \pr_z \mathcal{H} \Big[ \frac{T_\perp^{(1)}}{T_\parallel^{(0)}} + a_p \Big(a_p-1\Big)\frac{B_z}{B_0} \Big].
\end{eqnarray}
The $R_{3,0}(\zeta)$ has precision $o(\zeta^3)$, $o(1/\zeta^2)$.

\newpage
\subsection{Table of moments $(T_\perp,q_\perp,\widetilde{r}_{\parallel\perp})$ for various Pad\'e approximants}
The following summarizing table for quantities $\mathcal{T}_\perp(T_\perp^{(1)}), q_\perp^{(1)}, \widetilde{r}_{\parallel\perp}^{(1)}$ is created to clearly see the possibilities of a closure. 
All the proportionality constants (including the minus signs) and including the common denominator of $R(\zeta)$, are suppressed here.
The approximants $R_{2,1}, R_{4,5}, R_{6,9}, R_{8,13}$ are marked with an asterisk ``*'',  because these do not account for the Landau residue and
are not well-behaved. These approximants are provided only for completeness and should be disregarded.\\\\
1-pole and 2-pole approximants\\
\begin{tabular}{| c | c | c | c |}
  \hline
  & $R_1$   & $R_{2,0}$     &  $R_{2,1}^*$ \\
  \hline
  $\mathcal{T}_\perp$ & $\zeta$  & $\zeta^2,\zeta$      &  $\zeta^2$ \\
  $q^{(1)}_\perp$ & $\zeta$      & $\zeta$ & $\zeta$ \\
  $\widetilde{r}_{\parallel\perp}^{(1)}$ & $\zeta^2,\zeta$   &$\zeta$ & $0$ \\
  \hline
\end{tabular}
\\
3-pole approximants\\
\begin{tabular}{| c | c | c | c |}
  \hline
  & $R_{3,0}$     &  $R_{3,1}$ &  $R_{3,2}$ \\
  \hline
  $\mathcal{T}_\perp$ & $\zeta^3,\zeta^2,\zeta$    &  $\zeta^3,\zeta^2,\zeta$   &  $\zeta^3,\zeta^2,\zeta$ \\
  $q^{(1)}_\perp$ & $\zeta^2,\zeta$ & $\zeta^2,\zeta$  & $\zeta^2,\zeta$\\
  $\widetilde{r}_{\parallel\perp}^{(1)}$ & $\zeta^2,\zeta$ & $\zeta$ & $\zeta$\\
  \hline
\end{tabular}
\\
4-pole approximants\\
\begin{tabular}{| c | c | c | c | c | c | c |}
  \hline			
          & $R_{4,0}$                                            & $R_{4,1}$               & $R_{4,2}$                & $R_{4,3}$                & $R_{4,4}$                & $R_{4,5}^*$ \\
  \hline
$\mathcal{T}_\perp$ & $\zeta^4\cdots\zeta$                             & $\zeta^4\cdots\zeta$         & $\zeta^4\cdots\zeta$         &  $\zeta^4\cdots\zeta$        & $\zeta^4\cdots\zeta$         & $\zeta^4,\zeta^2$ \\
$q^{(1)}_\perp$ & $\zeta^3\cdots\zeta$                     & $\zeta^3\cdots\zeta$  & $\zeta^3\cdots\zeta$         &  $\zeta^3\cdots\zeta$               & $\zeta^3\cdots\zeta$         & $\zeta^3,\zeta$ \\
$\widetilde{r}^{(1)}_{\parallel\perp}$ & $\zeta^3\cdots\zeta$ & $\zeta^2,\zeta$ & $\zeta^2,\zeta$ & $\zeta^2, \zeta$ & $\zeta^2,\zeta$                 & $\zeta^2$ \\
  \hline  
\end{tabular}
\\
5-pole and 6-pole approximants\\
\begin{tabular}{| c | c | c | c | c || c | c | c | c | c |}
  \hline			
        & $R_{5,0}$ &   $R_{5,1}$    & $\cdots$      & $R_{5,6}$       & $R_{6,0}$          & $R_{6,1}$        & $\cdots$   & $R_{6,8}$ & $R_{6,9}^*$\\
  \hline
$\mathcal{T}_\perp$ & $\zeta^5\cdots\zeta$ & $\zeta^5\cdots\zeta$  & $\cdots$  & $\zeta^5\cdots\zeta$   &  $\zeta^6\cdots\zeta$   &$\zeta^6\cdots\zeta$ & $\cdots$ &$\zeta^6\cdots\zeta$ &$\zeta^6,\zeta^4,\zeta^2$\\
$q^{(1)}_\perp$ & $\zeta^4\cdots\zeta$ & $\zeta^4\cdots\zeta$  & $\cdots$  & $\zeta^4\cdots\zeta$           &  $\zeta^5\cdots\zeta$         & $\zeta^5\cdots \zeta$       & $\cdots$ & $\zeta^5\cdots \zeta$ &$\zeta^5,\zeta^3,\zeta$\\
$\widetilde{r}^{(1)}_{\parallel\perp}$ & $\zeta^4\cdots\zeta$ & $\zeta^3\cdots\zeta$ & $\cdots$     & $\zeta^3\cdots\zeta$   & $\zeta^5\cdots\zeta$          & $\zeta^4\cdots\zeta$     & $\cdots$ & $\zeta^4\cdots\zeta$ & $\zeta^4,\zeta^2$\\
  \hline  
\end{tabular}
\\
7-pole and 8-pole approximants\\
\begin{tabular}{| c | c | c | c | c || c | c | c | c | c |}
  \hline			
  & $R_{7,0}$ &   $R_{7,1}$    & $\cdots$      & $R_{7,10}$       & $R_{8,0}$          & $R_{8,1}$        & $\cdots$
  & $R_{8,12}$          & $R_{8,13}^*$\\
  \hline
  $\mathcal{T}$ & $\zeta^7\cdots\zeta$ & $\zeta^7\cdots\zeta$  & $\cdots$  & $\zeta^7\cdots\zeta$   &  $\zeta^8\cdots\zeta$   &$\zeta^8\cdots\zeta$ & $\cdots$
  &  $\zeta^8\cdots\zeta$   &$\zeta^8,\zeta^6,\zeta^4,\zeta^2$ \\
  $q^{(1)}_\perp$ & $\zeta^6\cdots\zeta$ & $\zeta^6\cdots\zeta$  & $\cdots$  & $\zeta^6\cdots\zeta$           &  $\zeta^7\cdots \zeta$         & $\zeta^7\cdots\zeta$       & $\cdots$
  &  $\zeta^7\cdots \zeta$         & $\zeta^7,\zeta^5,\zeta^3,\zeta$\\
  $\widetilde{r}_{\parallel\perp}^{(1)}$ & $\zeta^6\cdots\zeta$& $\zeta^5\cdots\zeta$ & $\cdots$ & $\zeta^5\cdots\zeta$   & $\zeta^7\cdots\zeta$   & $\zeta^6\cdots\zeta$     & $\cdots$  &
  $\zeta^6\cdots\zeta$ & $\zeta^6,\zeta^4,\zeta^2$\\
  \hline  
\end{tabular}\\\\

By observing the table, there are altogether 2 possible quasi-static closures:
\begin{eqnarray}
  R_{1}: \quad q^{(1)}_\perp &=& \alpha_T \mathcal{T}_\perp;\nn\\
  R_{2,0}: \quad \widetilde{r}_{\parallel\perp}^{(1)} &=& \alpha_q q_\perp^{(1)}, \label{eq:RparPerpF1}
\end{eqnarray}
and 6 time-dependent closures:
\begin{eqnarray}
  R_{2,0}: \quad (\zeta+\alpha_q) q_{\perp}^{(1)} &=& \alpha_T \mathcal{T}_\perp; \qquad
     (\zeta+\alpha_r)\widetilde{r}_{\parallel\perp}^{(1)} = \alpha_T \mathcal{T}_\perp; \qquad \cancel{\zeta\widetilde{r}_{\parallel\perp}^{(1)} = \alpha_q q_\perp^{(1)} + \alpha_T \mathcal{T}_\perp};\nn\\
  R_{3,0}: \quad (\zeta+\alpha_r)\widetilde{r}^{(1)}_{\parallel\perp} &=& \alpha_q q_\perp^{(1)} +\alpha_T \mathcal{T}_\perp ;\nn\\
  R_{3,1}: \quad (\zeta+\alpha_r)\widetilde{r}^{(1)}_{\parallel\perp} &=& \alpha_q q_\perp^{(1)};\nn\\
  R_{3,2}: \quad (\zeta+\alpha_r)\widetilde{r}^{(1)}_{\parallel\perp} &=& \alpha_q q_\perp^{(1)}. \label{eq:RparPerpF2}
\end{eqnarray}

We briefly checked dispersion relations that these closures yield for parallel propagation (proton species only, electrons cold),
where the $q^{(1)}_\perp$ and $\widetilde{r}^{(1)}_{\parallel\perp}$ closures produce only higher-order modes. This eliminated one $R_{2,0}$ closure that produced a growing mode.
The $R_1$ closure yields $\zeta=-i/\sqrt{\pi}$; the remaining $R_{2,0}$ closures yield $\zeta=\pm\sqrt{8-\pi}/4-i\sqrt{\pi}/4$ (result reported also in the Appendix of
\cite{Hunana2011}), the $R_{3,0}$ closure yields $\zeta=\pm 0.92-0.91i;\zeta=-1.02i$; the $R_{3,1}$ closure yields $\zeta=\pm 0.96-0.64i;\zeta=-0.78i$; and the $R_{3,2}$ closure
yields $\zeta=\pm1.04-0.33i;\zeta=-0.47i$.\\

\clearpage
\section{Conclusions} \label{section:Conclusions}

We offer a brief summary of the major results discussed throughout the text.

\begin{itemize} 
\item The kinetic Vlasov equation implicitly contains ``singularities'' in velocity space, referred to as wave-particle resonances.
  These resonances occur, because particles of a given species traveling along magnetic field lines with a velocity component $v_{\parallel}$ interact
  with plasma waves propagating in that system with a parallel phase speed $(\omega+n\Omega)/\kpar$,
  where $\Omega$ is the cyclotron frequency for that given species and $n=0,\pm 1,\pm 2\ldots$ is an integer.
  Wave-particle resonances can be separated  
  into Landau resonances ($n=0$) and cyclotron resonances ($n\neq 0$).

\item The presence of wave-particle resonances in the Vlasov equation is revealed by considering perturbations $f^{(1)}=f-f_0$
  around an equilibrium distribution function $f_0$, and by obtaining an explicit expression for $f^{(1)}$ that satisfies the Vlasov equation.
  For example, in a simplified 1D electrostatic geometry (which can be viewed as electrostatic propagation along $\bb_0$),
  the perturbations read $f^{(1)}=-\frac{iqE_\parallel}{m} \frac{\pr f_0/\pr v_\parallel}{\omega-v_\parallel\kpar}$, and contain Landau resonances.
  
\item  Obtaining $f^{(1)}$ in a general 3D electromagnetic geometry requires quite complicated procedure of
  integration along an unperturbed orbit (zero-order trajectory, from time $t'=-\infty$ to $t'=t$), see eq. (\ref{eq:f1Int}).
  The procedure can be considered as a core of any plasma book and here it is summarized in Appendix \ref{sec:Orbit}.   
  General $f^{(1)}$ perturbations around a gyrotropic $f_0$ are given by eq. (\ref{eq:GrandF1}).   
  Prescribing bi-Maxwellian $f_0$ yields (\ref{eq:GrandMax}), and prescribing bi-Kappa $f_0$ yields (\ref{eq:GrandKappa}).
  Obviously, perturbations $f^{(1)}\sim\frac{1}{\omega-v_\parallel k_\parallel - n\Omega}$, and contain Landau resonances and cyclotron resonances.
  
\item After an $f^{(1)}$ is obtained, integration over velocity space can be performed, eventually yielding an infinite hierarchy of ``kinetic'' moments.
  Combining Maxwell's equations
  $\nabla\times\bb = \frac{4\pi}{c}\bj+\frac{1}{c}\frac{\pr\bE}{\pr t}$ and $\nabla\times\bE=-\frac{1}{c}\frac{\pr\bb}{\pr t}$ yields the
  following wave equation
  \begin{equation} \label{eq:Wave}
  \boldsymbol{k}\times(\boldsymbol{k}\times\bE) +\frac{\omega^2}{c^2}\Big( \frac{4\pi i}{\omega} \bj +\bE \Big) =0.
  \end{equation}  
  Therefore, to obtain full dispersion relation of kinetic theory, it is sufficient to stop the hierarchy at the 1st-order (velocity) moment, which determines
  the current $\bj=\sum_r q_r n_r \bu_r=\boldsymbol{\sigma}\cdot\bE$. Calculations of pressure or higher-order kinetic moments
  are not necessary and thus typically omitted (provided that the full non-gyrotropic $f^{(1)}$ is considered, so that the perpendicular velocity moments
  $\bu_\perp$ are non-zero). In addition to the conductivity tensor $\boldsymbol{\sigma}$,
  one can also use the susceptibility tensor $\boldsymbol{\chi}=\frac{4\pi i}{\omega}\boldsymbol{\sigma}$, and the dielectric
  tensor $\boldsymbol{\epsilon}=\boldsymbol{\chi}+\boldsymbol{I}$ (the $\boldsymbol{I}$ is a unit matrix and here it represents contributions
  of the displacement current). The definitions of $\boldsymbol{\chi}$ and $\boldsymbol{\epsilon}$ are naturally motivated
  by the wave equation (\ref{eq:Wave}).

\item In Landau fluid models, the kinetic hierarchy has to be calculated at least up to the 3rd-order (heat flux) moment, or
  preferably, the 4th-order moment $\boldsymbol{r}$ (or beyond). Importantly, a closure has to be found where the last retained moment is
  expressed through lower-order moments. Subsequently, a simplification of $f^{(1)}$ is necessary, and in general one needs to
  impose low-frequency limit $\omega/\Omega\ll 1$, which eliminates the $n\neq 0$ cyclotron resonances. The exception is the 1D electrostatic geometry,
  where the low-frequency restriction is not required, and closures for arbitrary frequencies (and wavelengths) can be obtained.

\item In the 3D electromagnetic geometry, we restricted our attention to perturbations $f^{(1)}$ in the gyrotropic limit, see eq. (\ref{eq:f1_gyrotropic}).
  In this geometry, in addition to the low-frequency limit, one also assumes that the gyroradius is small, which corresponds to the limit $k_\perp v_\perp/\Omega \ll 1$
  (the gyroradius is defined as $v_{\textrm{th}\perp}/\Omega$, but here the limit is applied directly on $f^{(1)}$ before integration over velocity space).
  It is rather mind boggling that to obtain the correct $f^{(1)}$ in the laboratory reference frame, one needs to first calculate the complicated
  integration around the unperturbed orbit, and only then prescribe the gyrotropic limit.
  
\item Alternatively, the $f^{(1)}$ in the gyrotropic limit can be derived by using the guiding-center reference frame,
  and by imposing the conservation of the magnetic moment in the Vlasov equation from the beginning.
  Then, it is possible to show that various terms in $f^{(1)}$ correspond to the conservation of magnetic
  moment, electrostatic Coulomb force (which yields Landau damping), and magnetic mirror force (which yields transit-time damping, also called Barnes damping),
  see eq. (\ref{eq:f1_gyrotropic2}) and Section \ref{sec:Forces}.
  
\item We considered Landau fluid closures only for a bi-Maxwellian $f_0$ (which in the 1D geometry simplifies to Maxwellian $f_0$), even though one
  should be able to construct closures for a different $f_0$ with a similar technique.
  
\item In the 1D electrostatic geometry, the kinetic hierarchy calculated up to the 4th-order moment is given by eq. (\ref{eq:RefPica-n})-(\ref{eq:RefPica-rr}).
  All the moments contain the plasma response function $R(\zeta)=1+\zeta Z(\zeta)$, where $Z(\zeta)$ is the plasma dispersion function defined by eq. (\ref{eq:PDF}),
  and the variable $\zeta=\frac{\omega}{|\kpar|v_{\textrm{th}\parallel}}$. Importantly, the $\zeta$ variable is here defined with $|\kpar|=\sign(\kpar)\kpar$.
  If the $\zeta$ variable is defined with $\kpar$, the plasma dispersion function has to be redefined to $Z_0(\zeta)$, eq. (\ref{eq:PDFZ0}).
  The $R(\zeta)$ in the kinetic hierarchy can be quickly interpreted according to eq. (\ref{eq:LandauX}).  

\item It is impossible to find any ``direct'' rigorously exact fluid closure in the kinetic hierarchy of moments.
  In other words, it is impossible to take the last retained n-th order moment,
  and directly express it through lower-order moments by using exact un-approximated $R(\zeta)$ function,
  in such a way that the closure eliminates the $R(\zeta)$ function.    
  Technically, such a closure is possible only when $n\to \infty$.
  
\item To find a closure, the $R(\zeta)$ in the kinetic hierarchy needs to be analytically approximated, for example by a suitable Pad\'e approximant $R_{n,n'}(\zeta)$
  (as a ratio of two polynomials in $\zeta$).
  Approximants $R_{n,n'}(\zeta)$ are constructed by matching power series expansions $|\zeta|\ll 1$ of $R(\zeta)$, see eq. (\ref{eq:RzetaSMA}),
  and asymptotic series expansions $|\zeta|\gg 1$, see eq. (\ref{eq:RzetaL}). Perhaps the most convenient is to expand (\ref{eq:ZzetaDefX}).
  
\item  Importantly, contributions from the Landau residue $\sim\zeta e^{-\zeta^2}$ in $R(\zeta)$ are
  retained in the power series expansion, however, the contributions are eliminated in the
  asymptotic series expansion (since there is no asymptotic expansion of $e^{-\zeta^2}$).
  The same procedure is used in the kinetic solver WHAMP.
  Consequently, deeply down in the lower complex plane where damping becomes very large, Pad\'e approximants of $R(\zeta)$ become less accurate.
  
\item  Another example is the Langmuir mode, see Section \ref{sec:Langmuir}, where in the long-wavelength limit the frequency $\omega$ does not decrease,
  but is equal to the plasma frequency. Thus, $|\zeta|\gg 1$, and the Landau damping of the Langmuir mode in the long-wavelength limit typically disappears much more
  rapidly in kinetic theory than in Landau fluid models (see Figure  \ref{fig:Langmuir1}),
  which is a direct consequence of the missing $\zeta e^{-\zeta^2}$ in the asymptotic expansions of $R(\zeta)$. Nevertheless, at spatial scales that are shorter than five Debye lengths,
  the damping of the Langmuir mode can be captured very accurately in a fluid framework, see closure (\ref{eq:R75closure}) and Figure \ref{fig:Lmode}.
  Notably, it was indeed the example of the
  Langmuir mode that was used by \cite{Landau1946} to predict this collisionless damping phenomenon.

\item  We introduced a new classification scheme, 
  that we consider more natural than previous classifications. 
  The $n$ index in $R_{n,n'}(\zeta)$ represents the number of poles, and
  the ``basic'' approximant $R_{n,0}(\zeta)$ is defined as having the correct (leading-order) asymptote $-1/(2\zeta^2)$, see eq. (\ref{eq:Rn0}).
  The $R_{n,0}(\zeta)$ therefore correctly captures the asymptotic profile of the 0th-order (density) moment, and 
  approximants with less asymptotic points should be avoided if possible. 
  The $R_{n,n'}(\zeta)$ is defined as using $n'$ additional points in the asymptotic series expansion in comparison to $R_{n,0}(\zeta)$.
  The exception is the 1-pole approximant
  $R_1(\zeta) = \frac{1}{1-i\sqrt{\pi}\zeta}$, which obviously does not have the correct asymptote.

\item  Approximant $R_{n,n'}(\zeta)$ has power series precision
  $o(\zeta^{2n-3-n'})$ and asymptotic series precision $o(\zeta^{-2-n'})$.
  Analytic forms of 2-pole approximants of $R(\zeta)$ and $Z(\zeta)$ are given in Section \ref{section:2poleC}, 3-pole approximants in Section \ref{section:3poleC} and
  4-pole approximants in Section \ref{section:4poleC}. In Appendix \ref{section:Pade}, we provide valuable tables of 5-, 6-, 7- and 8-pole
  approximants of $R(\zeta)$, many in an analytic form. The precision of all approximants is compared in Section \ref{sec:PrecissionR}.
  
\item The limit $|\zeta|\ll 1$ can be viewed as isothermal limit, and $|\zeta|\gg 1$ can be viewed as adiabatic limit. Therefore, classical adiabatic fluid
  models discussed in Part 1 can be obtained by considering a high phase-speed limit $|\frac{\omega}{\kpar}|\gg v_{\textrm{th}\parallel}$.
  The exception is the generalized isothermal (``static'') closure used to capture the mirror instability, where a low phase-speed limit
  $|\frac{\omega}{\kpar}|\ll v_{\textrm{th}\parallel}$ must be used.

\item In many instances, solely expanding in $|\zeta|\ll 1$ or $|\zeta|\gg 1$ is not appropriate, and the $R(\zeta)$ together with $Z(\zeta)$ can be viewed as
  the most important functions of kinetic theory. For example, considering proton-electron plasma 
  at scales that are much longer than the Debye length, the dispersion relation of the parallel ion-acoustic mode is given by eq. (\ref{eq:soundKin}),
  and for equal proton and electron temperatures it reads $R(\zeta_p)+R(\zeta_e)=0$.
  No expansion of $R(\zeta)$ is possible, since the numerical solution is $\zeta_p=\pm 1.46-0.63i$.
  Only when electrons are hot and $T_{\parallel e}^{(0)}\gg T_{\parallel p}^{(0)}$, a simplified dispersion relation for the ion-acoustic mode can be obtained
  by prescribing $|\zeta_p|\gg 1$ and $|\zeta_e|\ll 1$.
  By employing Pad\'e approximants $R_{n,n'}(\zeta)$ in Landau fluid closures,
  the $R(\zeta)$ function is analytically approximated for all $\zeta$ values, see Figures \ref{fig:6} and \ref{fig:7}.

\item For the 1D electrostatic geometry, all the Landau fluid closures that can be constructed for the heat flux $q$ and the 4th-order moment perturbation
  $\widetilde{r}=r-3p^2/\rho$, are summarized in eq.
  (\ref{eq:Table1S})-(\ref{eq:Table1T}). The same closures are obtained in the 3D electromagnetic geometry for parallel moments
  $q_\parallel$ and $\widetilde{r}_{\parallel\parallel}$. These closures do not have any restrictions for frequencies and wavenumbers, and are therefore
  valid from the largest astrophysical scales down to the Debye length. 

\item Landau fluid closures can be separated into two categories.
  1) A closure is called static (or quasi-static), when the last retained moment $X_l$ is directly expressed through lower-order moments.
     2) A closure is called dynamic (or time-dependent), when $\zeta X_l+\alpha X_l$ is expressed through lower-order moments (where $\alpha$ is a coefficient). 
     After a dynamic closure is transformed to real space, $\pr/\pr t$ is replaced by the convective derivative $d/dt$ to preserve Galilean invariance.

\item  In real space, all the closures contain the negative Hilbert transform operator $\mathcal{H}$, defined according to 
  $\mathcal{H} f(z) = -\frac{1}{\pi z} * f(z) = - \frac{1}{\pi} V.P. \int_{-\infty}^\infty \frac{f(z-z')}{z'} dz'$, where $*$ represents convolution.
  The $\mathcal{H}$ operator in closures comes from Fourier space, where it is equal to $i\sign(\kpar)$. In real space, the $\mathcal{H}$ operator represents
  non-locality of closures, and ideally, the integrals in $\mathcal{H}f(z)$ should be calculated along magnetic field lines. The effect is
  pronounced in numerical simulations, where calculating the Hilbert transform along the ambient magnetic field $B_0$ can cause instabilities,
  see \cite{Passot2014}.   

\item For example, the simplest closure for the heat flux $q_\parallel$ is given by eq. (\ref{eq:Qcollisionless})
  of \cite{HammettPerkins1990}  (or equivalently by (\ref{eq:Qpar_Again}) when written in the 3D geometry).
  The simplest closure for the heat flux $q_\perp$ is given by eq. (\ref{eq:QperpClosureX}) of \cite{Snyder1997}. Both
  closures are proportional to the Hilbert transform of
  temperatures $T_\parallel$, $T_\perp$. Therefore, Landau fluid closures yield gyrotropic heat fluxes  $q_\parallel$, $q_\perp$ that are non-local, and
  influenced by temperatures along the entire magnetic field line. Notably, this is in contrast to ``classical'' non-gyrotropic heat flux
  vectors $\boldsymbol{S}^\parallel_\perp$,  $\boldsymbol{S}^\perp_\perp$ discussed in Part 1, which were local and proportional to the gradient of temperatures. 

\item In the 3D electromagnetic geometry in the gyrotropic limit, the closure for the perturbation $\widetilde{r}_{\perp\perp}$ is simply
  $\widetilde{r}_{\perp\perp}=0$. One needs to consider only closures for $q_\perp$ and $\widetilde{r}_{\parallel\perp}$, which are given
  in Section \ref{sec:3DclosuresX} and summarized in
  eq. (\ref{eq:RparPerpF1})-(\ref{eq:RparPerpF2}).

\item  Only one static closure for $q_\perp$ is available, the closure (\ref{eq:QperpClosureX}) of \cite{Snyder1997}.
  However, the closure is obtained with the $R_1(\zeta)$ approximant. Since $q_\perp\sim \zeta R(\zeta)$, see eq. (\ref{eq:QperpPP}) or (\ref{eq:Qperp_R1}),
  using $R_1(\zeta)$ implies that for large $\zeta$ values the heat flux does not disappear and instead converges to a constant value, which is erroneous. 
  Additionally, for $\zeta>1$ the real part of $R_1(\zeta)$ even has a wrong sign, see Figure \ref{fig:6}. The $R_1(\zeta)$ is still a valuable
  approximant for small $|\zeta|\ll 1$ values, and a Landau fluid model with static
  heat flux closures (\ref{eq:Qpar_Again}), (\ref{eq:QperpClosureX}) recovers the correct mirror threshold. 

\item If one comes to the conclusion that the $R_1(\zeta)$ approximant is unsatisfactory, then no static closure for $q_\perp$ is available.
  Consequently, 3D Landau fluid simulations are possible only if the heat fluxes $q_\parallel$, $q_\perp$ are described by time-dependent
  equations. Of course, one could possibly consider a model with a static $q_\parallel$ closure and time-dependent $q_\perp$ closure.

\item Perhaps, the most natural way to perform 3D Landau fluid simulations is to keep the ``classical'' nonlinear evolution equations for
  $q_\parallel$ and $q_\perp$ obtained in Part 1, and use static Landau fluid closures for the perturbations of the 4th-order moment.
  Of course, it is easy to imagine that in some numerical simulations the heat flux equations might be ``too much nonlinear'', i.e. responsible for instabilities.
  In such a case, the dynamic (linear) heat flux closures might be useful to verify the instability.

\item For the $\widetilde{r}_{\parallel\parallel}$ moment, there are 3 static closures available: the $R_{4,2}$ closure
  (\ref{eq:Static_R42_F}), the $R_{4,3}$ closure (\ref{eq:Static_R43_F}) of \cite{HammettPerkins1990}, and the  $R_{4,4}$ closure (\ref{eq:Static_R44_F}).
  In real space, the $R_{4,2}$ closure is given by (\ref{eq:Static_R42}), the $R_{4,3}$ closure by (\ref{eq:Static_R43}) and the $R_{4,4}$ closure by (\ref{eq:Static_R44}).
  The $R_{4,2}$ closure has the highest power-series precision $o(\zeta^3)$, and the $R_{4,4}$ closure has the highest asymptotic-series precision $o(\zeta^{-6})$.
  It is of course difficult to recommend which closure is clearly better without considering a specific situation.

\item  We considered the example of the ion-acoustic mode, see Figures \ref{fig:IAdamping1}, \ref{fig:IAdamping2} and
  associated discussion. The $R_{4,4}$ closure can be useful for simulations with sufficiently high electron temperatures,
  namely $\tau=T_e/T_p>15$, which corresponds to $\zeta_p>3$. However, such simulations will be perhaps not performed very frequently.
  In the most interesting regime with comparable proton and electron temperatures (or $\tau\in[1,5]$) the most precise static closure
  is by far the $R_{4,2}$ closure. Nevertheless, the $R_{4,3}$ closure is still a globally precise closure. We can only recommend to use
  both the $R_{4,2}$ closure (\ref{eq:Static_R42}) of \cite{HunanaPRL2018} and the $R_{4,3}$ closure (\ref{eq:Static_R43}) of  \cite{HammettPerkins1990},
  and clarify possible differences in numerical simulations. The differences might be more pronounced during
  nonlinear dynamics. 
  
\item As an example, Landau fluid simulations of turbulence typically show a curious behavior (see e.g. \cite{PerronePassot2018} and references therein),
  that at sub-proton scales, the spectrum of the parallel velocity field $u_\parallel$ is much steeper in kinetic simulations than in Landau fluid
  simulations. In contrast to the $R_{4,3}$ closure, our $R_{4,2}$ closure contains the parallel velocity $u_\parallel$. It would be interesting
  to explore if the $R_{4,2}$ closure influences the $u_\parallel$ spectrum. 

\item For the $\widetilde{r}_{\parallel\perp}$ moment, there is only one static closure, the $R_{2,0}$ closure (\ref{eq:R20_boring1}) of
  \cite{Snyder1997}.

\item If higher precision is desired, one can use dynamic closures for the $\widetilde{r}_{\parallel\parallel}$ and $\widetilde{r}_{\parallel\perp}$ moments,
  which however introduces two additional evolution equations. Of course, it is possible to use dynamic closure only for the $\widetilde{r}_{\parallel\perp}$ moment. 
  As discussed above, it appears that closures with the highest power-series precision (p.s.p.) are the most desirable (at least for $T_e\sim T_p$).      
  Concerning the $\widetilde{r}_{\parallel\parallel}$ moment, the static $R_{4,2}$ closure has p.s.p. $o(\zeta^3)$. Thus, it is possible to have a view
  that a worthy dynamic closure for $\widetilde{r}_{\parallel\parallel}$ should have a p.s.p. $o(\zeta^4)$.
  There is only one such closure, the $R_{5,3}$ closure (\ref{eq:R53_closure}) of \cite{HunanaPRL2018}.

\item Concerning dynamic closures for the $\widetilde{r}_{\parallel\perp}$ moment, the static $R_{2,0}$ closure (\ref{eq:R20_boring1}) has a p.s.p. $o(\zeta)$.
  Therefore, a worthy dynamic closure for the $\widetilde{r}_{\parallel\perp}$ moment should have a p.s.p. $o(\zeta^2)$, or higher. There are only two such closures.
  One with a p.s.p. $o(\zeta^2)$, the $R_{3,1}$ closure (\ref{eq:PS1closure}) of \cite{PassotSulem2007}; and one with a p.s.p. $o(\zeta^3)$, the
  $R_{3,0}$ closure (\ref{eq:3D-R30closure}) of \cite{HunanaPRL2018}.
  
\item To summarize, if one desires the highest power-series precision that is available at the 4th-order moment level,
  one should use the dynamic $R_{5,3}$ closure (\ref{eq:R53_closure}) for the $\widetilde{r}_{\parallel\parallel}$ moment, and
  the dynamic $R_{3,0}$ closure (\ref{eq:3D-R30closure}) for the $\widetilde{r}_{\parallel\perp}$ moment. Nevertheless, the dynamic closures might not
  be worth the computational cost, and it is possible to have a view that the static closures are sufficiently precise. In that case,
  for the $\widetilde{r}_{\parallel\parallel}$ moment one should use either the $R_{4,2}$ closure (\ref{eq:Static_R42}), or
  the $R_{4,3}$ closure (\ref{eq:Static_R43}) (see the discussion above),
  and for the $\widetilde{r}_{\parallel\perp}$ moment the $R_{2,0}$ closure (\ref{eq:R20_boring1}).
  Alternatively, one can use a dynamic closure only for the $\widetilde{r}_{\parallel\perp}$ moment. In that case, it is possible to match the
  power-series precision of $\widetilde{r}_{\parallel\parallel}$ and $\widetilde{r}_{\parallel\perp}$ moments. The precision $o(\zeta^2)$ is achieved by
  using the $R_{4,3}$ closure (\ref{eq:Static_R43}) for the $\widetilde{r}_{\parallel\parallel}$ moment
  and the $R_{3,1}$ closure (\ref{eq:PS1closure}) for the $\widetilde{r}_{\parallel\perp}$ moment. The precision $o(\zeta^3)$ is achieved by
  using the $R_{4,2}$ closure (\ref{eq:Static_R42}) for the $\widetilde{r}_{\parallel\parallel}$ moment and the $R_{3,0}$ closure (\ref{eq:3D-R30closure})
  for the $\widetilde{r}_{\parallel\perp}$ moment.
  
\item The most surprising result discussed in Part 2 is the observation that some closures reproduce a considered kinetic dispersion
  relation exactly, after $R(\zeta)$ is replaced by the approximant $R_{n,n'}(\zeta)$ used to obtain that fluid closure.
  We consider this observation as highly non-trivial and not obvious. For example, a 1D fluid model described by eq.
  (\ref{eq:sound_N2})-(\ref{eq:sound_RparparQ4}) that uses the $R_{4,3}$ closure for the $\widetilde{r}_{\parallel\parallel}$ moment, has
  a dispersion relation that is equivalent to the kinetic dispersion relation (\ref{eq:soundKin}), after the $R(\zeta)$ is replaced
  by the $R_{4,3}(\zeta)$. The results are equivalent only after the $R_{4,3}(\zeta_p)$ and $R_{4,3}(\zeta_e)$ terms in (\ref{eq:soundKin}) are transferred 
  to the common denominator and the resulting numerator is made to be equal to zero. That example concerns the ion-acoustic mode, but the same
  observation is true for the Langmuir mode as well, see Section \ref{sec:Langmuir}, dispersion relation (\ref{eq:Langmuir_kin}).
  We called such closures ``reliable'', or physically-meaningful.  

\item We only verified which closures are ``reliable'' on dispersion relations of the ion-acoustic mode and the Langmuir mode in the 1D electrostatic
  geometry, see closures marked with ``\checkmark'' in (\ref{eq:Table1S})-(\ref{eq:Table1T}). Nevertheless, it is expected that the same closures
  will remain ``reliable'' when the full 1D electrostatic dispersion relation of proton-electron plasma (\ref{eq:SoundPEls}) is considered,
  and which can be further generalized to multi-species, see eq. (\ref{eq:Estatic}).  

\item In the 1D electrostatic geometry, for a given n-th order moment $X_n$, a closure with the highest possible power series precision appears
  to be the dynamic closure constructed with the approximant $R_{n+1,n-1}(\zeta)$. For example, for the 3rd-order (heat flux) moment it is  
  the $R_{4,2}$ closure (\ref{eq:R42_closure}), for the 4th-order moment the $R_{5,3}$ closure (\ref{eq:R53_closure}),
  for the 5th-order moment the $R_{6,4}$ closure (\ref{eq:R64_closure}),
  and for the 6th-order moment the $R_{7,5}$ closure (\ref{eq:R75closure}). It was verified that all of these closures are ``reliable''.

\item Similarly, for a given n-th order moment $X_n$, a static closure with the highest power series precision is contructed with $R_{n,n-2}(\zeta)$. 

\item Importantly, by observing the summary of closures  (\ref{eq:Table1S})-(\ref{eq:Table1T}), it appears that closures that are ``unreliable'' can
  be constructed only if there are several possibilities in constructing the closure. The dynamic closure with $R_{n+1,n-1}(\zeta)$ approximant
  expresses $\zeta X_n+\alpha X_n$ through all the available lower-order moments $X_{m}$ where $m=1\cdots n-1$ (for even $m$, deviations $\widetilde{X}_m$ are used). 
  Thus, the $R_{n+1,n-1}$ closure for $\zeta X_n+\alpha X_n$ is unique, and it is expected to be ``reliable''. 

\item Curiously, it appears that the summary (\ref{eq:Table1S})-(\ref{eq:Table1T}) suggests, that all the dynamic closures
  $\zeta X_n+\alpha X_n$ with $\alpha=0$ are ``unreliable''. Construction of such closures is therefore discouraged. In other words, 
  the $\zeta X_n$ must be expressed through lower-order moments, including the moment $X_n$ itself, in order to construct a dynamic closure. 
  
\item To summarize, it appears that for a given n-th order moment $X_n$, the dynamic closure with the approximant $R_{n+1,n-1}(\zeta)$ is indeed ``reliable''.
  Therefore, one can go higher and higher in the hierarchy of moments and construct ``reliable'' closures with approximants $R_{n+1,n-1}(\zeta)$
  that converge to $R(\zeta)$ with increasing precision. In other words, one can reproduce linear Landau damping in the fluid
  framework to any desired precision. This establishes the convergence of fluid and collisionless kinetic descriptions. 
  
\item It is difficult to imagine that such a convergence of fluid and collisionless kinetic descriptions can be ever established in a general 3D electromagnetic
  geometry, since both kinetic and fluid systems must be obviously derived by using the same perturbations $f^{(1)}$.
  The exception is the 3D electromagnetic geometry in the gyrotropic limit, where such a convergence should exist.
  However, for a given moment $X_n$, the number of its gyrotropic moments is equal to $1+\textrm{int}[n/2]$, and increases with $n$. 
  It will be thefore much more difficult to show such a convergence. Nevertheless, one should at least use the kinetic dispersion relation
  in the gyrotropic limit (see for example \cite{FerriereAndre2002,Tajiri1967}), and establish if closures for the
  $\widetilde{r}_{\parallel\perp}$ moment summarized in (\ref{eq:RparPerpF1})-(\ref{eq:RparPerpF2}) are ``reliable'', which we did not do.
  It is expected that all of them are ``reliable''. 

\item We considered closures for the $\widetilde{r}_{\parallel\perp}$ and $\widetilde{r}_{\perp\perp}$ moments only in the gyrotropic limit (closures
  for $\widetilde{r}_{\parallel\parallel}$ have general validity). However,
  it is possible to keep the low-frequency restriction, but make the size of the gyroradius in $f^{(1)}$ unrestricted. 
  Such closures for the $\widetilde{r}_{\parallel\perp}$ and $\widetilde{r}_{\perp\perp}$ moments were obtained by \cite{PassotSulem2007}.
  In this geometry, it is also possible to obtain the non-gyrotropic (FLR) pressure tensor $\boldsymbol{\Pi}$
  (and other FLR contributions such as the non-gyrotropic heat flux vectors $\boldsymbol{S}^\parallel_\perp$,  $\boldsymbol{S}^\perp_\perp$ and $r^{\textrm{ng}}$),
  by integrating over the $f^{(1)}$ and by finding appropriate closures. The final model is rather complicated, but for sufficiently slow dynamics  
  such as the highly-oblique kinetic Alfv\'en waves (KAWs) or the mirror instability, the model reproduces linear kinetic theory very accurately on all spatial scales, 
  see \cite{PassotSulem2007,PSH2012,Hunana2013,SulemPassot2015} and references therein. Our new $R_{3,0}$ closure (\ref{eq:3D-R30closure})
  for the $\widetilde{r}_{\parallel\perp}$ moment has a higher $o(\zeta^3)$ precision than the $R_{3,1}$ closure (\ref{eq:PS1closure}) of \cite{PassotSulem2007},
  and it should be relatively easy to generalize the $R_{3,0}$ closure with FLR effects.
  By also employing our new more precise closures for the $\widetilde{r}_{\parallel\parallel}$ moment
  (which can not be generalized with FLR effects), the kinetic theory should be reproduced to a new level of precision.  
  
\item Another good example worth exploring might be the electromagnetic propagation along the magnetic field (the slab geometry),  
where $k_\perp=0$, but where no restriction on the frequency is imposed. In this case, the full kinetic $f^{(1)}$ enormously simplifies to the following form
\begin{equation}
  f^{(1)}_r = - \frac{q_r}{m_r} \bigg\{ \frac{1}{2} \bigg[  \frac{ (i E_x+E_y) e^{i\phi}}{\omega-k_\parallel v_\parallel + \Omega_r}
    +\frac{(iE_x-E_y) e^{-i\phi}}{\omega-k_\parallel v_\parallel - \Omega_r}  \bigg] \bigg[ \Big( 1-\frac{k_\parallel v_\parallel}{\omega}\Big) \frac{\pr f_{0r}}{\pr v_\perp}
    +\frac{k_\parallel v_\perp}{\omega}\frac{\pr f_{0r}}{\pr v_\parallel} \bigg] 
   + \frac{i E_z}{\omega-k_\parallel v_\parallel} \frac{\pr f_{0r}}{\pr v_\parallel} \bigg\}. \nn
\end{equation}
By prescribing a bi-Maxwellian $f_0$, integration over velocity space yields a hierarchy of moments. In this geometry,
the electrostatic dynamic $(\sim E_z)$ can be completely separated from the electromagnetic dynamics $(\sim E_x, E_y)$. 
The electromagnetic dynamics with cyclotron resonances $n=\pm 1$ yields a hierarchy of non-gyrotropic moments  
containing $Z(\zeta_{\pm})$ and $R(\zeta_\pm)$, where $\zeta_\pm = (\omega\pm\Omega)/(|\kpar|v_{\textrm{th}\parallel})$.
The $Z(\zeta_{\pm})$ and $R(\zeta_\pm)$ functions can be approximated with the same Pad\'e approximants as discussed here,
and by going sufficiently high in the hierarchy, simple closures might become available. Such closures should capture the collisionless
cyclotron damping in the fluid framework, even though only in the slab geometry. It should also be possible to verify,
if such closures are ``reliable'', i.e. if the kinetic dispersions of the ion-cyclotron and whistler modes are reproduced
exactly, after the $Z(\zeta_{\pm})$ and $R(\zeta_\pm)$ are replaced by the corresponding Pad\'e approximant. 
  
\end{itemize}

\vspace{1cm}  
\section{Acknowledgments}
We acknowledge support of the NSF EPSCoR RII-Track-1 Cooperative Agreement No. OIA-1655280 ``Connecting the Plasma Universe to Plasma
Technology in Alabama'', led by Gary P. Zank.  This work was supported by the European Research Council in the frame
of the Consolidating Grant ERC-2017-CoG771310-PI2FA ``Partial Ionisation: Two-Fluid Approach'', led by Elena Khomenko.  
Anna Tenerani acknowledges support of the NASA Heliophysics Supporting Research Grant \#80NSSC18K1211.
PH thanks Thierry Passot, Monica Laurenza, Nikola Vitas, Petr Hellinger and S. Peter Gary for many useful discussions. 
We are also very thankful to two anonymous referees whose comments and suggestions had a great impact on this text.
Significant effort has been made to eliminate all the misprints from the equations.
However, we will amend possible misprints, if found, in a corrigendum.

\clearpage
\appendix

\section{Higher order Pad\'e approximants of $R(\zeta)$} \label{section:Pade}
\subsection{5-pole approximants of $R(\zeta)$}
A general 5-pole approximant of the plasma response function that is worth considering is written as
\begin{equation}
R_5(\zeta)=\frac{1+a_1\zeta+a_2\zeta^2+a_3\zeta^3}{1+b_1\zeta+b_2\zeta^2+b_3\zeta^3+b_4\zeta^4+b_5\zeta^5}.
\end{equation}
Additionally, the minimum choice that we consider interesting, and that is defined as $R_{5,0}(\zeta)$, is to match the asymptotic expansion
for $|\zeta|\gg 1$ up the first $-1/(2\zeta^2)$ term, that requires $b_5=-2a_3$. The matching with the asymptotic expansion then proceeds step by step, according to
\begin{equation}
\begin{tabular}{ l  l  l }
  $R_{5,0}(\zeta):$ & $b_5=-2a_3$;                           &        $o(\zeta^{-2});$\\
  $R_{5,1}(\zeta):$ & $b_4=-2a_2$;                           &        $o(\zeta^{-3});$\\
  $R_{5,2}(\zeta):$ & $b_3=3a_3-2a_1$;                       &        $o(\zeta^{-4});$\\
  $R_{5,3}(\zeta):$ & $b_2=3a_2-2$;                       &        $o(\zeta^{-5});$\\
  $R_{5,4}(\zeta):$ & $b_1=3(a_1+a_3)$;                     &   $o(\zeta^{-6});$\\
  $R_{5,5}(\zeta):$ & $a_2=-\frac{2}{3}$;                     &   $o(\zeta^{-7});$\\
  $R_{5,6}(\zeta):$ & $a_3=-\frac{2}{7}a_1$;                  &   $o(\zeta^{-8}),$\\
\end{tabular} \label{eq:ApxR5expand}
\end{equation}
the $R_{5,7}(\zeta)$ does not make sense and is not defined. The matching with the power series is performed according to
\begin{eqnarray}
  R_{5,0}(\zeta) &=& 1+i\sqrt{\pi}\zeta -2\zeta^2 -i\sqrt{\pi}\zeta^3 +\frac{4}{3}\zeta^4 +i\frac{\sqrt{\pi}}{2}\zeta^5 -\frac{8}{15}\zeta^6 -i\frac{\sqrt{\pi}}{6}\zeta^7;\nn\\
  R_{5,1}(\zeta) &=& 1+i\sqrt{\pi}\zeta -2\zeta^2 -i\sqrt{\pi}\zeta^3 +\frac{4}{3}\zeta^4 +i\frac{\sqrt{\pi}}{2}\zeta^5 -\frac{8}{15}\zeta^6; \nn\\
  R_{5,2}(\zeta) &=& 1+i\sqrt{\pi}\zeta -2\zeta^2 -i\sqrt{\pi}\zeta^3 +\frac{4}{3}\zeta^4 +i\frac{\sqrt{\pi}}{2}\zeta^5 ;\nn\\
  R_{5,3}(\zeta) &=& 1+i\sqrt{\pi}\zeta -2\zeta^2 -i\sqrt{\pi}\zeta^3 +\frac{4}{3}\zeta^4; \nn\\
  R_{5,4}(\zeta) &=& 1+i\sqrt{\pi}\zeta -2\zeta^2 -i\sqrt{\pi}\zeta^3 ;\nn\\
  R_{5,5}(\zeta) &=& 1+i\sqrt{\pi}\zeta -2\zeta^2;\nn\\
  R_{5,6}(\zeta) &=&  1+i\sqrt{\pi}\zeta;
\end{eqnarray}
and the results are
\begin{eqnarray}
  R_{5,0}(\zeta): \qquad a_1 &=& i\sqrt{\pi}\frac{(621\pi^2-3927\pi+6208)}{(801\pi^2-5124\pi+8192)}; \quad
  a_2 = \frac{(900\pi^3-10665\pi^2+40268\pi-49152)}{5(801\pi^2-5124\pi+8192)}; \nn\\
  a_3 &=& i\sqrt{\pi} \frac{(450\pi^2-2799\pi+4352)}{10(801\pi^2-5124\pi+8192)}; \quad
  b_1 = -i\sqrt{\pi} \frac{(180\pi^2-1197\pi+1984)}{(801\pi^2-5124\pi+8192)}; \nn\\
  b_2 &=& \frac{2(1665\pi^2-10446\pi+16384)}{5(801\pi^2-5124\pi+8192)}; \quad
  b_3 = -i\sqrt{\pi}\frac{(1800\pi^2-11685\pi+18944)}{10(801\pi^2-5124\pi+8192)}; \nn\\
  b_4 &=& \frac{(7065\pi^2-43056\pi+65536)}{30(801\pi^2-5124\pi+8192)},
\end{eqnarray}
\begin{eqnarray}
  R_{5,1}(\zeta): \qquad a_1 &=& \frac{i}{\sqrt{\pi}} \frac{(360\pi^3-2445\pi^2+4780\pi-2048)}{5(72\pi^2-435\pi+656)};\quad
  a_2 = -\frac{(180\pi^2-1197\pi+1984)}{10(72\pi^2-435\pi+656)}; \nn\\
  a_3 &=& \frac{i}{\sqrt{\pi}} \frac{(801\pi^2-5124\pi+8192)}{30(72\pi^2-435\pi+656)}; \quad
  b_1 = -\frac{i}{\sqrt{\pi}}\frac{2(135\pi^2-750\pi+1024)}{5(72\pi^2-435\pi+656)}; \nn\\
  b_2 &=& \frac{(720\pi^2-4503\pi+7040)}{10(72\pi^2-435\pi+656)}; \quad
  b_3 = -\frac{i}{\sqrt{\pi}}\frac{2(495\pi^2-2859\pi+4096)}{15(72\pi^2-435\pi+656)},
\end{eqnarray}
\begin{eqnarray}
  R_{5,2}(\zeta): \qquad a_1 &=& i\sqrt{\pi}\frac{3(12\pi^2-81\pi+136)}{4(9\pi^2-69\pi+128)};\quad
  a_2 = -\frac{(135\pi^2-750\pi+1024)}{12(9\pi^2-69\pi+128)};\nn\\
  a_3 &=& i\sqrt{\pi} \frac{(72\pi^2-435\pi+656)}{12(9\pi^2-69\pi+128)}; \quad
  b_1 = i\sqrt{\pi} \frac{(33\pi-104)}{4(9\pi^2-69\pi+128)}; \nn\\
  b_2 &=& \frac{(90\pi^2-609\pi+1024)}{6(9\pi^2-69\pi+128)},
\end{eqnarray}  
\begin{eqnarray}
  R_{5,3}(\zeta): \qquad a_1 &=& \frac{i}{\sqrt{\pi}}\frac{(27\pi^2-126\pi+128)}{3(9\pi-28)}; \quad
  a_2 = \frac{(33\pi-104)}{3(9\pi-28)}; \nn\\
  a_3 &=& \frac{i}{\sqrt{\pi}} \frac{2(9\pi^2-69\pi+128)}{3(9\pi-28)}; \quad
  b_1 = -\frac{i}{\sqrt{\pi}}\frac{2(21\pi-64)}{3(9\pi-28)},  \label{eq:ApxR53}
\end{eqnarray}  
\begin{eqnarray}
  R_{5,4}(\zeta): \qquad a_1 &=& i\sqrt{\pi}\frac{(9\pi-26)}{(9\pi-32)}; \quad a_2 = \frac{(21\pi-64)}{(9\pi-32)};
  \quad a_3 = -i\sqrt{\pi}\frac{(9\pi-28)}{(9\pi-32)}; \label{eq:ApxR54}\\
  R_{5,5}(\zeta): \qquad a_1 &=& -i\frac{(16-3\pi)}{3\sqrt{\pi}}; \quad a_3 = i\frac{(32-9\pi)}{9\sqrt{\pi}}; \label{eq:ApxR55}\\
  R_{5,6}(\zeta): \qquad a_1 &=& -i\sqrt{\pi}\frac{7}{8},
\end{eqnarray}
so that for example
\begin{equation} \boxed{
R_{5,3}(\zeta)=\frac{1+\frac{i}{\sqrt{\pi}}\frac{(27\pi^2-126\pi+128)}{3(9\pi-28)}\zeta+\frac{(33\pi-104)}{3(9\pi-28)}\zeta^2+\frac{i}{\sqrt{\pi}} \frac{2(9\pi^2-69\pi+128)}{3(9\pi-28)}\zeta^3}{1-\frac{i}{\sqrt{\pi}}\frac{2(21\pi-64)}{3(9\pi-28)}\zeta+\frac{3(5\pi-16)}{(9\pi-28)}\zeta^2-\frac{i}{\sqrt{\pi}}\frac{2(81\pi-256)}{3(9\pi-28)}\zeta^3-\frac{2(33\pi-104)}{3(9\pi-28)}\zeta^4-\frac{i}{\sqrt{\pi}} \frac{4(9\pi^2-69\pi+128)}{3(9\pi-28)}\zeta^5};}
\end{equation}
\begin{equation}
R_{5,4}(\zeta)=\frac{1+i\sqrt{\pi}\frac{(9\pi-26)}{(9\pi-32)}\zeta+\frac{(21\pi-64)}{(9\pi-32)}\zeta^2-i\sqrt{\pi}\frac{(9\pi-28)}{(9\pi-32)}\zeta^3}{1+i\sqrt{\pi}\frac{6}{(9\pi-32)}\zeta+\frac{(45\pi-128)}{(9\pi-32)}\zeta^2-i\sqrt{\pi}\frac{(45\pi-136)}{(9\pi-32)}\zeta^3-\frac{2(21\pi-64)}{(9\pi-32)}\zeta^4+i\sqrt{\pi}\frac{2(9\pi-28)}{(9\pi-32)}\zeta^5};
\end{equation}
\begin{eqnarray}  
  R_{5,5}(\zeta) &=& \frac{1-i\frac{(16-3\pi)}{3\sqrt{\pi}}\zeta-\frac{2}{3}\zeta^2+i\frac{(32-9\pi)}{9\sqrt{\pi}}\zeta^3}
  {1-i\frac{16}{3\sqrt{\pi}}\zeta-4\zeta^2+i\frac{(64-15\pi)}{3\sqrt{\pi}}\zeta^3+\frac{4}{3}\zeta^4-i\frac{2(32-9\pi)}{9\sqrt{\pi}}\zeta^5};
\end{eqnarray}
\begin{eqnarray}
  R_{5,6}(\zeta) &=& \frac{1-i\sqrt{\pi}\frac{7}{8}\zeta-\frac{2}{3}\zeta^2+i\frac{\sqrt{\pi}}{4}\zeta^3}
  {1-i\sqrt{\pi}\frac{15}{8}\zeta-4\zeta^2+i\sqrt{\pi}\frac{5}{2}\zeta^3+\frac{4}{3}\zeta^4-i\frac{\sqrt{\pi}}{2}\zeta^5}.\label{eq:ApxR56}
\end{eqnarray}
\clearpage
\subsection{6-pole approximants of $R(\zeta)$}
A general 6-pole Pad\'e approximant to $R(\zeta)$ that we consider is
\begin{equation}
 R_{6}(\zeta)=\frac{1+a_1\zeta+a_2\zeta^2+a_3\zeta^3+a_4\zeta^4}{1+b_1\zeta+b_2\zeta^2+b_3\zeta^3+b_4\zeta^4+b_5\zeta^5+b_6\zeta^6},
\end{equation}
where as a minimum choice, we match the first asymptotic term by $b_6=-2a_4$, which defines $R_{6,0}(\zeta)$. The procedure of matching with
the asymptotic expansion yields step by step
\begin{equation}
\begin{tabular}{ l  l  l }
  $R_{6,0}(\zeta):$ & $b_6=-2a_4$;                           &        $o(\zeta^{-2});$\\
  $R_{6,1}(\zeta):$ & $b_5=-2a_3$;                           &        $o(\zeta^{-3});$\\
  $R_{6,2}(\zeta):$ & $b_4=3a_4-2a_2$;                       &        $o(\zeta^{-4});$\\
  $R_{6,3}(\zeta):$ & $b_3=3a_3-2a_1$;                       &        $o(\zeta^{-5});$\\
  $R_{6,4}(\zeta):$ & $b_2=3(a_2+a_4)-2$;                     &   $o(\zeta^{-6});$\\
  $R_{6,5}(\zeta):$ & $b_1=3(a_1+a_3)$;                     &   $o(\zeta^{-7});$\\
  $R_{6,6}(\zeta):$ & $a_4=-\frac{4}{21}-\frac{2}{7}a_2$;        &  $o(\zeta^{-8});$ \\
  $R_{6,7}(\zeta):$ & $a_3=-\frac{2}{7}a_1$;               &  $o(\zeta^{-9});$\\
  $R_{6,8}(\zeta):$ & $a_2=-\frac{8}{5}$;                  & $o(\zeta^{-10});$\\
  $R_{6,9}(\zeta):$ & $a_1=0$;                             &  $o(\zeta^{-11}),$
\end{tabular}
\end{equation}
where the approximant $R_{6,9}(\zeta)$ is not a good approximant (no imaginary part for real $\zeta$), and is eliminated.
Matching with the power series is performed according to
\begin{eqnarray}
   R_{6,0}(\zeta) &=& 1+i\sqrt{\pi}\zeta-2\zeta^2-i\sqrt{\pi}\zeta^3+\frac{4}{3}\zeta^4+i\frac{\sqrt{\pi}}{2}\zeta^5-\frac{8}{15}\zeta^6-i\frac{\sqrt{\pi}}{6}\zeta^7
   +\frac{16}{105}\zeta^8 +i\frac{\sqrt{\pi}}{24}\zeta^9;\nn\\
   &\vdots& \nn\\
   R_{6,8}(\zeta) &=& 1+i\sqrt{\pi}\zeta.
\end{eqnarray} 
Even though analytic results can be obtained with Maple, they are too long to write down, additionally, as we accidentally found out,
they are also tricky to evaluate. For example, if the default precision (of 10 digits) is used in Maple, the analytic $a_1$ in $R_{6,0}(\zeta)$ is
evaluated with command evalf as $-0.57i$, whereas the correct value is $-0.69i$. Alternatively, the system can be solved numerically from the onset.
We almost erroneously concluded that $R_{6,0}(\zeta)$ is not a very precise approximant,
even though its relative precision (for real valued $\zeta$) is better than $0.7\%$ for both real and imaginary parts of $R(\zeta)$. 
We provide results with 10 correct significant digits, which is a sufficient precision introducing relative numerical errors of less than $3\times10^{-7}\%$,
i.e. negligible in comparison with the $R_{6,0}(\zeta)$ relative precision to $R(\zeta)$. The results are
\begin{eqnarray}
R_{6,0}(\zeta): \qquad  
a_1 &=& -i0.6916731200; \quad
a_2 = -0.2854457889; \quad
a_3 =  i0.05976861370; \quad
a_4 =  0.005619524175; \nn\\
b_1 &=&  -i2.464126971; \quad
b_2 = -2.652997128; \quad
b_3 = i 1.606283498; \quad
b_4 = 0.5809066463; \nn\\
b_5 &=& -i 0.1201024988,
\end{eqnarray}
\begin{eqnarray}
R_{6,1}(\zeta): \qquad
a_1 &=& -i0.7895801201; \quad
a_2 =  -0.3391528628; \quad
a_3 = i0.07728246365; \quad
a_4 =  0.007840755018; \nn\\
b_1 &=& -i2.562033971; \quad
b_2 =  -2.880239841;\quad
b_3 = i1.830760570;\quad
b_4 = 0.7000533404,
\end{eqnarray}
\begin{eqnarray}
R_{6,2}(\zeta): \qquad
a_1 &=& -i0.8965446682; \quad
a_2 =   -0.4102783438; \quad
a_3 =   i0.1015110114; \quad
a_4 =   0.01132035970; \nn\\
b_1 &=&  -i2.668998519; \quad
b_2 =   -3.140955047; \quad
b_3 =   i2.103165693, 
\end{eqnarray}
\begin{eqnarray}
R_{6,3}(\zeta): \qquad
a_1 &=&  -i1.012753086;\quad
a_2 =  -0.5024864543;\quad
a_3 =  i0.1361229028;\quad
a_4 =  0.01700049686;\nn\\
b_1 &=&  -i2.785206937;\quad
b_2 =  -3.439137216,
\end{eqnarray}

\begin{eqnarray}
  R_{6,4}(\zeta): \qquad
a_1 &=& i \sqrt{\pi} \frac{270 \pi^2-1653 \pi+2528}{2(135 \pi^2-750 \pi+1024)};\quad
a_2 = \frac{9\pi(7 \pi-22)}{2(135 \pi^2-750 \pi+1024)};\nn\\
a_3 &=& i \sqrt{\pi}\frac{180 \pi^2-1197 \pi+1984}{2(135 \pi^2-750 \pi+1024)};\quad
a_4 =\frac{801 \pi^2-5124 \pi+8192}{6(135 \pi^2-750 \pi+1024)};\nn\\
b_1 &=& -i \sqrt{\pi} \frac{3(51 \pi-160)}{2(135 \pi^2-750 \pi+1024)},
\end{eqnarray}
\begin{eqnarray}
  R_{6,5}(\zeta): \qquad
a_1 &=& i\sqrt{\pi}\frac{4(9\pi-28)}{(81\pi-256)}; \quad
a_2 = \frac{3\pi(15\pi-47)}{(81\pi-256)}; \nn\\
a_3 &=& -i\sqrt{\pi}\frac{(51\pi-160)}{(81\pi-256)}; \quad
a_4 = -\frac{135\pi^2-750\pi+1024}{3(81\pi-256)},
\end{eqnarray}
\begin{eqnarray}
  R_{6,6}(\zeta): && \qquad a_1=i\sqrt{\pi}\frac{(45\pi-152)}{(45\pi-128)}; \quad a_2=\frac{(159\pi-512)}{(45\pi-128)}; \quad a_3=-9i\sqrt{\pi}\frac{(5\pi-16)}{(45\pi-128)};\\
  R_{6,7}(\zeta): && \qquad a_1=-i\sqrt{\pi}\frac{7}{8}; \quad a_2=4-\frac{105}{64}\pi;\\
  R_{6,8}(\zeta): && \qquad a_1=-i\sqrt{\pi}\frac{7}{8}.
\end{eqnarray}
\subsection{7-pole approximants of $R(\zeta)$}
\begin{equation}
  R_{7}(\zeta)=\frac{1+a_1\zeta+a_2\zeta^2+a_3\zeta^3+a_4\zeta^4+a_5\zeta^5}{1+b_1\zeta+b_2\zeta^2+b_3\zeta^3+b_4\zeta^4+b_5\zeta^5+b_6\zeta^6+b_7\zeta^7},
\end{equation}  
and the procedure of matching with asymptotic expansion yields
\begin{equation}
\begin{tabular}{ l  l  l }
  $R_{7,0}(\zeta):$ & $b_7=-2a_5$;                           &        $o(\zeta^{-2});$\\
  $R_{7,1}(\zeta):$ & $b_6=-2a_4$;                           &        $o(\zeta^{-3});$\\
  $R_{7,2}(\zeta):$ & $b_5=3a_5-2a_3$;                       &        $o(\zeta^{-4});$\\
  $R_{7,3}(\zeta):$ & $b_4=3a_4-2a_2$;                       &        $o(\zeta^{-5});$\\
  $R_{7,4}(\zeta):$ & $b_3=3a_5+3a_3-2a_1$;                  &   $o(\zeta^{-6});$\\
  $R_{7,5}(\zeta):$ & $b_2=3a_4+3a_2-2$;                     &   $o(\zeta^{-7});$\\
  $R_{7,6}(\zeta):$ & $b_1=\frac{21}{2}a_5+3a_3+3a_1$;        &  $o(\zeta^{-8});$ \\
  $R_{7,7}(\zeta):$ & $a_4=-\frac{4}{21}-\frac{2}{7}a_2$;     &  $o(\zeta^{-9});$\\
  $R_{7,8}(\zeta):$ & $a_5=-\frac{14}{69}a_3-\frac{4}{69}a_1$; & $o(\zeta^{-10});$\\
  $R_{7,9}(\zeta):$ & $a_2=-\frac{8}{5}$;                     &  $o(\zeta^{-11});$\\
  $R_{7,10}(\zeta):$ &$a_3=-\frac{12}{19}a_1$;                &   $o(\zeta^{-12}).$
\end{tabular}
\end{equation}
The $R_{7,11}(\zeta)$ is not defined because it would require $a_1\to\infty$.
Matching with the power series is performed according to
\begin{eqnarray}
   R_{7,0}(\zeta) &=& 1+i\sqrt{\pi}\zeta-2\zeta^2-i\sqrt{\pi}\zeta^3+\frac{4}{3}\zeta^4+i\frac{\sqrt{\pi}}{2}\zeta^5-\frac{8}{15}\zeta^6-i\frac{\sqrt{\pi}}{6}\zeta^7
   +\frac{16}{105}\zeta^8 +i\frac{\sqrt{\pi}}{24}\zeta^9-\frac{32}{945}\zeta^{10} -i\frac{\sqrt{\pi}}{120}\zeta^{11};\nn\\
   &\vdots& \nn\\
   R_{7,10}(\zeta) &=& 1+i\sqrt{\pi}\zeta.
\end{eqnarray}
The results are
\begin{eqnarray}
R_{7,0}(\zeta): \qquad 
a_1 &=&  -i0.8324695834; \quad
a_2 =     -0.4049799755; \quad
a_3 =     i0.1121082796; \quad
a_4 =     0.01799681258; \nn\\
a_5 &=&-i0.001293708127; \quad
b_1 =      -i2.604923434; \quad
b_2 =     -3.022086548; \quad
b_3 =    i2.031224201; \nn\\
b_4 &=&     0.8578481138; \quad
b_5 =   -i0.2288461173; \quad
b_6 =    -0.035945334608,
\end{eqnarray}
\begin{eqnarray}
R_{7,1}(\zeta): \qquad
a_1 &=&    -i0.9178985928; \quad
a_2 =     -0.4640689249; \quad
a_3 =     i0.1364936305; \quad
a_4 =     0.02310278605; \nn\\
a_5 &=&   -i0.001773778511; \quad
b_1 =     -i2.690352444; \quad
b_2 =      -3.232594474; \quad
b_3 =     i2.257867118; \nn\\
b_4 &=&      0.9950713218; \quad
b_5 =    -i0.2784723967,
\end{eqnarray}
\begin{eqnarray}
R_{7,2}(\zeta): \qquad
a_1 &=&     -i1.010198516; \quad
a_2 =     -0.5369471092; \quad
a_3 =    i 0.1677974137; \quad
a_4 =     0.03023595150; \nn\\
a_5 &=&   -i0.002497479595; \quad
b_1 =     -i2.782652367; \quad
b_2 =      -3.469070012; \quad
b_3 =    i 2.523713033; \nn\\
b_4 &=&      1.164050381,
\end{eqnarray}
\begin{eqnarray}
R_{7,3}(\zeta): \qquad
a_1 &=&     -i1.109722119; \quad
a_2 =     -0.6261744648; \quad
a_3 =   i  0.2086297926; \quad
a_4 =     0.04033869308; \nn\\
a_5 &=&   -i0.003624122579; \quad
b_1 =     -i2.882175970; \quad
b_2 =      -3.734698361; \quad
b_3 =    i2.836312196,
\end{eqnarray}
\begin{eqnarray}
R_{7,4}(\zeta): \qquad
a_1 &=&     -i1.216585782; \quad
a_2 =     -0.7344009695; \quad
a_3 =     i0.2623273358; \quad
a_4 =     0.05488528512; \nn\\
a_5 &=&   -i0.005440857949; \quad
b_1 =     -i2.989039633; \quad
b_2 =      -4.032335777,
\end{eqnarray}
\begin{eqnarray}
R_{7,5}(\zeta): \qquad
a_1 &=&     -i1.330549030; \quad
a_2 =     -0.8640648164; \quad
a_3 =     i0.3328884746; \quad
a_4 =     0.07606674237; \nn\\
a_5 &=&   -i0.008482851988; \quad
b_1 =     -i3.103002881,
\end{eqnarray}
\begin{eqnarray}
R_{7,6}(\zeta): \qquad
a_1 &=&     -i1.450931895; \quad
a_2 =      -1.016999244; \quad
a_3 =     i0.4247000792; \quad
a_4 =      0.1068986701; \nn\\
a_5 &=&    -i0.01378002846, 
\end{eqnarray}
\begin{eqnarray}
R_{7,7}(\zeta): \qquad
a_1 &=&     -i1.576631991; \quad
a_2 =      -1.194087585; \quad
a_3 =     i0.5420816788; \quad
a_5 =    -i0.02337475294.
\end{eqnarray}
We later found that the most precise (power-series) closure on 6th-order moment is a dynamic closure constructed with approximant $R_{7,5}(\zeta)$, and
therefore, starting with this approximant, we also provide analytic coefficients. The results are 
\begin{eqnarray}
  R_{7,5}(\zeta): \qquad
  a_1 &=& i\frac{2(3375\pi^3-24525\pi^2+54168\pi-32768)}{15(450\pi^2-2799\pi+4352)\sqrt{\pi}}; \quad
  a_2 = \frac{(6030\pi^2-37197\pi+57344)}{30(450\pi^2-2799\pi+4352)}; \nn\\
  a_3 &=& i \frac{3(600\pi^2-3805\pi+6032)\sqrt{\pi}}{10(450\pi^2-2799\pi+4352)}; \quad
  a_4 = \frac{(1545\pi^2-9743\pi+15360)}{5(450\pi^2-2799\pi+4352)};\nn\\
  a_5 &=& i\frac{(10800\pi^3-120915\pi^2+440160\pi-524288)}{90(450\pi^2-2799\pi+4352)\sqrt{\pi}}; \quad
  b_1:=-i \frac{(7065\pi^2-43056\pi+65536)}{15(450\pi^2-2799\pi+4352)\sqrt{\pi}},
\end{eqnarray}
\begin{eqnarray}
  R_{7,6}(\zeta): \qquad
  a_1 &=& i \sqrt{\pi} \frac{(1350\pi^2-8601\pi+13696)}{2(675\pi^2-4728\pi+8192)}; \quad
  a_2 = -\frac{3(135\pi-424)\pi}{2(675\pi^2-4728\pi+8192)}; \nn\\
  a_3 &=&  i\sqrt{\pi}\frac{(1800\pi^2-10707\pi+15872)}{2(675\pi^2-4728\pi+8192)}; \quad
  a_4 = \frac{(7065\pi^2-43056\pi+65536)}{6(675\pi^2-4728\pi+8192)}; \nn\\
  a_5 &=& -i\sqrt{\pi} \frac{(450\pi^2-2799\pi+4352)}{(675\pi^2-4728\pi+8192)},
\end{eqnarray}
\begin{eqnarray}
  R_{7,7}(\zeta): \qquad
  a_1 &=& i \frac{(675\pi^2-3432\pi+4096)}{3(-704+225\pi)\sqrt{\pi}}; \quad
  a_2 = \frac{(1545\pi-4864)}{3(-704+225\pi)}; \nn\\
  a_3 &=& i \frac{4(225\pi^2-2010\pi+4096)}{3(-704+225\pi)\sqrt{\pi}}; \quad
  a_5 = - i \frac{2(675\pi^2-4728\pi+8192)}{9(-704+225\pi)\sqrt{\pi}},
\end{eqnarray}
\begin{eqnarray}
R_{7,8}(\zeta): \qquad
  a_1 &=& -i\sqrt{\pi}\frac{3(25\pi-72)}{256-75\pi}; \quad
  a_2 = -\frac{335\pi-1024}{256-75\pi}; \quad
  a_3 = i\sqrt{\pi}\frac{5(165\pi-512)}{4(256-75\pi)}, \\
R_{7,9}(\zeta): \qquad
  a_1 &=& -i\frac{32-5\pi}{5\sqrt{\pi}}; \quad
  a_3 = i\frac{1024-275\pi}{100\sqrt{\pi}}, \\
R_{7,10}(\zeta): \qquad
  a_1 &=& -i \sqrt{\pi}\frac{19}{16}.  
\end{eqnarray}
\subsection{8-pole approximants of $R(\zeta)$}
\begin{equation} \label{eq:R8_nice}
  R_{8}(\zeta)=\frac{1+a_1\zeta+a_2\zeta^2+a_3\zeta^3+a_4\zeta^4+a_5\zeta^5+a_6\zeta^6}{1+b_1\zeta+b_2\zeta^2+b_3\zeta^3+b_4\zeta^4+b_5\zeta^5+b_6\zeta^6+b_7\zeta^7+b_8\zeta^8},
\end{equation}
and the procedure of matching with the asymptotic expansion step by step
\begin{equation}
  R(\zeta) = - \frac{1}{2\zeta^2}-\frac{3}{4\zeta^4} -\frac{15}{8\zeta^6} -\frac{105}{16\zeta^8}-\frac{945}{32\zeta^{10}}
  -\frac{10395}{64\zeta^{12}} -\frac{135135}{128\zeta^{14}} \cdots; \qquad |\zeta|\gg 1.
\end{equation}
yields the following table
\begin{equation}
\begin{tabular}{ l  l  l }
  $R_{8,0}(\zeta):$ & $b_8=-2a_6$;                                  & $o(\zeta^{-2});$ \\
  $R_{8,1}(\zeta):$ & $b_7=-2a_5$;                                  & $o(\zeta^{-3});$ \\
  $R_{8,2}(\zeta):$ & $b_6=3a_6-2a_4$;                              &  $o(\zeta^{-4});$\\
  $R_{8,3}(\zeta):$ & $b_5=3a_5-2a_3$;                              & $o(\zeta^{-5});$ \\
  $R_{8,4}(\zeta):$ & $b_4=3a_6+3a_4-2a_2$;                         & $o(\zeta^{-6});$ \\
  $R_{8,5}(\zeta):$ & $b_3=3a_5+3a_3-2a_1$;                         &  $o(\zeta^{-7});$\\
  $R_{8,6}(\zeta):$ & $b_2=\frac{21}{2}a_6+3a_4+3a_2-2$;            & $o(\zeta^{-8});$ \\
  $R_{8,7}(\zeta):$ & $b_1=\frac{21}{2}a_5+3a_3+3a_1$;              & $o(\zeta^{-9});$ \\
  $R_{8,8}(\zeta):$ & $a_6=-\frac{14}{69}a_4-\frac{4}{69}a_2-\frac{8}{207}$; & $o(\zeta^{-10});$\\
  $R_{8,9}(\zeta):$ & $a_5=-\frac{14}{69}a_3-\frac{4}{69}a_1$;      & $o(\zeta^{-11});$\\
  $R_{8,10}(\zeta):$ & $a_4=-\frac{12}{19}a_2-\frac{212}{285}$;     & $o(\zeta^{-12});$ \\
  $R_{8,11}(\zeta):$ & $a_3=-\frac{12}{19}a_1$;                     &  $o(\zeta^{-13});$ \\
  $R_{8,12}(\zeta):$ & $a_2=-\frac{94}{35}$;                        &  $o(\zeta^{-14});$ \\
  $R_{8,13}(\zeta):$ & $a_1=0$;                                     & $o(\zeta^{-15}),$
\end{tabular}
\end{equation}
where the approximant $R_{8,13}(\zeta)$ is not well behaved and is eliminated.
Matching with the power series is performed according to
\begin{eqnarray}
   R_{8,0}(\zeta) &=& 1+i\sqrt{\pi}\zeta-2\zeta^2-i\sqrt{\pi}\zeta^3+\frac{4}{3}\zeta^4+i\frac{\sqrt{\pi}}{2}\zeta^5-\frac{8}{15}\zeta^6-i\frac{\sqrt{\pi}}{6}\zeta^7
   +\frac{16}{105}\zeta^8 +i\frac{\sqrt{\pi}}{24}\zeta^9 -\frac{32}{945}\zeta^{10} \nn\\
   && \quad -i\frac{\sqrt{\pi}}{120}\zeta^{11} +\frac{64}{10395}\zeta^{12}+i\frac{\sqrt{\pi}}{720}\zeta^{13} ; \qquad |\zeta|\ll 1; \nn\\
   &\vdots& \nn\\
   R_{8,12}(\zeta) &=& 1+i\sqrt{\pi}\zeta.
\end{eqnarray}
Such a high-order Pad\'e approximants are very precise, and to retain the accuracy, we provide solutions with 16 correct significant digits
(even though this is actually not necessary and 10 digits is still fully sufficient). The approximant $R_{8,3}(\zeta)$ is a bit special,
since its corresponding $Z_{8,3}(\zeta)$ \emph{should} be the approximant that is used in the WHAMP code. This is inferred from a sentence on page 12 of the WHAMP manual
\cite{Ronnmark1982}, where it is stated that an 8-pole approximant was derived, using 10 equations from the power series expansion and 6 equations from the asymptotic series
expansion. However, the Pad\'e coefficients in the WHAMP manual are given in a different form than we use here, and an alternative Pad\'e approximation is used where for example
an 8-pole approximant is given by $Z_8(\zeta)=\sum_{j=0}^8 b_j/(\zeta-c_j)$, and the coefficients $b_j,c_j$ are obtained. We did not bother to re-derive the coefficients in that
form, instead, we compare the precision of various approximants in Section \ref{sec:PrecissionR}.

\begin{eqnarray} R_{8,0}(\zeta): \quad
a_1 &=&  -i0.9690248260959390; \quad
a_2 =     -0.5368540729623971; \quad
a_3 =     i0.1799961104391385; \nn\\
a_4 &=&      0.03849976076674387; \quad
a_5 =  -i0.004838817622209550; \quad
a_6 =    -0.0002789155539114067; \nn\\
b_1 &=&    -i2.741478677001455; \quad
b_2 =     -3.395998511188985; \quad
b_3 =    i2.488743246168061; \nn\\
b_4 &=&     1.183496393867702; \quad
b_5 =    -i0.3752177277401555; \quad
b_6 =    -0.07776565572655091; \nn\\
b_7 &=&    i0.009681326596560459,
\end{eqnarray}
\begin{eqnarray} R_{8,1}(\zeta): \quad
a_1 &=&  -i1.045465281824923; \quad
a_2 =     -0.6004884272987368; \quad
a_3 =     i0.2109529643239577; \nn\\
a_4 &=&       0.04706936874656537; \quad
a_5 =    -i0.006214502177680713; \quad
a_6 =     -0.0003778071927517807; \nn\\
b_1 &=&     -i2.817919132730439; \quad
b_2 =      -3.595120045647135; \quad
b_3 =     i2.719752919143475; \nn\\
b_4 &=&       1.338764097514731; \quad
b_5 =    -i0.4407920285051489; \quad
b_6 =     -0.0952587572277513,
\end{eqnarray}
\begin{eqnarray} R_{8,2}(\zeta): \quad
a_1 &=&  -i1.127283578226963; \quad
a_2 =     -0.6755893264302076; \quad
a_3 =     i0.2489222931730291; \nn\\
a_4 &=&       0.05823704506630824; \quad
a_5 =   -i0.008104732774508430; \quad
a_6 =   -0.0005229347287036976; \nn\\
b_1 &=&    -i2.899737429132479; \quad
b_2 =     -3.815240099310931; \quad
b_3 =    i2.984238291966390; \nn\\
b_4 &=&     1.523498938607364; \quad
b_5 =    -i0.5222070688557393,
\end{eqnarray}
\begin{eqnarray} R_{8,3}(\zeta): \quad
 a_1 &=& - i 1.214803859035098; \quad
 a_2  = -0.7640021842041184; \quad
 a_3  =  i 0.2959160549490394;\nn\\
 a_4 &=&  0.07292272182826132;\quad
 a_5  = -i 0.01075099173987222; \quad
 a_6  = -0.0007415148441966772; \nn\\
 b_1 &=& -i 2.987257709940614;\quad
 b_2  = -4.058778615835553;\quad
 b_3  =  i 3.287852273584013;\nn\\ 
 b_4 &=&  1.744375011977697,
\end{eqnarray}
\begin{eqnarray} R_{8,4}(\zeta): \quad
a_1 &=&  -i1.308257217643640; \quad
a_2 =     -0.8677094433207613; \quad
a_3 =    i 0.3544341617560842; \nn\\
a_4 &=&     0.09241987665571996; \quad
a_5 =    -i0.01452077809048251; \quad
a_6 =    -0.001080201271194285; \nn\\
b_1 &=&     -i3.080711068549157; \quad
b_2 =      -4.328127640297961; \quad
b_3 =     i3.636872378820012,
\end{eqnarray}
\begin{eqnarray} R_{8,5}(\zeta): \quad
a_1 &=&     -i1.407720282460896; \quad
a_2 =     -0.9887147938014795; \quad
a_3 =     i0.4274799329839440; \nn\\
a_4 &=&      0.1185117574638203; \quad
a_5 =    -i0.01997983676237579; \quad
a_6 =     -0.001621365678069347; \nn\\
b_1 &=&     -i3.180174133366412; \quad
b_2 =      -4.625426683036889,
\end{eqnarray}
\begin{eqnarray} R_{8,6}(\zeta): \quad
a_1 &=&  -i1.513048776977928; \quad
a_2 =      -1.128859520019374; \quad
a_3 =     i0.5184905792641649; \nn\\
a_4 &=&      0.1535743993372947; \quad
a_5 =    -i0.02799183221943639; \quad
a_6 =     -0.002514851707175031; \nn\\
b_1 &=&     -i3.285502627883444, \label{eq:R86dig}
\end{eqnarray}
\begin{eqnarray} R_{8,7}(\zeta): \quad
a_1 &=&  -i1.623826833670546; \quad
a_2 =      -1.289590935716420; \quad
a_3 =     i0.6311517791766421; \nn\\
a_4 &=&      0.2006218487856471; \quad
a_5 =    -i0.03983385915184296; \quad
a_6 =     -0.004041376481615575,
\end{eqnarray}
\begin{eqnarray} R_{8,8}(\zeta): \quad
a_1 &=&  -i1.739359630417800; \quad
a_2 =      -1.471743639038102; \quad
a_3 =     i0.7691080574071934; \nn\\
a_4 &=&      0.2632500611991985; \quad
a_5 =    -i0.05724369164680910,
\end{eqnarray}
\begin{eqnarray} R_{8,9}(\zeta): \quad
a_1 &=&  -i1.858726543442496; \quad
a_2 =      -1.675414915338742; \quad
a_3 =     i0.9356405409666494; \nn\\
a_4 &=&      0.3458159069196990,
\end{eqnarray}
and we provide analytic results for the last 3 approximants:
\begin{eqnarray}
  R_{8,10}(\zeta):  \qquad a_1 &=& i\sqrt{\pi}\frac{175\pi-592}{175\pi-512};\quad   a_2 = \frac{955\pi-3072}{175\pi-512};\quad
  a_3 = i\sqrt{\pi}\frac{6144-1925\pi}{4(175\pi-512)};\\
  R_{8,11}(\zeta):\qquad  a_1 &=& -i\sqrt{\pi}\frac{19}{16};\quad  a_2 = 6-\frac{665}{256}\pi;\\
  R_{8,12}(\zeta):\qquad  a_1 &=& -i\sqrt{\pi}\frac{19}{16}.
\end{eqnarray}
We also provide analytic coefficients for $R_{8,6}(\zeta)$, since this approximant can be used to construct the most precise dynamic closure 
for the 7th-order moment, which we will not do, however, an enthusiastic reader is encouraged to do the calculation as an exercise ! The $R_{8,6}(\zeta)$
coefficients read
\begin{eqnarray} R_{8,6}(\zeta): \qquad
  a_1 &=& i \sqrt{\pi} \frac{(189000\pi^3-1707165\pi^2+5130216\pi-5128192)}{(189000\pi^3-1612215\pi^2+4534656\pi-4194304)}; \nn\\
  a_2 &=& -\frac{2(46125\pi^3-715200\pi^2+3126720\pi-4194304)}{5(189000\pi^3-1612215\pi^2+4534656\pi-4194304)}; \nn\\
  a_3 &=& i \sqrt{\pi} \frac{(378000\pi^3-3424725\pi^2+10324380\pi-10354688)}{5(189000\pi^3-1612215\pi^2+4534656\pi-4194304)}; \nn\\
  a_4 &=& - \frac{(221400\pi^3-4788045\pi^2+23537664\pi-33554432)}{30(189000\pi^3-1612215\pi^2+4534656\pi-4194304)};\nn\\
  a_5 &=& i \sqrt{\pi} \frac{(252000\pi^3-2506275\pi^2+8286200\pi-9109504)}{5(189000\pi^3-1612215\pi^2+4534656\pi-4194304)};\nn \\
  a_6 &=& \frac{(1028700\pi^3-9863235\pi^2+31514112\pi-33554432)}{15(189000\pi^3-1612215\pi^2+4534656\pi-4194304)};\nn \\
  b_1 &=& -i \sqrt{\pi} \frac{6(15825\pi^2-99260\pi+155648)}{(189000\pi^3-1612215\pi^2+4534656\pi-4194304)}.
\end{eqnarray}  
We advise to be very careful when evaluating the above analytic expressions, since for example when the default 10-digit precision is used in Maple,
yields $a_1=-i0.63$, whereas the correct value provided in (\ref{eq:R86dig}) is $a_1=-i1.51$.

\clearpage
\section{Operator $(\bE+\frac{1}{\lowercase{c}}\bV\times\bb) \cdot\nabla_{\lowercase{v}} \lowercase{f}_0$ for gyrotropic $\lowercase{f}_0$} \label{section:B}
The magnetic field is transformed to the electric field with
induction equation $\pr \bb^{(1)} / \pr t=-c\nabla\times\bE^{(1)}$ that in Fourier space reads $\omega\bb^{(1)} = c\bk\times \bE^{(1)}$. From now on, for the electric and
magnetic field we drop the superscript (1), so in general
\begin{eqnarray}
  \bE + \frac{1}{c}\bV\times\bb &=& \bE + \frac{1}{\omega}\bV\times(\bk\times\bE) \nn\\
  &=& \bE + \frac{1}{\omega} \Big( \bk (\bV\cdot\bE) - \bE(\bV\cdot\bk) \Big) \nn\\
  &=& \bE\Big( 1-\frac{\bV\cdot\bk}{\omega}\Big) + \frac{\bk}{\omega}(\bV\cdot\bE).
\end{eqnarray}
For any general vector $\boldsymbol{A}=(A_x,A_y,A_z)$, the expression
\begin{eqnarray}
  \boldsymbol{A}\cdot\nabla_{v} f_0 =  A_x \frac{\pr f_0}{\pr v_x} + A_y \frac{\pr f_0}{\pr v_y}+ A_z\frac{\pr f_0}{\pr v_z},
\end{eqnarray}
so a general expression 
\begin{eqnarray}
  \Big( \bE + \frac{1}{c}\bV\times\bb \Big) \cdot \nabla_{v} f_0
  &=& \bigg[ E_x \Big( 1-\frac{\bV\cdot \bk}{\omega}\Big) + \frac{k_x}{\omega}(\bV\cdot\bE) \bigg] \frac{\pr f_0}{\pr v_x}
  + \bigg[ E_y \Big( 1-\frac{\bV\cdot \bk}{\omega}\Big) + \frac{k_y}{\omega}(\bV\cdot\bE) \bigg] \frac{\pr f_0}{\pr v_y}\nn\\
 &&  +\bigg[ E_z \Big( 1-\frac{\bV\cdot \bk}{\omega}\Big) + \frac{k_z}{\omega}(\bV\cdot\bE) \bigg] \frac{\pr f_0}{\pr v_z},
\end{eqnarray}
and by straightforward grouping of electric field components together 
\begin{eqnarray}
  \Big( \bE + \frac{1}{c}\bV\times\bb \Big) \cdot \nabla_{v} f_0
  &=& E_x \bigg[ \Big( 1-\frac{v_yk_y+v_zk_z}{\omega}\Big)\frac{\pr f_0}{\pr v_x}  + \frac{v_x}{\omega} \Big( k_y \frac{\pr f_0}{\pr v_y}
    + k_z \frac{\pr f_0}{\pr v_z}\Big) \bigg] \nn\\
  &+& E_y \bigg[ \Big( 1-\frac{v_x k_x+v_z k_z}{\omega}\Big)\frac{\pr f_0}{\pr v_y} + \frac{v_y}{\omega} \Big( k_x \frac{\pr f_0}{\pr v_x}
   + k_z \frac{\pr f_0}{\pr v_z} \Big) \bigg]   \nn\\
  &+& E_z \bigg[ \Big( 1-\frac{v_x k_x + v_y k_y}{\omega}\Big) \frac{\pr f_0}{\pr v_z} + \frac{v_z}{\omega} \Big( k_x \frac{\pr f_0}{\pr v_x}
   + k_y \frac{\pr f_0}{\pr v_y}\Big) \bigg]. 
\end{eqnarray}
Since nothing was essentially calculated, the above expression is of general validity and correct for any distribution function $f_0$.
The expression simplifies by considering gyrotropic $f_0(v_\perp,v_\parallel)$, that depend only on $v_\perp=|\bV_\perp|=\sqrt{v_x^2+v_y^2}$,
and which allows us to calculate
\begin{eqnarray}
  \frac{\pr v_\perp}{\pr v_x} &=& \frac{\sqrt{v_x^2+v_y^2}}{\pr v_x} = \frac{v_x}{\sqrt{v_x^2+v_y^2}} = \frac{v_x}{v_\perp};\nn\\
  \frac{\pr f_0}{\pr v_x} &=& \frac{\pr v_\perp}{\pr v_x} \frac{\pr f_0}{\pr v_\perp} = \frac{v_x}{v_\perp} \frac{\pr f_0}{\pr v_\perp} ; \qquad
  \frac{\pr f_0}{\pr v_y} = \frac{\pr v_\perp}{\pr v_y} \frac{\pr f_0}{\pr v_\perp} = \frac{v_y}{v_\perp}  \frac{\pr f_0}{\pr v_\perp}.
\end{eqnarray}
Or in another words, in cylindrical co-ordinate system the $f_0$ is $\phi$ independent and $\pr f_0/\pr \phi=0$, so that the velocity gradient
\begin{eqnarray}
  \nabla_v f_0 =
\left( \begin{array}{c}
    \frac{v_x}{v_\perp} \frac{\pr}{\pr v_\perp} \\
    \frac{v_y}{v_\perp} \frac{\pr}{\pr v_\perp} \\
    \frac{\pr}{\pr v_\parallel}
  \end{array} \right) f_0
  =\left( \begin{array}{c}
    \cos \phi \frac{\pr}{\pr v_\perp} \\
    \sin \phi \frac{\pr}{\pr v_\perp} \\
    \frac{\pr}{\pr v_\parallel}
  \end{array} \right) f_0.
\end{eqnarray}
This simplification for $f_0$ being gyrotropic therefore yields 
\begin{eqnarray}
  \Big( \bE + \frac{1}{c}\bV\times\bb \Big) \cdot \nabla_{v} f_0
  &=& E_x \bigg[ \Big( 1-\frac{v_\parallel \kpar}{\omega}\Big) \frac{v_x}{v_\perp} \frac{\pr f_0}{\pr v_\perp} + \frac{v_x}{\omega}
    \kpar \frac{\pr f_0}{\pr v_\parallel} \bigg] \nn\\
  &+& E_y \bigg[ \Big( 1-\frac{v_\parallel \kpar}{\omega}\Big)\frac{v_y}{v_\perp} \frac{\pr f_0}{\pr v_\perp} + \frac{v_y}{\omega}
    \kpar \frac{\pr f_0}{\pr v_\parallel} \bigg]   \nn\\
  &+& E_z \bigg[ \Big( 1-\frac{v_x k_x + v_y k_y}{\omega}\Big) \frac{\pr f_0}{\pr v_\parallel} + \frac{v_\parallel}{\omega} \Big( k_x  \frac{v_x}{v_\perp} \frac{\pr f_0}{\pr v_\perp}
   + k_y \frac{v_y}{v_\perp} \frac{\pr f_0}{\pr v_\perp}\Big) \bigg], 
\end{eqnarray}
that is conveniently re-arranged as
\begin{eqnarray}
  \Big( \bE + \frac{1}{c}\bV\times\bb \Big) \cdot \nabla_{v} f_0
  &=& ( E_x v_x + E_y v_y ) \bigg[ \Big( 1-\frac{v_\parallel \kpar}{\omega}\Big) \frac{1}{v_\perp} \frac{\pr f_0}{\pr v_\perp} + \frac{\kpar}{\omega}
    \frac{\pr f_0}{\pr v_\parallel} \bigg] \nn\\
  &+& E_z \bigg[ \Big( 1-\frac{v_x k_x + v_y k_y}{\omega}\Big) \frac{\pr f_0}{\pr v_\parallel} + \frac{v_\parallel}{\omega v_\perp} \Big( k_x  v_x 
   + k_y v_y \Big) \frac{\pr f_0}{\pr v_\perp} \bigg], 
\end{eqnarray}
or alternatively as
\begin{eqnarray} 
  \Big( \bE + \frac{1}{c}\bV\times\bb \Big) \cdot \nabla_{v} f_0
  &=& ( E_x v_x + E_y v_y ) \bigg[ \Big( 1-\frac{v_\parallel \kpar}{\omega}\Big) \frac{1}{v_\perp} \frac{\pr f_0}{\pr v_\perp} + \frac{\kpar}{\omega}
    \frac{\pr f_0}{\pr v_\parallel} \bigg] \nn\\
  &+& E_z \bigg[ \frac{\pr f_0}{\pr v_\parallel}  -\frac{v_x k_x + v_y k_y}{\omega}\Big( \frac{\pr f_0}{\pr v_\parallel}
    - \frac{v_\parallel}{v_\perp} \frac{\pr f_0}{\pr v_\perp} \Big) \bigg]. \label{eq:Gyro_pic} 
\end{eqnarray}
In cylindrical co-ordinate system $d^3 v=v_\perp dv_\perp dv_\parallel d\phi$.
\clearpage
\section{General kinetic $\lowercase{f}^{(1)}$ distribution (effects of non-gyrotropy)} \label{sec:Orbit}
The calculation is actually not that difficult once the coordinate change is figured out, as elaborated in the plasma physics
books by Stix, Swanson, Akheizer etc. In the general equation (\ref{eq:f1Int}) the (1) quantities must be Fourier transformed according to
\begin{eqnarray}
  f^{(1)}(\bx,\bV,t) &=& f^{(1)} e^{i\bk\cdot\bx-i\omega t};\nn\\
  \bE^{(1)}(\bx',t') &=& \bE^{(1)} e^{i\bk\cdot\bx'-i\omega t'};\nn\\
  \bb^{(1)}(\bx',t') &=& \bb^{(1)} e^{i\bk\cdot\bx'-i\omega t'},
\end{eqnarray}
and the equation (\ref{eq:f1Int}) rewrites
\begin{equation} \label{eq:f1Int2}
  f^{(1)} e^{i\bk\cdot\bx-i\omega t} = - \frac{q_r}{m_r}\int_{-\infty}^t e^{i\bk\cdot\bx'-i\omega t'}
  \Big[\bE^{(1)} +\frac{1}{c}\bV'\times\bb^{(1)} \Big]\cdot\nabla_{v'} f_0 (\bV') \;  dt'.
\end{equation}
In the cylindrical coordinate system with velocity (\ref{eq:KinVel}) and the wave-vector
\begin{eqnarray}
\bk =  \left( \begin{array}{c}
    k_\perp \cos \psi \\
    k_\perp \sin \psi \\
    k_\parallel
  \end{array} \right).
\end{eqnarray}
The integration is changed to be done with respect to variable
\begin{equation}
  \tau=t-t'.
\end{equation}  
The time $t$ is a constant here and since $d\tau=-dt'$, the integration reads $\int_{-\infty}^t dt' = \int_{\infty}^0 (-d\tau) = \int_0^\infty d\tau$.
The variable transformation is performed according to
\begin{eqnarray}
\bV' =  \left( \begin{array}{c}
    v_x' \\
    v_y' \\
    v_z'
\end{array} \right) =
\left( \begin{array}{c}
    v_\perp \cos(\phi+\Omega \tau) \\
    v_\perp \sin(\phi+\Omega \tau)\\
    v_\parallel
\end{array} \right)
\end{eqnarray}
\begin{eqnarray}
\bx' =  \left( \begin{array}{c}
    x' \\
    y' \\
    z'
\end{array} \right) =
 \left( \begin{array}{c}
    x \\
    y \\
    z
\end{array} \right) + 
\left( \begin{array}{c}
    -\frac{v_\perp}{\Omega}\big[ \sin(\phi+\Omega \tau)-\sin\phi\big] \\
    +\frac{v_\perp}{\Omega}\big[ \cos(\phi+\Omega \tau)-\cos\phi\big] \\
    -v_\parallel\tau
\end{array} \right), 
\end{eqnarray}
which at time $\tau=0$ satisfies the initial condition.
Now by straightforward calculation (and by using $\sin(a)\cos(b)-\cos(a)\sin(b)=\sin(a-b)$), the exponential factor is transformed as
\begin{eqnarray}
  \bk\cdot\bx'-\omega t' = \bk\cdot\bx-\omega t
  -\frac{k_\perp v_\perp}{\Omega}\big[ \sin(\phi-\psi+\Omega\tau)-\sin(\phi-\psi) \big] + (\omega-k_\parallel v_\parallel) \tau, 
\end{eqnarray}  
so that
\begin{eqnarray}
  e^{i\bk\cdot\bx'-i\omega t'} = e^{i\bk\cdot\bx-i\omega t} e^{-i\frac{k_\perp v_\perp}{\Omega} \sin(\phi-\psi+\Omega\tau)}
  e^{+i\frac{k_\perp v_\perp}{\Omega} \sin(\phi-\psi) } e^{i(\omega-k_\parallel v_\parallel) \tau}. \label{eq:expFact}
\end{eqnarray}
The complicated expressions encountered in the kinetic dispersion relations originate in using identity
\begin{equation} \label{eq:Bessel}
e^{i z \sin\phi} = \sum_{n=-\infty}^\infty e^{i n\phi} J_n (z).
\end{equation}  
There are two such exponents, and therefore the linear kinetic theory contains two independent summations, usually one through ``n'' and one
through ``m'' (which should not be confused with mass), i.e.
\begin{eqnarray}
  e^{-i\frac{k_\perp v_\perp}{\Omega} \sin(\phi-\psi+\Omega\tau)} &=& \sum_{n=-\infty}^\infty e^{-i n (\phi-\psi+\Omega\tau)} J_n (\frac{k_\perp v_\perp}{\Omega}); \label{eq:StixB1}\\
  e^{+i\frac{k_\perp v_\perp}{\Omega} \sin(\phi-\psi) } &=& \sum_{m=-\infty}^\infty e^{+i m (\phi-\psi)} J_m (\frac{k_\perp v_\perp}{\Omega}), \label{eq:StixB2}
\end{eqnarray}
and together
\begin{eqnarray}
  e^{-i\frac{k_\perp v_\perp}{\Omega} \sin(\phi-\psi+\Omega\tau)} e^{+i\frac{k_\perp v_\perp}{\Omega} \sin(\phi-\psi) } =
  \sum_{n=-\infty}^\infty \sum_{m=-\infty}^\infty e^{+i (m-n)(\phi-\psi)} e^{-i n \Omega \tau} J_n (\frac{k_\perp v_\perp}{\Omega}) J_m (\frac{k_\perp v_\perp}{\Omega}).
\end{eqnarray}  
It is obvious that the quantity $k_\perp v_\perp /\Omega$ will be always present and it is useful to use some abbreviation. Each book chooses different
notation, Swanson uses ``b'', Stix uses ``z'', etc. Since we are interested in Landau fluid models, we choose to follow the notation of
Passot and Sulem 2006 and call this quantity for $r$-species $\lambda_r$, so 
\footnote{Note that this notation should not be confused with notation in Peter Gary's book where $\lambda$ is reserved for quantities encountered in
the final dispersion relation and is $\sim k_\perp^2$}
\begin{equation}
\lambda_r \equiv \frac{k_\perp v_\perp}{\Omega_r},  
\end{equation}
where for clarity of calculations, we again drop the species index $r$. The transformation of the full exponential factor (\ref{eq:expFact}) therefore yields
\begin{eqnarray}
  e^{i\bk\cdot\bx'-i\omega t'} = e^{i\bk\cdot\bx-i\omega t} \sum_{n=-\infty}^\infty \sum_{m=-\infty}^\infty
  e^{+i (m-n)(\phi-\psi)} e^{+i(\omega-k_\parallel v_\parallel -n \Omega)\tau} J_n (\lambda) J_m (\lambda).
\end{eqnarray}
Using this result in (\ref{eq:f1Int2}) allows the usual cancellation of the exponential factor $e^{i\bk\cdot\bx-i\omega t}$ on both sides
of the Fourier transformed equation, a step that we omitted to explicitly write down many times before. The partially transformed equation (\ref{eq:f1Int2})
therefore reads
\begin{eqnarray} \label{eq:f1Int3}
  f^{(1)} &=& - \frac{q_r}{m_r}\int_{0}^{\infty} \bigg[ \sum_{n=-\infty}^\infty \sum_{m=-\infty}^\infty
  e^{+i (m-n)(\phi-\psi)} e^{+i(\omega-k_\parallel v_\parallel -n \Omega)\tau} J_n (\lambda) J_m (\lambda)
  \Big(\bE^{(1)} +\frac{1}{c}\bV'\times\bb^{(1)} \Big)\cdot\nabla_{v'} f_0 (\bV') \bigg]  d\tau, \nn\\
\end{eqnarray}
where we still did not perform the coordinate change  in the operator at the end of the equation. From now on, for the electric and
magnetic field we drop the superscript (1). Let's first calculate the gradient $\nabla_{v'} f_0 (\bV')$. 

It is useful to emphasize a very important property
\begin{equation}
|\bV_\perp'|^2 = v_x'^2+v_y'^2 = v_\perp^2 \cos^2 (\phi+\Omega\tau) + v_\perp^2 \sin^2 (\phi+\Omega\tau) = |\bV_\perp|^2, 
\end{equation}  
or in another words $|\bV_\perp'|=|\bV_\perp|$ that is often abbreviated with non-bolded $v_\perp'=v_\perp$ (and since $v_\parallel'=v_\parallel$ also $|\bV'|=|\bV|$).
At first, it can be perhaps a bit confusing when one writes that the non-bolded $v_\perp'=v_\perp$,
since $v_x'\neq v_x$, $v_y'\neq v_y$ and also the bolded $\bV'\neq \bV$. The above identity implies
that for the gyrotropic $f_0$ (which is a strict requirement for $f_0$) 
\begin{equation}
f_0(|\bV_\perp'|^2,v_\parallel') = f_0(|\bV_\perp|^2,v_\parallel),
\end{equation}
further implying that
\begin{equation}
  \frac{\pr f_0}{\pr v_\perp'} \equiv  \frac{\pr f_0}{\pr |\bV_\perp'|} = \frac{\pr f_0}{\pr |\bV_\perp|} \equiv \frac{\pr f_0}{\pr v_\perp}.
\end{equation}  
The $\nabla_{v'} f_0$ can now be calculated easily, since
\begin{eqnarray}
\frac{\pr f_0}{\pr v_x'} &=& \frac{\pr f_0}{\pr |\bV_\perp'|} \frac{\pr |\bV_\perp'|}{\pr v_x'} = \frac{\pr f_0}{\pr v_\perp} \frac{v_x'}{v_\perp};\qquad 
\frac{\pr f_0}{\pr v_y'} = \frac{\pr f_0}{\pr v_\perp} \frac{v_y'}{v_\perp},
\end{eqnarray}  
and the gradient is written as (for gyrotropic $\pr f_0/\pr \phi=0$)
\begin{eqnarray}
  \nabla_{v'} f_0 =
\left( \begin{array}{c}
   \frac{v_x'}{v_\perp} \frac{\pr}{\pr v_\perp} \\
   \frac{v_y'}{v_\perp} \frac{\pr}{\pr v_\perp} \\
    \frac{\pr}{\pr v_\parallel}
  \end{array} \right) f_0
=  \left( \begin{array}{c}
   \cos (\phi+\Omega\tau) \frac{\pr}{\pr v_\perp} \\
   \sin (\phi+\Omega\tau) \frac{\pr}{\pr v_\perp} \\
    \frac{\pr}{\pr v_\parallel}
  \end{array} \right) f_0.
\end{eqnarray}
It is actually simpler to postpone the introduction of angles $\phi,\psi$ and for a moment keep a general notation $\bV'=(v_x',v_y',v_z')$ and $\bk=(k_x,k_y,k_z)$.
To transform the $( \bE + \frac{1}{c}\bV'\times\bb ) \cdot \nabla_{v'} f_0$, one can do the completely same operations as were done in the previous subsection
where the operator $( \bE + \frac{1}{c}\bV\times\bb ) \cdot \nabla_{v} f_0$ was considered. One can just use the result (\ref{eq:Gyro_pic}), add primes
to all velocities, and delete those on $v_\perp'=v_\perp$, $v_\parallel'=v_\parallel$, finally yielding
\begin{eqnarray} 
  \Big( \bE + \frac{1}{c}\bV'\times\bb \Big) \cdot \nabla_{v'} f_0
  &=& ( E_x v_x' + E_y v_y' ) \bigg[ \Big( 1-\frac{v_\parallel \kpar}{\omega}\Big) \frac{1}{v_\perp} \frac{\pr f_0}{\pr v_\perp} + \frac{\kpar}{\omega}
    \frac{\pr f_0}{\pr v_\parallel} \bigg] \nn\\
  &+& E_z \bigg[ \frac{\pr f_0}{\pr v_\parallel}  -\frac{v_x' k_x + v_y' k_y}{\omega}\Big( \frac{\pr f_0}{\pr v_\parallel}
    - \frac{v_\parallel}{v_\perp} \frac{\pr f_0}{\pr v_\perp} \Big) \bigg],  
\end{eqnarray}
which is equivalent to equation (4.83) in Swanson. Only now we introduce the angles and finish the transformation. Since
\begin{eqnarray}
  v_x'k_x + v_y'k_y &=& v_\perp \cos(\phi+\Omega\tau)k_\perp\cos\psi + v_\perp \sin(\phi+\Omega\tau)k_\perp\sin\psi \nn\\
  &=& v_\perp k_\perp \cos(\phi+\Omega\tau-\psi),
\end{eqnarray}  
the transformation yields
\begin{eqnarray}
  \Big( \bE + \frac{1}{c}\bV'\times\bb \Big) \cdot \nabla_{v'} f_0 &=&
  \Big(E_x \cos(\phi+\Omega\tau)+E_y \sin(\phi+\Omega\tau) \Big) \bigg[ \Big( 1-\frac{v_\parallel k_\parallel}{\omega}\Big) \frac{\pr f_0}{\pr v_\perp}
    +\frac{k_\parallel v_\perp}{\omega}\frac{\pr f_0}{\pr v_\parallel} \bigg]\nn\\
  && + E_z \bigg[  \frac{\pr f_0}{\pr v_\parallel} + \frac{k_\perp}{\omega} \cos(\phi+\Omega\tau-\psi) \Big(v_\parallel \frac{\pr f_0}{\pr v_\perp}-v_\perp \frac{\pr f_0}{\pr v_\parallel} \Big)
     \bigg]. \label{eq:Stix39}
\end{eqnarray}
To be clear, lets write down the complete result (\ref{eq:f1Int3}) that we have for now
\begin{eqnarray}
  f^{(1)} &=& - \frac{q_r}{m_r}\int_{0}^{\infty} \sum_{n=-\infty}^\infty \sum_{m=-\infty}^\infty
  e^{+i (m-n)(\phi-\psi)} e^{+i(\omega-k_\parallel v_\parallel -n \Omega)\tau} J_n (\lambda) J_m (\lambda) \nn\\
&& \times \bigg\{ \Big(E_x \cos(\phi+\Omega\tau)+E_y \sin(\phi+\Omega\tau) \Big) \bigg[ \Big( 1-\frac{v_\parallel k_\parallel}{\omega}\Big) \frac{\pr f_0}{\pr v_\perp}
    +\frac{k_\parallel v_\perp}{\omega}\frac{\pr f_0}{\pr v_\parallel} \bigg]\nn\\
  && + E_z \bigg[  \frac{\pr f_0}{\pr v_\parallel} + \frac{k_\perp}{\omega} \Big(v_\parallel \frac{\pr f_0}{\pr v_\perp}-v_\perp \frac{\pr f_0}{\pr v_\parallel} \Big)
    \cos(\phi+\Omega\tau-\psi) \bigg] \bigg\}  d\tau, \label{eq:NoStix}
\end{eqnarray}
where the $\times$ at the beginning of the second line is just a multiplication and not a cross product (the equation is not written in the vector form anyway). 
The result agrees with Stix's expressions (10.38) and (10.39), even though Stix at this stage did not use the Bessel expansion yet.
Stix now does not proceed with the evaluation of the integral along $\tau$, and instead goes ahead and already
starts to partially calculate the 1st-order velocity moment with integrals $\int \bV f^{(1)} d^3v$ (with first integrating over $\int_0^{2\pi} d\phi$)
to eventually obtain the kinetic current $\boldsymbol{j}=\sum_r q_r n_r \bu_r=\sum_r q_r \int \bV_r f^{(1)}_r d^3 v_r$ and the conductivity matrix $(\boldsymbol{\sigma})_{ij}$
(through $\boldsymbol{j}=\boldsymbol{\sigma}\cdot\bE$) that leads to the kinetic dispersion relation.
Stix actually first derives (\ref{eq:Stix39}) plugged into (\ref{eq:f1Int2}).
After introducing the Bessel expansion, Stix immediately performs the integration over $d\phi$. The integration over $\tau$ is done later during other calculations, and
this somewhat simplifies the amount of algebra that needs to be written down. The simplified algebra is beneficial and surely appreciated by experienced kinetic researchers,
however, especially for new researchers, it somewhat blurs the main point, how the kinetic dispersion relation is derived. The kinetic dispersion relation is
derived by obtaining the $f^{(1)}$, and by calculating the current $\boldsymbol{j}$.
Moreover, we later want to obtain higher order moments of $f^{(1)}$
than just the 1st-order velocity moment. We therefore follow Swanson, Akheizer, Passot and Sulem, and finish the calculation of $f^{(1)}$
by evaluating the $\int_0^\infty d\tau$ integral in (\ref{eq:NoStix}).    

By examining equation (\ref{eq:NoStix}), there is only one factor that is $\tau$ dependent in the first line that needs to be integrated, $e^{i(\omega-k_\parallel v_\parallel -n \Omega)\tau}$,
and the factor is multiplied by four different possibilities, $\cos(\phi+\Omega\tau)$, $\sin(\phi+\Omega\tau)$, 1, and $\cos(\phi+\Omega\tau-\psi)$.
We first need to examine the following integral
\begin{equation}
\int_0^\infty e^{iax}dx = \frac{i}{a}; \qquad \textrm{if}\quad Im(a)>0.
\end{equation}
This perhaps surprising integral can be easily verified since an indefinite integral $e^{iax}/(ia)$ exists and the curious limit
\begin{equation}
\lim_{x\to\infty}  e^{iax} = \lim_{x\to\infty}  e^{i[Re(a)+i Im(a)]x} = \lim_{x\to\infty}  e^{iRe(a)x} e^{-Im(a)x} = 0; \qquad \textrm{if}\quad Im(a)>0.
\end{equation}  
Obviously, the $Im(a)>0$ is a strict requirement. If $Im(a)=0$, the limit is undefined since $\cos(x)$ and $\sin(x)$ always oscillate,
and if $Im(a)<0$ the limit diverges to $\pm\infty$. Therefore, one of the four needed integrals is   
 \begin{equation}
\int_0^\infty e^{i(\omega-k_\parallel v_\parallel -n \Omega)\tau} d\tau = \frac{i}{\omega-k_\parallel v_\parallel -n \Omega}; \qquad \textrm{if}\quad Im(\omega)>0,
\end{equation}
 where the $Im(\omega)>0$ requirement is obtained, because $k_\parallel$, $v_\parallel$, $\Omega$ are real numbers, $n$ is an integer, and none
 of these can have an imaginary part. For the other 3 integrals we need
\begin{eqnarray}
   \int_0^\infty e^{ia\tau}\cos(\phi+\Omega\tau)d\tau &=& \frac{1}{2}\int_0^\infty e^{ia\tau} \Big( e^{i(\phi+\Omega\tau)} + e^{-i(\phi+\Omega\tau)} \Big) d\tau
   = \frac{e^{i\phi}}{2}\int_{0}^\infty e^{i(a+\Omega)\tau}d\tau + \frac{e^{-i\phi}}{2}\int_{0}^\infty e^{i(a-\Omega)\tau}d\tau \nn\\
   &=& \frac{e^{i\phi}}{2}\frac{i}{a+\Omega} + \frac{e^{-i\phi}}{2}\frac{i}{a-\Omega}
   = \frac{i}{2}\Big(\frac{e^{i\phi}}{a+\Omega}+ \frac{e^{-i\phi}}{a-\Omega} \Big); \qquad \textrm{if}\quad Im(a)>0,
\end{eqnarray}
and similarly
\begin{eqnarray}
   \int_0^\infty e^{ia\tau}\sin(\phi+\Omega\tau)d\tau &=& \frac{1}{2i}\int_0^\infty e^{ia\tau} \Big( e^{i(\phi+\Omega\tau)} - e^{-i(\phi+\Omega\tau)} \Big) d\tau
   = \frac{e^{i\phi}}{2i}\int_{0}^\infty e^{i(a+\Omega)\tau}d\tau - \frac{e^{-i\phi}}{2i}\int_{0}^\infty e^{i(a-\Omega)\tau}d\tau \nn\\
   &=& \frac{e^{i\phi}}{2i}\frac{i}{a+\Omega} - \frac{e^{-i\phi}}{2i}\frac{i}{a-\Omega}
   = \frac{1}{2}\Big(\frac{e^{i\phi}}{a+\Omega}- \frac{e^{-i\phi}}{a-\Omega} \Big); \qquad \textrm{if}\quad Im(a)>0.
\end{eqnarray}
The 3 required integrals therefore calculate
\begin{eqnarray}
  \int_0^\infty e^{i(\omega-k_\parallel v_\parallel -n \Omega)\tau} \cos(\phi+\Omega\tau) d\tau
  &=& \frac{i}{2}\Big(\frac{e^{i\phi}}{\omega-k_\parallel v_\parallel -(n-1) \Omega}
    + \frac{e^{-i\phi}}{\omega-k_\parallel v_\parallel -(n+1) \Omega} \Big);\\
  \int_0^\infty e^{i(\omega-k_\parallel v_\parallel -n \Omega)\tau} \cos(\phi-\psi+\Omega\tau) d\tau
  &=& \frac{i}{2}\Big(\frac{e^{i(\phi-\psi)}}{\omega-k_\parallel v_\parallel -(n-1) \Omega}
    + \frac{e^{-i(\phi-\psi)}}{\omega-k_\parallel v_\parallel -(n+1) \Omega} \Big);\\
  \int_0^\infty e^{i(\omega-k_\parallel v_\parallel -n \Omega)\tau} \sin(\phi+\Omega\tau) d\tau
  &=& \frac{1}{2}\Big(\frac{e^{i\phi}}{\omega-k_\parallel v_\parallel -(n-1) \Omega}
    - \frac{e^{-i\phi}}{\omega-k_\parallel v_\parallel -(n+1) \Omega} \Big),
\end{eqnarray}
and all 3 results require $Im(\omega)>0$. This strictly appearing restriction is removed later by the analytic continuation, once
the fluid integrals over the $f^{(1)}$ are calculated.
We therefore managed to finish the integration of (\ref{eq:NoStix}) along the unperturbed orbit and our latest full result
for $f^{(1)}$ reads
\begin{eqnarray}
  f^{(1)} &=& - \frac{q_r}{m_r} \sum_{n=-\infty}^\infty \sum_{m=-\infty}^\infty
  e^{+i (m-n)(\phi-\psi)} J_n (\lambda) J_m (\lambda) \bigg\{ \bigg[\frac{i E_x}{2}\Big(\frac{e^{i\phi}}{\omega-k_\parallel v_\parallel -(n-1) \Omega}
    + \frac{e^{-i\phi}}{\omega-k_\parallel v_\parallel -(n+1) \Omega} \Big)\nn\\
 && \qquad +\frac{E_y}{2}\Big(\frac{e^{i\phi}}{\omega-k_\parallel v_\parallel -(n-1) \Omega}
  - \frac{e^{-i\phi}}{\omega-k_\parallel v_\parallel -(n+1) \Omega} \Big) \bigg] \bigg[ \Big( 1-\frac{v_\parallel k_\parallel}{\omega}\Big) \frac{\pr f_0}{\pr v_\perp}
    +\frac{k_\parallel v_\perp}{\omega}\frac{\pr f_0}{\pr v_\parallel} \bigg]\nn\\
  && + E_z \bigg[\frac{i}{\omega-k_\parallel v_\parallel -n \Omega} \frac{\pr f_0}{\pr v_\parallel}
    + \frac{k_\perp}{\omega} \Big(v_\parallel \frac{\pr f_0}{\pr v_\perp}-v_\perp \frac{\pr f_0}{\pr v_\parallel} \Big)
   \frac{i}{2}\Big(\frac{e^{i(\phi-\psi)}}{\omega-k_\parallel v_\parallel -(n-1) \Omega}
    + \frac{e^{-i(\phi-\psi)}}{\omega-k_\parallel v_\parallel -(n+1) \Omega} \Big) \bigg] \bigg\}.\nn \label{eq:NoStix2}\\
\end{eqnarray}
Obviously, the result is not very pretty, and we would like to pull somehow out the denominator $\omega-k_\parallel v_\parallel -n\Omega$ from all the expressions,
so that the ``resonances'' are grouped together. The trouble is the shifted $(n-1)\Omega$ and $(n+1)\Omega$. However, all expressions are preceded by $\sum_{n=-\infty}^\infty$.
It is therefore easy to shift the summation by one index, where terms that contain $(n-1)\Omega$ require shift $n=l+1$, and terms that contain $(n+1)\Omega$ require shift
$n=l-1$. The transformation is easy to calculate, for example the terms proportional to $E_x$ transform as
\begin{eqnarray}
  \sum_{n=-\infty}^\infty e^{+i (m-n)(\phi-\psi)} J_n (\lambda) \frac{e^{i\phi}}{\omega-k_\parallel v_\parallel -(n-1) \Omega}
  &=& \sum_{l=-\infty}^\infty e^{+i (m-l-1)(\phi-\psi)} J_{l+1} (\lambda) \frac{e^{i\phi}}{\omega-k_\parallel v_\parallel -l \Omega}\nn\\
  &=& \sum_{l=-\infty}^\infty e^{+i (m-l)(\phi-\psi)} J_{l+1} (\lambda) \frac{e^{-i(\phi-\psi)} e^{i\phi}}{\omega-k_\parallel v_\parallel -l \Omega}\nn\\
  &=& \sum_{l=-\infty}^\infty e^{+i (m-l)(\phi-\psi)} J_{l+1} (\lambda) \frac{e^{+i\psi}}{\omega-k_\parallel v_\parallel -l \Omega}; \label{eq:ExKin1}\\
  \sum_{n=-\infty}^\infty e^{+i (m-n)(\phi-\psi)} J_n (\lambda) \frac{e^{-i\phi}}{\omega-k_\parallel v_\parallel -(n+1) \Omega}
  &=& \sum_{l=-\infty}^\infty e^{+i (m-l+1)(\phi-\psi)} J_{l-1} (\lambda) \frac{e^{-i\phi}}{\omega-k_\parallel v_\parallel -l \Omega}\nn\\
  &=& \sum_{l=-\infty}^\infty e^{+i (m-l)(\phi-\psi)} J_{l-1} (\lambda) \frac{e^{-i\psi}}{\omega-k_\parallel v_\parallel -l \Omega}. \label{eq:ExKin2}
\end{eqnarray}
Adding the two equations together, both $E_x$ terms therefore transform as
\begin{eqnarray}
  (\ref{eq:ExKin1}) + (\ref{eq:ExKin2}) = \sum_{l=-\infty}^\infty \frac{e^{+i (m-l)(\phi-\psi)}}{\omega-k_\parallel v_\parallel -l \Omega}
  \Big( J_{l+1} (\lambda) e^{+i\psi} + J_{l-1} (\lambda) e^{-i\psi} \Big). 
\end{eqnarray}
The transformation of terms proportional to $E_y$ is almost identical since the 2 terms are just subtracted, and it is equivalent to
\begin{eqnarray}
  (\ref{eq:ExKin1}) - (\ref{eq:ExKin2}) = \sum_{l=-\infty}^\infty \frac{e^{+i (m-l)(\phi-\psi)}}{\omega-k_\parallel v_\parallel -l \Omega}
  \Big( J_{l+1} (\lambda) e^{+i\psi} - J_{l-1} (\lambda) e^{-i\psi} \Big). 
\end{eqnarray}
The terms proportional to $E_z$ are now very easy to transform and 
\begin{eqnarray}
  \sum_{n=-\infty}^\infty e^{+i (m-n)(\phi-\psi)} J_n (\lambda) \frac{e^{i(\phi-\psi)}}{\omega-k_\parallel v_\parallel -(n-1) \Omega}
  &=& \sum_{l=-\infty}^\infty e^{+i (m-l)(\phi-\psi)} J_{l+1} (\lambda) \frac{1}{\omega-k_\parallel v_\parallel -l \Omega}; \label{eq:ExKin3}\\
  \sum_{n=-\infty}^\infty e^{+i (m-n)(\phi-\psi)} J_n (\lambda) \frac{e^{-i(\phi-\psi)}}{\omega-k_\parallel v_\parallel -(n+1) \Omega}
  &=& \sum_{l=-\infty}^\infty e^{+i (m-l)(\phi-\psi)} J_{l-1} (\lambda) \frac{1}{\omega-k_\parallel v_\parallel -l \Omega}, \label{eq:ExKin4}
\end{eqnarray}
and together
\begin{eqnarray}
 (\ref{eq:ExKin3}) + (\ref{eq:ExKin4}) = \sum_{l=-\infty}^\infty \frac{e^{+i (m-l)(\phi-\psi)}}{\omega-k_\parallel v_\parallel -l \Omega}
  \Big( J_{l+1} (\lambda) + J_{l-1} (\lambda) \Big).
\end{eqnarray}
We therefore managed to rearrange the summation and equation (\ref{eq:NoStix2}) for $f^{(1)}$ transforms to
\begin{eqnarray}
  f^{(1)} &=& - \frac{q_r}{m_r} \sum_{l=-\infty}^\infty \sum_{m=-\infty}^\infty
  \frac{e^{+i (m-l)(\phi-\psi)}}{\omega-k_\parallel v_\parallel -l \Omega} J_m (\lambda) \bigg\{ \bigg[\frac{i E_x}{2}\Big(J_{l+1} (\lambda) e^{+i\psi} + J_{l-1} (\lambda) e^{-i\psi} \Big)\nn\\
 && \qquad +\frac{E_y}{2}\Big(J_{l+1} (\lambda) e^{+i\psi} - J_{l-1} (\lambda) e^{-i\psi}  \Big) \bigg] \bigg[ \Big( 1-\frac{v_\parallel k_\parallel}{\omega}\Big) \frac{\pr f_0}{\pr v_\perp}
    +\frac{k_\parallel v_\perp}{\omega}\frac{\pr f_0}{\pr v_\parallel} \bigg]\nn\\
  && + E_z \bigg[i J_l(\lambda) \frac{\pr f_0}{\pr v_\parallel}
    + \frac{k_\perp}{\omega} \Big(v_\parallel \frac{\pr f_0}{\pr v_\perp}-v_\perp \frac{\pr f_0}{\pr v_\parallel} \Big)
   \frac{i}{2}\Big( J_{l+1} (\lambda) + J_{l-1} (\lambda) \Big) \bigg] \bigg\}.\label{eq:NoStix3}
\end{eqnarray}
This is much prettier result than (\ref{eq:NoStix2}) since all the cyclotron resonances of the same order are nicely grouped together.
We are essentially done, however, there is one more step that allows further simplification and that is the use of Bessel identities
\begin{eqnarray}
  J_{l-1}(z) + J_{l+1}(z) &=& \frac{2l}{z} J_l(z);\\
  J_{l-1}(z) - J_{l+1}(z) &=& 2 J_l'(z),
\end{eqnarray}
where the prime represents a derivative, so $J_l'(z)=\pr J_l(z) / \pr z$. The contributions proportional to $E_x$, $E_y$ are rewritten as
\begin{eqnarray}
  J_{l+1} (\lambda) e^{+i\psi} + J_{l-1} (\lambda) e^{-i\psi} &=& J_{l+1} (\lambda) (\cos\psi+i\sin\psi) +  J_{l-1} (\lambda)(\cos\psi-i\sin\psi) \nn\\
  &=&  \Big( J_{l+1} (\lambda) + J_{l-1} (\lambda) \Big)\cos\psi  + i \Big( J_{l+1} (\lambda) - J_{l-1} (\lambda) \Big)\sin\psi \nn\\
  &=&  \frac{2l}{\lambda} J_l(\lambda)\cos\psi - 2i J_l'(\lambda)\sin\psi;\\
  J_{l+1} (\lambda) e^{+i\psi} - J_{l-1} (\lambda) e^{-i\psi} &=& J_{l+1} (\lambda) (\cos\psi+i\sin\psi) -  J_{l-1} (\lambda)(\cos\psi-i\sin\psi) \nn\\
  &=&  \Big( J_{l+1} (\lambda) - J_{l-1} (\lambda) \Big)\cos\psi  + i \Big( J_{l+1} (\lambda) + J_{l-1} (\lambda) \Big)\sin\psi \nn\\
  &=&  -2 J_l'(\lambda)\cos\psi + i\frac{2l}{\lambda} J_l(\lambda)\sin\psi,
\end{eqnarray}  
and the contributions proportional to $E_z$ are trivial. The expression for $f^{(1)}$ reads
\begin{eqnarray}
  f^{(1)} &=& - \frac{q_r}{m_r} \sum_{l=-\infty}^\infty \sum_{m=-\infty}^\infty
  \frac{e^{+i (m-l)(\phi-\psi)}}{\omega-k_\parallel v_\parallel -l \Omega} J_m (\lambda) \bigg\{ \bigg[i E_x
    \Big(\frac{l}{\lambda} J_l(\lambda)\cos\psi - i J_l'(\lambda)\sin\psi \Big)\nn\\
    && \qquad +iE_y\Big(i J_l'(\lambda)\cos\psi + \frac{l}{\lambda} J_l(\lambda)\sin\psi  \Big) \bigg]
  \bigg[ \Big( 1-\frac{v_\parallel k_\parallel}{\omega}\Big) \frac{\pr f_0}{\pr v_\perp}
    +\frac{k_\parallel v_\perp}{\omega}\frac{\pr f_0}{\pr v_\parallel} \bigg]\nn\\
  && + i E_z J_l(\lambda) \bigg[\frac{\pr f_0}{\pr v_\parallel}
    + \frac{k_\perp}{\omega} \frac{l}{\lambda} \Big(v_\parallel \frac{\pr f_0}{\pr v_\perp}-v_\perp \frac{\pr f_0}{\pr v_\parallel} \Big)
    \bigg] \bigg\}. \label{eq:GrandAlmost}
\end{eqnarray}
Pulling the $i$ out to the front, re-grouping the $E_x,E_y$ terms together and renaming back $l\to n$ (since it is somewhat nicer and
cannot be confused with imaginary $i$, even though it can be confused with density $n_r$),
together with reintroducing back the species index $r$ for $f^{(1)}_r$, $f_{0r}$, $\lambda_r$ and $\Omega_r$, yields the ``\emph{grand-finale}'' result of this section,
in the form 
\begin{eqnarray}
  f^{(1)}_r &=& - \frac{iq_r}{m_r} \sum_{n=-\infty}^\infty \sum_{m=-\infty}^\infty
  \frac{e^{+i (m-n)(\phi-\psi)}}{\omega-k_\parallel v_\parallel -n \Omega_r} J_m (\lambda_r) \bigg\{ \bigg[
    \frac{nJ_n(\lambda_r)}{\lambda_r}  \Big(E_x\cos\psi +E_y\sin\psi\Big)  \nn\\
&&\qquad    +i J_n'(\lambda_r) \Big( -E_x\sin\psi +E_y\cos\psi\Big)  \bigg]
  \bigg[ \Big( 1-\frac{k_\parallel v_\parallel}{\omega}\Big) \frac{\pr f_{0r}}{\pr v_\perp}
    +\frac{k_\parallel v_\perp}{\omega}\frac{\pr f_{0r}}{\pr v_\parallel} \bigg]\nn\\
  && + E_z J_n(\lambda_r) \bigg[ \frac{\pr f_{0r}}{\pr v_\parallel}
    - \frac{n\Omega_r}{\omega}\Big(\frac{\pr f_{0r}}{\pr v_\parallel} -\frac{v_\parallel}{v_\perp} \frac{\pr f_{0r}}{\pr v_\perp}  \Big)
   \bigg] \bigg\}. \label{eq:GrandF1}
\end{eqnarray}
The quantity $\lambda_r \equiv k_\perp v_\perp/\Omega_r$, and $\Omega_r=q_r B_0/(m_r c)$.
The expression is equivalent to equation (4.88) in Swanson.\footnote{Swanson and others use notation
  $\frac{\pr f_0}{\pr v_\perp}\equiv f_{0\perp}$ and $\frac{\pr f_0}{\pr v_\parallel}\equiv f_{0\parallel}$,
also in Swansons notation $\lambda_r=b$.}

\subsection{Case $\psi=0$, propagation in the x-z plane}
If we are interested only in linear dispersion relations (and not in the development of higher order fluid hierarchy suitable for numerical simulations),
we can restrict ourselves to the propagation in the x-z plane, as we have done many times before when solving dispersion relations.
In the x-z plane, the wavenumber $\bk=(k_x,0,k_z)=(k_{\perp},0,k_\parallel)$, or equivalently the angle $\psi=0$.
In this case, the expression (\ref{eq:GrandF1}) simplifies to
\begin{eqnarray}
  f^{(1)}_r &=& - \frac{iq_r}{m_r} \sum_{n=-\infty}^\infty \sum_{m=-\infty}^\infty
  \frac{e^{i (m-n)\phi}}{\omega-k_\parallel v_\parallel -n \Omega_r} J_m (\lambda_r) \bigg\{ \bigg[
    \frac{nJ_n(\lambda_r)}{\lambda_r}  E_x 
+i J_n'(\lambda_r) E_y  \bigg]
  \bigg[ \Big( 1-\frac{k_\parallel v_\parallel}{\omega}\Big) \frac{\pr f_{0r}}{\pr v_\perp}
    +\frac{k_\parallel v_\perp}{\omega}\frac{\pr f_{0r}}{\pr v_\parallel} \bigg]\nn\\
  && + E_z J_n(\lambda_r) \bigg[ \frac{\pr f_{0r}}{\pr v_\parallel}
    - \frac{n\Omega_r}{\omega v_\perp}\Big( v_\perp \frac{\pr f_{0r}}{\pr v_\parallel} - v_\parallel\frac{\pr f_{0r}}{\pr v_\perp}  \Big)
   \bigg] \bigg\},
\end{eqnarray}
which is equivalent to the equation (10.3.12) in Gurnett and Bhattacharjee.  
In this case, the coupling of the electric field components with the sum over index $m$ disappears, and the sum can be left in its original
form $\sum_{m=-\infty}^\infty e^{+im\phi} J_m (\lambda)=e^{+i\lambda\sin\phi}$, yielding
\begin{eqnarray}
  f^{(1)}_r &=& - \frac{iq_r}{m_r} e^{i\lambda_r\sin\phi} \sum_{n=-\infty}^\infty 
  \frac{e^{-in\phi}}{\omega-k_\parallel v_\parallel -n \Omega_r} \bigg\{ \bigg[
    \frac{nJ_n(\lambda_r)}{\lambda_r}E_x  +i J_n'(\lambda_r) E_y\bigg]
  \bigg[ \Big( 1-\frac{k_\parallel v_\parallel}{\omega}\Big) \frac{\pr f_{0r}}{\pr v_\perp}
    +\frac{k_\parallel v_\perp}{\omega}\frac{\pr f_{0r}}{\pr v_\parallel} \bigg]\nn\\
  && + E_z J_n(\lambda_r) \bigg[ \frac{\pr f_{0r}}{\pr v_\parallel}
    - \frac{n\Omega_r}{\omega v_\perp}\Big(v_\perp \frac{\pr f_{0r}}{\pr v_\parallel} - v_\parallel \frac{\pr f_{0r}}{\pr v_\perp}  \Big)
   \bigg] \bigg\}.
\end{eqnarray}
If the last term proportional to $E_z$ is compared with the expression (5.2.1.9) of Akhiezer, it appears that Akhiezer has a typo,
where instead of the correct $v_\perp$ there is a typo $v_\parallel$.
\clearpage
\subsection{General $f^{(1)}$ for a bi-Maxwellian distribution}
Prescribing $f_0$ to be a bi-Maxwellian distribution function, the general expression (\ref{eq:GrandF1}) for $f^{(1)}$ further simplifies.
Since in the Vlasov expansion the gyrotropic $f_0$ was assumed to dependent only on $\bV$, i.e. $f_0(v_\perp^2,v_\parallel^2)$ and be $\bx,t$
independent, the fluid velocity $\bu$ is removed from the distribution function and the ``pure'' bi-Maxwellian is
\begin{eqnarray}
f_0 = n_{0r} \sqrt{\frac{\alpha_\parallel}{\pi}} \frac{\alpha_\perp}{\pi} e^{-\alpha_\parallel v_\parallel^2 - \alpha_\perp v_\perp^2},
\end{eqnarray}  
where $\alpha_\parallel \equiv m/(2T_\parallel^{(0)})$, $\alpha_\perp = m/(2T_\perp^{(0)})$, or in the language of thermal speeds,
$v_{th\parallel}^2=2T^{(0)}_{\parallel}/m=\alpha_\parallel^{-1}$ and $v_{th\perp}^2=2T^{(0)}_{\perp}/m=\alpha_\perp^{-1}$. We prefer the $\alpha$ notation instead
of the thermal speed $v_{\textrm{th}}$, since in long analytic calculations, there is a less chance of an error.

It is straightforward to calculate that for a bi-Maxwellian 
\begin{eqnarray}
  \frac{\pr f_0}{\pr v_\parallel} &=& -2 \alpha_\parallel v_\parallel f_0 = -\frac{m_r}{T_{\parallel}^{(0)}} v_\parallel f_0;\\ 
  \frac{\pr f_0}{\pr v_\perp} &=& -2 \alpha_\perp v_\perp f_0 = -\frac{m_r}{T_{\perp}^{(0)}} v_\perp f_0.
\end{eqnarray}  
The bi-Maxwellian distribution $f_0$ therefore can be pulled out (together with $-m_r$) and the general expression (\ref{eq:GrandF1}) rewrites
\begin{eqnarray}
  f^{(1)}_r &=&  iq_r f_{0r} \sum_{n=-\infty}^\infty \sum_{m=-\infty}^\infty
  \frac{e^{+i (m-n)(\phi-\psi)}}{\omega-k_\parallel v_\parallel -n \Omega_r} J_m (\lambda_r) \bigg\{ \bigg[
    nJ_n(\lambda_r)\frac{\Omega_r}{k_\perp}  \Big(E_x\cos\psi +E_y\sin\psi\Big)  \nn\\
&&\qquad    +i J_n'(\lambda_r) v_\perp \Big( -E_x\sin\psi +E_y\cos\psi\Big)  \bigg]
  \bigg[ \frac{1}{T_{\perp r}^{(0)}} +\frac{k_\parallel v_\parallel}{\omega}\Big( \frac{1}{T_{\parallel r}^{(0)}} - \frac{1}{T_{\perp r}^{(0)}} \Big) \bigg]\nn\\
  && + E_z J_n(\lambda_r) v_\parallel \bigg[ \frac{1}{T_{\parallel r}^{(0)}}
    - \frac{n\Omega_r}{\omega}\Big(\frac{1}{T_{\parallel r}^{(0)}} -\frac{1}{T_{\perp r}^{(0)}} \Big)
   \bigg] \bigg\}. \label{eq:GrandMax}
\end{eqnarray}
\subsection{General $f^{(1)}$ for a bi-Kappa distribution}
A bi-Kappa distribution function (as used previously in part I of the manuscript) reads 
\begin{equation}
  f_0 = n_{0r} \frac{\Gamma(\kappa+1)}{\Gamma(\kappa-\frac{1}{2})}
   \sqrt{\frac{\alpha_\parallel}{\pi}} \frac{\alpha_\perp}{\pi}  \bigg[ 1+\alpha_\parallel v_\parallel^2 + \alpha_\perp v_\perp^2 \bigg]^{-(\kappa+1)},
\end{equation}
where the abbreviated $\alpha_\parallel=1/(\kappa\theta_\parallel^2)$, $\alpha_\perp=1/(\kappa\theta_\perp^2)$, and the thermal speeds are
$\theta_\parallel^2 = (1-\frac{3}{2\kappa}) \frac{2T_{\parallel r}^{(0)}}{m_r}$, $\theta_\perp^2 = (1-\frac{3}{2\kappa}) \frac{2T_{\perp r}^{(0)}}{m_r}$. 
We again emphasized the species index $r$ only where necessary, even though in the final expression for $f^{(1)}$ we will use the proper
$\alpha_{\parallel r}$, $\alpha_{\perp r}$. Also, the $\kappa$-index should be written as $\kappa_r$, since the index will be different for
each particle species. The derivatives of $f_0$ are  
\begin{eqnarray}
  \frac{\pr f_0}{\pr v_\parallel} = -\frac{2(\kappa+1) \alpha_\parallel v_\parallel}{1+\alpha_\parallel v_\parallel^2+\alpha_\perp v_\perp^2}f_0;\\
   \frac{\pr f_0}{\pr v_\perp} = -\frac{2(\kappa+1) \alpha_\perp v_\perp}{1+\alpha_\parallel v_\parallel^2+\alpha_\perp v_\perp^2}f_0,
\end{eqnarray}
which yields the $f^{(1)}$ for a bi-Kappa distribution
\begin{eqnarray}
  f^{(1)}_r &=& \frac{iq_r}{m_r} \frac{2(\kappa_r+1)f_{0r}}{(1+\alpha_{\parallel r} v_\parallel^2+\alpha_{\perp r} v_\perp^2)} \sum_{n=-\infty}^\infty \sum_{m=-\infty}^\infty
  \frac{e^{+i (m-n)(\phi-\psi)}}{\omega-k_\parallel v_\parallel -n \Omega_r} J_m (\lambda_r) \bigg\{ \bigg[
    nJ_n(\lambda_r)\frac{\Omega_r}{k_\perp}  \Big(E_x\cos\psi +E_y\sin\psi\Big)  \nn\\
&&\qquad    +i J_n'(\lambda_r) v_\perp \Big( -E_x\sin\psi +E_y\cos\psi\Big)  \bigg]
  \bigg[ \alpha_{\perp r} +\frac{k_\parallel v_\parallel}{\omega}(\alpha_{\parallel r} -\alpha_{\perp r}) \bigg]\nn\\
  && + E_z J_n(\lambda_r) v_\parallel \bigg[ \alpha_{\parallel r}
    - \frac{n\Omega_r}{\omega} \Big(\alpha_{\parallel r} -\alpha_{\perp r}  \Big)
   \bigg] \bigg\}. \label{eq:GrandKappa}
\end{eqnarray}
In the limit $\kappa\to\infty$, the (\ref{eq:GrandKappa}) should ``obviously'' converge to the bi-Maxwellian (\ref{eq:GrandMax}).
\clearpage
\subsection{Formulation with scalar potentials $\Phi,\Psi$} \label{sec:PhiPsi}
In kinetic theory and especially in formulation of Landau fluid models, instead of electric fields, it is often useful to
work with scalar potentials $\Phi,\Psi$, that should not be confused with azimuthal angles $\phi,\psi$ for the velocity and wavenumber
in the cylindrical coordinate system. The usual decomposition employs the scalar potential $\Phi$ and the vector potential $\boldsymbol{A}$, according to
\begin{eqnarray}
  \bb &=& \bb_0 + \nabla\times\boldsymbol{A};\\
  \bE &=& -\nabla\Phi -\frac{1}{c}\frac{\pr \boldsymbol{A}}{\pr t}, \label{eq:EfieldGauge}
\end{eqnarray}
and it is useful to choose the Coulomb gauge $\nabla\cdot\boldsymbol{A}=0$. By exploring the equation for $f^{(1)}$, 
it is noteworthy that the perpendicular electric fields $E_x, E_y$ are ``coupled'' through the azimuthal angle $\psi$. In contrast, the parallel electric field component $E_z$
is on its own. Of course, this is partially a consequence of using the cylindrical coordinate system, which are natural coordinates to describe gyrating particle. 
It turns out, that in this case the calculations can be simplified, if the (\ref{eq:EfieldGauge}) is kept for the perpendicular components $E_x,E_y$,
but the $E_z$ field is rewritten with another scalar potential $\Psi$ according to
\begin{eqnarray}
  E_x &=& -\pr_x\Phi -\frac{1}{c}\frac{\pr A_x}{\pr t};\\
  E_y &=& -\pr_y\Phi -\frac{1}{c}\frac{\pr A_y}{\pr t};\\
  E_z &=& -\pr_z \Psi. 
\end{eqnarray}
Here we follow the notation of \cite{PassotSulem2006,PassotSulem2007}.
 Note that in Section 3 and Section 4, we used variable $\Phi$ for the potential of the parallel electric field $E_z$, which is here referred to as $\Psi$.
This transformation enables the elimination of vector potential $\boldsymbol{A}$, as we will see shortly.
Since for the $E_z$ component the eq. (\ref{eq:EfieldGauge}) is still valid, implying
\begin{equation}
E_z = -\pr_z\Phi -\frac{1}{c}\frac{\pr A_z}{\pr t} = -\pr_z \Psi; \quad => \quad \frac{\pr A_z}{\pr t} = -c\pr_z (\Phi-\Psi), 
\end{equation}  
or in Fourier space
\begin{equation}
A_z = \frac{ck_\parallel}{\omega}(\Phi-\Psi).
\end{equation}
Using the Coulomb gauge in Fourier space $\bk\cdot\boldsymbol{A}=0$ implies \footnote{The Coulomb gauge is sometimes called ``perpendicular gauge'' since
the vector potential $\boldsymbol{A}$ is obviously perpendicular to the direction of propagation $\bk$.}
\begin{eqnarray}
  k_x A_x + k_y A_y + k_z A_z &=& 0;\\
  A_x \cos\psi + A_y\sin\psi &=& -\frac{k_\parallel}{k_\perp} A_z = - \frac{c k_\parallel^2}{\omega k_\perp}(\Phi-\Psi).
\end{eqnarray}
The electric field components in Fourier space read
\begin{eqnarray}
  E_x &=& i\Big[ -\Phi k_\perp \cos\psi +\frac{\omega}{c}A_x\Big];\\
  E_y &=& i\Big[ -\Phi k_\perp \sin\psi +\frac{\omega}{c}A_y\Big];\\
  E_z &=& i\Big[ -k_\parallel \Psi \Big],
\end{eqnarray}
and the expression with $E_x,E_y$ components at the first line of equation (\ref{eq:GrandMax}) for $f^{(1)}$ is
\begin{eqnarray}
  E_x \cos\psi +E_y\sin\psi &=& i \Big[ -\Phi k_\perp +\frac{\omega}{c}\big(A_x\cos\psi+A_y\sin\psi\big) \Big]\nn\\
  &=&  i \Big[ -\Phi k_\perp -\frac{\omega}{c} \frac{c k_\parallel^2}{k_\perp \omega} (\Phi-\Psi) \Big] = i\Big[ -\Phi k_\perp -\frac{k_\parallel^2}{k_\perp} (\Phi-\Psi) \Big]\nn\\
  &=&  i k_\perp \Big[ -\Big(1+\frac{k_\parallel^2}{k_\perp^2} \Big)\Phi + \frac{k_\parallel^2}{k_\perp^2} \Psi \Big].
\end{eqnarray}
Furthermore, since $\pr B_z/\pr t=-c(\pr_x E_y-\pr_y E_x)$, which in Fourier space rewrites $\frac{\omega}{c} B_z = k_x E_y - k_y E_x$, implying
\begin{equation} \label{eq:Induct_Bz_end}
-  E_x \sin\psi +E_y\cos\psi = \frac{\omega}{c k_\perp} B_z.
\end{equation}  
The bi-Maxwellian equation (\ref{eq:GrandMax}) then reads
\begin{eqnarray}
  f^{(1)}_r &=&  q_r f_{0r} \sum_{n=-\infty}^\infty \sum_{m=-\infty}^\infty
  \frac{e^{+i (m-n)(\phi-\psi)}}{\omega-k_\parallel v_\parallel -n \Omega_r} J_m (\lambda_r) \nn\\
&&  \times \bigg\{ \bigg[
    nJ_n(\lambda_r) \Omega_r \Big( \big(1+\frac{k_\parallel^2}{k_\perp^2} \big)\Phi - \frac{k_\parallel^2}{k_\perp^2} \Psi  \Big)
  -J_n'(\lambda_r) \frac{\omega v_\perp}{c k_\perp} B_z  \bigg] 
  \bigg[ \frac{1}{T_{\perp r}^{(0)}} +\frac{k_\parallel v_\parallel}{\omega}\Big(\frac{1}{T_{\parallel r}^{(0)}}- \frac{1}{T_{\perp r}^{(0)}} \Big)\bigg]\nn\\
  && \quad + \Psi J_n(\lambda_r)k_\parallel v_\parallel \bigg[ \frac{1}{T_{\parallel r}^{(0)}}
    - \frac{n\Omega_r}{\omega}\Big(\frac{1}{T_{\parallel r}^{(0)}} -\frac{1}{T_{\perp r}^{(0)}} \Big)
   \bigg] \bigg\}, 
\end{eqnarray}
which verifies eq. (7) of \cite{PassotSulem2006} (their preceding eq. (6) contains a small misprint, and on the r.h.s. should have $f_r^{(0)}$ instead of $f_r^{(1)}$).

\bibliographystyle{jpp}
\bibliography{hunana_mhd}

\end{document}